\documentclass[pdflatex,sn-basic]{sn-jnl}
\UseRawInputEncoding 
\usepackage{graphicx}%
\usepackage{multirow}%
\usepackage{amsmath,amssymb,amsfonts}%
\usepackage{amsthm}%
\usepackage{mathrsfs}%
\usepackage[title]{appendix}%
\usepackage{xcolor}%
\usepackage{textcomp}%
\usepackage{manyfoot}%
\usepackage{booktabs}%
\usepackage{algorithm}%
\usepackage{algorithmicx}%
\usepackage{algpseudocode}%
\usepackage{listings}%
\usepackage[utf8]{inputenc}   
\usepackage[T1]{fontenc}      
\usepackage{xcolor}
\usepackage{float}
\usepackage[section]{placeins}

\theoremstyle{thmstyleone}%
\theoremstyle{thmstyletwo}%

\theoremstyle{thmstylethree}%

\usepackage{enumitem}
\usepackage{natbib}
\setcitestyle{aysep={}}
\begin{document}
\def\apj{ApJ}
\def\mnras{MNRAS}
\def\aap{A\&A}
\def\apjl{ApJ}
\def\pasj{PASJ}
\def\nat{Nature}
\def\iaucirc{IAU Circ.}
\def\skytel{S\&T}
\def\aaps{A\&A S.S.}
\def\apss{Ap\&SS}
\def\apjs{ApJS}
\def\araa{ARA\&A}
\def\pasp{PASP}
\def\rmxaa{Rev. Mex. Astron. Astrofis.}
\def\aapr{Astron.~Astrophys.~Rev.}
\def\nar{New~Astron.~Rev.}
\def\apspr{Astrophys.~Space~Phys.~Res.}
\def\aj{AJ}
\newcommand{\ha}{H$\alpha$}
\newcommand{\hb}{H$\beta$}
\newcommand{\hg}{H$\gamma$}
\newcommand{\hd}{H$\delta$}
\newcommand{\hep}{H$\epsilon$}
\newcommand{\pab}{Pa$\beta$}   
\newcommand{\pag}{Pa$\gamma$}  
\newcommand{\pad}{Pa$\delta$}  
\newcommand{\pae}{Pa$\epsilon$} 
\newcommand{\brg}{Br$\gamma$}  
\newcommand{\hei}{He~{\sc i}}
\newcommand{\heii}{He~{\sc ii}}
\newcommand{\Feii}{Fe~{\sc ii}}
\newcommand{\niii}{N~{\sc iii}}
\newcommand{\ciii}{C~{\sc iii}}

\newcommand{\Civ}{C~{\sc iv}}
\newcommand{\Nv}{N~{\sc v}}
\newcommand{\Siiv}{Si~{\sc iv}}

\newcommand{\Nvi}{N~{\sc vi}}
\newcommand{\Nvii}{N~{\sc vii}}
\newcommand{\Oiv}{O~{\sc iv}}
\newcommand{\Ov}{O~{\sc v}}
\newcommand{\Ovii}{O~{\sc vii}}
\newcommand{\Oviii}{O~{\sc viii}}
\newcommand{\Neix}{Ne~{\sc ix}}
\newcommand{\Nex}{Ne~{\sc x}}
\newcommand{\Mgxi}{Mg~{\sc xi}}
\newcommand{\Mgxii}{Mg~{\sc xii}}
\newcommand{\Sixiii}{Si~{\sc xiii}}
\newcommand{\Sixiv}{Si~{\sc xiv}}
\newcommand{\Sxvi}{S~{\sc xvi}}
\newcommand{\Caxx}{Ca~{\sc xx}}
\newcommand{\FexxivL}{Fe~{\sc xxiv(L)}}
\newcommand{\Fexxiv}{Fe~{\sc xxiv}}
\newcommand{\Fexxv}{Fe~{\sc xxv}}
\newcommand{\Fexxvi}{Fe~{\sc xxvi}}
\newcommand{\Fexix}{Fe~{\sc xix}}
\newcommand{\kms}{\mbox{km\,s$^{-1}$}}          

\newcommand{\ergs}{\mbox{erg\,s$^{-1}$}}        

\newcommand{\flux}{\mbox{erg\,s$^{-1}$\,cm$^{-2}$}} 
\newcommand{\lum}{\mbox{erg\,s$^{-1}$}}         

\newcommand{\degree}{\mbox{$^{\circ}$}}         


\newcommand{\msun}{\mbox{$\mathrm{M}_\odot$}}   

\newcommand{\rsun}{R$_{\odot}$}                 

\newcommand{\lsun}{L$_{\odot}$}                 


\newcommand{\lx}{$L_\mathrm{X}$}                

\newcommand{\logxi}{$\log\xi$}  

\newcommand{\ledd}{$L_\mathrm{Edd}$}            


\newcommand{\ang}{\AA{}}                        

\newcommand{\arcsec}{\ensuremath{^{\prime\prime}}} 

\newcommand{\arcmin}{\ensuremath{^{\prime}}}    

\newcommand{\FigPCyg}{
\begin{figure}[t]
\centering
\includegraphics[width=1\textwidth]{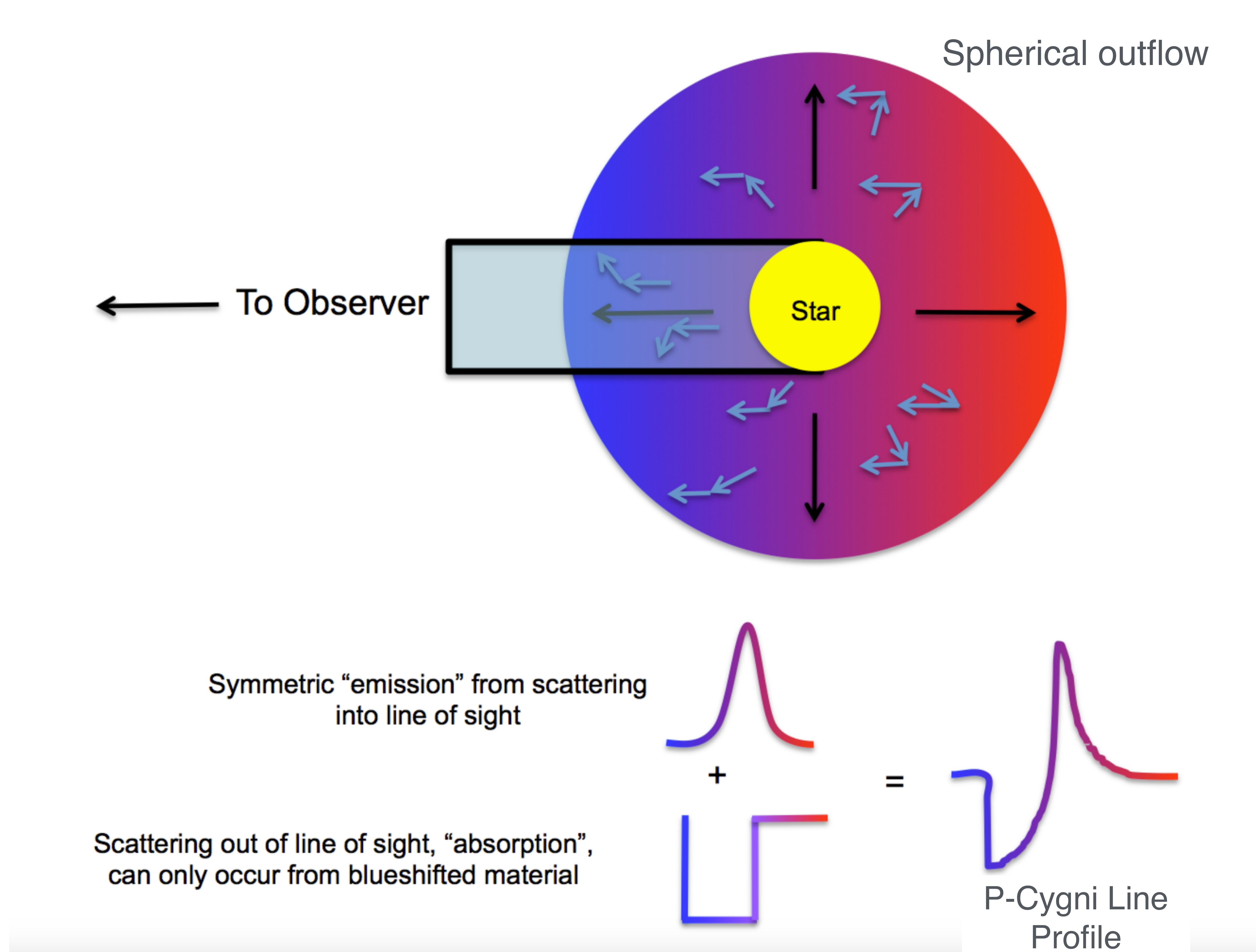}
\caption{Diagram showing the spectral signatures from an expanding spherical wind. The outflow presents significant line opacity around a continuum source (a star in this case, but it can also be a disc), leading to the formation of P-Cygni profiles (sketched at the bottom in an intensity-versus-wavelength representation). These are characterised by an emission component superimposed on a blue-shifted absorption. Black arrows indicate the direction of the outflow, while blue arrows illustrate typical scattering interactions. Adapted from \citet{Matthews2016}.}\label{fig:P-Cyg_scheme}
\end{figure}
}

\newcommand{\FigSimProfiles}{
\begin{figure}[t]
\centering
\includegraphics[width=1\textwidth]{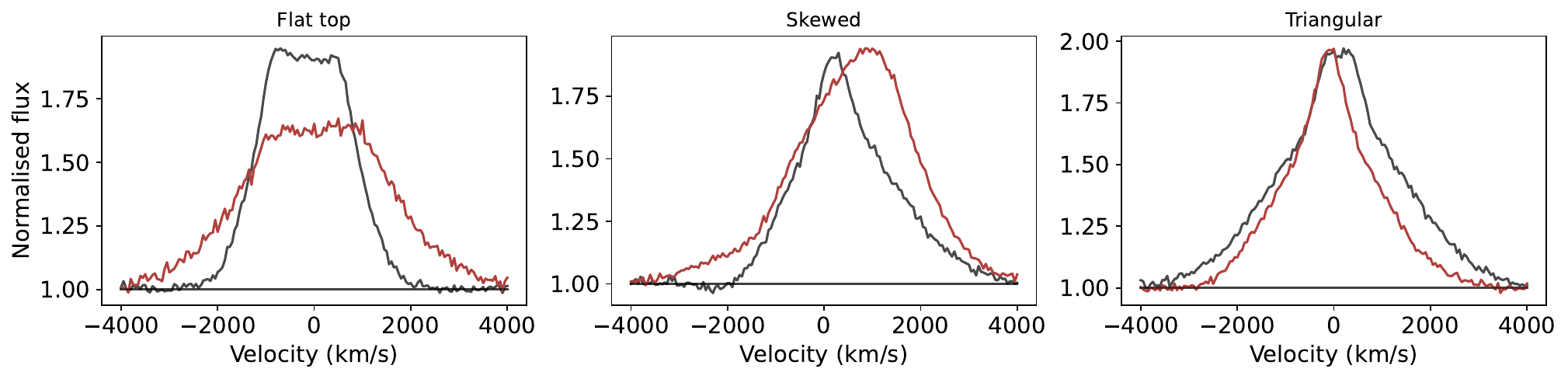}
\caption{Simulated line profiles, distinct from P-Cyg profiles, resulting from the combination of (non-physical) absorption, emission, and double-peaked components expected to play a role in outflows arising in accretion discs. Adapted from \citet{MataSanchez2023}.}
\label{fig:SimProfiles}
\end{figure}
}

\newcommand{\FigNebularPhase}{
\begin{figure}[t]
\centering
\includegraphics[width=1\textwidth]{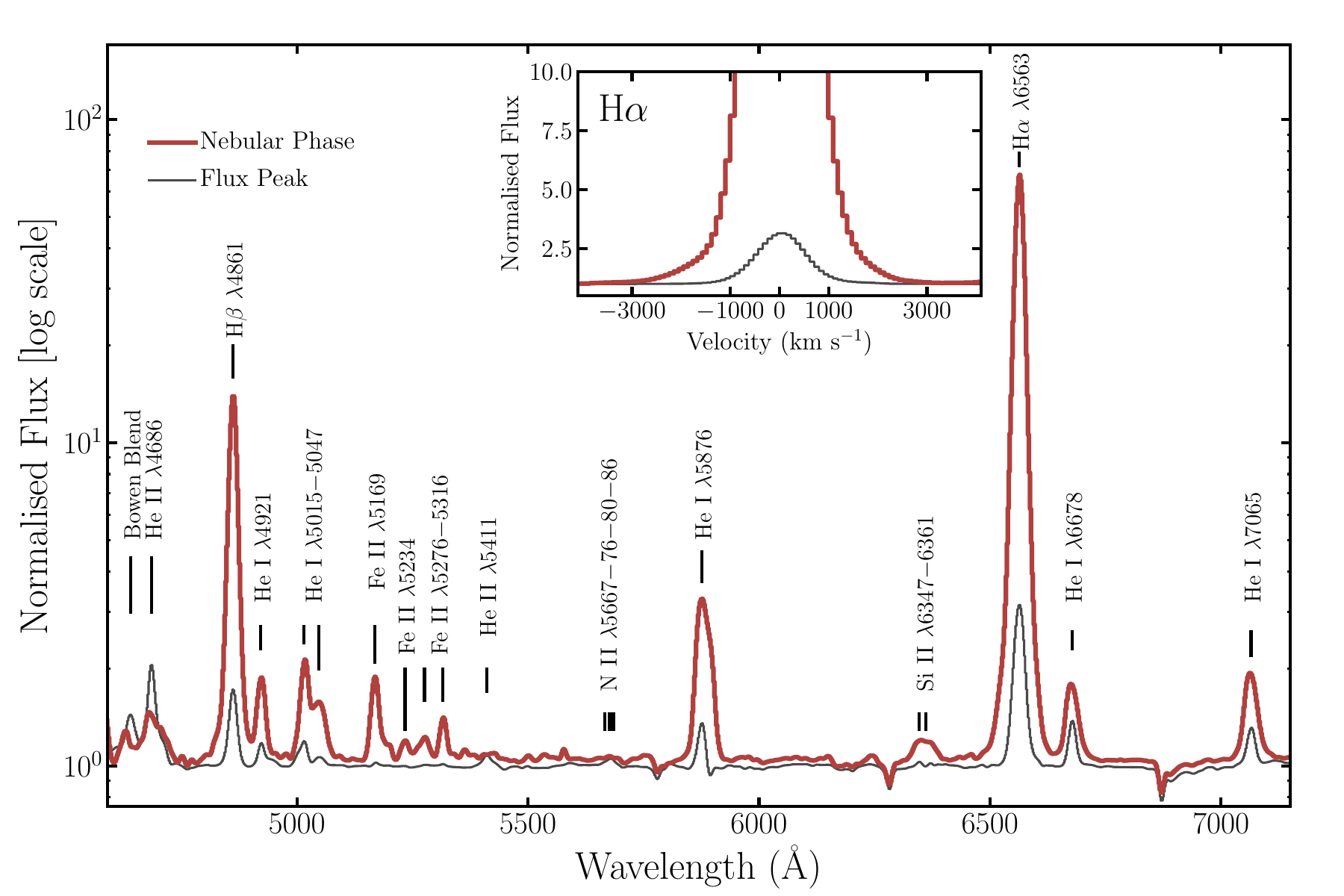}
\caption{The nebular phase of V404~Cyg. Normalised spectra corresponding to the flux peak (black) and the maximum intensity of the nebular phase (red), two days later and following an abrupt drop in the X-ray, OIR and radio fluxes.  During this phase, \hei\ and Balmer lines become intense and broad, while additional transitions such as Si\,\textsc{ii} and \Feii\ emerge. A logarithmic scale is used to represent the intense \ha\ emission, which reached an equivalent width of $\sim$ 2000~\AA. A vertical offset of 0.1 has been added to the nebular phase spectrum for clarity.
The inset shows the \ha\ region, where broad wings extending to $\pm$3000~\kms\ are visible. Adapted from \citet{Munoz-Darias2016}.}\label{fig:nebular_phase}
\end{figure}
}

\newcommand{\FigTrail}{
\begin{figure}[t]
\centering
\includegraphics[width=1.0\textwidth]{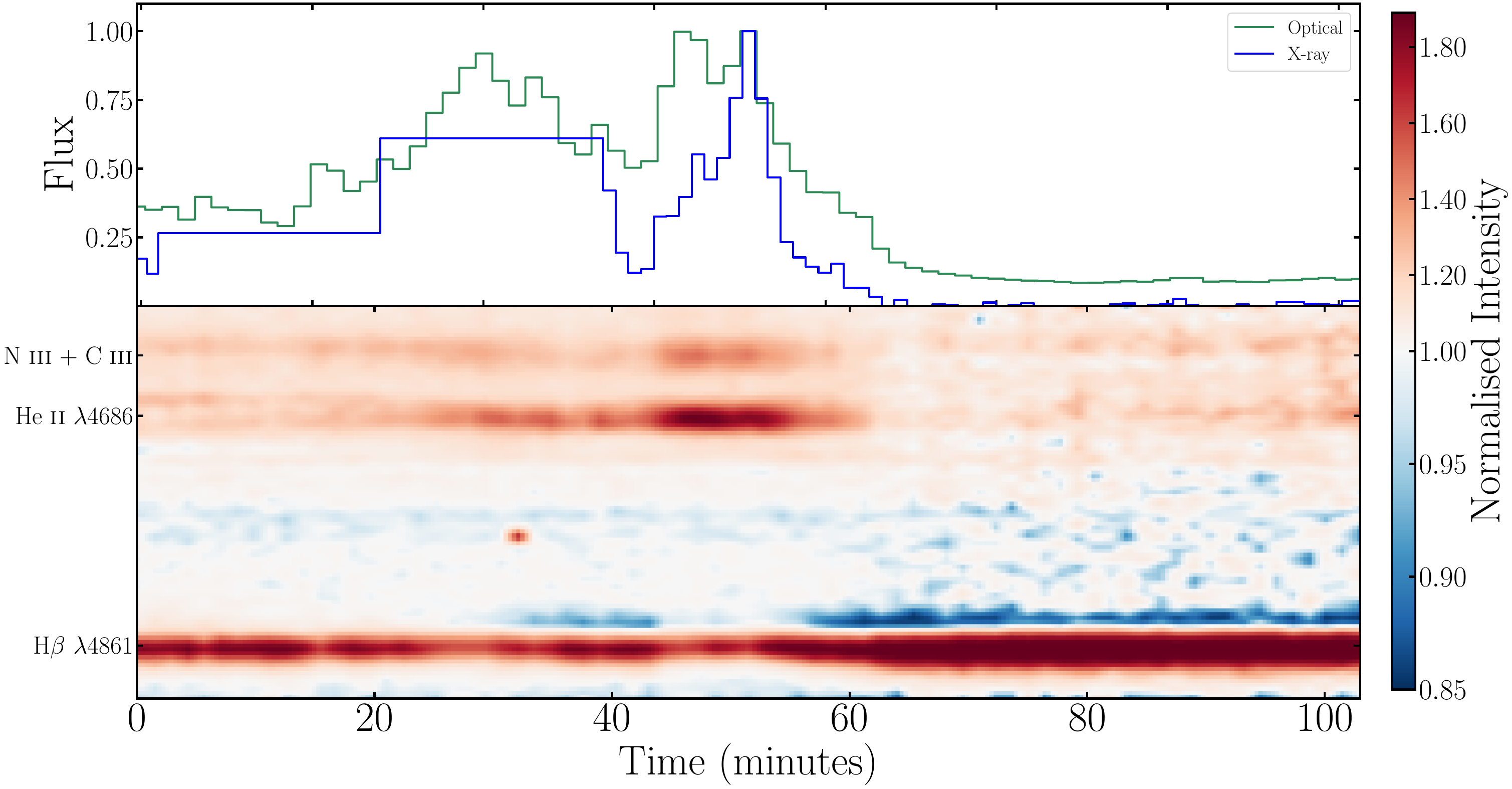}
\caption{Minute time-scale evolution of wind features in V404~Cyg. Top panel: Optical and X-ray light curves (normalised to the peak) during a 103~min GTC spectroscopic window. Bottom panel: Trailed spectrum (from 75 individual spectra) showing the evolution of \hb\ (bottom), as well as \heii~$\lambda4686$ and the Bowen blend (approximately \niii\ + \ciii\ at $\lambda \sim 4640$ \AA). The \hb\ blue-shifted absorption (blue horizontal band) disappears during the main flare, when the higher ionisation features (i.e. \heii, \niii\ and \ciii) become stronger. From data originally presented in \citealt{Munoz-Darias2016}.} \label{fig:trail_hb}
\end{figure}
}

\newcommand{\FigNIRwinds}{
\begin{figure}[t]
\centering
\includegraphics[width=1.0\textwidth]{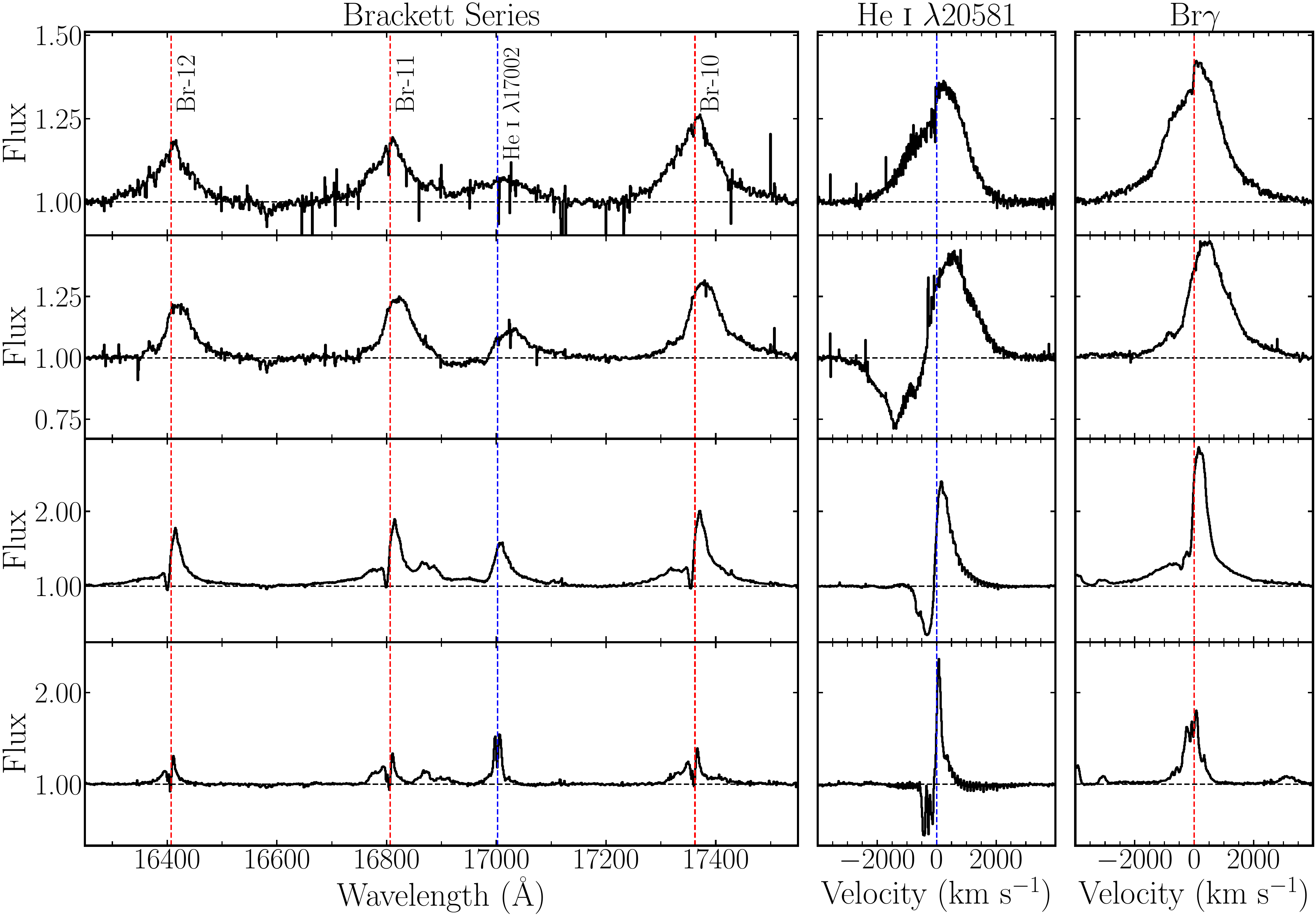}
\caption{NIR wind signatures in GRS~1915+105. Each row corresponds to a different observing epoch, taken from top to bottom in 2019 (19A and 19C) and 2023 (23A and 23D; see \citealt{SanchezSierras2023b}). Dashed vertical lines mark the rest positions of the different helium (blue) and hydrogen Brackett (red) lines. Different wind signatures are observed depending on the epoch and the specific transition. Brackett lines (left and right columns) are more prone to show emission components (top two panels), which are sometimes accompanied by sharp absorption troughs (bottom two panels). In contrast, \hei~$\lambda20581$ (and $\lambda10830$, not shown) have (typically) higher optical depth and often display strong P-Cyg profiles (middle column; see Sec.~\ref{liwinds:lines}). Adapted from \citet{SanchezSierras2023b}.}\label{fig:NIR_lines}
\end{figure}
}

\newcommand{\FigTEC}{
\begin{figure}[h]
\centering
\includegraphics[width=0.9\textwidth]{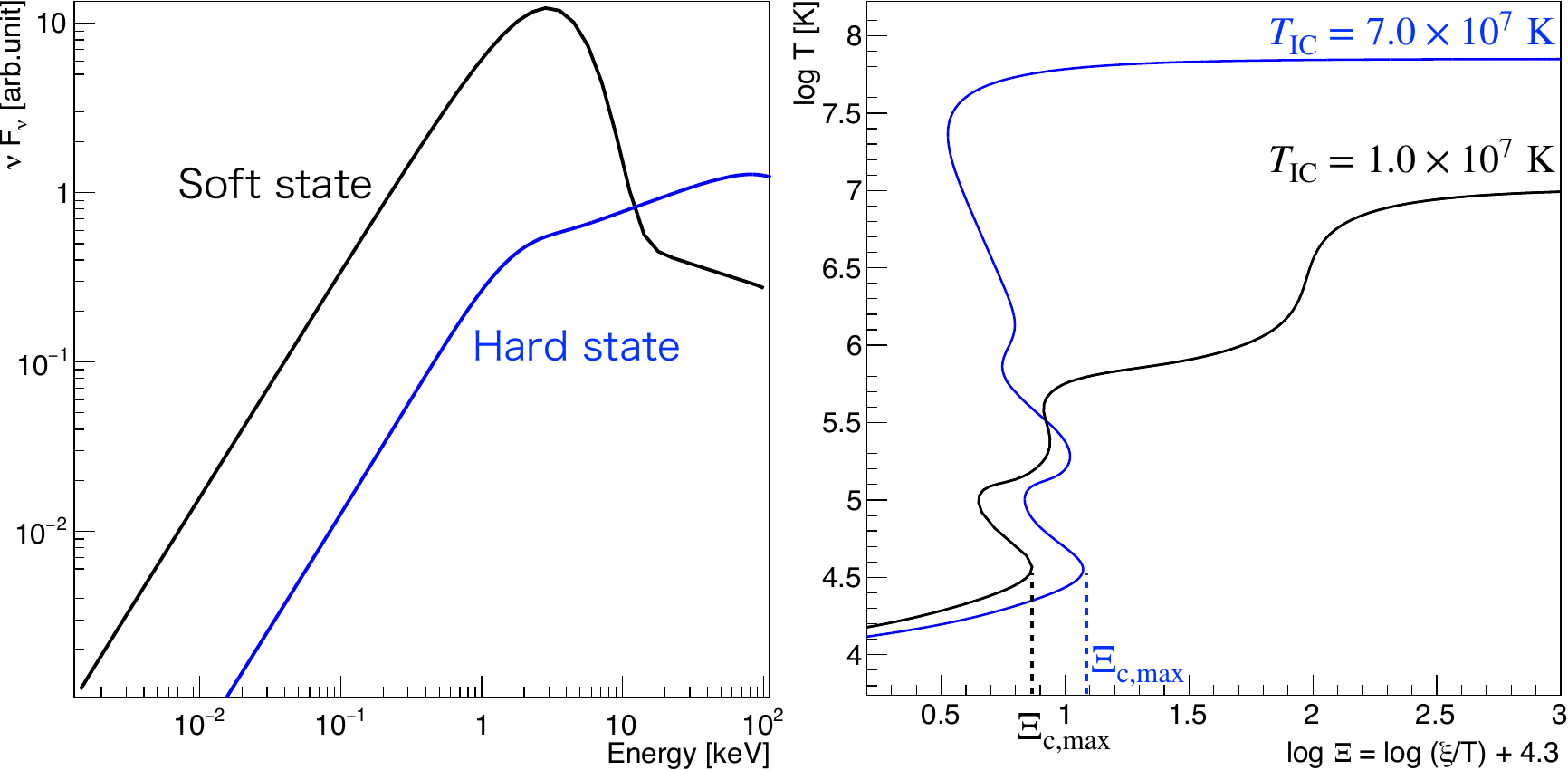}
\caption{Typical SEDs of soft and hard states in BH LMXBs illuminating ionised plasma (left) and the corresponding thermal equilibrium curves (right). Adapted from \citet{Tomaru2019}.}
\label{fig:TEC}
\end{figure}
}
\newcommand{\TRWmodel}{
\begin{figure}[h]
\centering
\includegraphics[width=0.9\textwidth]{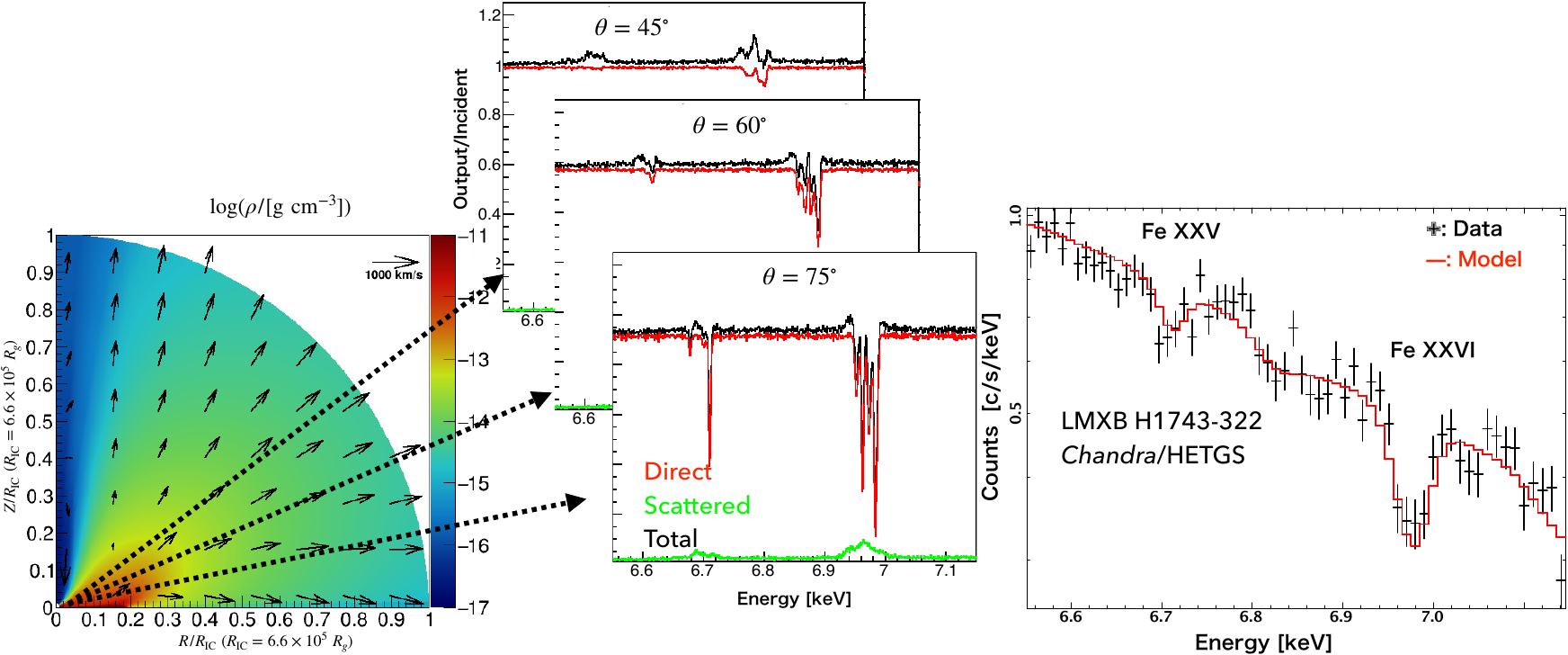}
\caption{Thermal--radiative wind model, which can explain the observed X-ray spectrum of the BH LMXB H~1743$-$322. Panels depict the density distribution from the radiation hydrodynamics simulation, including radiative heating and acceleration (left), the line profiles from Monte Carlo radiative transfer at different inclination angles (middle), and the comparison with observations (right). Adapted from \citet{Tomaru2019}.}
\label{fig:TRWmodel}
\end{figure}
}
\newcommand{\MHDWmodel}{
\begin{figure}[h]

\includegraphics[width=0.4\textwidth]{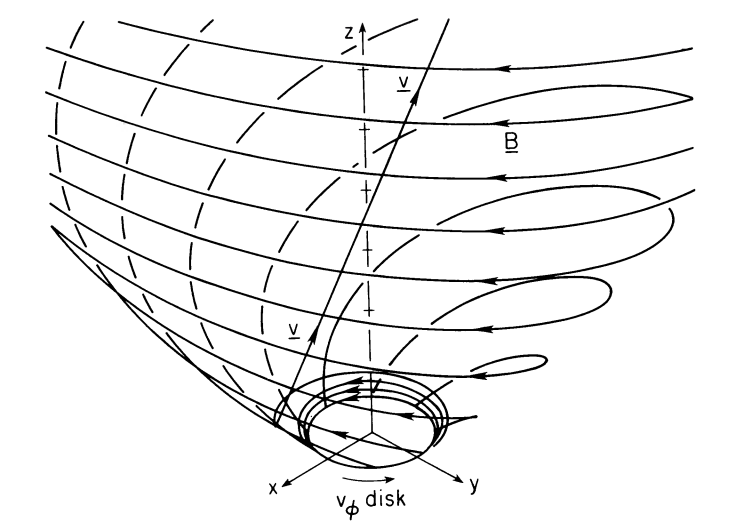}
\includegraphics[width=0.6\textwidth]{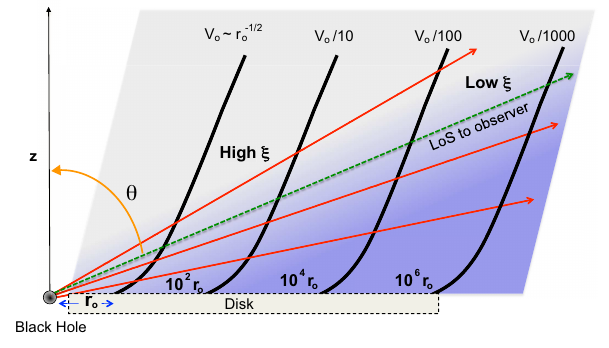}
\caption{Schematic diagram of the MHD wind model, with 3D streamlines on the left. The  right shows their projection onto 2D, where 
the solid black lines show the magnetic field lines and streamlines of the wind. Composite figure from \citealt{Lovelace1991} and \citealt{Fukumura2017}, respectively}
\label{fig:MHDmodel}
\end{figure}
}
\newcommand{\ADCmodel}{
\begin{figure}[h]
\centering
\includegraphics[width=0.9\textwidth]{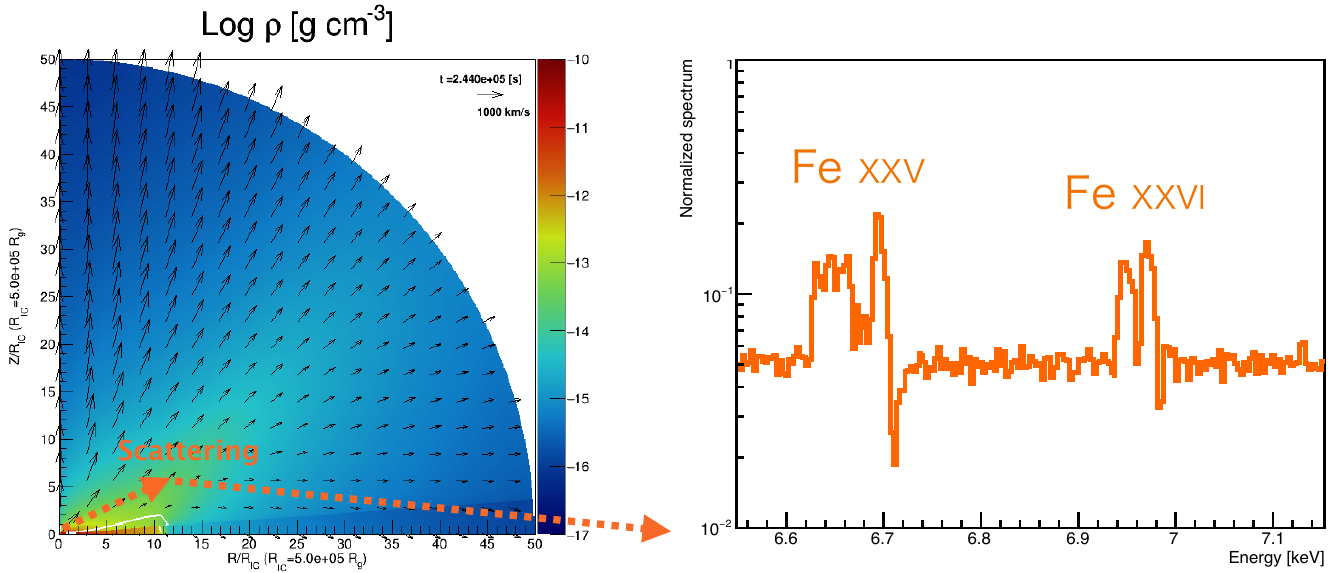}
\caption{Schematic view of an accretion disc corona source. The density and velocity distributions (left) are taken from an RHD simulation \citep[adapted from][]{Tomaru2020b}, while the line profiles (right) are obtained from a Monte Carlo radiative transfer simulation \citep[adapted from][]{Tomaru2023b}.}
\label{fig:ADC_source}
\end{figure}
}

\newcommand{\MHDlines}{
\begin{figure}[h]
\centering
\includegraphics[width=0.45\textwidth]{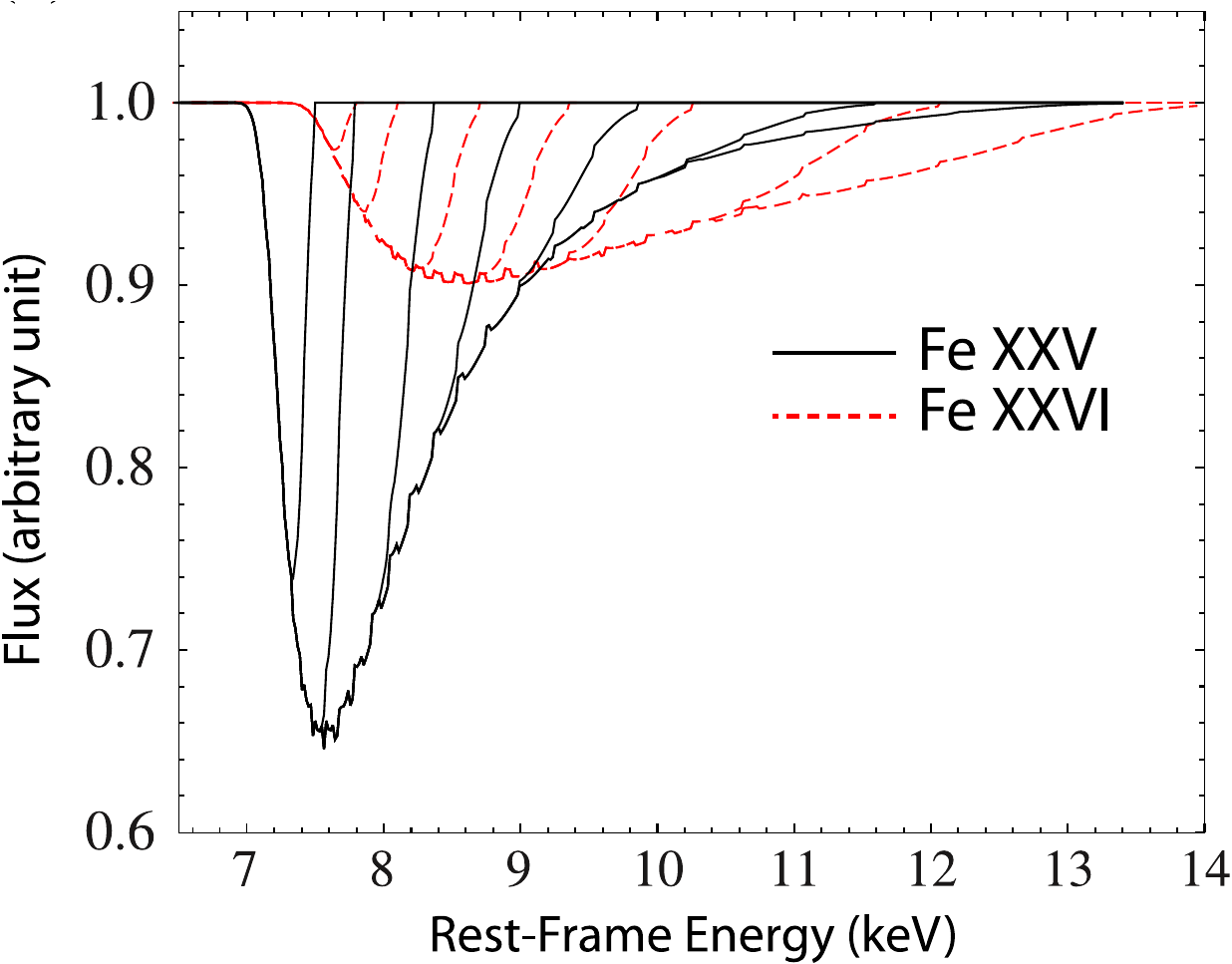}
\caption{
Typical line profiles of the MHD wind model adapted from \citet{Fukumura2015}. 
The highly ionised Fe {\sc xxvi} is faster than the lower ionisation Fe {\sc xxv}. 
Both show strong blue wing components. 
}
\label{fig:line_MHD}
\end{figure}
}

\newcommand{\Windstructure}{
\begin{figure}[h]
\centering
\includegraphics[width=\textwidth]{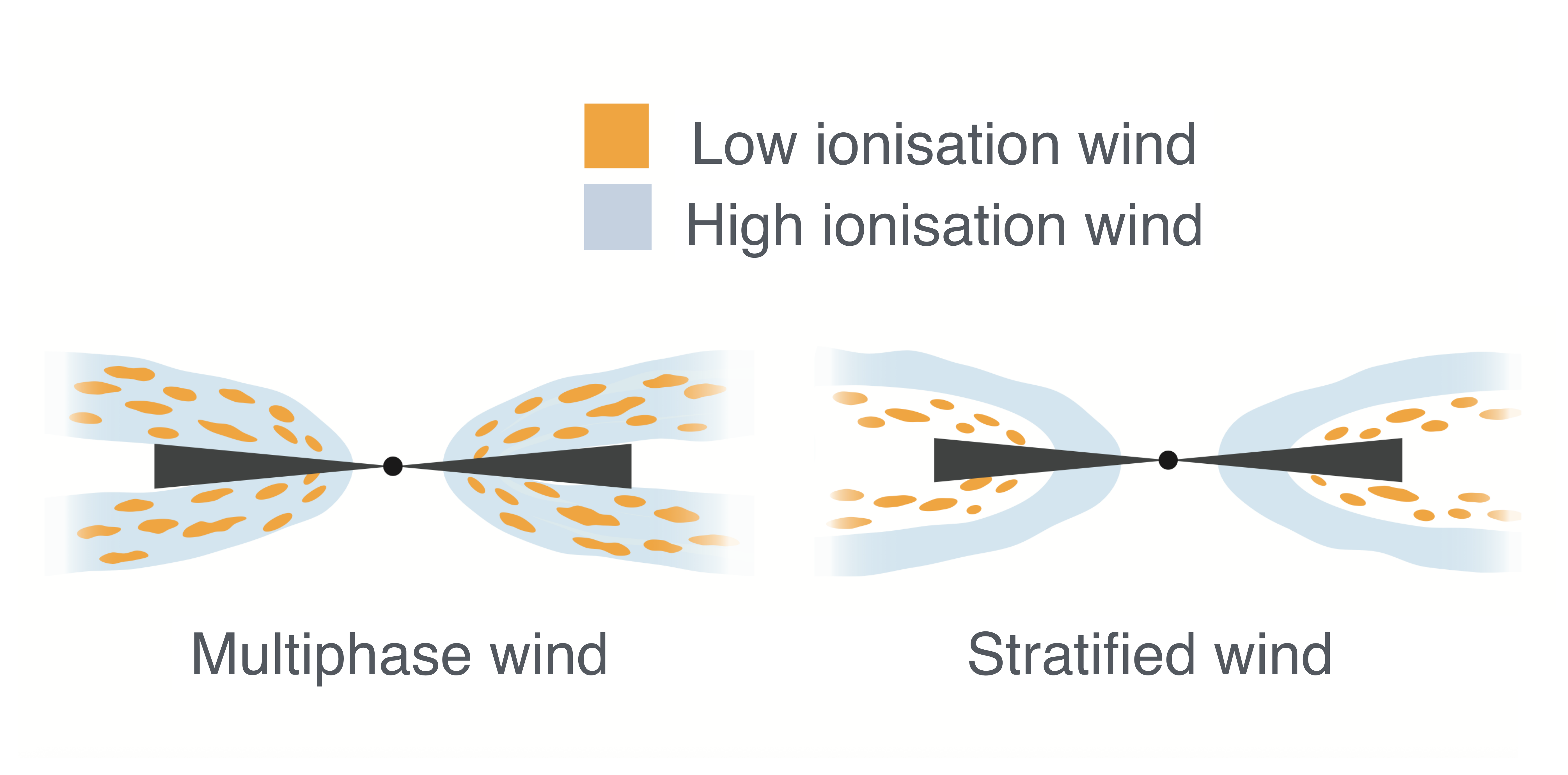}
\caption{
Wind structure. Sketch illustrating alternative configurations for a clumpy, low-ionisation wind and a lower-density, highly ionised outflow. An equatorial geometry is assumed in both cases. Adapted from \citet{Munoz-Darias2022}.
}
\label{fig:structure_sketch}
\end{figure}
}

\newcommand{\liwindsstates}{
\begin{figure}[t]
\centering
\includegraphics[scale=0.5]{Our_Files/Images/JSS-MD2020_sketch.pdf}
\caption{
Detection of low ionisation winds across accretion states. 
Schematic illustration adapted from \citet{SanchezSierras2020}, showing optical and NIR wind signatures during different phases of the outburst of MAXI~J1820+070. \tmd{this my go away if we found it repetitive when adding the in prep. sketch on the observability of winds across the states (in the Discussion).}
}
\label{fig:liwinds_states}
\end{figure}
}

\newcommand{\FigGXxrism}{

\begin{figure}
\includegraphics[width=\linewidth]{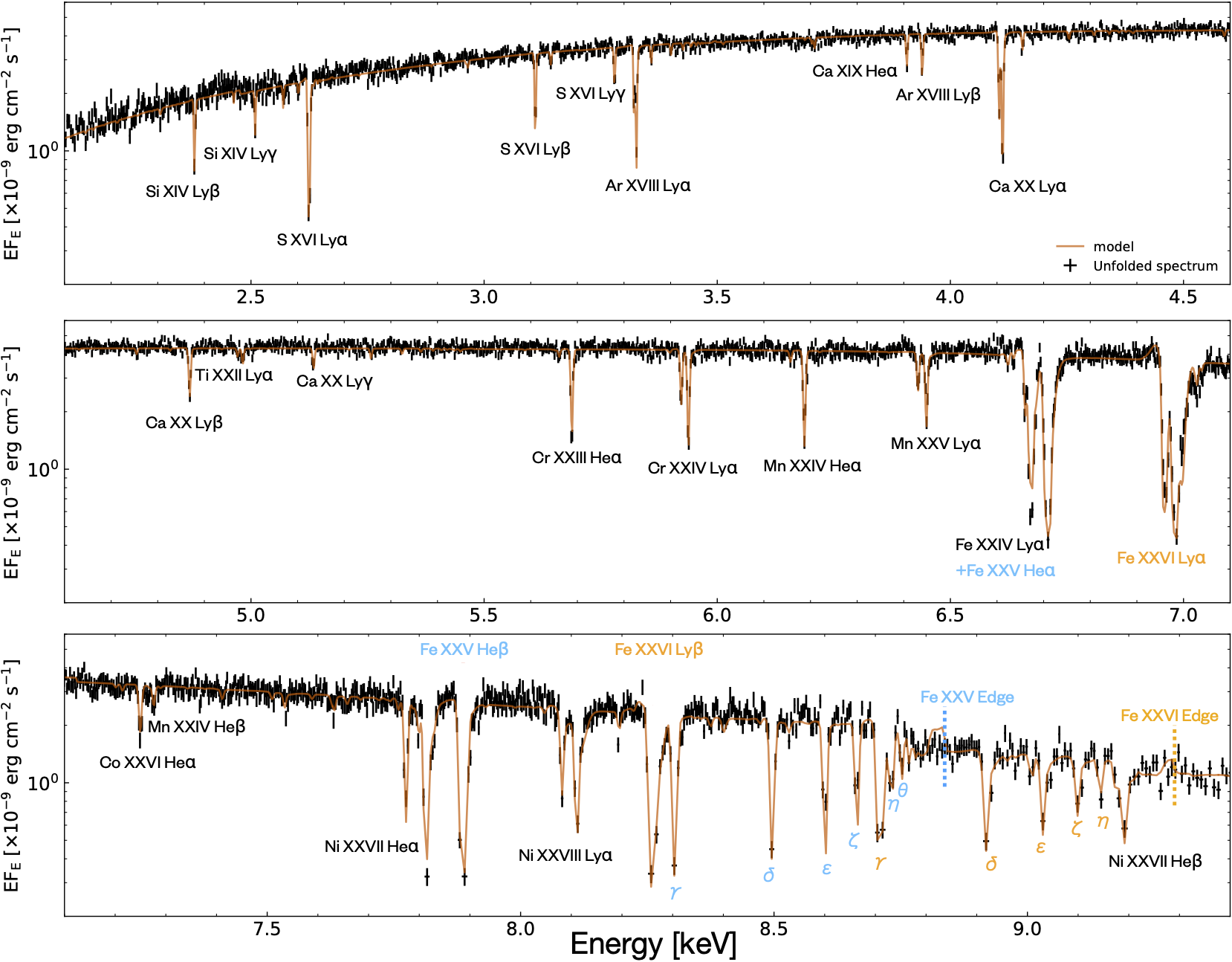}
    \caption{XRISM spectrum (black) of GX~13+1 together with the fitted model (orange). The main transitions are labelled, with the \Fexxv\ and \Fexxvi\ lines indicated in cyan and orange, respectively. The data were taken during a luminous, likely super-Eddington, epoch. A slower, wind component ($\sim300$~\kms) and a faster ($\sim700$~\kms), highly ionised one are detected. Adapted from \citet{XRISM2025GX13}.}
\label{fig:gx13spectrumxrism}
\end{figure}
}

\newcommand{\FigUVwindSimu}{
\begin{figure}
    \centering
    \includegraphics[width=0.5\linewidth]{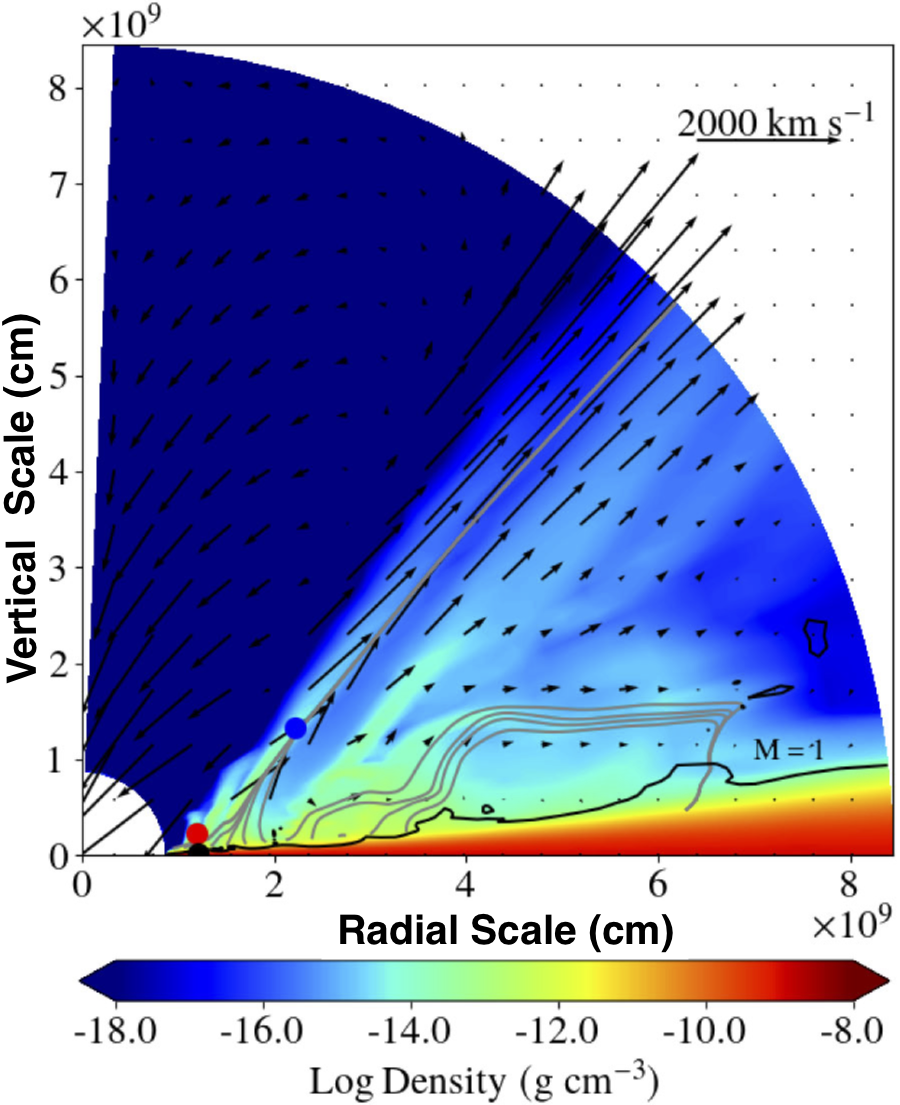}
    \caption{Density distribution of an ultraviolet line-driven wind from an accreting white dwarf, taken from a state-of-the-art radiation hydrodynamics simulation (adapted from \citealt{Higginbottom2024}).}
    \label{fig:sim_UV}
\end{figure}
}

\newcommand{\FigSketch}{
\begin{figure}[t]
\centering
\includegraphics[width=0.9\linewidth]{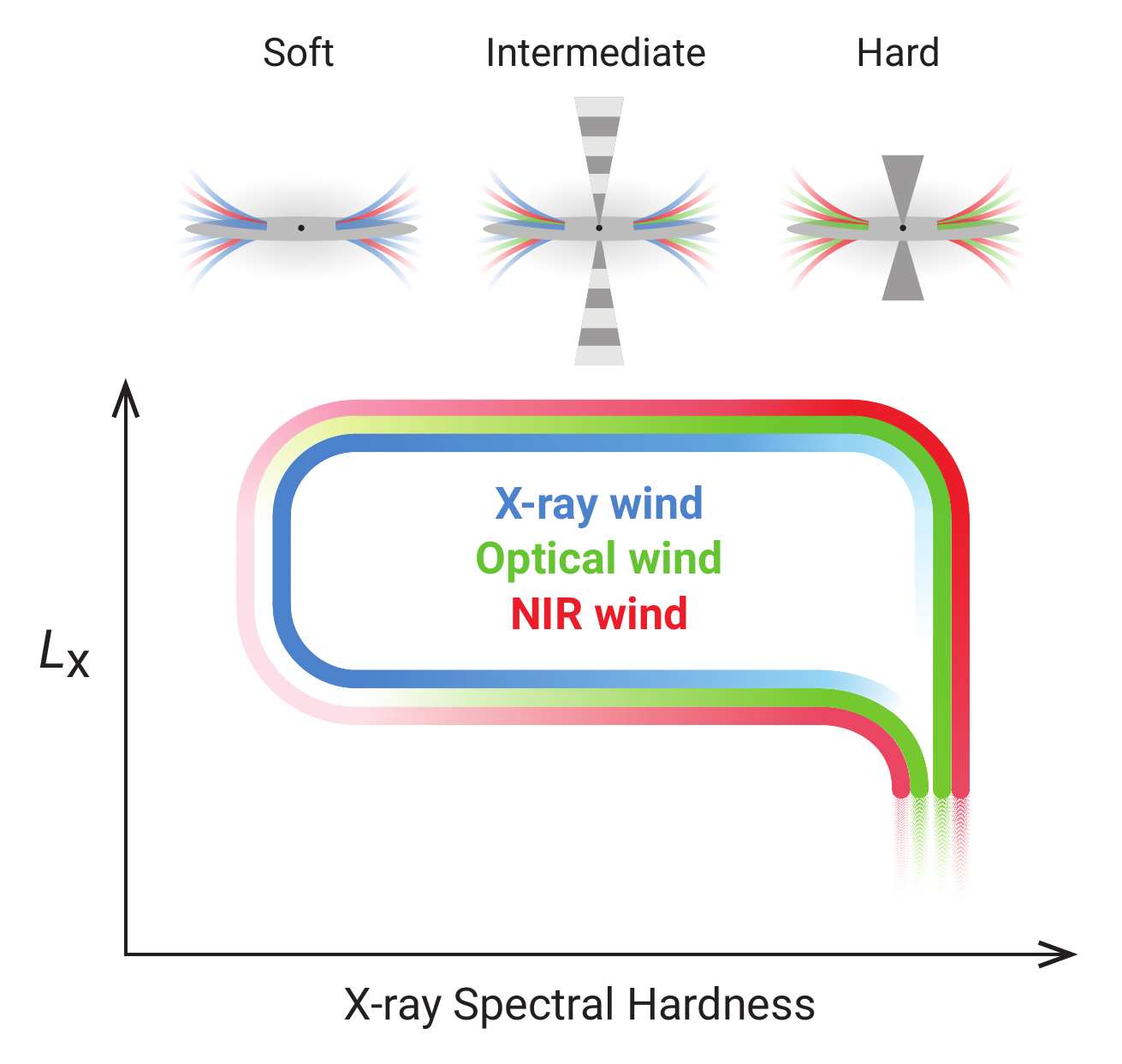}
\caption{
Sketch summarising the current observational evidence for both high (X-ray) and low (optical and NIR) ionisation winds across the BH accretion states. The coloured tracks illustrate wind detectability for the different spectral domains, where colour intensity follows the typical absorption strength for a high-inclination source. The tracks follow the standard outburst evolution of BHs in the X-ray luminosity versus spectral hardness plane (commonly plotted as the hardness–intensity diagram; e.g. \citealt{Homan2001, Dunn2010}). The amount of fast X-ray variability can also be used as an alternative to X-ray hardness \citep{Munoz-Darias2011, Munoz-Darias2014}. This approach is particularly useful for NS systems, for which limited observations suggest that they follow a similar pattern (see Sec.~\ref{sec:discussion:states}). The full disc–outflow configuration, following the same colour scheme and showing the jet behaviour (e.g. \citealt{Fender2004}), is also depicted in the top panel.
}
\label{fig:sketch}
\end{figure}
}

\newcommand{\FigTsujimoto}{
\begin{figure}[t]
\centering
\vspace{-1cm}
\includegraphics[width=\linewidth]{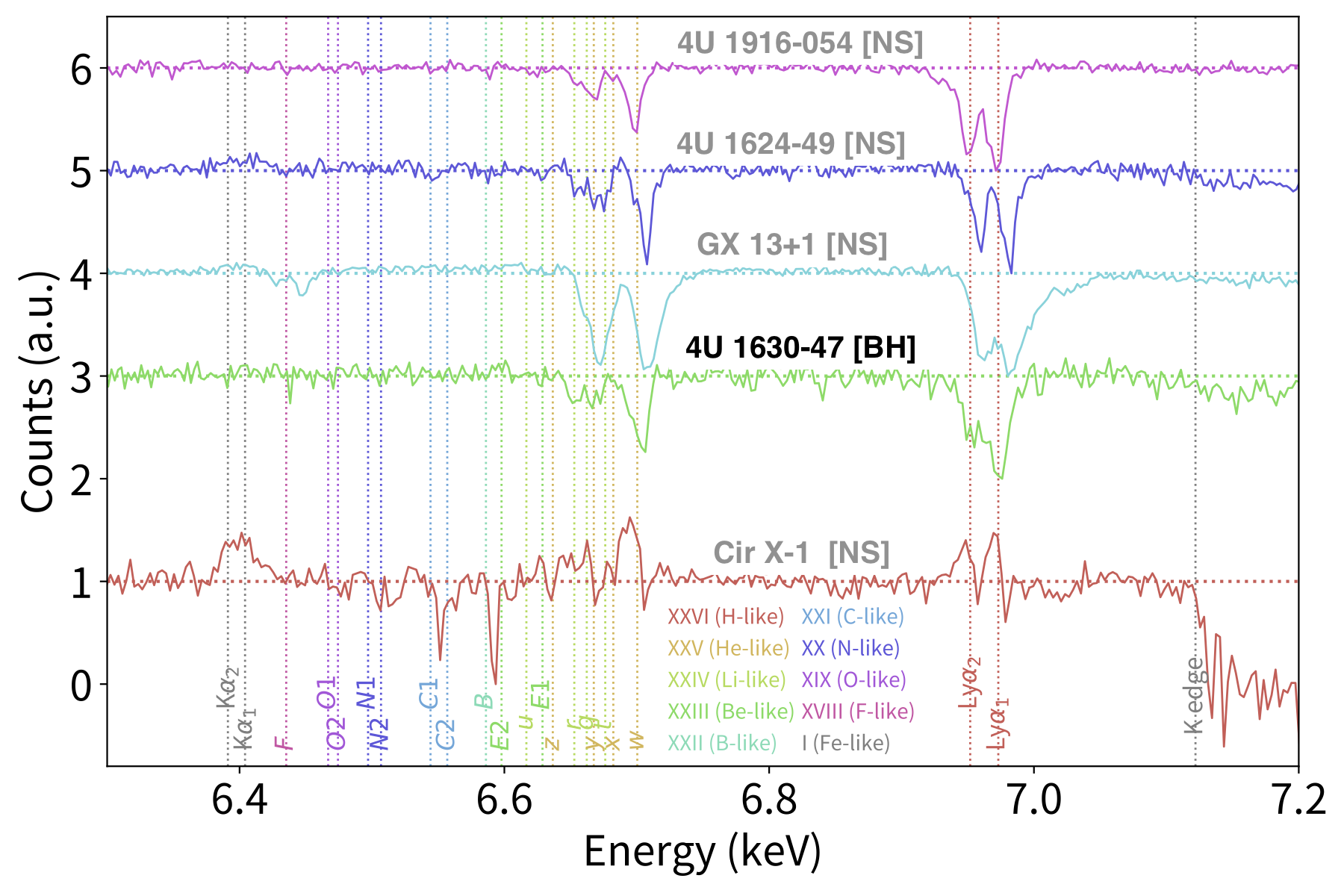}
\caption{\textit{XRISM} view of Fe wind signatures in four NS LMXBs (from top to bottom: 4U~1916$-$054, 4U~1624$-$49, GX~13+1, and Cir~X$-$1) and in the BH transient 4U~1630$-$47. Vertical dotted lines mark the rest-frame positions of transitions associated with different species of Fe, from neutral Fe (Fe-like) to \Fexxvi\ (H-like). Adapted from \citealt{Tsujimoto2025}.} 
\label{fig:XRISM_Fe}
\end{figure}
}
\newcommand{\Tableli}{
\begin{table}[h!]
    \centering
    \caption{Low-ionisation winds: observational evidence}
\begin{tabular}{lccccc}
    \toprule
    Source (nature) & Signature$^{a}$ & Cold$^{b}$ & Cool$^{c}$ & Vel$^{d}$ (\kms) & References \\
    \midrule
    \multicolumn{6}{l}{\textbf{Confident detections}} \\
    \midrule
   
    V404~Cyg (BH) & P-Cyg ($>$10\%), Emi & Optical &  & 1500--3000 & 1--1f \\
    V4641~Sgr (BH) & P-Cyg ($>$10\%) & Optical &  & 900--1600 & 2--2c \\
    & Emi& Optical &  & $\sim$3000 & \\
    GRS~1915+105 (BH) & P-Cyg ($>$10\%), Emi & NIR &  & 1000--3000 & 3 \\
    Swift J1858.6--0814 (NS) & P-Cyg ($>$10\%) & Optical & Ultraviolet & 1700--2400 & 4, 4b \\
   
    \hline
   
    MAXI~J1820+070 (BH) & P-Cyg & Optical, NIR &  & 1200--1800 & 5--5c \\
    & Emi& Optical, NIR &  & $\sim$1800 & \\
    GRS~1716--249 (BH) & P-Cyg & Optical & \heii(?) & $\sim$2000 & 6 \\
    Swift~J1727.8$-$162 (BH) & P-Cyg & Optical & \heii(?) & 1150 & 7 \\
    MAXI~J1803--298 (BH) & P-Cyg & Optical &  & $\sim$1250 & 8 \\
    
    \hline
  
    MAXI~J1348--630 (BH) & Emi& Optical, NIR &  & 1500--1700 & 9 \\
    & Abs & Optical, NIR &  & 500--900 & \\
    GX~339--4 (BH) & Emi, Other & Optical, NIR &  & $\sim$1000 & 10--10c \\
    
    \hline
   
    Swift J1357.2--0933 (BH) & P-Cyg & Optical & \heii & 1600--4000 & 11, 11b \\
    UW CrB (NS) & P-Cyg & Optical (?)  & Ultraviolet & 1500 & 12--12c \\  
    \midrule
    \multicolumn{6}{l}{\textbf{Tentative detections}} \\
    \midrule
    GX~13+1 (NS) & P-Cyg [SL] & NIR &  & $\sim$2400 & 13 \\
    Sco~X-1 (NS) & P-Cyg [SL] & NIR &  & $\sim$2600 & 14 \\
    GRO~J1655--40 (BH) & Other [T] & Optical &  &  & 15 \\
    4U~1543-47 (BH) & Other [T] & NIR &  &  & 16 \\
    Aql~X-1 (NS) & Emi [SL] & Optical &  & 800 & 17 \\
    \bottomrule
\end{tabular}
\textbf{Notes.} Horizontal lines separate sources according to the wind detection categories described in Sec.~\ref{sec:liwinds:obs}.
\textbf{(a)} P-Cyg line profiles, indicating whether the blue-shifted absorption dips below 10\% the continuum level ($>10\%$); Broad emission components, including flat-top and nebular phases (Emi); blue-shifted absorption troughs (Abs.). 'Other' refers to skewed profiles, asymmetries in several emission lines. In cases of tentative detections, the reason is indicated in brackets: SL (single line) and T (the features themselves are only tentative or there are alternative explanations). These are further explained in Sec. \ref{sec:liwinds:tentative}.  
\textbf{(b)} Detection of cold winds [mainly H (Balmer, Paschen, Brackett series) and \hei\ transitions]. The detection's spectral range (optical, NIR or both) is indicated.
\textbf{(c)} Detection of of cool winds [\heii\ and ultraviolet (mainly \Civ\ and \Nv) transitions]. 
\textbf{(d)} Outflow velocity (blue-edge for P-Cyg and Emi). Question marks (?) indicate tentative detections for the corresponding line or spectral range.

    \textbf{References.}  
    (1) \citet{Munoz-Darias2016}; (1b) \citet{Munoz-Darias2017}; (1c) \citet{MataSanchez2018};  
    (1d) \citet{Rahoui2017};
    (1e) \citet{Vincentelli2025}; (1f) \citet{Casares1991};
    (2) \citet{Munoz-Darias2018}; (2b) \citet{Lindstrom2005}; (2c) \citet{Chaty2003};  
    (3) \citet{SanchezSierras2023b}; 
    (4) \citet{Munoz-Darias2020}; (4b) \citet{CastroSegura2022};  
    (5) \citet{Munoz-Darias2019}; (5b) \citet{SanchezSierras2020}; (5c) \citet{Baglio2025}
    (6) \citet{Cuneo2020};  
    (7) \citet{MataSanchez2024a};  
    (8) \citet{MataSanchez2022};  
    (9) \citet{PanizoEspinar2022};  
    (10) \citet{Ambrifi2025}; (10b) \citet{Rahoui2014}; (10c) \citet{Soria1999};  
    (11) \citet{Jimenez-ibarra2019b}; (11b) \citet{Charles2019};
    (12) \citet{Fijma2023};  (12b) \citet{Kennedy2025}; (12c) \citet{Fijma2025};
    (13) \citet{Bandyopadhyay1999};  
    (14) \citet{Bandyopadhyay1999};
    (15) \citet{Soria2000};  
    (16) \citet{SanchezSierras2023};  
    (17) \citet{PanizoEspinar2021}.  

\label{tab:liwinds}    
\end{table}
}

\newcommand{\TablehiBH}{
\begin{table}[]
    \centering
    \caption{High-ionisation winds in BH systems: observational evidence.}
    \begin{tabular}{lccccc}
        \toprule
        Source & Signature$^{a}$ & Warm$^{b}$ & Hot$^{c}$ & Vel$^{d}$ (\kms) & References \\
        \midrule
        \multicolumn{5}{l}{\textbf{Confident detections}} \\
        \midrule
        4U~1630$-$47 & Abs &  &\checkmark & 200--400 & 1--1m \\
        GRO~J1655$-$40 & Abs & \Mgxii, \FexxivL, \Sixiv &\checkmark &200--1200 & 2--2p \\
        GRS~1915$+$105 & Abs &  &\checkmark & 200--1100 & 3--3m \\
                       & Emi &  &\checkmark &  & 3m \\
        H~1743$-$322 & Abs &  &\checkmark & 200--400 & 4--4j \\
        V404~Cyg & P-Cyg, Emi & \Mgxii, \Sixiv, \Sxvi &\checkmark & 1500--4000 & 5, 5b\\
        MAXI~J1803$-$298 & Abs &  &\checkmark & few $\times 10^2$ & 6 \\
        IGR~J17451$-$3022$^{\dagger}$ & Abs &  &\checkmark &  & 7 \\
        IGR~J17091$-$3624 & Abs & \Nex, \Mgxii, \Sixiii\ &\checkmark &  & 8a--8c \\
        \midrule
        \multicolumn{5}{l}{\textbf{Tentative detections}} \\
        \midrule
        GX~339$-$4 & Emi [T] &  &\checkmark &  & 9  \\
        EXO~1846$-$031 & Abs [SL, BC] &  &\checkmark & 18000 & 10 \\
        MAXI~J1348$-$630 & Abs [SL, BC] &  &\checkmark & 10000 & 11, 11b\\
        MAXI~J1631$-$479 & Abs [SL, BC] &  &\checkmark & 20000 & 12 \\
        MAXI~J1820$+$070 & Abs [LR, BC] &  &\checkmark & 4000 & 13 \\
        4U~1543$-$47 & Abs [SL, BC] &  &\checkmark &  & 14, 14b \\
        MAXI~J1810$-$222 & Abs [SL, LR] &  &\checkmark & 15000--45000 & 15 \\
        \bottomrule
    \end{tabular}
    \vspace{0.5em}
    \footnotesize
    \textbf{Notes.}  
\textbf{(a)} P-Cyg line profile (P-Cyg); Absorption (Abs); Emission (Emi). In cases of tentative detections, the reason is indicated in brackets: SL (single line), BC (broad continuum on reflection), LR (low-resolution spectra), and T (the features themselves are only tentative or there are alternative explanations). These are further explained in Sec. \ref{sec:hiwinds:bh:tentative}. 
\textbf{(b)} Detection of warm winds (\logxi\ $\sim 1.5-3$). 
\textbf{(c)} Detection of hot winds (\logxi\ $\gtrsim 3$; mainly \Fexxv\ and \Fexxvi).
\textbf{(d)} Outflow velocity (blue-edge for P-Cyg). We only quote velocities above $\sim$200 \kms for gratings  (\textit{XMM-Newton}/RGS, \textit{Chandra}/HETGS or Chandra/LETGS) and above a few tens of km/s for \textit{XRISM}/Resolve. \textbf{($\dagger$)} The nature of the compact object, either BH or NS, is unclear.\\
    \textbf{References.}  
    (1) \citet{Ponti2012}; (1b) \citet{King2013}; (1c) \citet{King2014}; (1d) \citet{Neilsen2014}; (1e) \citet{DiazTrigo2014}; (1f) \citet{Miller2015}; (1g) \citet{Wang2016}; (1h) \citet{Pahari2018}; (1i) \citet{Hori2018}; (1j) \citet{Gatuzz2019}; (1k) \citet{Zeegers2019}; (1l) \citet{Trueba2019}; (1m) \citet{Miller2025};
     (2) \citet{Miller2006}; (2b) \citet{DiazTrigo2007}; (2c) \citet{Sala2007}; (2d) \citet{Miller2008}; (2e) \citet{Kallman2009}; (2f) \citet{Reis2009}; (2g) \citet{Zhang2012}; (2h) \citet{Neilsen2012}; (2i) \citet{Luo2014}; (2j) \citet{Madej2014}; (2k) \citet{Miller2015}; (2l) \citet{Gatuzz2016}; (2m) \citet{Shidatsu2016}; (2n) \citet{Fukumura2017}; (2o) \citet{Higginbottom2018}; (2p) \citet{Tomaru2023}; 
    (3) \citet{Ponti2012}; (3b) \citet{Neilsen2012}; (3c) \citet{King2013}; (3d) \citet{Madej2014}; (3e) \citet{Miller2015}; (3f) \citet{Lee2002}; (3g) \citet{Neilsen2009}; (3h) \citet{Ueda2009}; (3i) \citet{Miller2016}; (3j) \citet{Zoghbi2016}; (3k) \citet{Rogantini2020}; (3l) \citet{Ratheesh2021}; 
    (3m) \citet{Miller2020};
   (4) \citet{Miller2006}; (4b) \citet{Blum2010}; (4c) \citet{Miller2012}; (4d) \citet{Tetarenko2018}; (4e) \citet{Shidatsu2019}; (4f) \citet{Tomaru2020}; (4g) \citet{Ponti2012}; (4h)\citet{King2013}; (4i) \citet{Madej2014}; (4j) \citet{Miller2015};
   (5) \citet{King2015}; (5b) \citet{Munoz-Darias2022}; 
   (6) \citet{Zhang2024}; 
   (7) \citet{Bozzo2016};
   (8) \citet{King2012}; (8b) \citealt{Gatuzz2020}; (8c) \citet{Wang2024}; 
   (9) \citet{Miller2015}; 
   (10) \citet{Wang2021}; 
   (11) \citet{Wu2023}; (11b) \citet{Chakraborty2021}; 
   (12) \citet{Xu2020}; 
   (13) \citet{Fabian2020} 
   (14) \citet{Prabhakar2023}; (14b) \citet{Shidatsu2013}; 
   (15) \citet{DelSanto2023};  
\label{tab:hiwinds:bh}
\end{table}
}

\newcommand{\TablehiNS}{
\begin{table}[]
\caption{Highly ionised winds and atmospheres in NS systems: observational evidence.}
\begin{tabular}{lccccc}
    \hline
    Source & Signature$^a$ & Warm$^b$ & Hot$^c$ & Vel$^d$  & References \\
    \hline
    \textbf{Confident Detections} \\
    \hline
    4U~1916$-$054 & Abs & \Nex, \Mgxii\ & \checkmark & atm & 1--1g \\
    1A~1744$-$361 & Abs &  & \checkmark & atm & 2--2c \\
    4U~1323$-$62 & Abs &  & \checkmark & atm & 3--3c \\
    EXO~0748$-$676 & Abs, Emi & \Ovii, \Neix, \Nex\ & \checkmark & atm & 4--4g \\
    4U~1254$-$69 & Abs &  & \Fexxvi & atm & 5--5d \\
    4U~1822$-$37 & Emi & \Ovii,  \Neix, \Nex\ & \checkmark & atm & 6--6d \\
    MXB~1659$-$298 & Abs & \Oviii, \Nex, \FexxivL\ & \checkmark & atm & 7--7e \\
    AX~J1745.6$-$2901 & Abs &  & \checkmark & $\sim$ 200 & 8--8e \\
    4U~1624$-$49 & Abs &  & \checkmark & $\sim$300 & 9--9e\\
    Her~X$-$1 & Abs, Emi & \Nvi, \Nvii, \Ovii & \checkmark & 300--1000 & 10--10e \\
    2S~0921$-$63 & Emi & \Ovii, \Oviii, \Nex & \checkmark & $<$1000 & 11--11c \\
    Cir~X$-$1 & Abs, Emi, P-Cyg & \Nex, \Mgxi, \Mgxii\ & \checkmark & 300--2000 & 12--12g \\
    GX~13+1 & Abs, Emi & \Mgxii, \Sixiv, \Sxvi\ & \checkmark & 300--1000 & 13--13i \\
    \hline
    \textbf{Tentative Detections} \\
    \hline 
    4U~1820$-$30 & Abs [T] & \Oiv, \Ov\ &  & 1200 & 14 \\
    XTE~J1710$-$281 & Abs [SL, LR] &  & \Fexxv? & atm & 15, 15b \\
    Swift~J1858.6$-$0814 & Emi [T] & \Nvii, \Ovii\ &  & atm & 16 \\
    IGR~17062$-$6143 & Abs [T] & \Oviii? &  & 2000--3500 & 17, 17b \\
    1RXS~J180408.9$-$342058 & Abs [SL] &   & \Fexxvi? & $\sim$26000 & 18 \\
    IGR~J17591$-$2342 & Abs [BC] & \Sixiii\ &  & $\sim$3000 & 19, 19b \\
    IGR~J17480$-$2446 & Abs [SL] &  & \Fexxv? & $\sim$3000 & 20 \\
    GRO~J1744$-$28 & Abs [SL] &  & \Fexxv? & $\sim$8000 & 21, 21b \\
    GX~340+1 & Abs [SL] & \Caxx? & \Fexxv? & $\sim$12000, $\sim$ 2700 & 22, 22b \\
    \hline
\end{tabular}
\textbf{Notes.} \textbf{(a)} P-Cyg line profile (P-Cyg); Absorption (Abs); Emission (Emi). In cases of tentative detections, the reason is indicated in brackets: SL (single line), BC (broad continuum on reflection), LR (low-resolution spectra), and T (the features themselves are only tentative or there are alternative explanations). These are further explained in Sec. \ref{sec:hiwinds:ns:tentative}. 
\textbf{(b)} Detection of warm winds (\logxi\ $\sim 1.5$-$3$). Some of the transitions associated with the lowest \logxi\ are given. 
\textbf{(c)} Detection of hot winds (\logxi\ $\gtrsim 3$; \Fexxv\ and \Fexxvi\ unless indicated otherwise).
\textbf{(d)} Outflow velocity (blue-edge for P-Cyg). Low-velocities consistent with disc atmospheres are indicated marked as \textit{atm}. We only quote velocities above $\sim$200 \kms for gratings  (\textit{XMM-Newton}/RGS, \textit{Chandra}/HETGS or Chandra/LETGS) and above a few tens of km/s for \textit{XRISM}/Resolve. \\
\textbf{References:}
(1) \citet{Boirin2004}; (1b) \citet{DiazTrigo2006}; (1c) \citet{Juett2006};  (1d) \citet{Iaria2006}; (1e) \citet{Zhang2014};  (1f) \citet{Gambino2019}; (1g) \citet{Trueba2020}; 
(2) \citet{Gavriil2012}; (2b) \citet{Mondal2024}; (2c) \citet{Ng2024};
(3) \citet{Boirin2005}; (3b) \citet{Church2005}; (3c) \citet{Balucinska-Church2009};
(4) \citet{Cottam2001}; (4b) \citet{Jimenez2003}; (4c) \citet{DiazTrigo2006}; (4d) \citet{vanPeet2009}; (4e) \citet{Ponti2014}; (4f) \citet{Psaradaki2018}; (4g) \citet{Bhattacharya2024};
(5) \citet{Boirin2003}; (5b) \citet{DiazTrigo2006}; (5c) \citet{Iaria2007}; (5d) \citet{DiazTrigo2009};
(6) \citet{Cottam2001}; (6b) \citet{Ji2011}; (6c) \citet{Iaria2013}; (6d) \citet{Sasano2014};
(7) \citet{Sidoli2001}; (7b) \citet{DiazTrigo2006}; (7c) \citet{Ponti2018}; (7d) \citet{Ponti2019}; (7e) \citet{Iaria2019}; 
(8) \citet{Hyodo2009}; (8) \citet{Ponti2015}; (8) \citet{Bianchi2017}; (8) \citet{Trueba2022}; (8) \citet{Tanaka2026}; 
(9) \citet{Parmar2002}; (9b) \citet{DiazTrigo2006}; (9c) \citet{Iaria2007b}; (9d) \citet{Xiang2009}; (9e) \citet{DiazTrigo2026}; 
(10) \citet{Jimenez-Garate2002}; (10b) \citet{Jimenez-Garate2005}; (10c) \citet{Ji2009}; (10d) \citet{Kosec2020}; (10e) \citet{Kosec2023};
(11) \citet{Kallman2003}; (11b) \citet{Yoneyama2023}; (11c) \citet{Tomaru2023b};
(12) \citet{Brandt2000}; (12b) \citet{Schulz2002}; (12c) \citet{D'Ai2007}; (12d) \citet{Iaria2008}; (12e) \citet{Schulz2008}; (12f) \citet{Schulz2020} ; (12g) \citet{Tsujimoto2025}; 
(13) \citet{Ueda2001}; (13b) \citet{Sidoli2002}; (13c) \citet{Ueda2004}; (13d) \citet{DiazTrigo2012}; (13e) \citet{Madej2014}; (13f) \citet{Allen2018}; (13g) \citet{Tomaru2018}; (13h) \citet{Rogantini2025}; (13i)  \citet{XRISM2025GX13}; 
(14) \citet{Costantini12}
(15) \citet{Raman2018}; (15b) \citet{Trueba2022};
(16) \citet{Buisson2020}; 
(17) \citet{Degenaar2017}; (17b) \citet{vandenEijnden2018}; 
(18) \citet{Degenaar2016}; 
(19) \citet{Nowak2019}; (19b) \citet{Manca2023}; 
(20) \citet{Miller2011};
(21) \citet{Degenaar2014}; (21b) \citet{Younes2015}; 
(22) \citet{Miller2016}; (22b) \citet{Chakraborty2025}.

\label{tab:hiwinds:ns}
\end{table}
}


\newcommand{\TabMASTER}{
\begin{table}[h!]
    \centering
    \caption{Systems with confident detections of either low-ionisation (cold–cool) winds or high-ionisation (warm–hot) atmosphere/wind.}
\begin{tabular}{lccccccc}
    \toprule
    Source & ${P_{\textsc{orb}}}^a$ & Inclination$^b$ & ${v_{\rm max}}^c$ & Cold$^d$ & Cool$^e$ & Warm$^f$ & Hot$^g$ \\
     & (h) &  & (\kms) &  &  &  &  \\
    \midrule
    \midrule
\multicolumn{8}{c}{\textbf{Black Hole Systems [15]}} \\
\midrule
        IGR~J17451$-$3022   &  ?    &  & -- & &  &  & \textbf{\checkmark} (no vel.) \\
        IGR~J17091$-$3624 &   ?   &  & -- & &  &  & \textbf{\checkmark} (no vel.) \\
        \textbf{4U~1630$-$47} &  ?    & High [Dips] & 400 & &  &  & \textbf{\checkmark} \\
        \textbf{H~1743$-$322} &   ?   &  High [Dips] & 400 & & &  & \textbf{\checkmark} \\
        \textbf{GX~339--4} & 42.1 & Mid & 1000 \texttt{(b-e)} &\textbf{\checkmark} &  &  & \checkmark? \\
        \textbf{Swift~J1727.8$-$162} & 10.8    & Mid & 1150 \texttt{(b-e)} & \textbf{\checkmark} & \textbf{\checkmark}? &  &  \\ 
        \textbf{GRO~J1655$-$40} &  62.9  & High [Dips] & 1200 &\textbf{\checkmark}?  &  & \textbf{\checkmark} & \textbf{\checkmark} \\
        \textbf{MAXI~J1803--298} &  7.0   & High [Dips] & 1250 \texttt{(b-e)} & \textbf{\checkmark} &  &  & \textbf{\checkmark} \\
        \textbf{MAXI~J1348--630}  &  ?   & Mid-to-low & 1700 \texttt{(b-e)} & \textbf{\checkmark} & &  & \checkmark? \\
        \textbf{MAXI~J1820+070} &  16.5   & High [67--81$^{\circ}$] & 1800 \texttt{(b-e)} & \textbf{\checkmark} &  &  &  \\
        \textbf{GRS~1716--249} &   6.7  & Mid & 2000 \texttt{(b-e)} & \textbf{\checkmark} & \checkmark? &  &  \\
        \textbf{V4641~Sgr} &  67.6   & High [60--70$^{\circ}$]  & 3000 \texttt{(b-e)} & \textbf{\checkmark} &  &  &  \\
        \textbf{GRS~1915+105} &  812   & High [60--70$^{\circ}$] & 3000 \texttt{(b-e)} & \textbf{\checkmark} &  &  & \textbf{\checkmark} \\
        \textbf{V404~Cyg} &  155   & High [60--70$^{\circ}$] & 4000 \texttt{(b-e)} & \textbf{\checkmark} &  & \textbf{\checkmark} & \textbf{\checkmark} \\
        \textbf{Swift~J1357.2--0933} &  2.8  & High [$\geq$80$^{\circ}$] & 4000 \texttt{(b-e)} &  \checkmark & \checkmark &  &  \\
    \midrule
\multicolumn{8}{c}{\textbf{Neutron Star Systems [15]}} \\
\midrule
        4U~1916$-$054 & 0.8 & High [Dips] & -- &  &  & \textbf{\checkmark}-atm & \textbf{\checkmark}-atm \\
        1A~1744$-$361 &  1.62  & High [Dips] & -- & &  &  & \checkmark-atm \\
        4U~1323$-$62 &  2.94    & High [Dips] & -- & &  &  & \checkmark (no vel.) \\
        EXO~0748$-$676 &   3.82   & High [Eclipses] & -- & & &  \checkmark-atm & \checkmark-atm \\
        4U~1254$-$69 &   3.93   & High [Dips] & -- & &  &  & \checkmark-atm \\
        4U~1822$-$37 &   5.57   & High [Eclipses] & -- & & & \checkmark-atm   & \checkmark-atm \\
        MXB~1659$-$298 &   7.11   & High [Eclipses] & -- & &  & \checkmark-atm   & \checkmark-atm \\
        \textbf{AX~J1745.6$-$2901} &  8.35    & High [Eclipses] & 200 & &  &  & \checkmark \\
        \textbf{4U~1624$-$49} &  20.89    & High [Dips] & 350 & &  &  & \checkmark \\
        \textbf{2S~0921$-$63} &  216.2    & High [Eclipses] & <1000 & &  &\checkmark   & \checkmark \\
        \textbf{Her~X$-$1} &  40.8    & High [Eclipses] & 1000 & &  & \checkmark   & \checkmark \\
        
        \textbf{GX~13+1} &   577.6   & High [Dips] & 1000 &\checkmark? &  &  \checkmark   & \checkmark \\
        \textbf{UW CrB} &     1.8 & High [Eclipses] & 1500 \texttt{(b-e)}& \checkmark? & \checkmark &  &  \\
        \textbf{Cir~X$-$1 }&   398.4  & High [Eclipses] & 2000 \texttt{(b-e)} & &  & \checkmark   & \checkmark \\
        \textbf{Swift~J1858.6--0814} &  21.3   & High [Eclipses] & 2400 \texttt{(b-e)} & \checkmark & \checkmark & \checkmark? &  \\

        \bottomrule
    \end{tabular}

\vspace{0.5em}

\textbf{Notes}. References are provided in Tables \ref{tab:liwinds}, \ref{tab:hiwinds:bh} and \ref{tab:hiwinds:ns}. Sources with confident wind detections (i.e. a blueshift) are highlighted in boldface. \textbf{(a)} Orbital period of the system. \textbf{(b)} Estimated inclination of the system. Values in brackets indicate either the approximate inclination angle in degrees or observational constraints (e.g. eclipses or dips). A low inclination refers to $\lesssim$30$^\circ$, mid to 30--60$^\circ$, and high to $\gtrsim $60$^\circ$ (see catalogues by \citealt{Corral-Santana2016} and \citealt{Avakyan2023} for details on the fundamental parameters).
\textbf{(c)} Maximum outflow velocity measured across all spectral ranges. Values derived from the blue edge of P-Cygni line profiles or broad emission components are indicated with \texttt{(b-e)}. Confident detections with a measured outflow velocity are indicated by \checkmark . Detections consistent with a disc atmosphere (i.e. velocity consistent with zero) are marked as \checkmark-atm, while those with no solid velocity constraints are labelled \checkmark(no-vel). Tentative detections are indicated by \checkmark?.
\textbf{(d)} Low-ionisation wind detected in optical and NIR transitions of mainly H and \hei\ ($\log\xi \lesssim 1$).  
\textbf{(e)} Low-ionisation winds detected in \heii\ and ultraviolet transitions (mainly \Civ, \Nv) reaching up to $\log\xi \sim 1.5$. 
\textbf{(f)} High-ionisation (X-ray) winds with $1.5 \lesssim \log\xi \lesssim 3$.  
\textbf{(g)} High-ionisation winds detected mainly via \Fexxv\ and \Fexxvi, indicating $\log\xi \gtrsim 3$. 
\label{tab:master}      
\end{table}
}


\title[Accretion Disc Winds]{Accretion disc winds in X-ray binaries}

\author*[1,2]{\fnm{Teo} \sur{Mu\~noz-Darias}}\email{teo.munoz-darias@iac.es}

\affil*[1]{\orgname{Instituto de Astrofísica de Canarias}, \orgaddress{\street{Via Lactea s/n}, \city{La Laguna}, \postcode{E-38205}, \state{Tenerife}, \country{Spain}}}

\affil[2]{\orgdiv{Departamento de Astrofísica}, \orgname{Universidad de La Laguna}, \orgaddress{ \city{La Laguna}, \postcode{E-38206}, \state{Tenerife}, \country{Spain}}}

\author[3]{\fnm{María} \sur{Díaz Trigo}}\equalcont{These authors contributed equally to this work.}
\affil[3]{\orgname{European Southern Observatory (ESO)}, \orgaddress{\street{Karl-Schwarzschild-Str.\ 2}, \city{Garching bei München}, \postcode{85748}, \country{Germany}}}

\author[4]{\fnm{Chris} \sur{Done}}\equalcont{These authors contributed equally to this work.}
\affil[4]{\orgname{Centre for Extragalactic Astronomy, Department of Physics, University of Durham}, \orgaddress{\city{Durham}, \postcode{DH1 3LE}, \country{UK}}}

\author*[5,6,7]{\fnm{Gabriele} \sur{Ponti}}\email{gabriele.ponti@inaf.it}\equalcont{These authors contributed equally to this work.}
\affil[5]{\orgname{INAF--Osservatorio Astronomico di Brera}, \orgaddress{\street{Via Emilio Bianchi 46}, \city{Merate}, \postcode{I-23807}, \country{Italy}}}
\affil[6]{\orgname{Max-Planck-Institut für Extraterrestrische Physik}, \orgaddress{\street{Gießenbachstraße 1}, \city{Garching}, \postcode{85748}, \country{Germany}}}

\affil[7]{\orgname{Como Lake Center for Astrophysics (CLAP), DiSAT, Università degli Studi dell’Insubria}, \orgaddress{\street{via Valleggio 11}, \city{Como}, \postcode{I-22100}, \country{Italy}}}

\author[8]{\fnm{Ryota} \sur{Tomaru}}\equalcont{These authors contributed equally to this work.}
\affil[8]{\orgname{Department of Earth and Space Science, Graduate School of Science, Osaka University}, \orgaddress{\street{1-1 Machikaneyama}, \city{Toyonaka}, \state{Osaka}, \postcode{560-0043}, \country{Japan}}}



\abstract{Despite early theoretical expectations that large-scale, massive outflows would be triggered by accretion onto black holes and neutron stars, their presence was not firmly established until the 2000s. Since then, these accretion disc winds have been recognised as a common, perhaps ubiquitous, feature of accretion discs in X-ray binaries. Over the past two decades, our understanding of these outflows has expanded significantly, with their associated phenomenology now observed across the X-ray, ultraviolet, optical, and near-infrared regimes.

In this review, we provide a comprehensive summary of the observational properties of both low- and high-ionisation winds, treating each separately as well as part of a broader phenomenon, and place these findings in the context of current theoretical modelling. We discuss their close connection with disc atmospheres, their impact on the accretion process, and their role within the broader framework that includes the radio jet and the different accretion flow configurations and states. We also address current challenges and outline some of the anticipated developments, particularly those linked to upcoming observational facilities.}




\maketitle

\section{Introduction}\label{sec:intro}
The association between accretion onto compact objects and the production of outflows was recognised from the early days of accretion theory. In fact, they are already described in the model by \citet{Shakura1973}, where a strong outflow of matter is expected to occur at high accretion rates. At that stage, the local accretion rate approaches the Eddington limit and, consequently, radiation pressure due to electron scattering surpasses the gravitational force, leading to the launch of ejecta. Such high accretion rates and ensuing luminosities (approaching the Eddington Luminosity, hereafter \ledd) are possible if the accretor is a neutron star (NS) or a black hole (BH), and thus the presence of massive outflows was naturally anticipated under these critical conditions. However, as we will see, additional launching mechanisms are required to explain the broad observational evidence available today.

In this review, we will present the current observational and theoretical evidence for the presence of non-collimated outflows -- hereafter referred to as winds -- associated with accretion discs onto stellar-mass BHs and NSs, a topic that we will generally refer to as accretion disc winds in low-mass X-ray binaries (LMXBs). In these stellar systems (see~\citealt{Bahramian2023} for a recent review), a low-mass star, typically less massive than our Sun, transfers material onto a BH or a NS via an accretion disc, making LMXBs excellent laboratories to simultaneously study accretion and its associated outflow processes. Galactic LMXBs can be bright targets evolving over timescales from milliseconds to years, which makes them particularly well suited for detailed observational studies within human timescales. The same is true for some systems with high-mass companions (i.e. high-mass X-ray binaries), which are, in some cases, discussed in this review. In many of them, however, accretion is mainly supplied by stellar winds or by a decretion disc associated with the massive donor star, and the role of a persistent accretion disc is therefore reduced. Additionally, qualitatively similar winds are also observed, to some extent, in accreting white dwarfs and quasars, and will be briefly mentioned in several sections of the text.

Winds and jets are the two generic classes of outflows observed in LMXBs. The presence of jets associated with X-ray binaries was discovered already in the 1970s (e.g.~\citealt{Spencer1979}), as they can be spatially extended and are bright in the radio band, where the rest of the components of an LMXB do not emit significant radiation. Jets are collimated outflows, often reaching relativistic speeds (e.g.~\citealt{Mirabel1999}). They can extract a significant amount of energy from the system, although they are not expected to carry away much mass. In contrast, winds typically have lower velocities (up to a few thousand \kms) and kinetic luminosities, but can carry substantial mass -- sometimes even exceeding that ultimately accreted by the BH or the NS (see \citealt{Fender2016} for a review on the mass and energetic balance of jets and winds). Winds, therefore, can have a strong impact on the accretion process itself (e.g.~\citealt{Neilsen2011, Munoz-Darias2016}). In addition, winds might carry away a large amount of angular momentum (e.g.~\citealt{Tetarenko2018}), which is key to understanding how matter is ultimately accreted, and more generally, how accretion discs operate. Likewise, this angular momentum removal might also impact the orbital evolution of the LMXB itself (e.g.~\citealt{GallegosGarcia2024}).

BH LMXBs are, in general, transient sources that go into outburst with recurrence times of years to decades. Their transient nature results from a combination of their relatively long orbital periods -- a few hours to several days -- and their masses, typically in the range of $\sim 6-12$~M$_\odot$ (e.g.~\citealt{Corral-Santana2016, Tetarenko2016}). As a consequence, they tend to have larger accretion discs than their NS counterparts. This implies that higher accretion rates are required to keep the entire disc fully ionised and, therefore, to remain in the active state according to the disc instability model \citep{Lasota2001}. For this reason, NS LMXBs account for the vast majority of persistent systems (i.e. systems permanently in outburst). During outburst, BH systems are known to show two distinctive accretion states (\citealt{Miyamoto1992,Miyamoto1993}; see \citealt{Done2007, Belloni2010} for reviews). The hard state is displayed at the beginning and end of the outburst. It is characterised by the presence of a relatively cold, likely truncated accretion disc emitting seed photons that are up-Comptonised in a corona of hot electrons, producing hard X-ray emission with a power-law shape (e.g.~\citealt{Gilfanov2010}). In contrast, during the soft state, the X-ray spectrum is dominated by a thermal component, which can, in many cases, be well fitted by a multicolour disc following the Shakura--Sunyaev prescription. These two states are observed following a relatively precise hysteresis pattern, in which a q-shaped track is described in the hardness--intensity diagram (e.g.~\citealt{Homan2001, Belloni2005}). The presence of jets is tightly correlated with these accretion states, with a compact jet being a defining property of hard accretion states, and more extended jet ejections observed during the transitions between hard and soft states (e.g.~\citealt{Fender2004}). Phenomenology is arguably more complex in systems harbouring NS accretors, at least regarding spectral analysis, due to the additional emission from the NS surface (and/or boundary layer; e.g.~\citealt{Lin2007, Armaspadilla2017}). However, when considering the accretion phenomenology as a whole, including jets (e.g.~\citealt{Migliari2006}) and fast variability properties, a similar accretion scheme -- which also includes hysteresis patterns -- can be built (e.g.~\citealt{Munoz-Darias2014}).

Winds entered the accretion-ejection picture in the late 1990s, with the discovery of X-ray absorption features by the \textit{ASCA} mission \citep{Ueda1998,Kotani2000}. While these early detections hinted at the presence of ejecta, it was only with the advent of higher resolution spectroscopy from \textit{Chandra} and \textit{XMM-Newton} that significant line blueshifts were securely measured and attributed to material being expelled from the system (e.g.~\citealt{Miller2006}). In the following years, X-ray signatures of these winds were detected in over twenty systems (see e.g.~\citealt{Ponti2016, DiazTrigo2016,Neilsen2023, Parra2024}). This effort has entered a new phase with the advent of the \textit{X-Ray Imaging and Spectroscopy Mission (XRISM)}, whose microcalorimeter (\textit{Resolve}) delivers unprecedented spectral resolution in the X-ray band (\citealt{Tashiro2025}). 

Observations at lower energies -- initially in the optical, and increasingly in the near-infrared (NIR) and ultraviolet domains -- have opened a new window into the study of BH winds. This progress followed the 2015 outburst of V404~Cyg, during which prominent wind signatures (see Sec.~\ref{sec:signatures}) were detected in a dozen optical lines, including a conspicuous nebular phase \citep{Munoz-Darias2016}. However, wind signatures, although more subtle, were also seen in individual lines during the 1989 outburst of the same source \citep{Casares1991}, and in the BH transient V4641~Sgr \citep{Lindstrom2005}.
\smallskip
\smallskip

In this review, we approach accretion disc winds as a multiwavelength phenomenon, observed from the X-ray to the NIR. Since the spectral range does not necessarily reflect the physical state of the ejecta responsible for a given transition, we classify winds into two categories based on their ionisation parameter ($\xi$), defined as:

\begin{equation}
\label{eq:xi}
\xi = \frac{L}{nR^2}   
\end{equation}

where $L$ is the source luminosity, $n$ is the wind density, and $R$ is the distance of the wind from the illuminating source.

In this context, we consider low-ionisation winds as those with \logxi\ $\lesssim 1.5$ (detected in optical, NIR, and ultraviolet lines), while high-ionisation winds refer to those detected through X-ray lines tracing ejecta with \logxi\ $\gtrsim 1.5$. We note, however, that this division in \logxi\ is only approximate and should be treated with caution (see \citealt{Kallman1982} for a detailed description of ionic species and the ionisation parameter). To better organise the observational evidence within each category, we often chose to divide low-ionisation winds into \textit{cold} and \textit{cool} types, and high-ionisation winds into \textit{warm} and \textit{hot}, reflecting the different degrees of ionisation inferred from the data.

This review is organised as follows. Sec.~\ref{sec:signatures} describes the main observational signatures from which the presence of low- and high-ionisation winds can be inferred, whose main properties are presented and discussed in Sec.~\ref{sec:liwinds} and \ref{sec:hiwinds}, respectively. Theoretical background and modelling efforts are presented in Sec.~\ref{sec:theory}. Finally, accretion disc winds in X-ray binaries are globally discussed in Sec.~\ref{sec:discussion}, while a broad summary and a look at expected developments during the next decade are provided in Sec.~\ref{sec:next}.

\section{Observational signatures}
\label{sec:signatures} 

Accretion disc winds are revealed by a plethora of characteristic spectral features, primarily emission and absorption line profiles or a combination of both. These features are not necessarily weak; however, they coexist with other spectral components that veil and contaminate them. Additionally, their correct identification, as well as the extraction of key encoded information, such as kinematic properties, requires velocity resolutions of a few hundred \kms\ at most, ideally less. This level of precision is achievable at low energies (ultraviolet, optical and NIR; see Sec.~\ref{sec:liwinds}), typically in combination with the largest ground-based telescopes [e.g.~the \textit{Gran Telescopio Canarias (GTC)}, the \textit{Very Large Telescope (VLT)}, and the \textit{Southern African Large Telescope (SALT)}] and space-based facilities such as the \textit{James Webb} and \textit{Hubble} space telescopes. However, achieving this in the X-ray regime has traditionally been challenging, with only \textit{Chandra} and, to some extent, \textit{XMM-Newton} providing relatively precise measurements (see Sec.~\ref{sec:hiwinds}). Fortunately, the recent advent of \textit{XRISM} is enabling, for the first time, X-ray studies at resolutions comparable to those at lower energies.

The observational signatures of disc winds are broadly similar across the different spectral regimes, although, as we will see below, the combination of physical conditions, atomic properties, and observational limitations makes some signatures more frequent or easier to detect in certain spectral domains.

\subsection{P-Cygni line profiles} 
\FigPCyg
P-Cygni (P-Cyg) line profiles are arguably the most reliable indicators of wind-type outflows. These features, named after the supergiant star where they were first observed \citep{Maury1897,Menzel1929,Beals1929}, are common in massive stars and have been extensively studied in various astrophysical contexts. A P-Cyg profile consists of a blueshifted absorption feature overlapping with a broad emission component (see Fig.~\ref{fig:P-Cyg_scheme}). When correctly identified, such profiles are often considered a definitive signature of an outflowing wind.

Taking the simplest case -- photon scattering in a spherically symmetric outflow composed of two-level atoms -- regions outside our line of sight absorb photons from the central source and re-emit them in different directions. Some of these photons reach the observer, producing broad emission lines centred around zero velocity. The blueshifted absorption occurs when approaching gas scatters photons out of the observer’s line of sight. The blue-edge of this absorption is usually taken as the characteristic outflow velocity, sometimes referred to as the wind terminal velocity. 

The balance between absorption and emission determines the final shape of the P-Cyg profile, which in turn reflects the physical properties of the ejecta. In this simplified scenario, altering the line opacity and plasma conditions can produce profiles without absorption components, flat-top emissions, and other types of distorted profiles, such as skewed ones (e.g.~\citealt{Castor1970,Proga2002a}; see also Fig.~\ref{fig:SimProfiles}). As already shown by these early works, it is clear that a physically motivated fit to recover the parameters of the outflow is not possible without additional constraints on the properties of the ejecta and the irradiation field. Moreover, assuming scattering of photons by two-level atoms may be valid for resonance lines (i.e., transitions from a ground to an upper level; resonant scattering), such as those associated with some metallic ions in the ultraviolet (e.g., \Civ\ and \Nv). However, wind-related profiles are frequently observed in recombination lines (i.e. lines produced when a free electron recombines with an ion), making their modelling significantly more complex (see Sec.~\ref{liwinds:lines}).

Things become even more complicated when considering the geometry of the outflow. In the context of LMXBs and other compact binaries, the primary mass source is an accretion disc. Consequently, line-of-sight effects are expected to play a crucial role.  The orbital inclination ($i$), defined as the angle between the orbital plane and the observer, strongly impacts the wind's observational signatures. For example, equatorial outflows (e.g. bi-conical winds from the disc)  should produce stronger blueshifted absorptions in systems observed close to edge-on ($i \approx 90^\circ$), while the absorption weakens or even disappears at lower inclinations ($i \approx 0^\circ$).

In addition, interpreting wind-related profiles is further complicated by the presence of emission from other spectral components, and in particular the accretion flow itself. In X-rays, the strong disc continuum veils any emission feature, which are often revealed during dips and eclipses (Sec.~\ref{sec:hiwinds}), when part of the disc's emission is blocked by the bulge\footnote{The region of the disc where the incoming stream of material impacts, increasing its vertical size.} or the companion star, respectively. In the optical, strong emission lines originating in the disc atmosphere are commonly observed (see also Sec. \ref{sec:discussion:atmospheres}). These take the shape of double-peaked lines and are relatively well understood \citep{Marsh1988,Casares2015}. However, when combined with P-Cyg profiles, this can result in complex line shapes. A similar situation occurs in the NIR, with an additional challenge: the presence of jet continuum veiling during the hard state, which dilutes --and in some cases completely overshadows -- any spectral feature arising from the disc wind or atmosphere (e.g.~\citealt{Ambrifi2025} and references therein). Nevertheless, conspicuous examples of P-Cyg line profiles have been observed in X-rays (e.g.~\citealt{Brandt2000}), ultraviolet (e.g.~\citealt{CastroSegura2022}), optical (e.g.~\citealt{Munoz-Darias2016}), and the NIR (e.g.~\citealt{SanchezSierras2023b}; see Fig.~\ref{fig:NIR_lines}).

\FigSimProfiles

\subsection{Additional wind signatures}
As described above, the actual shape of a P-Cyg line profile depends on many factors, including (but not limited to) the geometry, density, and velocity of the expanding outflow (e.g.~\citealt{Castor1979}), as well as possible additional contributions from the accretion flow. Thus, extreme variations of the P-Cyg profile -- particularly when one of the components is absent or overwhelmingly dominant -- naturally result in characteristic wind signatures. These wind profiles, which we label as ‘additional’ to P-Cyg profiles, can be grouped into the categories described below, with some representative examples shown in Fig.~\ref{fig:SimProfiles}. It is worth mentioning that, already in the early days of spectroscopic studies of P-Cygni and other massive stars, up to eight different classes of wind-related profiles were identified. These ranged from the ‘standard’ P-Cyg line profile to those completely dominated by either an emission or an absorption component \citep{Beals1953}. Therefore, it is not surprising that similar, if not more complex, phenomenology is also present in LMXBs.

\begin{itemize}[leftmargin=0cm, itemsep=1pt] \smallskip

\item \textbf{Absorption profiles:} The presence of a blueshifted absorption feature can be taken as robust evidence for the existence of a wind. This is especially true if two or more lines are detected at a consistent blueshift, allowing for a confident identification of the spectral transitions involved. This is the signature observed in some of the most reliable detections in the X-ray regime (e.g., \citealt{Miller2006}).  

In particular, the presence of \Fexxv\ and \Fexxvi\ blueshifted absorptions within the Fe complex at $\sim$6.7--7 keV has been the most prolific method for detecting high-ionisation winds (see Sec.~\ref{sec:hiwinds}). These detections are associated with systems seen at high inclination (\citealt{DiazTrigo2006, Ponti2012}) and generally lack the (redshifted) emission component. This may be related to an equatorial wind geometry, which reduces the scattering of photons moving away from the orbital plane. However, the mere fact that the material originates from a disc geometry can explain stronger blueshifts at edge-on lines of sight (e.g., \citealt{Higginbottom2019}).

Blueshifted absorptions with no associated emission component (i.e. the equivalent of a P-Cyg line profile lacking its emission component; see Fig.~\ref{fig:P-Cyg_scheme}) are rare in the optical and NIR domains. This can be explained by several factors. On the one hand, emission lines from the disc atmosphere are strong, and when observed together with a blueshifted absorption, they can mimic a P-Cyg profile. Indeed, there are cases of ‘P-Cyg line profiles’ with double-peaked emission, as well as evidence for multiple emission components (\citealt{Munoz-Darias2016,Munoz-Darias2020}). On the other hand, optical and NIR spectroscopy can be highly sensitive, making it possible to detect emission just a few percent above the continuum level. Additionally, most of the relevant lines in these domains are recombination transitions between excited levels of hydrogen, which are prone to strong emission components (see Sec.~\ref{liwinds:lines}).

\FigNIRwinds

\item \textbf{Emission profiles:} In massive stars, broad emission components are often detected in the ultraviolet and optical domains simultaneously with P-Cyg profiles in other spectral transitions (e.g., \citealt{Beals1953,Prinja1994}). These are also common during certain phases of nova evolution (e.g., \citealt{iijima2003}). As discussed above, these components can be accounted for by the same principles as P-Cyg profiles (e.g., \citealt{Castor1970}; see Fig.~\ref{fig:SimProfiles}).  

Clear examples of this phenomenon are also found in LMXBs, particularly in the optical and NIR regimes (see Fig.~\ref{fig:NIR_lines} and Sec.~\ref{sec:liwinds}). These lines have full widths at zero intensity of several thousand \kms\ and are approximately at their rest wavelength, as expected for a P-Cyg profile lacking the blueshifted absorption. Likewise, P-Cyg profiles and broad emission components are seen to alternate in the same spectral transitions over time-scales of just a few days. In these cases, the broad emission profiles meet the blue continuum at roughly the same velocities as those indicated by the blue-edge of the P-Cyg profile (e.g., \citealt{MataSanchez2022}).  

An interesting feature often observed in these optical and NIR broad components is the presence of an extended blue wing (or a lack of flux in the red wing). Interestingly, this feature is commonly found in the ionised (e.g., \Civ) emission of luminous quasars and is generally thought to be associated with the ubiquitous presence of winds (e.g., \citealt{Richards2002,Richards2011,Matthews2023}).

\item \textbf{Nebular phase:} An extreme example of broad emission lines occurs during the so-called nebular phase. This is characterised by intense emission components with equivalent widths of several hundred to a few thousand~\AA\ and high Balmer decrement (\ha-to-\hb) values, typically $\gtrsim 5$.

This phase was first identified during the 2015 outburst of V404 Cyg (\citealt{Munoz-Darias2016}; see Fig.~\ref{fig:nebular_phase}) and was associated with a cooling, expanding nebula created by a previously launched wind. It was observed after the sudden drop in luminosity that marked the end of the outburst (see also \citealt{Casares2019, Rahoui2017}), although scaled-down versions of this phenomenon may be present in earlier stages (see \citealt{MataSanchez2018} for details). A similar phase, though much less extreme, was observed in V4641 Sgr (\citealt{Chaty2003, Munoz-Darias2018}) and tentatively in the NS transient Aql X-1.  (\citealt{PanizoEspinar2021}; see Sec. \ref{sec:liwinds:tentative}).
\FigNebularPhase

\item \textbf{Flat-top emission profiles}  
Flat-top, or box-like, emission profiles are commonly observed in outflowing stellar objects, particularly in Wolf-Rayet stars and novae (e.g., \citealt{Payne1930}). These broad profiles exhibit nearly constant intensity around their central wavelengths and can be reproduced by considering an optically thin spherical shell of plasma moving at a constant velocity (or with a small spread that avoids zero velocities; \citealt{Beals1931}). As such, and in the absence of contamination, they should be considered a reliable signature of the presence of an outflow.  
These features have also been found in the optical and NIR spectra of LMXBs, particularly in sources with intermediate inclination and sometimes accompanied by blueshifted absorption features (e.g., \citealt{Cuneo2020, PanizoEspinar2022, Ambrifi2025}).  

\item \textbf{Distorted and single-peaked emission profiles:}  
The wind and disc line profiles discussed above can be considered \textit{clean profiles}. However, a greater variety of wind-related profiles can be obtained simply by combining them, even empirically (e.g., \citealt{MataSanchez2023}; see Fig.~\ref{fig:SimProfiles}). This includes red-skewed profiles and fully redshifted emission lines, which can be attributed to absorption of the blue part of the emission feature, though not enough to dip below the continuum level. Some clear cases have been detected in LMXBs in the optical regime, appearing simultaneously with P-Cyg profiles (e.g., \citealt{Munoz-Darias2020}), the latter occurring in transitions more prone to blueshifted absorption due to atomic properties (see Fig.~\ref{fig:NIR_lines} and Sec.~\ref{liwinds:lines}). 

Finally, single-peaked profiles that are not necessarily particularly broad (i.e. unlike those described above) may also be associated with outflows. In this case, a wind could modify an intrinsically double-peaked profile through radiative-transfer effects. Alternatively, these single-peaked profiles may originate high up in the disc atmosphere, where rotational velocities are significantly lower due to angular momentum conservation (see e.g.~\citealt{Murray1996, Matthews2015}).

\end{itemize}


\section{Low ionisation winds}\label{sec:liwinds}
Over the last decade, and in particular since the 2015 outbursts of the BH transient V404 Cygni, high-quality spectroscopy -- mainly at optical wavelengths but increasingly in the NIR and ultraviolet -- has revealed signatures of accretion disc winds in LMXBs (see Table \ref{tab:liwinds} and Sec.~\ref{sec:liwinds:obs}). However, earlier detections exist (see Sec.~\ref{sec:intro}), although in some cases they are either restricted to a single transition or the signatures were inconclusive. 

Low ionisation winds are primarily detected in transitions of hydrogen and helium, which trace cold gas. Including the detections in \heii, as well as the resonance lines in the far-ultraviolet (e.g.~\Civ\ and \Nv) we can ascribe these low-ionisation winds to \textit{cold} and \textit{cool} gas with \logxi\ $\lesssim 1.5$ (e.g.~\citealt{Kallman1982}).

\subsection{Main spectral lines for low-ionised winds}
\label{liwinds:lines}

Most low-ionisation wind signatures are found in a common set of spectral transitions. In the optical, the range with the highest number of detections, the strongest signatures are observed in \hei\ at 5876 \AA\ (hereafter \hei~$\lambda5876$) and \ha. However, additional features, particularly in sources with strong P-Cyg profiles, are also seen in other \hei\ lines such as \hei~$\lambda6678$ and \hei~$\lambda7065$, as well as in other Balmer transitions. Weak \Feii\ P-Cyg profiles have also been reported in V404~Cyg \citep{Munoz-Darias2016} and V4641~Sgr \citep{Lindstrom2005}.

The NIR has been less explored, with significantly fewer observational campaigns. In addition, some sources exhibit very weak NIR lines during the hard state, particularly at the reddest wavelengths. This is likely due to contamination from a strong jet continuum, which can dilute emission features originating in the disc or disc wind (e.g.~\citealt{ Rahoui2012, Ambrifi2025}). Focusing on P-Cyg line profiles in the NIR, the best case study is, by far, that of GRS~1915+105 (\citealt{SanchezSierras2023b}).

Still within the low-ionised regime, and focussing on wind signatures in cool material, a handful of detections probing slightly higher ionisation conditions have been reported in optical and NIR transitions of \heii\ (see Table \ref{tab:liwinds}). In addition to these optical and NIR diagnostics, wind signatures have also been detected in the ultraviolet via classical resonance lines such as \Civ\ and \Nv\ (e.g. \citealt{CastroSegura2022}).

\subsubsection{Lines tracing cold winds}
\label{sec:liwinds:lines:cold}
Balmer, Paschen and Brackett hydrogen lines, together with optical and NIR \hei\ transitions, trace relatively cold gas and often coexist in LMXBs and in other wind-relevant objects such as massive stars. However, not all lines are equally sensitive to outflows. The NIR transitions \hei~$\lambda10830$ ($2^3\mathrm{S}$--$2^3\mathrm{P}$) and \hei~$\lambda20581$ ($2^1\mathrm{S}$--$2^1\mathrm{P}$) originate from the two $2s$ levels of the triplet and singlet systems of \hei, respectively. These long-lived lower levels can accumulate significant populations, analogous to ground-state levels in ultraviolet resonance lines, making the corresponding transitions particularly responsive to outflows. Of the two, the $2^3\mathrm{S}$ level is much more long-lived and is effectively metastable.

The influence of these metastable levels extends to other transitions connected to them. Combined with the typically higher population of the triplet system compared to the singlet, this helps explain why \hei~$\lambda5876$ ($2^3\mathrm{P}$--$3^3\mathrm{D}$) often displays the strongest optical blueshifted absorption features. In contrast, the analogous singlet transition \hei~$\lambda6678$ ($2^1\mathrm{P}$--$3^1\mathrm{D}$) is generally weaker due to the lower population and the shorter effective lifetime associated with the $2^1\mathrm{S}$ level.

Hydrogen lines (particularly the Paschen, transitions to $n=3$, and Brackett, transitions to $n=4$, series), while commonly observed as strong emission features, are typically less sensitive to wind-driven absorption than their \hei\ counterparts. This is primarily because they originate from higher-$n$ atomic levels, which are less populated and therefore yield lower line opacity. By contrast, the Balmer series remains an excellent wind tracer because the $n=2$ level is often significantly populated. Under suitable conditions this level can effectively act as a ground state, allowing \ha\ ($n=3\rightarrow2$) to mimic, to some extent, the behaviour of a resonance transition, as observed in winds from massive stars (e.g. \citealt{Puls1998,Petrov2014}). In this regard, \citet{SanchezSierras2020} showed that the emission profiles of \ha\ and \hei~$\lambda10830$ are often similar in BH transients, as both transitions originate from highly populated lower levels and trace overlapping regions of either the disc wind or the disc atmosphere.

\subsubsection{Lines tracing cool winds}

Wind signatures in cool material, traced by the \heii~$\lambda4686$ line and ultraviolet resonance lines (e.g.~\Civ\ and \Nv; \citealt{CastroSegura2022, Fijma2023}), have been detected in a few objects (see Table \ref{tab:liwinds}). Ultraviolet resonance lines are primary wind tracers in massive stars (e.g.~\citealt{Kudritzki2000}) and cataclysmic variables (e.g.~\citealt{Drew1988, Prinja2000}). These lines are formed by resonant scattering -- that is, they involve a ground-state lower level -- and are particularly sensitive to outflows via blueshifted absorption, often producing conspicuous P-Cyg profiles.

In contrast, the \heii~$\lambda4686$ line is formed by recombination from $n=4$ to $n=3$ and tends to have lower optical depth, making it less prone to absorption. It primarily traces dense, ionised gas and typically appears in pure emission. However, although rare, P-Cyg profiles in \heii~$\lambda4686$ have also been reported in massive stars (e.g.~\citealt{Herrero2012}). In LMXBs, the best example is Swift~J1357.2$-$0933, where the modelling by \citet{Charles2019}  supports the presence of a relatively dense outflow, possibly favoured by the highly equatorial line-of-sight (see Sec.~\ref{sec:liwinds:unique}).

\Tableli

\subsection{Observations of low-ionised winds}
\label{sec:liwinds:obs}

Table \ref{tab:liwinds} presents the list of systems in which observational signatures consistent with the presence of winds have been detected in transitions associated with low ionisation gas. We have divided them into two groups, confident and tentative, according to the typology and confidence we assign to these detections. 

In general, we consider confident detections to be those associated with the presence of wind signatures in multiple transitions, especially those that include P-Cyg line profiles or blueshifted absorptions. In total, twelve sources meet these criteria and can be subdivided according to their observed phenomenology.

\subsubsection{Strong P-Cyg profile sources}

Four sources have shown blueshifted absorptions deeper than 10 per cent of the continuum level. These are found across multiple transitions and observing epochs. The sample includes three BH systems (V404~Cyg, V4641~Sgr, and GRS~1915$+$105), with variable associated velocities reaching up to 3000 ~\kms, and the NS transient Swift~J1858.6$-$0814, which exhibited blue-edge velocities exceeding 2000~\kms. Thus, these sources have not only shown the strongest P-Cyg profiles but also the highest velocities (see below for the case of Swift~J1357.2$-$0933). These systems have long orbital periods ($\gtrsim$1 day) and are seen at relatively high orbital inclinations ($\gtrsim$60$^\circ$). They also display complex outburst evolutions characterised by strong variability as well as variable, intrinsic X-ray absorption (e.g.~\citealt{Motta2017b, Motta2021, Miller2020}).

All sources except GRS~1915$+$105 show these P-Cyg profiles in the optical (hydrogen and \hei\ lines). GRS~1915$+$105 is well known for its X-ray winds (see Sec.~\ref{sec:hiwinds}), but high extinction prevents optical observations. However, the source is accessible in the NIR, where wind signatures have been seen simultaneously in multiple lines during the radio-loud, X-ray-obscured stages displayed since 2018 (Fig.~\ref{fig:NIR_lines}; \citealt{Motta2021, SanchezSierras2023b}). The deepest blueshifted absorptions appear in \hei$~\lambda10830$ and \hei~$\lambda20581$ (i.e. the strongest NIR \hei\ transitions) reaching down to 40 per cent of the continuum.

Finally, in all cases winds are observed simultaneously (or contemporaneously) with radio jets (e.g.~\citealt{Munoz-Darias2016, Munoz-Darias2018, SanchezSierras2023b}; see Sec.~\ref{sec:discussion:states}). Additional wind features (see Sec.~\ref{sec:signatures}), often observed simultaneously with P-Cyg profiles but in different transitions, have also been recurrently detected in these sources.

\subsubsection{Sources with multiple detections across the accretion states}

The BH transients MAXI~J1820+070, GRS~1716$-$249, Swift~J1727.8$-$162 and MAXI~J1803$-$298 have exhibited P-Cyg profiles along with broad emission profiles and other wind-related features in several optical transitions and across multiple observing epochs. The blueshifted absorptions have depths lower than 10 per cent the continuum level (typically between 2 and 5 per cent) and show (blue-edge) velocities between $\sim$ 1000 and 2000 ~\kms. With the exception of GRS~1716$-$249, these systems displayed outburst with both hard and soft states allowing to search and study winds across the different accretion regimes (\citealt{Munoz-Darias2019, SanchezSierras2020}). The overall picture that emerges is that low-ionisation winds are best detected in optical and NIR transitions during the hard state, while these signatures become much weaker or absent during the soft state. However, wind signatures with similar kinematic properties can still be found in NIR transitions during the latter (see Sec.~\ref{sec:liwinds:states} below).

\subsubsection{Lower inclination sources}
\label{liwinds:li}
This group includes two BH transients, MAXI J1348$-$630 and GX~339$-$4, which exhibited a roughly similar outburst evolution and observational properties to the previous group. However, they show a different wind phenomenology, and substantial evidence suggests that they are seen at lower inclinations than most, if not all, systems in the previous group (e.g.~\citealt{Munoz-Darias2013b,Carotenuto2021}).

MAXI~J1348$-$630 and GX~339$-$4 have shown compelling evidence for the presence of low-ionisation winds. These signatures primarily include broad and flat-top profiles in multiple transitions across different epochs, as well as red-skewed line profiles. Additionally, MAXI~J1348$-$630 exhibited blueshifted absorption simultaneously detected in several optical and NIR emission lines during the hard-to-soft transition; this feature was also observed in the NIR during the soft state. The absorption has a low velocity (centred at $\sim$ 500 ~\kms) and was seen to either dip below the continuum level (i.e. a classical P-Cyg profile) or appear as an absorption trough on the blue wing of emission components, depending on the transition \citep{PanizoEspinar2022}. MAXI~J1348$-$630 also shows broad lines with a prominent blue emission wing, which meets the continuum at a larger velocity, comparable to those seen in the previous group (up to 1700 ~\kms).

GX~339$-$4 has exhibited similar phenomenology to MAXI~J1348$-$630 (e.g. broad, flat-top, and asymmetric profiles; e.g.~\citealt{Ambrifi2025}) in several outbursts, albeit without any blueshifted absorption.

\subsubsection{Very high inclination sources with distinct detections} 
\label{sec:liwinds:unique}
Two sources have shown confident detections associated with unique properties. Interestingly, both have short orbital periods and are observed at very high orbital inclination ($i \gtrsim 80^\circ$).

Swift~J1357.2$-$0933, a BH transient with a 3-hour orbital period, is best known for its optical dips, which recur on minute time-scales (e.g.~\citealt{Corral-santana2013, Paice2018, Panizoespinar2024}). Dip-resolved optical spectroscopy has revealed strong blueshifted absorptions associated with these dips. They indicate blue-edge velocities up to 4000 ~\kms (e.g. \citealt{Jimenez-ibarra2019b}).

UW~CrB (1E~1603.6+2600) is one of the shortest-period LMXBs, with an orbital period of 111 minutes \citep{Morris1990, Armaspadilla2023}. This eclipsing NS system has exhibited transient ultraviolet P-Cyg profiles (\Siiv\ and \Nv) with blue-edge velocities of $\sim$ 1500 ~\kms. While possible hints of similar signatures have been reported at optical wavelengths, wind detections remain unconfirmed outside the ultraviolet domain (see \citealt{Fijma2023, Kennedy2025, Fijma2025}).

Interestingly, both sources exhibit wind signatures in lines tracing warmer material (i.e. cool winds). Swift~J1357.2$-$0933, in particular, has provided the clearest examples of \heii\ P-Cyg profiles during dips. The modelling by \citet{Charles2019} suggests that these detections favour hotter and denser ejecta than that typically associated with cold winds. Given the high inclinations of these sources, it is tempting to associate these detections with the high densities expected in equatorial winds, particularly when the line of sight is nearly parallel to the orbital plane. In this context, the high velocities observed in Swift~J1357.2$-$0933 could be better understood as part of an equatorial wind component. 

Finally, it is worth noting that in addition to UWCrB, the only other LMXB with ultraviolet wind detections, Swift~J1858.6$-$0814 (\citealt{CastroSegura2022}), is also an eclipsing NS LMXB. However, its orbital period is $\sim$21h (\citealt{Buisson2021}), much longer than that of UWCrB.

\subsubsection{Tentative detections}
\label{sec:liwinds:tentative}
At least five additional sources have shown signatures that we qualitatively classify as tentative. We note that in the case of low-ionisation winds, detections labelled this way do not generally mean that the observed features are non-significant or caused by artifacts, but rather that there are other, plausible interpretations.  
  
\begin{itemize}[leftmargin=0cm, itemsep=1pt] \smallskip
\item \textbf{GX~13+1 and Sco~X-1:} Strong Br$\gamma$ P-Cyg profiles were reported in these two luminous NS X-ray binaries \citep{Bandyopadhyay1999}. Both profiles show similar absorption depth ($\gtrsim$ 10 per cent below the continuum level) and blue-edge velocity ($\sim 2500$~\kms). However, they are not seen in other transitions that are typically more prone to P-Cyg profiles (e.g.~\hei; see Sec.~\ref{liwinds:lines}), nor in earlier spectroscopy of either object \citep{Bandyopadhyay1997}.

There are several reasons to treat these detections with caution. First, the P-Cyg profiles are only seen in Br$\gamma$, a line affected by telluric absorption on both the blue and red wings. At low spectral resolution, this is difficult to correct and may result in spurious absorption components. In fact, high-resolution studies of systems with strong NIR P-Cyg profiles have never found them in this transition, which instead often shows very broad emission profiles (see Fig.~\ref{fig:NIR_lines}; \citealt{SanchezSierras2023b}). It is also striking how similar the observed P-Cyg profiles are in both sources, especially given the significant difference in inclination. In particular, most modern studies of Sco~X-1 have failed to detect P-Cyg profiles in either \brg\ \citep{matasanchez2015} or other NIR transitions (Mu\~noz-Darias et al., in prep). Thus, wind detections in these objects remain tentative.

\item \textbf{GRO~J1655$-$40 and 4U~1543$-$47:} Both BH transients have exhibited peculiar emission profiles that might be linked to wind activity. However, their classification remains ambiguous. In the case of 4U~1543$-$47, a low-inclination source, sensitive NIR spectroscopy during its ultraluminous state is reminiscent of that seen in other low-inclination systems (Sec.~\ref{liwinds:li}), although the signatures are significantly more subtle \citep{SanchezSierras2023}. GRO~J1655$-$40 shows very complex optical line profiles \citep{Soria2000}. Given the high inclination of the system and the strong signatures of X-ray winds it has displayed (e.g.~\citealt{Miller2006}), these features may be linked to low-ionisation outflows. This system should be considered a prime candidate for low-ionisation wind searches in future outbursts.

\item \textbf{A nebular phase in Aql~X-1:} Detailed optical spectroscopy of the prototypical NS X-ray transient Aql X-1 revealed a highly ionised spectrum from the early phases of its 2016 outburst, with very weak hydrogen (Balmer) and \hei\ lines \citep{PanizoEspinar2021}. However, enhanced Balmer emission, especially in \ha, was observed at the end of this luminous event. The line profile, as well as other observables such as the Balmer decrement (i.e. the \ha-to-\hb\ ratio), are reminiscent of a nebular phase. This is, however, characterised by relatively low velocities (the lines meet the continuum at $\sim$800~\kms), which does not rule out an alternative scenario in which the nebular-like line profiles are formed in the disc atmosphere.
\end{itemize}

\subsection{Wind visibility across the states and ionisation effects}
\label{sec:liwinds:states}

As described above, the picture that emerges from observational studies of low-ionisation winds is that they are preferentially detected during the hard state (i.e. when the radio jet is active). These detections have been made primarily in the optical regime, most notably in the H$\alpha$ and \hei~$\lambda5876$ lines. In contrast, such optical wind features are typically absent during the soft state, with only occasional weak detections reported during soft-intermediate states in some sources. Observations in the NIR remain scarce, but conspicuous signatures have been found during the hard state and the hard-to-soft transition, similarly to what is observed in optical lines. However, growing evidence suggests that, in some cases (see below), these features persist—at least to some extent—during the soft state. (e.g.~\citealt{Munoz-Darias2019, SanchezSierras2020, PanizoEspinar2022, Ambrifi2025}). This observational picture contrasts markedly with that of high-ionisation winds (Sec.~\ref{sec:hiwinds}), as discussed further in Sec.~\ref{sec:discussion}.

There is solid evidence indicating that ionisation effects significantly impact the visibility of the wind, which might help explaining the above picture. In V404~Cyg, the system with arguably the richest low-ionisation wind dataset, wind signatures are only observed at low \heii/\hb\ ratios\footnote{This ratio, often referred to as an ionisation ratio, is a useful diagnostic because \heii\ traces higher excitation conditions than the Balmer lines. In addition, the two transitions are closely spaced in wavelength, making the ratio largely insensitive to interstellar reddening.} (\citealt{Munoz-Darias2016}). Likewise, high-cadence spectroscopy reveals how the blueshifted \hei\ components disappear as \heii\ emission becomes stronger during flare events (see Fig.~\ref{fig:trail_hb}). Interestingly, this complex variability in wind-affected line profiles can, in turn, be used for deep searches of wind signatures, as it imprints characteristic shapes in the optical variability spectrum \citep{Vincentelli2025}. Similar results have been obtained in AGN using the same technique \citep{Igo2020}.

\FigTrail

The optical spectra of BH transients during the soft state are characterised by strong high-excitation lines (e.g.~\heii, \ciii, \niii), consistent with the absence of wind signatures due to the ionisation effects described above (e.g.~\citealt{Munoz-Darias2019}). In contrast, NIR observations during this state have revealed cold wind signatures, particularly in the \pab\ line, but not in \hei\ transitions (e.g.~\hei~$\lambda10830$ and $\lambda20581$), which typically follow the behaviour of H$\alpha$ and the optical \hei\ lines. Although the available evidence is limited (see e.g.~\citealt{PanizoEspinar2022}), the presence of \pab\ absorption without a corresponding feature in \hei\ or H$\alpha$ allows us to speculate on the ionisation structure of the outflow.

As discussed in Sec.~\ref{sec:liwinds:lines:cold}, the transition \hei~$\lambda10830$ is a powerful tracer of stellar and disc winds due to its large oscillator strength and the long-lived metastable $2^3\mathrm{S}$ lower level (see e.g.~\citealt{Erkal2022} for applications in young stellar objects). However, its diagnostic power depends critically on the population of that level, which can be reduced, for instance, under intense irradiation. In such environments, helium is often significantly overionised (i.e.~\heii\ dominates), and \hei~$\lambda10830$ absorption may be absent. Although hydrogen is also typically ionised, its high abundance and fast recombination rate can sustain small reservoirs of neutral gas. In these regions, recombination cascades can populate the $n=3$ level, enabling \pab\ absorption ($n=3\rightarrow$5). This absorption component might be absent in \ha\ if the significantly ionised gas makes the required $n=2$ population hard to sustain. In addition, the high \ha\ optical depth results in strong emission that fills in any absorption. Thus, under these conditions, \pab, typically less optically thick, can retain a detectable blueshifted absorption component and act as a more sensitive wind tracer.

Br$\gamma$ ($n=7\rightarrow4$), another strong NIR line, shares a similar physical origin with Pa$\beta$, but it is almost exclusively observed in emission and only rarely shows wind-related absorption features. This is likely due to its intrinsically lower opacity (transition involving $n=4$ instead of $n=3$), which makes it typically less sensitive to absorption. It is worth noting that in particularly strong winds, such as those observed in GRS~1915+105, absorption components can appear superimposed on both the \brg\ and \pab\ emission profiles. However, in these cases, strong P-Cyg profiles are detected in \hei$\lambda10830$ and \hei~$\lambda20581$ (Fig.~\ref{fig:NIR_lines}; \citealt{SanchezSierras2023b}).


\section{High ionisation winds}\label{sec:hiwinds}

We consider high-ionisation winds to be those detected in the X-ray domain, typically characterised by \logxi\ $\gtrsim$ 1.5. They have been primarily identified through absorption lines of \Fexxv\ and \Fexxvi, which trace plasma with ionisation parameters of \logxi\ $\sim$ 3--6. We refer to these as \textit{hot winds}. However, as we will show, there are also detections of X-ray lines tracing lower-ionisation plasma, which we designate as \textit{warm}. While the first detections of such lines were made by \textit{ASCA} \citep{Ueda1998,Kotani2000}, determining significant line blueshifts ($\gtrsim$ 200–2000 \kms) only became possible with the resolution of gratings aboard \textit{Chandra} and \textit{XMM-Newton}. Since 2024, \textit{XRISM} has been pushing these boundaries down to just a few tens of \kms.

Tables~\ref{tab:hiwinds:bh} and \ref{tab:hiwinds:ns} list the BH and NS systems, respectively, with detections of highly ionised absorbers. Following Sect.~\ref{sec:liwinds}, we have divided the detections into two groups: confident and tentative. Confident detections are those in which (i) at least one absorption line is detected at more than $5\sigma$ in high-resolution spectra, (ii) multiple lines are identified with high significance in low-resolution data, or (iii) a single line is consistently detected in observations with different instruments.

We note that, even when the lines are confidently detected, their blueshift may not be reliable. This is often the case for observations performed at low resolution; when the \Fexxv\ and \Fexxvi\ transitions appear blended due to limited spectral quality; when spectra are modelled with a reflection component in addition to the absorber (introducing model degeneracy); or when two absorbing components are modelled but their lines are not spectrally resolved.

Given the resolution of the \textit{Chandra} and \textit{XMM-Newton} gratings, we conservatively quote only those shifts $\gtrsim$200 \kms, while those down to a few tens of \kms\ fall within the realm of \textit{XRISM}. Overall, the majority of the BH systems and one third of the NSs show significant blueshifts (Table~\ref{tab:hiwinds:bh} and  Table~\ref{tab:hiwinds:ns}, respectively). In many of these sources, lines have been detected on multiple occasions, and for BHs in particular, changes in velocity between observations have been reported.

The majority of sources show the lines in absorption. P-Cyg profiles have rarely been detected, with all cases appearing at high luminosities ($\gtrsim$0.5~\ledd; \citealt{Brandt2000, King2015}). On the other hand, emission lines\footnote{We list significant detections of narrow lines, up to widths of $\sim$1000–2000~\kms, thus excluding relativistically broadened lines.} are often observed in \textit{Accretion Disc Corona} sources. This is a phenomenological class of LMXBs, seen at very high inclination ($\gtrsim$80$^{\circ}$), for which scattered emission from the extended wind or atmosphere becomes apparent as the direct emission from the central source is blocked by the disc rim. However, a more general condition for seeing emission lines is likely to obscure the central engine (due to inclination or by other means), so that the emission lines can stand out above the continuum. As we discuss below, emission lines are more often seen in optical and NIR spectra (see Sec.~\ref{sec:discussion:atmospheres}).

Finally, we note that in what follows we do not discuss (unless explicitly indicated) absorption from plasmas observed during X-ray dips, particularly those associated with the impact of the accretion stream onto the disc and/or the presence of a thickened bulge at the disc rim (see, e.g., \citealt{White1982b}).

\subsection{Main spectral lines for high ionisation winds}
\label{sec:hiwinds:spectraltransitions}

The main signatures of a high-ionisation plasma are \Fexxv\ and \Fexxvi. Detection of \Fexxvi\ typically implies \logxi\ $\gtrsim 3$, and we will associate these signatures to hot winds. However, the Spectral Energy Distribution (SED) strongly affects the absolute value of $\xi$, and differences of up to one order of magnitude can arise for the same plasma when illuminated by different SEDs. In addition, the SED also influences whether plasma showing \Fexxv\ and \Fexxvi\ transitions may additionally exhibit lines down to \Oviii\ or \Nex. For instance, the more than 60 lines detected in the spectrum of MXB~1659–298, spanning from \Nex\ to \Fexxvi, were satisfactorily modelled with a single plasma component of \logxi $\sim 3.8$ and a column density of $10^{23.5}$cm$^{-2}$ \citep{Ponti2019}. In contrast, \citet{Allen2018} required at least two plasma layers to explain the presence of both \Mgxii\ and \Fexxvi\ in GX~13+1.

We note that the number of detections associated with \logxi $ \sim$ 2--3 and below may be underestimated as result of observational biases (see Sec. \ref{sec:hiwinds:states}). Fe~K transitions other than \Fexxv--\Fexxvi\ are blended at pre-\textit{XRISM} resolutions \citep[e.g.][]{Kallman2004}, while the strongest transitions for plasmas with \logxi $ \sim$ 2--3 and \logxi\ $\sim$ 1--2 correspond to the Fe~L and Fe~M shells, with prominent features at $\sim$ 1~keV (e.g. \citealt{Gu2019}) and $\sim$0.7--0.8~keV \citep{Behar2001}, respectively. Detecting such transitions in LMXBs is challenging, as these systems are typically located in the Galactic plane, which is often heavily obscured by interstellar gas and dust along the line of sight.  For example, among all LMXBs with confident line detections listed in Tables \ref{tab:hiwinds:bh} and \ref{tab:hiwinds:ns}, about 40 percent have column densities above $10^{22}$cm$^{-2}$, while only one source, EXO~0748$-$676, is observed through a column density below $10^{21}$~cm$^{-2}$. Overall, in Tables~\ref{tab:hiwinds:bh}–\ref{tab:hiwinds:ns}, only a few sources show detections spanning from \Oviii\ and \Nex\ to \Fexxvi.  We emphasise that modelling the full set of lines with a photoionised plasma and a broadband SED is required to robustly assign such features to one or more plasma components (see Sec.~\ref{sec:hiwinds:multiplecomponents}).

\subsection{Black hole systems}
\label{sec:hiwinds:bh}

Table~\ref{tab:hiwinds:bh} lists the 15 BH systems with both confident (8) and tentative (7) detections of high ionisation winds, together with the range of outflow velocities measured for each source. Most detections correspond to hot winds (i.e. \Fexxv\ and \Fexxvi\ lines), while observations of warm winds are restricted to a few cases, for which a selection of transitions associated with lower $\xi$ is given. 

\TablehiBH

\subsubsection{Confident detections}
The eight systems with confident detections indicate outflow velocities ranging from a few hundreds (i.e. pre-\textit{XRISM} resolutions) to $\sim$ 4000 \kms. We do not report the outflow velocity for the absorbing plasma of IGR~J17091-3624. In this source, \Fexxv\ and \Fexxvi\ absorption lines are detected at high significance in \textit{NICER} data, although the limited resolution (i.e. not a grating) does not enable to constrain the velocity \citep{Wang2024}. Likewise, \textit{Chandra} observations of the same source reveal an absorption feature that, if identified with \Fexxv, would correspond to a velocity of $\sim9\times10^3$~\kms. This may be accompanied by another absorption component, tentatively detected, indicating $\sim15\times10^3$~\kms. In addition to the possible uncertainty in line identification, the significance of these detections varies between 2 and 6$\sigma$, depending on the assumed underlying continuum, thus falling into the category of tentative detections \citep{King2012}.

Confident detections have been found in high-inclination systems, supporting an equatorial geometry for these outflows (see Sec.\ref{sec:discussion:obsproperties}). Furthermore, they are much more frequently detected during the softest states, as discussed in Sec.~\ref{sec:hiwinds:states} (see also Sec.~\ref{sec:discussion:states}).

\subsubsection{Tentative detections}
\label{sec:hiwinds:bh:tentative}
Table~\ref{tab:hiwinds:bh} also lists sources with tentative evidence for X-ray winds, a large fraction of which correspond to single-line detections in low-resolution spectra characterised by complex underlying continua. These tentative detections are generally associated with systems observed at low to intermediate inclinations, during or near the hard state.

\begin{itemize}[leftmargin=0cm, itemsep=1pt] \smallskip

\item \textbf{EXO~1846$-$031 and MAXI~J1631$-$479:} In these two sources, a single line was detected in low-resolution spectra after fitting complex broadband continua (see \citealt{Wang2021} and \citealt{Xu2020}, respectively). 

\item \textbf{MAXI~J1348$-$630:} The presence of absorption components around 7–7.3~keV has been reported in this source. The low-resolution spectra only allow the identification of either a single line \citep{Chakraborty2021} or an absorption trough \citep{Wu2023}, once the complex broadband continua is fitted.

\item \textbf{4U~1543$-$47:} \citet{Prabhakar2023} report the detection of a broad, dynamic absorption feature in the 8--10~keV range, interpreted as originating from a highly relativistic disc wind. However, the absorption trough is not resolved into individual lines, and is instead observed in a low-resolution spectrum with an underlying complex continuum.

\item \textbf{MAXI~J1820+070:} \citet{Fabian2020} found that adding a hot, outflowing, photoionised absorber to their model significantly improved the fit to complex broadband, low-resolution spectra taken during the soft state.

\item \textbf{MAXI~J1810$-$222:} \citet{DelSanto2023} report an absorption trough around 1~keV in the low-resolution \textit{NICER} spectra, which is interpreted as evidence for an ultrafast outflow. However, the limited energy resolution prevents the identification of individual absorption lines.

\item \textbf{GX~339--4:} \citet{Miller2015} detect tentative evidence for the presence of broad emission lines in third-order \textit{Chandra} spectra, suggesting the presence of an equatorial high-ionisation wind seen at low inclination. 

\end{itemize}

\subsection{Neutron star systems}
\label{sec:hiwinds:ns}
Table~\ref{tab:hiwinds:ns} lists the 13 NS LMXBs with confident detections of photoionised plasmas, together with nine systems showing tentative detections. Most correspond to hot components (i.e. \Fexxv\ and \Fexxvi\ lines), although a significant fraction also exhibit warm plasmas, for which a selection of transitions with lower \logxi\ values are recorded. As discussed below (Sec. \ref{sec:hiwinds:states}), the non-detection of soft X-ray lines is, in many cases, likely due to a combination of high interstellar absorption and/or relatively low flux at those energies.

\TablehiNS

\subsubsection{Confident detections}
Among the 13 NS systems with confident detections, outflow velocities ranging from 200 to 2000~\kms\ have been measured for six of them. For the remaining seven sources, no significant velocity (i.e. $\gtrsim200$~\kms) was determined, and they are thus consistent with disc atmospheres (see discussion in Sec.~\ref{sec:discussion:atmospheres}).

Confident detections are classified as such based on the observation of at least two lines (typically \Fexxv\ and \Fexxvi), identified through grating observations (with 4U~1323--62 being the exception). These are often detected across multiple epochs, which frequently reveal variability in the physical conditions of the plasma. For example, 1A~1744--361 \citep{Gavriil2012, Ng2024} shows detections of both \Fexxv\ and \Fexxvi, although sometimes only \Fexxvi\ is present. This mild variability is likely driven by small changes in the intrinsic luminosity or the spectral hardness of the irradiating SED. More abrupt variability -- such as the presence versus absence of both \Fexxv\ and \Fexxvi -- is observed in systems like AX~J1745.6--2901 \citep{Ponti2015} and MXB~1659--298 \citep{Ponti2019}, which appear to be correlated with the accretion state (see Sec.~\ref{sec:hiwinds:states}).

One source with detections labelled as confident, 4U~1254--69, shows only \Fexxvi. However, this line has been consistently observed across different epochs and with various instruments, which in some cases also reveal the presence of the \Fexxvi\ K$\beta$ transition, reinforcing the line identification and indicating a plasma with \logxi~$\gtrsim$~4 at rest (\citealt{Boirin2003, DiazTrigo2009}; see also \citealt{Iaria2007} for a tentative \textit{Chandra} detection). This high ionisation parameter likely explains the absence of soft X-ray features despite the relatively low interstellar absorption (2.3$\times$10$^{21}$~cm$^{-2}$) towards this source.

Winds have also been observed in Cir~X-1 and GX~13+1, which display \textit{Z}-source--like behaviour during their high-luminosity phases (see, e.g., \citealt{Hasinger1989, Kuulkers1994, Homan2010} for a description of \textit{Z}-sources, i.e. neutron stars accreting at $\gtrsim$0.5~\ledd, and their associated phenomenology). In GX~13+1, winds are consistently observed, showing moderate variability (\citealt{Ueda2004,Allen2018, XRISM2025GX13}; Fig. \ref{fig:gx13spectrumxrism}). On the other hand, Cir~X-1 exhibits strong, variable P-Cyg line profiles throughout its different \textit{Z}-source phases \citep{Brandt2000, Schulz2002, Tsujimoto2025}. However, different ionisation absorbers were instead observed during fainter epochs, when P-Cyg profiles were not detected \citep{Schulz2008}. In addition to variability in the plasma conditions driven by intrinsic changes in the source, orbital variability has also been detected in Cir~X-1 (\citealt{Tominaga2023}; see also \citealt{Xiang2009} for the case of 4U~1624$-$49).

Finally, we note that complex variability in the wind-related components can also be associated with super-orbital evolution. In particular, thanks to our high-inclination line of sight to the precessing disc of Her~X-1, strong super-orbital variability in the plasma properties is observed, which can be used to map the vertical structure of the disc wind (\citealt{Kosec2020, Kosec2023}; see Sec.~\ref{sec:discussion:structure}).

\subsubsection{Tentative detections}
\label{sec:hiwinds:ns:tentative}
In addition to the sources showing confident detections, nine other NS LMXBs have displayed possible X-ray wind signatures, which we classify as tentative detections:

\begin{itemize}[leftmargin=0cm, itemsep=1pt] \smallskip

   \item \textbf{4U~1820$-$30:} \citet{Costantini12} reported two features at 6.8 and 3.8$\sigma$ significance, identified as \Oiv\ and \Ov\ if blueshifted by $\sim$1200~\kms. However, the authors note that the inferred column densities of \Oiv\ and \Ov\ would imply a \Civ\ ultraviolet line at a level two orders of magnitude above the upper limits obtained from a \textit{Hubble Space Telescope} observation taken 8 years earlier. Moreover, the lines can be equally well fitted with either a photoionised or a collisionally ionised plasma.
    
   \item \textbf{XTE~J1710$-$281:} \citet{Raman2018} reported the detection of two absorption lines at 6.60 and 7.01~keV with \textit{Suzaku}, of which the first one -- identified as a blend of \Fexxv\ to \Fexix -- is highly significant. However, the line is detected in an aggregated spectrum that includes emission occurring during absorption dips, while for the phase-resolved persistent spectrum only a fit with a photoionised absorber (\logxi = 2.6) is reported. No lines could be resolved in \textit{Chandra} observations, likely due to the faintness of the source.

    \item \textbf{Swift~J1858.6$-$0814:} \citet{Buisson2020} report an emission line in a \textit{XMM-Newton} spectrum detected at the 99.7\% confidence level, which are identified as slightly redshifted \Nvii. A second line, consistent with \Ovii\ at rest, is also present at more than 95\% confidence level.
    
    \item \textbf{IGR~J17062--6143}: a handful of low-significance emission and absorption features at $\sim$0.5--1.5~keV were reported for this pulsar after modelling the continuum and reflection components. These features were tentatively identified as signatures of a potential outflow \citep{Degenaar2017}. However, a subsequent analysis modelled the features as collisionally ionised absorption from circumbinary material, effectively ruling out the disc outflow interpretation \citep{vandenEijnden2018}.

    \item \textbf{1RXS~J180408.9--342058}: an absorption feature at $\sim$7.64~keV was reported using \textit{Chandra} data in this low-inclination system. The detection has a significance of 4.8$\sigma$, while lower-significance features are also present in \textit{Swift} data \citep{Degenaar2016}.
    Identifying the line with \Fexxvi\ would imply an outflow velocity of $\sim$26\,000~\kms. In addition to the uncertainty in associating a single line with a specific transition, the unusually high velocity compared with other LMXB outflows and the low inclination of the system were also raised as concerns by the authors.
    
    \item \textbf{IGR~J17591--2342}:  an absorption feature is found at the Si~edge in the combined \textit{Chandra} spectrum of this pulsar \citep{Nowak2019}. When identified with \Sixiii, it indicates an outflow at $\sim$3000~\kms. However, the significance of the line only reaches 5$\sigma$ for one of the possible models and decreases to $\sim$ 3$\sigma$ for alternative models with similar goodness of fit. A modelling with a photoionised absorber is not performed. However, we note the rarity of detecting a transition from Si without corresponding transitions from other key elements. The line is not reported in low-resolution spectra \citep{Manca2023}, where potential \Oviii\ and \Neix\ features are instead mentioned.
    
    \item \textbf{IGR~J17480--2446}: in this pulsar, two absorption features at 6.77 and 6.98~keV were reported  by \citet{Miller2011} and identified with \Fexxv\ and \Fexxvi. However, this identification would imply different blueshifts for the lines of $\sim$3000 and 1000~\kms, respectively. Thus, modelling with photoionised absorption is challenging due to the different blueshifts implied by the two lines, while  including a relativistic disc line to account for the broad superposed Fe emission line results in model degeneracy. These facts, together with the 3$\sigma$ significance of the feature at 6.98~keV and the low inclination of the system justify our classification as tentative. 

    \item \textbf{GRO~J1744--28}: an absorption feature at 6.85~keV was reported after modelling the X-ray spectrum of this pulsar with continuum and disc reflection components \citep{Degenaar2014}. If identified with \Fexxv, this would imply an outflow at $\sim$8000~\kms. However, modelling the broad Fe emission with a relativistic disc line, instead of a broad Gaussian, reduces the significance of the line from $\sim$5$\sigma$ to $\sim$3$\sigma$. Moreover, the absorption feature is not reported by \citet{Younes2015}, who model the emission excess as a blend of neutral Fe, \Fexxv\ and \Fexxvi\ at rest, making the detection of the absorption line highly dependent on the modelling of the emission.

    \item \textbf{GX~340+1}:  an absorption line at 6.94~keV is reported at a 5$\sigma$ significance level and identified with \Fexxv, implying an outflow of $\sim$12000~\kms\ \citep{Miller2016}. However, this feature has only been observed in a relatively short ($\sim$ 6~ks) exposure and not in the numerous, longer observations of the source obtained with the same instrument. While variability is not unexpected, the other two \textit{Z} sources listed in Table~\ref{tab:hiwinds:ns} display quasi-persistent outflows with multiple transitions. On the other hand, a complex XRISM spectrum fitted with emission and absorption components favours a lower outflow velocity of $\sim$2700~\kms, largely based on a feature identified as \Caxx\ \citep{Tanaka2026}.
    
\end{itemize}

\FigGXxrism

\subsection{Multiple high-ionisation components}
\label{sec:hiwinds:multiplecomponents}

In the past two decades, the presence of multiple high-ionisation plasma components has been claimed in a handful of cases \citep[e.g.][]{Iaria2006,Schulz2008,Kallman2009, Allen2018, Tomaru2023b}.

These detections have mostly been reported in studies based on spectra with high spectral resolution and very good statistics, allowing the identification of multiple lines over a broad range of ionisation conditions. However, we note that, in general, the different components have not been resolved in velocity space and, in some cases, their presence strongly relies on the SED used to model the spectra and on whether it can produce a wide range of lines for a single ionisation parameter.

This picture is quickly changing with the arrival of \textit{XRISM}. The few studies published at the time of writing this review indicate the need for more than one plasma component in several cases \citep[e.g.][]{Miller2025, Tsujimoto2025,XRISM2025GX13}. However, how these components or plasma phases are structured within the wind has not yet been established. For example, the two wind components resolved in GX~13+1 point to wind stratification (see Fig.~\ref{fig:gx13spectrumxrism}; \citealt{XRISM2025GX13}). Conversely, the two plasma phases seen during a portion of a 4U~1630–47 observation, taken in a low-luminosity soft state, appear to indicate locally coexisting conditions, potentially related to temporary obscuration, such as that caused by a clumped medium. Wind stratification or multi-phase (i.e. spatially coexisting) components are expected in different wind scenarios (e.g. depending on the launching mechanism; see Sec.~\ref{sec:theory}). Therefore, resolving and characterising the different X-ray components is an important step towards understanding wind structure and launching mechanisms, not only in the context of high-ionisation winds but also in relation to the connection between low and high ionisation wind phases (see Secs.~\ref{sec:theory} and~\ref{sec:discussion}).

\subsection{High ionisation wind detection across accretion states}
\label{sec:hiwinds:states}

In contrast to low-ionisation winds (see Sec.~\ref{sec:liwinds:states}), the presence of X-ray wind signatures is prevalent during soft states and practically absent in hard states. 

\subsubsection{The observational picture}
The current picture of how wind signatures evolve with accretion state is mostly based on monitoring programmes of BH LMXBs across their outbursts. These studies reveal the presence of wind signatures during the softer states of high-inclination sources (see also Sec.~\ref{sec:discussion:obsproperties}), while such features are typically absent during the hard state \citep[e.g.][]{Neilsen2009, Ponti2012, Parra2024}.

Overall, NSs offer a consistent, albeit more complex, view than that inferred from BHs. Most of the NS LMXBs listed in Table~\ref{tab:hiwinds:ns} are persistent sources, and broadband SEDs, as well as timing and multiwavelength data, are not always simultaneously available to allow for meaningful state classification. In fact, our current knowledge relies primarily on three sources that sample hard and soft states in a way comparable to BH outbursts -- the \textit{atoll} sources EXO~0748$-$676, AX~J1745.6$-$2901, and MXB~1659$-$29 \citep[][]{Ponti2014, Ponti2015, Ponti2019} -- as well as the aforementioned luminous \textit{Z} sources (i.e. Cir~X$-$1 and GX~13+1; see also \citealt{Vanderklis2006} and \citealt{Munoz-Darias2014} for NS accretion states and their comparison to those of BHs). 

In order to explain the current observational picture, particularly the prevalence of high ionisation wind detections in the soft state, significant efforts have been made over the past decade. These have focused on determining whether radiative effects—either associated with an overall increase in luminosity or with changes in the SED—and/or a reconfiguration of the magnetic field resulting from accretion state transitions can account for the observations \citep[e.g.][]{Neilsen2009, DiazTrigo2014, Gatuzz2019, Tomaru2019}.

\FigTsujimoto

\subsubsection{Winds, ionisation, and the hard state}
The importance of characterising the broadband SED to understand the evolution of plasma conditions cannot be overstated. For instance, NSs exhibit an additional soft, thermal component from the boundary layer, which is absent in BHs. Among NSs themselves, the luminous \textit{Z}-sources are generally significantly softer than the fainter \textit{atolls} (e.g. \citealt{Vanderklis2006}). Moreover, variability in the physical conditions of the plasma may result from spectral changes not captured by observations limited to a narrow energy band, such as variations in the hard flux above the typical range of X-ray monitoring programmes.

X-ray spectral lines are strongly affected by both the amount (i.e. normalisation) and the shape of the broadband SED illuminating the plasma. Gradual changes in the ionisation state and column density have been observed in several cases \citep[e.g.][]{Miller2006b, Kubota2007, DiazTrigo2014}. In particular, plasmas dominated by \Fexxvi\ can become over-ionised and thus undetectable through line absorption. For instance, the disappearance of the wind when transitioning from a regular soft state to a very luminous one (sometimes referred to as a \textit{very high state}) might be explained by such ionisation effects (\citealt{DiazTrigo2014}; but see \citealt{Tomaru2019} for a counterexample).

Beyond BH transients, \Fexxv\ and \Fexxvi\ absorption has been reported during the soft state of the NS LMXBs EXO~0748$-$676, AX~J1745.6$-$2901, and MXB~1659$-$29, while it is absent during the hard state \citep[][]{Ponti2014, Ponti2015, Ponti2019}. However, the presence of a colder plasma (\logxi $<3$) in the latter state has been reported  for EXO~0748$-$676 \citep{DiazTrigo2006}, and cannot be excluded for AX~J1745.6$-$2901 and MXB~1659$-$29. Interestingly, EXO~0748$-$676 is the only NS in Table~\ref{tab:hiwinds:ns} with an interstellar medium column density below 10$^{21}$~cm$^{-2}$. This might suggest that the typically lower soft X-ray flux associated with the hard state, combined with high interstellar absorption, could conspire to obscure warm plasmas during NS hard states. Their presence, however, would be expected from the thermal instability operating in such states, which leads to a multi-phase medium with denser and cooler clumps embedded in a hot phase (e.g. \citealt{Krolik1981,chakravorty2013,Bianchi2017}).

Things becomes even more complex when considering \textit{Z}-sources, as winds have been consistently observed in the harder states of GX~13+1 and Cir~X-1 \citep{Homan2016, Allen2018}. However, it is important to bear in mind that these states do not correspond to the less luminous hard states of \textit{atoll sources} (such as EXO~0748$-$676), which are more alike to those seen in BH LMXBs. Instead, when considering the amount of Comptonised emission, as well as the timing and radio properties, they are likely more akin to (hard) intermediate states \citep[e.g.][]{Lin2007, Munoz-Darias2014}.

Finally, we note that while the BH picture seems significantly better established than for NSs, new evidence suggests that more sensitive observations with well-sampled SEDs across the different states are needed. For instance, in the case of 4U~1630–47, \textit{XRISM} has recently detected absorption from \Fexxv\ and \Fexxvi\ at rest during a very dim (0.05\ledd) soft state. This was observed shortly before the transition to the hard state during the decay of the outburst \citep{Miller2025}. This contrasts with the absence of lines in a brighter but slightly harder observation during a similar (soft-to-hard) transition previously obtained with \textit{Chandra} \citep{Gatuzz2019}. In another system, IGR~J17091–3624, a low ionisation absorber at rest (\logxi $\sim$2) has been reported prior to the hard-to-soft transition, simultaneously with the presence of jets \citep{Gatuzz2020}. This was interpreted as a potential precursor to the high ionisation winds observed in the soft state. If such low ionisation X-ray plasmas are indeed present during bright hard states, \textit{XRISM} should be able to detect them in the Fe~K band (see Fig. \ref{fig:XRISM_Fe}), overcoming the limitations associated with detecting them in the soft X-ray band. This highlights the exciting potential of combining \textit{XRISM} observations with optical/NIR spectroscopy during hard states, when low-ionisation winds are best traced (see Sec.~\ref{sec:discussion:states}).

\section{Theoretical wind models and numerical 
 simulations}\label{sec:theory}

Launching a wind requires a force which can overcome gravity. There are only three types of forces which can do this, from radiation pressure, hydrodynamic pressure or magnetic pressure. Whatever the driving force, the resulting wind velocity is typically of the order of the escape velocity at the launch radius, so the observed wind velocity can be used as a tracer of this radius via $v/c\sim (R_{\rm g}/R)^{1/2}$.

There is a broader context to understanding winds in X-ray binaries, which is that these may give us a higher signal to noise view of the winds in the supermassive black holes, where they link to AGN feedback. Some of the wind mechanisms described here connect quite naturally across the mass scale. For example, super-Eddington winds driven by radiation pressure on electrons should be similar for objects of different masses. Likewise, self-similar magnetic wind models are expected to scale with mass (e.g.~\citealt{Fukumura2017}).
Conversely, if the winds differ across systems of different masses, this would point to mechanisms that depend on the specific temperature of the radiation spectrum. In AGN, the accretion disc peaks in the ultraviolet, perhaps enhancing line driving winds, whereas in stellar-mass systems the disc peaks at X-ray temperatures, favouring the production of thermal winds.
Thus, understanding winds in LMXBs can also provide valuable clues for interpreting the winds in AGN.

There are several types of winds seen in AGN, but typically the fastest ones carry the most kinetic power. Ultra-fast outflows (UFOs) with $v \sim 0.1$–$0.3\,\mathrm{c}$, detected as X-ray winds in some AGN (e.g. PDS~456; \citealt{Reeves2009}, recently confirmed by \textit{XRISM}; \citealt{xrism_pds456}), are considered the most likely drivers of AGN feedback (e.g. \citealt{King2015a}).
 
While UFOs have not yet been firmly observed in X-ray binaries (although tentative detections exist; see Sec.~\ref{sec:hiwinds}), this does not imply that they are absent. They may simply be more highly ionised than in AGN, making them difficult to detect. Therefore, we use the observed, much slower X-ray winds in binaries to investigate the underlying wind mechanisms, and then examine what these mechanisms predict for faster material originating at smaller radii. We will focus on the high-ionisation (X-ray) winds, but we will also briefly discuss mechanisms for the low-ionisation (optical/NIR) winds at the end. 

\subsection{Radiatively driven wind} 

\subsubsection{Super-Eddington winds}

Bright accretion flows very naturally power winds. This was realised even in the earliest paper outlining the disc structure \citep{Shakura1973}. Radiation carries momentum, and the momentum is transferred to the gas in the photosphere as the radiation produced within the disc diffuses outwards towards the surface. 
For completely ionised material, the cross-section for momentum transfer is the electron cross-section, given by the Thomson cross-section ($\sigma_T$) in the classical limit.

This can launch a wind from $\sim 20R_{\rm g}$ for $\dot{m}=L/L_{\rm Edd}=2.4$ (higher than Eddington as it requires the local disc flux to be at Eddington, not just the total flux integrated over the entire disc, see e.g.~\citealt{kubota2019}).
The launching radius increases approximately linearly with higher $\dot{m}$, so any moderately super-Eddington flow should launch a wind from $20$–$50\,R_{\rm g}$, where the typical velocity is very fast ($0.15$–$0.2\,\mathrm{c}$; \citealt{Shakura1973,Poutanen2007}).

A pioneering radiation hydrodynamic (RHD) simulation of super-Eddington winds was carried out by \citet{Eggum1988}. Several others have followed, with increasing resolution driven by the exponential growth of computational power \citep{Ohsuga2005, Sadowski2014, Jiang2014, Zhang2025}. These studies consistently demonstrate the emergence of outflows with velocities of $\sim$0.1–0.2$c$. Such velocities are comparable to those observed in ultra-fast outflows (UFOs) in AGN (see, e.g., \citealt{xrism_pds456}), making super-Eddington winds a prime candidate for AGN feedback \citep[e.g.][]{King2015a}.

\subsubsection{Ultraviolet line driven winds}
Radiatively driven winds can occur even at luminosities below the Eddington limit if the gas is subject to additional sources of opacity beyond pure electron scattering. This reduces the effective Eddington limit at which radiation pressure can launch a wind to
$L_{\rm Eff, Edd} \sim L_{\rm Edd} /(1+\sigma/\sigma_T)$, where $\sigma$ represents an effective opacity that encapsulates the contribution of these additional processes. Hence, winds can be launched for $(1+{\cal M})\,L/L_{\rm Edd}>1$, where ${\cal M} = \sigma/\sigma_T$ is known as the force multiplier. Crudely, its effect can be understood as an enhancement of the effective cross-section relevant for radiative acceleration.

Bound-bound (line) transitions introduce very large but narrowly peaked cross sections that, in static gas, quickly saturate at the line centre and therefore imparts little net acceleration.
However, in an accelerating flow, the Doppler shift may move the line centres sufficiently to shift them out of their own absorption cores and into unattenuated continuum\footnote{Formally, the condition is given in terms of the 
Sobolev length $ l_{\rm Sob} \;\equiv\;v_{\rm th}/(dv/dl)$, where $v_{th}$ is the thermal speed, and $l$ is the path length along the radiation direction.}. As a result, the line continues to absorb photons and impart momentum, yielding the well-known ultraviolet line-driving mechanism.
This process is effective in gases with low ionisation, where a multitude of bound-bound transitions remain available. It is intrinsically non‐linear since the local radiative acceleration depends on the velocity gradient via the Sobolev optical depth\footnote{The Sobolev optical depth is the line optical depth across $ l_{\rm Sob}$.}. That gradient is, in turn, determined by the radiative force, and the force itself depends on the opacity distribution \citep{Castor1975}, but at its peak it can result
in ${\cal M}\sim 4000$.

The above means that ultraviolet line-driven disc winds can be produced in objects that are inherently less luminous, 
those with $L/L_{\rm Edd} > 10^{-3}$, if the disc peaks in the ultraviolet. This is why the 
first two-dimensional RHD accretion-disc wind simulations were performed for white dwarfs. \citep{Proga1998}. Any disc luminosity is set by the mass-transfer rate from the companion, giving $L \sim GM\dot{M}/(2R_{\rm in})$, but $R_{\rm in}$ is about $2000\times$ larger for a white dwarf than for a NS. Thus, the same mass-transfer rate that yields \ledd\ from a NS produces only $\sim10^{-3}$\ledd\ in a white dwarf. Hence, the observed winds cannot be powered by radiation pressure alone. However, the white-dwarf disc luminosity peaks in the ultraviolet, with an additional, hotter (ultraviolet/extreme-ultraviolet) boundary layer where the disc impacts the stellar surface. These first RHD simulations showed that ultraviolet line-driven winds could indeed be produced from disc-accreting white dwarfs.

However, the above simulations necessarily made some approximations in order to treat the complex non-linear dependencies of the matter and radiation coupling. In particular, the wind itself filters the radiation field which illuminates any point in the flow. But this radiation field sets the ionisation of the material, which determines how many ultraviolet line transitions can contribute to the force multiplier, which sets the dynamics of the wind, which sets its density, affecting the radiation field. These first RHD simulations of accreting white dwarfs assumed an isothermal outflow with fixed ionisation state to make the calculations tractable. More recent and complete calculations which couple the RHD with Monte Carlo radiation transfer find much weaker winds from accreting white dwarfs (Fig.~\ref{fig:sim_UV}). The issue appears quite fundamental: discs which are sufficiently ultraviolet luminous to produce a line-driven wind are also hot enough to overionise the wind, potentially suppressing it entirely \citep{Higginbottom2024}.

\FigUVwindSimu

AGN similarly have discs which peak in the ultraviolet, but are intrinsically much more luminous, with $L/L_{\rm Edd}\sim 0.1-1$. Thus these are systems where the requirement is only $\cal{M}$~>~$10-1$. However, AGN ubiquitously have an X-ray corona in addition to the ultraviolet emitting disc, and X-rays are very effective at overionising the wind material. Nontheless, the first simulations showed that the ultraviolet-luminous disc in bright AGN can launch a funnel-shaped wind due to self-shielding \citep{Proga2000}. Ultraviolet line driving lifts material off the disc photosphere, where it is illuminated by X-ray emission from the central source. This overionises it almost immediately, causing the material to fall back down. However, this produces a slightly vertically extended region that casts a shadow, allowing the ultraviolet line-driven wind from further out to be accelerated until it rises above this shadow, reaching higher speeds before it becomes overionised. This, in turn, increases the height of the shadow, and a wind can form if low-ionisation gas is accelerated by ultraviolet line driving within the shadowed region, reaching escape velocity before being exposed to ionising X-rays \citep{Proga2000}. 

The above provides an alternative mechanism to super-Eddington winds for driving the UFOs discussed earlier \citep{Nomura2016, mizumoto2021}. UFOs are highly ionised, far beyond the range where ultraviolet line driving is effective. Yet, in a two-dimensional disc-wind geometry, ultraviolet line driving can accelerate much less ionised gas within the shadowed region. This gas becomes overionised only after reaching escape velocity, so it is no longer accelerated but follows a ballistic trajectory, producing a highly ionised, high-velocity wind.

These simulation results, though, again rely on approximate radiation transfer to estimate the ionisation structure of the wind. In particular, shadows are crucial for accelerating the outflow, and scattering within the wind can enhance illumination, suppressing the wind \citep{Higginbottom2014}, and better treatment of ionisation within the wind also reduces the predicted outflow rates  \citep{Higginbottom2024}. Nevertheless, despite this significant progress, current radiative transfer codes remain incomplete in some respects.
In particular, they do not yet incorporate the full clumpy structure known to arise in ultraviolet line-driven winds in massive stars (but see \citealt{Mosallanezhad2026} for a recent effort in this direction). All radiation-driven winds become clumpy once they turn optically thick, as they cast shadows on gas further out, preventing efficient acceleration. Faster material from inner regions collides with slower, shadowed gas, producing a clumpy outflow. In the case of ultraviolet line driving, shadowing also occurs in velocity space through the line deshadowing instability, which intrinsically leads to highly clumped winds (see e.g.~\citealt{Elmellah2018}). However, the microphysical aspects of this instability are not yet fully captured in current simulations.

Modelling this is challenging even in the spherically symmetric O-star simulations, and no accretion-disc wind codes can yet handle this level of complexity. Secondly, the inner disc may puff up, especially at accretion rates close to Eddington. This would cast a shadow over the disc, shielding the critical acceleration region from the X-ray flux and enabling the wind to form. A better understanding of AGN disc physics is therefore required to place ultraviolet line-driving simulations—and hence our understanding of the origin of UFOs and AGN feedback—on a more secure footing.

In the above, line-driven winds have been discussed in the context of accreting white dwarfs and AGN, since they have cooler discs than those of LMXBs, which peak in the X-ray domain. This makes the overionisation issues even more challenging than in the case of AGN (see e.g.~\citealt{Proga2002}). Thus, other mechanisms, such as the thermal and magnetic winds (see below), are likely better suited to explain (sub-Eddington) winds in LMXBs, unless  stronger-than-anticipated shielding and/or clumps are present. 

\subsection{Thermally driven winds (Thermal winds)}

The second type of wind is where the thermal energy of the gas powers the outflow via a force from the pressure gradient. The bright inner disc 
heats the photosphere at larger radii by illumination, producing a hot skin 
as was clearly recognised in \cite{Shakura1973}. The electrons in the outer disc photosphere
are illuminated by the spectrum from the central regions, and are heated by scattering high energy photons, and cooled by scattering low energy photons. 
This gives an equilibrium Compton temperature as a weighted sum over the spectral energy distribution $F(E)$ as:  

$$T_{\rm IC}=0.25\frac{\int EF(E) dE}{\int F(E) dE}$$ 

\noindent Typically $T_{\rm IC} \sim 10^{7}-10^{8}$  K in X-ray binaries. Thus X-ray irradiation of the outer disc forms an X-ray heated atmosphere whose temperature depends only on the spectrum of the radiation. This stays constant with radius, but the local gravity decreases with it. Hence, the scale height of the atmosphere increases with radius until it reaches $H \sim R$ and can escape as a wind. This occurs at radii where the temperature implies a sound speed greater than the escape velocity; that is, $kT_{\rm IC}\ge GM/R_{\rm IC}$, defining the Compton radius as:

$$R_{\rm IC}\sim 
6.4 \times 10^4/T_{\rm IC, 8} R_{\rm g}$$ 

\noindent where $R_{\rm g} = GM/c^2$ and $T_{\rm IC, 8} = T_{\rm IC}/10^8$ \citep{Begelman1983, Done2018}]. More accurate estimates (see below) show that the wind can be launched at $\sim 0.2R_{\rm IC}$, but this is still a large radius, at which the sound speed is modest, leading to relatively low wind velocities (i.e. up to $\sim$ a few thousand \kms). The behaviour and mass-loss rate of thermal winds are closely linked to the thermal structure of photoionised plasma under X-ray illumination, including the effects of thermal instability. As we will discuss below, this relation is complex, since the equilibrium curve -- which depends sensitively on the illuminating SED and the properties of the gas -- shapes the thermodynamic conditions from which the wind is launched.

Fig.~\ref{fig:TEC} illustrates the incident SEDs for typical hard (blue, $T_{\rm IC}\sim 10^8$~K) and soft (black, $T_{\rm IC}\sim 10^7$~K) state BH LMXBs, with its corresponding thermal equilibrium (heating=cooling) curve to determine the resulting gas temperature for different illuminating X-ray flux
$F_x=L/4\pi R^2$ (right). 
This is calculated using photoionised plasma codes, such as {\sc cloudy} \citep{Gunasekera2023, Ferland2017} and  {\sc xstar} \citep{Kallman2001}, as the cooling processes include atomic processes such as lines and recombination radiation as well as bremsstrahlung and Compton scattering.
Thus, cooling is dependent on metallicity and density, so here we assume solar abundances and that the gas is at constant pressure, $P_{\rm gas}$.
The most physically revealing way to show the behaviour is to plot the gas temperature, $T$, 
not simply versus flux, but rather versus the pressure ionisation parameter $\Xi=P_{\rm rad}/P_{\rm gas}$. 
Radiation pressure is simply $F_{\rm x}/c=L_{\rm x}/(4\pi R^2 c)$
while gas pressure is $nkT$. Hence, in c.g.s. units:

$$\Xi=L_{\rm x}/(nR^2 T) \times 1/(4\pi k c) = 10^{4.3} \xi/T $$ 

\noindent where $\xi (=L_{\rm x}/nR^2$; Eq. \ref{eq:xi}) is the more usual density version of the photoionisation parameter.

\FigTEC

\subsubsection{Hard spectra mass loss rates}

The physics of the heating/cooling process is simpler for a hard illuminating spectrum, as here the gas is completely ionised when there is strong illumination of low-density material (i.e. high $\Xi$, the blue lines in Fig.~\ref{fig:TEC}). 
The temperature is set by the balance of Compton heating and cooling, so it reaches $T_{\rm IC}$. 
As the gas becomes denser ($\Xi$ decreases), thermal bremsstrahlung (free-free) cooling becomes larger than the Compton process at certain $\Xi$, and the temperature drops below $T_{\rm IC}$. 
Once the gas cools, the electrons can recombine with atoms, and line emission becomes more significant, further enhancing the cooling process.
This leads to a very rapid transition from the high-ionisation state at the Compton temperature to the much lower gas temperatures of $\sim 10^4$~K typical of atomic cooling by hydrogen and helium radiative recombination. 

This rapid transition is seen in Fig.~\ref{fig:TEC} (right panel), where the slope of the equilibrium curves changes sign.
Those curves indicate heating = cooling, and divide the $\Xi-T$ plane into regions above/left of the curve where cooling dominates, and below/right where heating dominates.
A small perturbation in temperature with constant $\Xi$ (the gas is isobaric) shifts the gas parcel vertically on the plot.
The system is stable if an increase(decrease) in temperature pushes the gas into a cooling(heating) dominated region, as it will return to the equilibrium curve.
This happens if the equilibrium curve has a positive gradient. 
Conversely, in the regions with a negative slope, a small temperature increase drives the system into a heating-dominated regime, causing further heating until the gas reaches the high Compton temperature (or a smaller intermediate stable branch). 

This rapid transition of the thermal instability enables analytic estimation of the mass loss rate from the disc. 
Material on the disc surface is at low $\Xi$ as it has high density, so it is at typical atomic temperatures of $10^4$~K (or the disc surface temperature).
If the hard X-ray illumination is bigger than $\Xi_{\rm c,max}$ (Fig.~\ref{fig:TEC}; right panel) then this triggers the thermal instability and the material moves vertically up to the upper branch at this $\Xi$, increasing in temperature by a factor of $1000$, producing a hot disc atmosphere.
The maximum ionisation parameter $\Xi_{\rm c, max}$ ($\approx 12$ for the hard state SED in Fig.~\ref{fig:TEC}a) determines the density $n_c$ at the base of the hot atmosphere/wind at radius $R$ via the pressure ionisation parameter:
    
$$\Xi_{\rm c, max}= L/(4\pi n_c ck_B T_{\rm IC} R^2)$$ 

This material becomes unbound and escapes as a wind for $R > R_{\rm in} = 0.2 R_{\rm IC}$, whereas it remains bound as an atmosphere for $R < R_{\rm in}$. In the wind region near the disc, the gas outflow is predominantly vertical with velocity given by the isothermal sound speed:

$$c_{\rm IC} = \sqrt{k_B T_{\rm IC}/(\mu m_p)}$$ 

This velocity is reached as the temperature approaches $T_{\rm IC}$ due to rapid heating. The mass loss rate per unit area is given by 

$$\dot{m} = 1/2 \mu m_p n_{\rm c} c_{\rm IC}$$

\noindent with the the total mass loss rate: 

$$\dot{M} = 2\times 2\pi \int^{R_{\rm in}}_{R_{\rm out}}  \dot{m} RdR=\frac{L}{(2c c_{\rm IC}\Xi_{\rm c, max} )}\ln{(R_{\rm out}/R_{\rm in})}$$ 

\noindent Here, $R_{\rm out}$ is the outer disc radius, which can be constrained from the orbital period of the system, while $R_{\rm in}$ is the inner wind radius. This can be simply estimated as $R_{\rm IC}$, or calculated more precisely from the 
energy equation for a Keplerian rotating hot atmosphere at $T=T_{\rm IC}$: 
$$\epsilon = 1/2 v^{2}_{\rm K}+\frac{p}{(\gamma-1)\rho}+p/\rho-GM/R$$
This gives $R_{\rm in} = (\gamma-1)/(2\gamma) R_{\rm IC} = 0.2 R_{\rm IC}$, where $\gamma = 5/3$ (see \citealt{Woods1996} for further details). Therefore, the total mass loss rate is then larger for larger luminosities and large disc radii.

On the other hand, the hardness of the spectrum has a complex impact, as harder spectra increase $T_{\rm IC}$, which means the material flows away faster, decreasing the density ($\propto T_{\rm IC}^{-1}$), and the total mass loss rate ($\propto T_{\rm IC}^{-1/2}$). However, this is partly compensated by the increase in the range of disc radii over which the wind can be produced ~\citep{Begelman1983, Woods1996,Done2018}.

\subsubsection{Soft spectra mass loss rates}

In cases where the irradiated SED is softer, e.g. in the soft state of BHs (black in Fig.~\ref{fig:TEC}), 
the simple arguments outlined above change slightly since not all elements are completely stripped for $\Xi>\Xi_{c,max}$ and optical depth effects become important in illuminating the disc. Fig \ref{fig:TEC} (right panel) is calculated assuming that the X-ray illuminated material is optically thin. However, this is not the case for a soft state SED. 

Fig.~\ref{fig:TEC}b shows the thermal equilibrium curve for illumination by the 
 soft-state SED. This is no longer a simple 
 S-shaped profile, where the gas jumps from atomic temperatures to $T_{\rm IC}$; instead, 
there is an intermediate stable branch at 
$T\sim 10^6$~K for $\Xi\sim 10-100$.
In this regime, heating resulting from Compton scattering and photoelectric absorption by oxygen and iron ions is balanced by cooling via bremsstrahlung and re-emission from these elements, leading to a plateau. This gas 
typically has \logxi\ $\sim 2$, and has considerable absorption opacity, so this gas shields the disc below from illumination. Hence the wind emerges not from $\Xi_{c,max}\sim 10$ from the disc surface itself, but at the transition between the hot and intermediate branch, at $\Xi_{h,min}\sim 100$ \citep{Tomaru2019}. Material in the intermediate branch forms a partially ionised atmosphere, transitioning to a very slow wind at larger radii where material at the lower temperature of $T\sim 10^6$~K can escape. 

Simulations, which include only radiative heating and cooling, do indeed show that the slow, outer wind from material in this plateau at $\sim 10^6$~K can dominate in some soft states \citep{Higginbottom2017, Tomaru2019}.
However, because the gas is not completely ionised in this regime, 
it can also experience radiative forces due to line absorption and photoionisation.
These forces are approximately ten times stronger than those from electron scattering for gas at the plateau ionisation state (standard $\xi\sim 100$),
and RHD simulations show that this is important in giving the material an extra push at luminosities $\gtrsim 0.1 L_{\rm Edd}$ \citep{Tomaru2019}.
Thus, in soft states, thermal winds are really thermal-radiative winds, where the radiation pressure is due to atomic absorption as well as electron scattering. 
Although there are some differences in the wind properties between soft states and hard states, especially the ionisation parameter, the mass loss rate of winds is still basically proportional to the luminosity of the central source \citep{Tomaru2019, Higginbottom2020}. 

\subsubsection{Predicting observables}

The analytic thermal wind models predict 
that the wind material should expand at approximately constant speed, $c_{\rm IC}$, so the wind lines should be blueshifted by this factor, and be quite narrow as there is little velocity spread. Both mean outflow velocity and velocity dispersion are directly observable from the line profile, so these can be directly tested. 

However, the mass loss rate, $\dot{M}$, is not directly observable. 
Instead, we observe column density and ionisation state along a single sight-line, so this depends on how the mass loss makes a 2D structure. 
Assuming the wind is radial and at constant velocity gives \citep{Hori2018}:  $N_{\rm H}\propto  \dot{M}/(R_{\rm in}c_{\rm IC})\propto L \ln{(R_{\rm out}/R_{\rm in})}$ and
$\xi\propto T_{\rm IC}\propto c_{\rm IC}^2$ (constant with radius as the wind density is approximately $\propto R^{-2}$).

This dependence of $\xi$ on $T_{\rm IC}$ means that the wind features are almost completely ionised in hard states, but not in the soft states. 
This naturally predicts that absorption lines of H- and He-like Fe are only seen in soft states. The wind may be present in the hard state, with a similar mass loss rate and $N_{\rm H}$ to the soft states at similar luminosities, but it can be invisible in terms of atomic features. 
Alternatively, the hard state wind may be suppressed by shadowing from an inner disc atmosphere~\citep{Begelman1983b,Tomaru2019}).

These analytic approximations can be used to track the wind in a given source as the spectral shape and luminosity change, where they are quite successful in quantitatively matching the observed trends~\citep{Done2018,Hori2018,Shidatsu2019}. 

\subsubsection{Numerical simulations and match to observations of hot winds}

The analytic approximations above only treat the region where the source luminosity is sufficient to heat the base of the wind to the Compton temperature.
There are multiple other regions that can be treated analytically \citep{Begelman1983,Begelman1983b,Done2018}.
However, these approaches can only take us so far until the neglect of the full 2D wind structure becomes important. 
Instead, the thermal (and thermal-radiative) winds can be numerically simulated in RHD codes, giving the full 2D behaviour at all luminosities \citep{Tomaru2019, Tomaru2020, Shidatsu2019, Higginbottom2020}.
Coupling these with radiation transfer allows detailed calculations of the line profiles for direct comparison with observables. 

\TRWmodel

These results show explicitly that the SED dependence of thermal (and thermal-radiative) winds can explain the observational absorption line in the soft states and disappearance in the hard states in H1743-322 (Fig.~\ref{fig:TRWmodel}), and the 
observed X-ray absorption line features in GX~13+1  \citep{Tomaru2020b} by changing the incident SED and outer disc size. Even the unusual high column, low ionisation absorption line features in a dim ($L\sim 0.05$\ledd) soft state of GRO~J1655-40
can be explained by these models \citep{Tomaru2023}. 
This is an important test case as the data include a density diagnostic metastable line which requires $n\sim 10^{14}$~cm$^{-2}$ for collisions to populate the level.
Combining this with the observed ionisation parameter requires a launching radius $<<0.2 R_{\rm IC}$, apparently ruling out thermal winds \citep{Miller2006}. 
However, for soft state spectra the metastable level is instead populated mostly from radiative recombinations, and the data can be reproduced if the source is much more luminous, with $L\sim L_{\rm Edd}$, resulting in more wind, which has gone optically thick, reducing the observed luminosity (\citealt{Tomaru2023}; see also \citealt{Shidatsu2016}).

These radiation transport codes enable the direct study of both wind emission and absorption.
The wind scatters some fraction of the illuminating flux, but the hot, photoionised gas also directly emits radiation as free-free and bound-free recombination continua and lines. 
Usually, those lines can not be observed since the direct source emission is much stronger than those lines.
However,  when the object is very highly inclined, so that the vertical extent of the disc and/or its atmosphere blocks the direct emission, then the observed flux is dominated by the wind. 
This is the case in \textit{Accretion Disc Corona} sources, and the previous model can well describe the emission line in the 2S 0921-630 (see Fig. \ref{fig:ADC_source}; \citealt{Tomaru2023b}).

\ADCmodel

\subsubsection{Numerical simulations and match to observations of cold winds}

Despite the success of thermal wind models in reproducing the X-ray line features observed in soft states (i.e. hot winds), they currently cannot easily account for the optical/NIR absorption and emission lines associated with the low-ionised winds predominantly detected during hard states (see Sec.~\ref{liwinds:lines}).

Focussing first on the optical absorption lines, the observed blueshifts require that this low ionisation gas is outflowing at $\sim 1000-2000$~\kms. 
These optical absorption lines require low ionisation gas, so might be expected to be at low $\Xi$, below the thermal instability.
But gas here has very low velocity, with sound speed $c_{\rm s} = \sqrt{k_B 10^4~{\rm K}/(\mu m_p)}= 10~{\rm km~s^{-1}} $. 

Instead, the observed velocity of the optical winds is quite close to the higher velocity ($\sim 1000~{\rm km~s^{-1}}$) of material at the high Compton temperature of the hard state ($\sim 10^8$~K). However, here the material it is far too highly ionised to make the optical transitions.

If these hard state optical winds are indeed thermal winds, then they could be produced if there is some clumping instability which operates after the wind is launched, such that the material returns below $\Xi_{\rm c,max}$ and triggers the thermal instability in reverse. Adiabatic expansion alone cannot do this as $\Xi$ {\it increases} in the expanding wind as the density drops as $R^{-2}$, far faster than the temperature. Instead, it requires some other mechanism to compress the gas to trigger the thermal instability, perhaps via interaction with higher inclination, slower material forming the disc photosphere. However, detailed models of this show that clumps would grow only at very large radii ($\sim 100-1000R_{\rm IC}$), resulting in velocities 
much smaller ($\leq 100$~\kms) than those observed (e.g.~\citealt{Waters2021}).  
Perhaps a more likely mechanism is that the hot, overionised wind expands out and sweeps up mass from previously ejected wind material, and that the subsequent shock gives rise to clumping, but no detailed simulations of this scenario exist as yet.

The emission lines are perhaps easier to explain, as these could be produced at the base of the wind, where the X-ray irradiation quickly heats the material from atomic cooling to the hot branch (see Fig \ref{fig:TEC}). There are simulations by \cite{Koljonen2023} which show that this can give a good match to all the emission lines seen in the soft state of the black hole MAXI J1820+070. However,
none of these lines have the obvious wind signatures of absorption, and/or broader wings on the emission lines. 
Also, this was a soft state observation, where the optical winds are typically non detected. Nonetheless, these simulations clearly show the potential to explain the standard optical emission lines seen from the irradiated disc at the base of the wind.

\subsection{Magnetically driven winds (MHD winds)}

Radiative and thermal (and thermal-radiative) winds can be calculated (with varying degrees of difficulty, see above) from the given conditions, 
but the MHD winds are currently completely undefined as the baseline magnetic field configuration in the accretion disc is not known. 
Magnetic fields are almost certainly present, as again recognised by \cite{Shakura1973}, as the viscosity required to transport angular momentum outwards so material in the disc can accrete inwards is most likely magnetic.
However, there is still debate about whether this is caused solely by a small-scale magnetic dynamo driven by the magneto-rotational instability (MRI) \citep{Chandrasekhar1961, Balbus1991}.

Simulations performed over the last decades have uncovered some of the properties of this process, and it is now clear that the initial conditions in terms of net magnetic flux imposed are very important in determining the non-linear outcome of the process.
Initialising with no net flux gives rise to SANE (standard and normal evolution) flows, where the small-scale dynamo gives an effective viscosity parameter $\alpha\sim 0.01$. 
These flows launch jets, but the jet power is low.
By contrast, with some initial net flux, the flow is magnetically arrested close to the innermost stable circular orbit (MAD: magnetically arrested disc;  \citealt{Narayan2003}), and the effective $\alpha$ and jet powers are both much higher. 

Thus, even for the known magnetic dynamo process, the properties of the resulting flow are not yet clear.
However, this need not be the dominant magnetic configuration.
Before the MRI was discovered, other ways of transporting angular momentum via large-scale, coherent fields threading the disc were explored.
The earliest were the Blandford-Payne self-similar models, where the magnetic field connects into the disc at all radii and torques it by accelerating some small amount of material up the field lines, converting azimuthal motion to radial outflow \citep{Blandford1982}.
After this pioneering work, different solutions with similar large-scale magnetic fields but different assumptions were studied (\citealt{Contopoulos1994, Ferreira1995, Ferreira1997}). This mechanism was also explored for the case of young stellar objects \citep{Pudritz1986}.

\MHDWmodel

The current  MHD wind model (see Fig. \ref{fig:MHDmodel}) for observational spectra developed by \citet{Fukumura2010, Fukumura2015, Fukumura2017} is based on the formalism of \citet{Contopoulos1994}, which itself generalises the pioneering centrifugally driven wind mechanism of \citet{Blandford1982}.
In these formulations, self-similar global magnetic fields thread the entire accretion disc, naturally driving an outflow across all radii.
By adjusting the field configuration, one may impose a radial density law of the form as $n(R) = n_0 (R/R_{\rm in})^{-\alpha}$,
where $n_0$ is the density at the inner radius of the wind, $R_{\rm in}$, and $\alpha~(0<\alpha<2)$ is a free parameter. 
At each radius, the wind speed have similar velocity as the Keplerian velocity, $v(R) \sim v_{\rm K}(R)$.
Using those parameters, one can calculate the line profile by a photoionisation code such as {\sc xstar} \citep{Kallman2001} or {\sc cloudy} \citep{Ferland2017,Gunasekera2023}.
The typical feature of this model is that the wind density, ionisation parameter $\xi = L/(nR^2) \propto R^{\alpha-2}$, and velocity $v \propto v_\phi(R) \propto R^{-1/2}$ decrease with radius. This causes the highly ionised ions to be faster than those at lower ionisation, producing asymmetric lines with a strong blue wing (Fig.~\ref{fig:line_MHD}). A different MHD formalism by \citet{Ferreira1995}, applied in models by \citet{Chakravorty2016, Chakravorty2023}, yields similar results, as also shown by \citet{Datta2024}.

MHD winds are then capable of explaining the wide range of outflow velocities observed -- from the highly relativistic velocities associated UFOs in AGN \citep{Fukumura2015} to the slower X-ray absorption lines seen in X-ray binaries, where the inner disc wind may be so highly ionised as to escape detection \citep{Fukumura2017}.

\MHDlines

However, a critical issue for these models 
is that they invoke a self-similar ordered  magnetic field structure which is not yet well physically justified. 
Although magnetic fields are a plausible driver for angular momentum transport via the small scale magnetorotational instability, the inherently turbulent nature of these fields casts doubt on the likelihood of establishing and sustaining the globally organised magnetic structures assumed in these models. This remains an important topic for future research. 


\section{Discussion}\label{sec:discussion}

In this review, we attempt to compile, discuss, and place within the context of theoretical work the current observational evidence for the presence of accretion disc winds in accreting BHs and NSs. We do so using a multiwavelength approach, arguably for the first time, given the growing number of detections at wavelengths other than X-rays in recent years. This allows us to study winds with significantly different physical conditions, such as ionisation and presumably densities. 

Since the focus is on accreting NSs and BHs with disc-fed accretion, this review is naturally centred on LMXBs. Nonetheless, we have included some systems with somewhat more massive donors, such as Her~X-1 and V4641~Sgr, which can be considered intermediate-mass X-ray binaries. However, we do not include other markedly different compact binary populations that also display disc winds, such as accreting white dwarfs (cataclysmic variables; e.g.~\citealt{Drew1988, Kafka2004, Cuneo2023}) and ultraluminous X-ray sources. The latter are thought to accrete in a supercritical regime where such winds are naturally produced and strongly influence their observational properties. For example, SS~433, the most promising Galactic candidate for this type of system, is known to display optical P-Cyg line profiles in addition to other features associated with its precessing jet (e.g.~\citealt{Fabrika2004, Blundell2011}).

Table~\ref{tab:master} includes the 30 systems in which, to the best of our knowledge and based on the criteria discussed above, signatures of disc outflows have been confidently detected in at least one spectral range. Systems with only tentative detections are excluded from this discussion, although an extensive list of such cases is provided in Tables~\ref{tab:liwinds},~\ref{tab:hiwinds:bh}, and~\ref{tab:hiwinds:ns}. The general discussion that follows focuses on the global properties of winds, from cold to hot outflows, as more specific issues related to low- and high-ionisation cases, as well as modelling, have already been addressed in their respective sections.   

\subsection{Atmospheres, winds and failed winds}
\label{sec:discussion:atmospheres}
Since LMXBs were first studied at low energies, optical and NIR spectroscopy has revealed the presence of emission lines. These features originate in the disc atmosphere, a vertically extended layer of gas surrounding the disc, and are often characterised by double-peaked profiles produced by the Doppler shifts from opposite sides of the outer accretion disc. However, such double-peaked profiles can persist even if the atmosphere is outflowing, as long as the opacity is such that the lines trace material close to the disc, where the velocity field is still dominated by rotation (e.g.~\citealt{Koljonen2023}). Single-peaked profiles are also observed, particularly during outbursts\footnote{Note, however, that at low velocity resolution, the double-peak cannot always be resolved during outburst, as the (cooler) optical emitting regions move further out to lower velocity zones, naturally forming narrower profiles (e.g.~\citealt{Casares2015}). This is particularly challenging for low inclination systems given their lower projected disc's velocities.} and are often interpreted as arising in an outflow or high up in the disc atmosphere (e.g.~\citealt{Matthews2015}).

In addition, broad absorption components with no significant velocity shifts are sometimes present in the optical spectra of LMXBs (particularly in the Balmer lines; see \citealt{MataSanchez2026} for a global study) and are also common in accreting white dwarfs. The origin of these features is uncertain, but they may arise in an optically thick layer of the disc photosphere under certain physical conditions \citep{Dubus2001, MataSanchez2026}. These absorptions are sometimes superimposed on emission lines, making their interpretation more complex (e.g.~\citealt{Miceli2024}). They can also be detected together with additional blueshifted components (i.e. outflows), as in the case of Swift~J1727.8$-$1624 (\citealt{MataSanchez2024a}) and possibly GRO~J1655$-$40 (\citealt{Soria2000}).

Most LMXBs accessible in the optical and NIR exhibit some form of emission and/or absorption features in their spectra. As such, in this review we have restricted ourselves to discussing only those that can be confidently (or at least tentatively) associated with winds (see Sec.~\ref{sec:signatures} and Sec.~\ref{sec:liwinds}).

\subsubsection{Static X-ray lines}
Relativistically broadened emission components aside (e.g.~\citealt{Miller2007}), the presence of emission and/or absorption lines in X-ray spectra is relatively rare. Out of the 30 systems included in Table~\ref{tab:master}, nine have shown confident detections of X-ray lines with no significant velocity shift, and are thus consistent with an origin in the disc atmosphere. Despite the lack of measured blueshifts, these features are in most cases similar to those associated with X-ray outflows. This supports a connection between disc winds and atmospheres, suggesting that the former may be understood as an outflowing version of the latter (see Sec \ref{sec:disussion:velocity}). For this reason, in this review we consider both static and shifted X-ray components. Furthermore, until the recent launch of \textit{XRISM}, instrumental capabilities for studying X-ray lines at high velocity resolution were limited. Future observations with this facility may therefore uncover low-velocity outflows in some of these systems

\subsubsection{Inflows and failed winds}
Redshifted absorption features that might trace the presence of infalling gas out of the disc plane have been observed in some LMXBs. These can be produced by material lifted from the disc (e.g. by one of the mechanisms described in Sec.~\ref{sec:theory}) but failing to accelerate beyond the escape velocity (i.e. failed winds).

This is a common phenomenology in other accreting objects. For instance, inverted P-Cyg profiles (i.e. blueshifted emission accompanied by redshifted absorption) are common in the NIR lines (e.g.~\pab\ and \brg) of protostars (e.g. T~Tauri stars; \citealt{Folha2001}). These are interpreted as signatures of infalling gas. However, the precise physical origin of the emission and absorption features remains uncertain, with magnetically driven accretion typically invoked.

In LMXBs, examples of this phenomenology exist, particularly for lines tracing low-ionised gas. These components can be narrow, pronounced, and transient on timescales significantly shorter than the orbital period (\citealt{Cuneo2020}; see also \citealt{Miceli2024}), or rather persistent across large phases of the outburst \citep{MataSanchez2024a}. Arguably, the main difficulty lies in ruling out a disc atmosphere origin or the possibility that the profile results from a combination of disc atmosphere and outflow features. Similar claims for high-ionisation gas exist (e.g.~\citealt{Miller2014}; see also \citealt{Shidatsu2013}), although the limited velocity resolution makes it difficult to distinguish these features from static disc-atmosphere lines. Even in the \textit{XRISM} era, the presence of multiple absorption components makes this distinction challenging \citep{Rozanska2014, Miller2025}.

The combination of \textit{XRISM} with optical and NIR spectroscopy at similarly high velocity resolution should lead to major improvements in our understanding of these components. In particular, a key question to address is whether winds, atmospheres, and failed winds represent different stages of the physical phenomenon, with winds being the material that escapes, and atmospheres or failed winds forming part of a cycle where material is lifted, forms an atmosphere, and eventually falls back.

\TabMASTER

\subsection{Disc winds: observational properties}
\label{sec:discussion:obsproperties}
It follows a summary of the most relevant observational properties that can be directly derived from Table \ref{tab:master}.
\subsubsection{Black holes versus neutron stars}
Outflow velocities are available for 21 X-ray binaries (sources in boldface in Table~\ref{tab:master}). These include most BH transients (top panel) and about half of the NS systems (lower panel). In the former, the majority of the detections come from the optical domain, specifically cold winds, while high ionisation winds vastly dominates in the case of NS. This might be the result of several combined factors: 

\begin{itemize}[leftmargin=0cm, itemsep=1pt] \smallskip

\item BH systems are mostly transient sources, and therefore observations rely on dedicated programmes such as \textit{Target-of-Opportunity} campaigns, which are more difficult to implement with high spectral resolution facilities\footnote{However, this is not the case for lower-resolution instruments such as \textit{NICER}, which has provided some tentative detections (see Table~\ref{tab:hiwinds:bh}).}. In the last decade, the advent of new large ground-based telescopes such as the \textit{GTC} and \textit{SALT}, in addition to existing facilities like the \textit{VLT}, has mitigated this limitation in the optical band. However, outburst observations with high-resolution X-ray instruments such as \textit{Chandra} and \textit{XMM-Newton} remain relatively scarce.

\item NS systems are mostly persistent sources, which makes it easier to schedule X-ray observations. However, many of them are faint or simply inaccessible in the optical, as they are located in the Galactic plane, where extinction is high. They also tend to exhibit hotter optical spectra (e.g. \citealt{PanizoEspinar2021}), often dominated by strong \heii\ lines -- a property that in BHs has been shown to be crucial for the detection of cold winds (see e.g. Fig.~\ref{fig:trail_hb}). This behaviour may stem from a combination of factors, including the smaller disc size in NS systems, the fact that luminous persistent sources typically remain in soft states, and the faster hard-to-soft transitions observed in NS transients (which prevents their observation in the former state; see e.g.~\citealt{Munoz-Darias2014}).
\end{itemize}

Nonetheless, several characteristics -- such as the range of observed wind velocities and the strong preference for detecting winds in high-inclination systems -- are common to both BHs and NSs. Hereafter, we discuss the wind properties of BH and NS LMXBs globally.

\subsubsection{The orbital inclination and the wind geometry}
As is evident from Table~\ref{tab:master}, inclination is a major factor in driving the detectability of winds. In fact, X-ray winds have so far only been confidently detected in high-inclination systems, which we define as those with inclinations greater than $\sim 60^{\circ}$. This behaviour has traditionally been interpreted as evidence for an equatorial wind geometry (e.g.~\citealt{Ponti2012}), where detection is naturally favoured when the line of sight lies close to the disc plane. While this remains the prevailing view, it is important to bear in mind that, in the context of disc winds, absorption will still appear stronger at high inclinations even for more isotropic geometries owing to the increased line-of-sight opacity (e.g., \citealt{higgionbottom2015, DiazTrigo2016}). In this context, observations taken at different phases of the precessing disc period of Her~X-1 offered a unique insight into the wind geometry, as the strongest and least clumpy outflow was detected when looking closer to the orbital plane \citep{Kosec2023}, supporting an equatorial geometry.

Although the situation is broadly similar in the optical and NIR domains, several low-ionised wind detections have been reported in lower-inclination systems. A key difference lies in the nature of the signatures. X-ray wind detections are associated with blueshifted absorption lines, as are optical-NIR P-Cygni line profiles (which are also always seen at high inclinations). However, detections at lower inclinations often arise from emission components (see Sec.~\ref{sec:signatures}). A unique case is that of MAXI~J1348$-$630 \citep{PanizoEspinar2022}. In this system, broad emission components were observed alongside an absorption trough detected in multiple high signal-to-noise optical and NIR spectra across several epochs. Interestingly, the system has a low-to-moderate orbital inclination (e.g.~\citealt{Carotenuto2021}), and the centroid of the absorption trough indicates a velocity of 500~\kms\ (with a blue-edge at $\sim$900~\kms), significantly lower than the velocities inferred in other systems (see below) and those of the emission components detected simultaneously in the same object (up to 1700~\kms). Since the emission components originate in material out of the line of sight (see Fig.~\ref{fig:P-Cyg_scheme}), their velocities should be less dependent on the inclination. The lower velocity of the absorption trough, on the other hand, could be naturally explained as a projection effect in an equatorial wind geometry. Additional detections in low-inclination systems are required to establish more robust conclusions based on these arguments. 
In summary, while more observations and detailed modelling are needed, the current observables from both low- and high-ionisation winds largely support a prevailing equatorial wind geometry.

\subsubsection{The wind velocity}
\label{sec:disussion:velocity}
The observed velocity is one of the most relevant wind properties to discuss. It can be directly derived from observations, with little to no modelling involved. However, distinguishing the specific wind signatures from which velocities are derived is fundamental for a meaningful comparison. In Table~\ref{tab:master}, we have sorted both NS and BH systems by the maximum velocity reported for each source, regardless of whether it is derived from lines tracing low- or high-ionisation material. Here, it is important to note that the maximum \textit{observed} velocity does not necessarily correspond to the maximum velocity of the wind, as higher-velocity components may remain undetected, for example if the relevant species becomes over-ionised or if the line opacity drops owing to decreasing densities. We also list the orbital period ($P_{\textsc{orb}}$), which serves as a proxy for accretion disc size\footnote{Note, however, that this also depends on the compact object mass. A typical BH mass implies a disc 2--3 times larger than that of a NS at the same $P_{\textsc{orb}}$; see e.g.~\citet{Frank1992}.}. 

In the NS sample (lower panel in Table~\ref{tab:master}), which is dominated by X-ray detections, there is a clear trend: atmospheres or very low-velocity winds are preferentially found at short orbital periods. This was already noted by \citet{DiazTrigo2016} and can be interpreted in the context of thermal winds. As discussed in Sec.~\ref{sec:theory}, these require a minimum disc size to launch an outflow ($\gtrsim$0.2$R_{\rm IC}$). Therefore, short-period systems -- particularly those hosting NSs -- might not reach the necessary outer disc radius. In these cases, the gas can still be lifted by the same mechanism but remains gravitationally bound to the system, forming a disc atmosphere. 

In general, the observed wind velocities range from $\lesssim200$~\kms\ (which we associate with X-ray atmospheres) to $\sim$4000~\kms. While the overall distributions for BHs and NSs are similar, the highest velocities are found in BH systems. The four targets with the largest values (above 3000~\kms) all harbour BHs. These also tend to be long-period systems, although exceptions exist. For instance, Swift~J1357.2$-$0933 exhibits very large velocities despite its short orbital period. However, this may be related to its extremely high inclination and possibly to its rather unique phenomenology (and perhaps outflow nature), since low ionisation winds are only detected during the peculiar optical dips displayed by the system (see e.g.~\citealt{Charles2019, Jimenez-ibarra2019b, Corral-santana2013, Panizoespinar2024}).

By comparing the detections listed in Table~\ref{tab:liwinds} with those in Tables~\ref{tab:hiwinds:bh} and~\ref{tab:hiwinds:ns}, one might conclude that X-ray winds generally show smaller velocities than cold winds. However, it is crucial to note that measurements in these regimes are typically obtained using different methods. In the optical and NIR (i.e. low-ionised winds), the absorption components are often well resolved, allowing for velocity measurements based on the blue-edge of the P-Cyg profile. In contrast, X-ray velocities are usually derived from the centroid of the absorption line. In Table~\ref{tab:master}, we indicate which measurements were obtained using the blue-edge (labelled \texttt{b-e}), and it is clear (and expected) that the highest velocities are found with this technique. Although rare, this also includes a few X-ray detections. Interestingly, when X-ray winds are measured using the blue-edge (e.g. V404~Cyg and Cir~X$-$1), the inferred velocities are as high as those derived from low-ionised winds. In fact, if we assume that the blue-edge velocity is typically about twice the velocity of the absorption centroid, then the characteristic velocities of low- and high-ionised winds become consistent. The arrival of \textit{XRISM}, combined with observations at lower energies, should shed light on this topic -- a key factor for understanding the nature of both low- and high-ionised winds. So far, the only simultaneous high-resolution study is that of V404~Cyg, which shows consistent kinematical properties across the different low- and high-ionisation components (\citealt{Munoz-Darias2022}).

Finally, it is worth noting the significant number of tentative X-ray wind detections reporting velocities an order of magnitude higher than those discussed here (see Tables~\ref{tab:hiwinds:bh} and~\ref{tab:hiwinds:ns}). The presence of these components -- approaching the UFOs regime observed in AGN -- is typically inferred from single lines, often detected using low-resolution instruments. This may reflect the intrinsic rarity of such features, as well as the need for high-cadence observations to capture them. While the existence of such high-velocity winds in LMXBs is intriguing and would have strong implications for constraining wind-launching mechanisms, conclusive detections are still lacking. Ideally, this would require several lines simultaneously displaying extreme blueshifts. Once again, the arrival of \textit{XRISM} may play a key role in resolving this matter. We also note that, so far, there is no evidence for such high-velocity components in low-ionisation winds.

\FigSketch

\subsection{Accretion states and winds}
\label{sec:discussion:states}
In addition to winds, LMXBs also launch collimated jets, best observed at radio wavelengths.  A close connection between the accretion regime and the presence and observational properties of the jet is well established, to the extent that steady, compact jets can be considered a defining element of the accretion state (e.g.~\citealt{Fender2012}). The discovery of X-ray winds, preferentially detected during the soft state (see Sec.~\ref{sec:hiwinds:states}), added further complexity to this picture since, in most cases, high-ionisation winds and jets appear to be mutually exclusive phenomena (e.g. \citealt{Neilsen2009, Ponti2012}). 

The extensive optical and, to some extent, NIR and ultraviolet spectroscopic campaigns conducted over the last decade have shown that low-ionisation winds are as common as X-ray winds. However, they are typically observed during the hard state (see Sec.~\ref{sec:liwinds:states}), that is, simultaneously with radio jets. As discussed above, both types of winds share key observational properties, such as similar velocities and a preference for being detected in high-inclination systems. The current observational evidence is sketched in Fig.~\ref{fig:sketch}, where the visibility of winds as a function of the accretion state (as traced by the commonly used hardness–intensity diagram; e.g. \citealt{Homan2001, Fender2004}), together with the jet properties, is depicted. It leads to some general conclusions, which can be summarised as follows:

\begin{itemize}[leftmargin=0cm, itemsep=1pt]
\smallskip
\item Winds, as a physical phenomenon, are present in all states, but their phenomenology changes dramatically from the soft to the hard state, with high-ionisation (X-ray) winds observed primarily in the soft state and low-ionisation (optical/NIR) winds in the hard state. This conclusion is primarily based on the hot and cold wind types, while the situation is less clear for the warm and cool winds owing to the smaller number of detections.

\item Low- and high-ionisation winds are typically not observed simultaneously. However, this observational picture is not absolute and should not be interpreted in a deterministic way. X-ray winds can also be observed simultaneously with the jet (i.e. during the hard and intermediate states), and thus may coexist with low-ionisation winds. Signatures of low-ionisation winds (both in the optical and NIR lines), although weaker than those found in the hard state, have also been found during the hard-to-soft transition, and even in the soft-intermediate state (see e.g. \citealt{PanizoEspinar2022, MataSanchez2024a}). In the NIR, these are also present during the soft state, supporting the presence of cold winds at this stage (see Sec.~\ref{sec:liwinds:states}). Therefore, the variability of wind signatures across the spectral range and accretion states may simply reflect how the physical conditions in the soft and hard states influence the detectability of high- and low-ionisation winds, respectively.

\item The current observational picture suggests that the relation between jets and winds is not direct. In addition, if jets and winds are linked by a common, albeit complex, mechanism, this would likely require them to originate from the same region of the system \footnote{It is worth noting, however, that general relativistic MHD models are able to simultaneously produce jets and winds (see \citealt{Liska2020}).}. For jets, this is generally thought to be the innermost few gravitational radii. However, most modelling of winds, as well as their kinematic properties, suggests that they are launched from the outer disc (see also Sec.~\ref{sec:theory}).

\end{itemize}

It is important to bear in mind that the observational view sketched in Fig.~\ref{fig:sketch} provides a simplified representation of the situation at the time of writing this review. For instance, not all regions of the hardness–intensity diagram have been equally sampled by observations, with the soft-to-hard transition being much less studied—at all wavelengths—than the more luminous hard-to-soft one. In the diagram, both are depicted in a similar way, but new observations may uncover significant differences. Also, as discussed in Sec.~\ref{sec:hiwinds:states}, future X-ray observations might reveal lower-ionisation (i.e. warm) X-ray winds during the harder states, while additional NIR studies are needed to establish the prevalence and properties of cold outflows in the softer states.

The above picture has been drawn mainly from BH transients. However, it seems to be broadly applicable to NS systems as well. Winds in these objects are primarily detected in persistent sources in the X-ray regime and show observational properties consistent with those seen in BHs (see Sec.~\ref{sec:hiwinds:states}). Considering both, high- and low-ionisation winds in NS LMXBs, the most complete comparison can be made against the NS transient Swift~J1858.6$-$0814. During its long and rather extreme 2020 outburst (see e.g.~\citealt{Vincentelli2023}), it displayed strong optical and ultraviolet P-Cyg line profiles (i.e. cold and cool winds), contemporaneously with jet emission (\citealt{vandenEijnden2020}; i.e. during the hard state). In particular, the optical features are remarkably similar to those seen in BH transients \citep{Munoz-Darias2020}. Unfortunately, no high-resolution optical or X-ray spectroscopy is available during its soft state. However, there are hints of static X-ray emission lines during the hard state \citep{Buisson2020}. Likewise, ultraviolet observations taken during the soft state do not show the P-Cyg profiles (i.e. cool winds) present during the bright hard-state epochs \citep{CastroSegura2022, CastroSegura2024}, suggesting that the observational picture of the cold and cool wind components across the accretion states is rather similar.

\subsection{Structure, mass content and impact}
\label{sec:discussion:structure}
\Windstructure
In the above, we have discussed how winds of different ionisation components are preferentially observed at high orbital inclination, and how this can be reasonably interpreted within the framework of an equatorial geometry. These gas components, observed through similar signatures in both NS and BH systems, show velocities of up to $\sim 4000$~\kms, although they can also appear static. We have also evaluated the main physical mechanisms proposed to launch winds (i.e. thermal, radiative, and magnetic; see Sec.~\ref{sec:theory}) and how they compare with current observational evidence.

\subsubsection{Wind structure}
Beyond understanding how winds are actually launched (see Sec. \ref{sec:theory}), some of the most pressing open questions concern their structure (i.e. whether different components coexist, and how) and their mass content. While the presence of winds has been established in more than 20 systems, confirmed detections of both low-ionised and high-ionised winds exist only for three systems (Table~\ref{tab:master}). Among them, simultaneous observations are available only in the case of V404~Cyg. A combined analysis of these data \citep{Munoz-Darias2022} shows that blueshifted absorption features at remarkably similar velocities are observed in both the optical and X-ray spectra. However, they do not appear simultaneously: the X-ray absorption components emerge during the luminous flares, whereas the optical absorptions occur during quieter periods. On the other hand, broad emission lines are observed in both regimes when the source exhibits intermediate behaviour. These components show consistent width between each other, strongly supporting a common (wind) origin for the optical and X-ray emitting regions.

Basic arguments based on the similar kinematical properties of hot and cold winds favour a multiphase wind structure, where both phases coexist spatially, as opposed to a stratified outflow (see sketch in Fig.~\ref{fig:structure_sketch}). In the simplest scenario, the multiphase approach explains the observed differences in ionisation parameter ($\xi$) as resulting from variations in the wind density ($n$ in Eq.~\ref{eq:xi}), that is, the presence of higher-density clumps accounting for the low-ionisation ejecta. In contrast, pure stratification would instead imply different distances from the irradiating source ($R$ in Eq.~\ref{eq:xi}). However, generalising these results is problematic, as both the 2015 outburst and the wind itself of V404~Cyg are at the extreme end of the population. For instance, the high luminosity of the source during some phases of the outburst suggests a strong contribution of radiation pressure in launching the wind (see e.g.~\citealt{Motta2017b,Casares2019}) -- a mechanism that is not applicable to other systems unless the line-driven mechanism plays a role (see Sec.~\ref{sec:theory}).

Connected to this, X-ray observations, particularly in the XRISM era, can offer the possibility to simultaneously study different highly-ionised outflow components (e.g. warm and hot winds) within the same dataset. One of the first studies of this class, that of GX~13+1 (Fig. \ref{fig:gx13spectrumxrism}; \citealt{XRISM2025GX13}), reveals the presence of a relatively slow wind ($\sim 330$ \kms) detected via blue-shifted absorption in dozens of transitions. While these can be generally modelled by a single absorber, a detailed look at the line profiles suggests the presence of different stratified components within the X-ray outflow. Likewise, the aforementioned study of Her~X-1 at different phases of its precession period also supports stratification in the wind: a less ionised and clumpy outflow is found when looking through lines-of-sight sampling the wind further away (vertically) from the equatorial plane \citep{Kosec2023}. On the other hand, a multiphase nature is invoked to explain the properties of the ultraviolet and optical wind (i.e. cold and cool components) detected in Swift~J1858.6-0814 \citep{CastroSegura2022}. All considered, both features -- that is, multiphase outflows (especially when looking across the full low to high ionisation picture) with some degree of stratification -- can likely account for the still limited observational evidence.

\subsubsection{Mass outflow rate and wind's impact}
Crude estimations, derived from X-ray observations, of the mass carried by the wind (i.e. the mass outflow rate) exist, particularly for some BH LMXBs. Typical values range from being comparable to the accretion rate, up to exceeding it by a factor of 10--20 (e.g.~\citealt{Lee2002, Neilsen2011, Ponti2012}). These figures are derived from the observed absorber column after properly constraining $\xi$ through spectral modelling. Likewise, it is necessary to assume a particular wind geometry (e.g. equatorial, with a 30$^{\circ}$ opening angle), as well as an accretion efficiency (i.e. the mass-to-radiation conversion; 10 percent is a typical value).

Proper modelling of low-ionised winds remains to be done, although new radiative transfer codes have recently become available to the community (e.g.~\textit{Sirocco}; \citealt{Matthews2025}). The first attempt (Ambrifi et al., in prep.) indicates that similarly large mass outflow rates, in addition to significant clumping, are likely required to reproduce the standard optical P-Cyg line profiles detected during regular hard states, such as those seen in MAXI~J1820+070 \citep{Munoz-Darias2019}.

Focussing on individual systems, mass outflow rate estimates have been calculated for the aforementioned case of V404~Cyg. This is possible thanks to the broad phenomenology displayed by the source, which allows the application of different methods. In particular, the evolution of the conspicuous optical nebular phase observed after the abrupt end of the outburst enabled an estimate of the ejected mass. This was done using both the diffusion and recombination timescales of the nebula (Fig.~\ref{fig:nebular_phase}) produced by the expanding shell of ejecta. The total mass was found to lie in the range 10$^{-8}$–10$^{-6}$~M$_{\odot}$, up to 100 times the mass accreted during the outburst \citep{Munoz-Darias2016, Casares2019}. This high outflow rate is able to explain the large intrinsic column densities (e.g.  $N_{\rm H} \sim 10^{23}$–-$10^{24}$~cm$^{-2}$; e.g.~\citealt{Motta2017b}) reported in this and other systems with strong outflow signatures. This also suggests that, at least for the four systems showing strong signatures of low-ionised winds (see Sec.~\ref{sec:liwinds:obs} and Table \ref{tab:liwinds}) -- which share some similar phenomenology -- these outflows are dynamically relevant in terms of mass content. For instance, in the case of the NS transient Swift~J1858.6$-$0814, an outflow rate larger than 0.2 times the accretion rate was derived from the ultraviolet P-Cyg profiles \citep{CastroSegura2022}, while the strong NIR wind detected in GRS~1915+105 offers a viable explanation for the heavily X-ray obscured phases displayed by the system \citep{SanchezSierras2023b}.

Although the above mass estimates for both high- and low-ionised winds carry significant uncertainties, they nonetheless suggest that a large fraction of the mass involved in the accretion process is carried away by the wind. This is expected to have a strong impact on regulating, in some way, the accretion process. At least for the most extreme cases, there is substantial observational evidence supporting this claim. For instance, the already widely discussed 2015 outburst of V404~Cyg was unusually short given the amount of mass stored in its large accretion disc. This has been attributed to a massive outflow that depleted a significant part of the disc and disrupted its structure, abruptly ending the outburst \citep{Munoz-Darias2016}. Likewise, in GRS~1915+105, besides the NIR wind detected during its X-ray obscured phases, strong X-ray winds were also observed on multiple occasions. The regulation of the inner accretion flow by such winds has been proposed as the origin of some of the different variability patterns detected in the system (\citealt{Neilsen2011}; see also \citealt{Vincentelli2023}). 

Even if winds do not display such extreme phenomenology (i.e. generally implying lower luminosities), their impact might still be significant. Some topics discussed in the literature include the possible role of winds in quenching the radio jet (i.e. triggering the transition to the soft state) and stabilising the disc during soft states (e.g.~\citealt{Neilsen2009, Ponti2012}). However, these arguments generally rely on a scenario in which winds are absent—or significantly weaker—during hard states (i.e. the X-ray wind view). Thus, they do not account for the low-ionisation winds observed in BH hard states. Winds have also been invoked to explain the persistent nature of some long orbital period NS systems, which would otherwise be expected to be transient. Scattering within a strong wind (typically associated with large accretion discs) would enhance the irradiation of the outer disc, keeping hydrogen fully ionised and hence preventing the transient behaviour \citep{Dubus2019}.

Another aspect worth bearing in mind is the possible role of winds in angular momentum removal—a key mechanism required for accretion discs to operate. In the \citet{Shakura1973} prescription, this is implicitly encoded in the viscosity parameter ($\alpha$), to which several physical processes may contribute. The fast outburst decays observed in some systems suggest high values of $\alpha$, which can be explained by different scenarios, including a strong contribution from state-independent winds (i.e. winds that remain active throughout the outburst) in removing angular momentum \citep{Tetarenko2018}. It is important to note, however, that not all winds are expected to remove angular momentum from the disc in the same way. Although more advanced modelling would be required to assess this in detail, thermal or radiative winds are, in general, not thought to exert a significant torque. Hence, while they carry angular momentum away from the system, they do so by removing mass while leaving the specific angular momentum of the remaining disc material largely unchanged. By contrast, MHD winds may extract angular momentum directly from the disc through magnetic torques, reducing the specific angular momentum of the accreting material and potentially contributing to the angular-momentum budget. Under this interpretation, magnetically launched winds may help to explain the aforementioned fast outburst decays (\citealt{Tetarenko2018, Dubus2019}).

Finally, the total angular momentum removed from the binary by the wind (i.e. irrespective of its driving mechanism) might also influence the long-term evolution of the system, as recent simulations suggest (see e.g.~\citealt{GallegosGarcia2024}). Likewise, on the observational side, wind-launched material has been proposed to exert ram pressure on the companion star, potentially helping to explain the rapid orbital decays seen in some LMXBs (e.g.~\citealt{ponti2017}). These two examples highlight the multifaceted role that disc winds might play. Winds are relevant not only for regulating accretion, but also for shaping the physical and observational properties of accreting binaries; a topic that is likely to be the focus of further research in the coming years.

\section{The next decade}
\label{sec:next}
Without forgetting earlier detections dating back to the final years of the 20th century (see Sec.~\ref{sec:intro}), we are now entering the third decade of intensive research on disc winds in accreting stellar-mass BHs and NSs, following the first blueshifted line detections discovered by \textit{Chandra}. The first decade (mid-2000s to mid-2010s) established the presence of high-ionisation winds, while observations with the largest ground-based telescopes (mainly \textit{GTC} and \textit{VLT}) in the second decade extended our understanding to lower ionisation outflows (optical and NIR), and even into the ultraviolet using the \textit{Hubble Space Telescope}. This period also saw significant efforts devoted to modelling, particularly in the X-ray regime.

After these two decades of intense study, winds emerge as a fundamental -- perhaps ubiquitous -- component of accretion onto BHs and NSs, being widely present during their active phases (i.e. outbursts and persistent sources). Their observational signatures depend critically on the spectral regime: high-ionisation outflows are seen during soft states, while low-ionisation winds are preferentially detected in hard states, when radio jets are also active. The physical properties reveal winds that can be relatively fast (though velocities significantly above $\sim 1$ per cent of the speed of light require confirmation) and geometrically constrained, with strong observational evidence supporting a preferentially equatorial configuration. Structurally, the limited available studies point towards a multi-phase, clumpy nature with some degree of stratification. Last but not least, winds can be massive compared to the accreted material, with compelling evidence that the most extreme cases substantially impact both the accretion disc dynamics and the source's observational properties.

The third decade, that we are starting at the time of writing, will most likely bring new surprises that we are not yet able to anticipate. However, we should be able to, at the very least, explore the following topics:

\begin{itemize}[leftmargin=0cm, itemsep=1pt]

    \item \textbf{\textit{XRISM} as a cornerstone for multiwavelength studies at high velocity resolution:} The arrival of \textit{XRISM} is allowing, for the first time, systematic X-ray studies at high velocity resolution. Among other things, this is expected to provide new insights into the presence of different high ionisation components (e.g. the warm and hot phases discussed in this review). It will also allow us to study the actual shapes of the line profiles, which encode information on the geometry of the wind and perhaps the launching mechanisms. Crucially, it also enables the combination of these X-ray studies with optical, NIR and ultraviolet spectroscopy at similarly high resolution, shedding light on the actual structure of the wind.

    \item \textbf{The NIR and beyond in the era of the \textit{James Webb Space Telescope}:} As discussed in this review, the NIR has proven to be an exciting window for studying winds, as it has revealed rich phenomenology (e.g. Fig.~\ref{fig:NIR_lines}) and allows one to track outflows during most parts of the outburst. We are still in an exploratory phase for these type of studies, as NIR spectroscopic campaigns remain scarce. The arrival of new instrumentation such as the \textit{Son-of-X-shooter} spectrograph \citep{Schipani2018} on the \textit{NTT-ESO} telescope can significantly improve this situation by dramatically increasing the number of observations. In addition, the exceptional spectroscopic capabilities of the \textit{James Webb Space Telescope} will allow us to study winds in lines that are not accessible from ground-based observatories.
    
   \item \textbf{The role of disc winds in polarisation measurements:} The Imaging X-ray Polarimetry Explorer \citep{Weisskopf2022} is opening a new window by systematically performing polarisation studies of X-ray binaries and AGN. As a potential source of X-ray scattering, accretion disc winds may contribute to the high levels of polarised emission observed in these systems (see e.g.~\citealt{Nitindala2025}). This means we not only need to understand the impact of winds on polarisation measurements, but also that such measurements could, in turn, help reveal the presence and properties of the winds themselves (see e.g.~\citealt{Tomaru2024,Tomaru2026}).
 \item \textbf{New modelling and simulations:} The past decade has seen major advances in the modelling and numerical simulations of disc winds, mostly focused on the high-ionisation components. Further progress, enhanced by the expected growth in computational power, will likely include the incorporation of low-ionisation components (e.g. the \textsc{Sirocco} code; \citealt{Matthews2025}), allowing us to achieve a more complete theoretical understanding of the physical properties of wind-type outflows. This, in turn, will enable us to test key aspects such as the range of possible launching mechanisms and their impact on the accretion process and the immediate environment of X-ray binaries.
\end{itemize}

\smallskip

All the above underscores that disc winds will remain a key focus of observational and theoretical studies in the foreseeable future, as we work towards understanding their main properties and role in the fundamental process of mass accretion onto compact objects.

\bmhead{Acknowledgements}
We thank the referees for their careful reading and constructive comments, which helped improve this paper, and the International Space Science Institute in Bern for hosting a workshop on the first 50 years of research on accretion discs that led to this review. We also acknowledge Jean-Pierre Lasota and Maurizio Falanga, who contributed to this workshop and unfortunately passed away while this review was being written. T.M.D. thanks Alessandra Ambrifi and Montserrat Armas Padilla for useful comments and suggestions, and Gabriel P\'erez (IAC) for developing the sketch presented in Fig.~\ref{fig:sketch}. T.M.D. acknowledges support by the Spanish \textit{Agencia estatal de investigaci\'on} via PID2021-124879NB-I00 and PID2024-161863NB-I00. C.D. acknowledges support from STFC via grant ST/T000244/1. G.P. acknowledges financial support from the European Research Council (ERC) under the European Union’s Horizon 2020 research and innovation program HotMilk (grant agreement No. 865637) and from the Framework per \textit{l’Attrazione e il Rafforzamento delle Eccellenze (FARE) per la ricerca in Italia} (R20L5S39T9). R.T. acknowledges JSPS KAKENHI Grant Number JP24KJ0152.

\smallskip
\noindent \textbf{Declarations.}
The authors declare no competing interests.

\let\oldbibitem\bibitem
\renewcommand{\bibitem}{\vspace{-7.1pt}\oldbibitem}

\providecommand{\bibfont}{}
\renewcommand{\bibfont}{\footnotesize}

\bibliography{References_winds}

@Article{Herrero2012,
  author    = {Herrero, A. and Garcia, M. and Puls, J. and Uytterhoeven, K. and Najarro, F. and Lennon, D. J. and Rivero-González, J. G.},
  journal   = {Astronomy \& Astrophysics},
  title     = {A peculiar Of star in the Local Group galaxy IC~1613},
  year      = {2012},
  issn      = {1432-0746},
  month     = jul,
  pages     = {A85},
  volume    = {543},
  doi       = {10.1051/0004-6361/201118383},
  publisher = {EDP Sciences},
}

@Article{Drew1988,
  author    = {Drew, Janet E. and Verbunt, Frank},
  journal   = {Monthly Notices of the Royal Astronomical Society},
  title     = {Regular orbital variations in the ultraviolet resonance lines of {YZ} {Cnc}.},
  year      = {1988},
  issn      = {0035-8711},
  month     = sep,
  note      = {ADS Bibcode: 1988MNRAS.234..341D},
  pages     = {341--351},
  volume    = {234},
  doi       = {10.1093/mnras/234.2.341},
  publisher = {OUP},
  url       = {https://ui.adsabs.harvard.edu/abs/1988MNRAS.234..341D},
  urldate   = {2025-04-07},
}

@Article{Tomaru2026,
  author    = {Tomaru, Ryota and Done, Chris and Odaka, Hirokazu},
  journal   = {Monthly Notices of the Royal Astronomical Society},
  title     = {The detection of high {X}-ray polarization from an accretion disc corona source and its modelling via {Monte} {Carlo} radiation transfer simulation},
  year      = {2026},
  issn      = {0035-8711},
  month     = apr,
  note      = {ADS Bibcode: 2026MNRAS.547ag498T},
  pages     = {stag498},
  volume    = {547},
  doi       = {10.1093/mnras/stag498},
  publisher = {OUP},
  url       = {https://ui.adsabs.harvard.edu/abs/2026MNRAS.547ag498T},
  urldate   = {2026-04-29},
}

@Article{Prinja2000,
  author    = {Prinja, Raman K. and Ringwald, F. A. and Wade, Richard A. and Knigge, Christian},
  journal   = {Monthly Notices of the Royal Astronomical Society},
  title     = {{HST} ultraviolet observations of rapid variability in the accretion-disc wind of {BZ} {Cam}},
  year      = {2000},
  issn      = {0035-8711},
  month     = feb,
  note      = {ADS Bibcode: 2000MNRAS.312..316P},
  pages     = {316--326},
  volume    = {312},
  doi       = {10.1046/j.1365-8711.2000.03111.x},
  publisher = {OUP},
  url       = {https://ui.adsabs.harvard.edu/abs/2000MNRAS.312..316P},
  urldate   = {2025-04-07},
}

@ARTICLE{Brandt2000,
       author = {{Brandt}, W.~N. and {Schulz}, N.~S.},
        title = "{The Discovery of Broad P Cygni X-Ray Lines from Circinus X-1 with the Chandra High-Energy Transmission Grating Spectrometer}",
      journal = {\apjl},
         year = 2000,
        month = dec,
       volume = {544},
       number = {2},
        pages = {L123-L127},
          doi = {10.1086/317313},
archivePrefix = {arXiv},
       eprint = {astro-ph/0007406},
 primaryClass = {astro-ph},
       adsurl = {https://ui.adsabs.harvard.edu/abs/2000ApJ...544L.123B}
}

@Article{Ambrifi2025,
  author	= {Ambrifi, A. and Mata Sánchez, D. and Muñoz-Darias, T. and
		  Sánchez-Sierras, J. and Armas Padilla, M. and Baglio, M. C.
		  and Casares, J. and Corral-Santana, J. M. and Cúneo, V. A.
		  and Fender, R. P. and Ponti, G. and Russell, D. M. and
		  Shidatsu, M. and Steeghs, D. and Torres, M. A. P. and Ueda,
		  Y. and Vincentelli, F.},
  journal	= {Astronomy and Astrophysics},
  title		= {State-dependent signatures of jets and winds in the
		  optical and infrared spectrum of the black hole transient
		  {GX} 339?4},
  year		= {2025},
  issn		= {0004-6361},
  month		= feb,
  note		= {ADS Bibcode: 2025A\&A...694A.109A},
  pages		= {A109},
  volume	= {694},
  doi		= {10.1051/0004-6361/202451024},
  publisher	= {EDP},
  url		= {https://ui.adsabs.harvard.edu/abs/2025A&A...694A.109A},
  urldate	= {2025-02-17}
}

@Article{Armaspadilla2017,
  author	= {Armas Padilla, M. and Ueda, Y. and Hori, T. and Shidatsu,
		  M. and Mu{\~n}oz-Darias, T.},
  journal	= {\mnras},
  title		= {Suzaku spectroscopy of the neutron star transient 4U
		  1608-52 during its outburst decay.},
  year		= {2017},
  month		= may,
  pages		= {290-297},
  volume	= {467},
  archiveprefix	= {arXiv},
  doi		= {10.1093/mnras/stx020},
  eprint	= {1701.02728},
  primaryclass	= {astro-ph.HE},
  url		= {http://adsabs.harvard.edu/abs/2017MNRAS.467..290A}
}

@Article{Armaspadilla2023,
  author	= {Armas Padilla, M. and Corral-Santana, J. M. and Borghese,
		  A. and C{\'u}neo, V. A. and Mu{\~n}oz-Darias, T. and
		  Casares, J. and Torres, M. A. P.},
  journal	= {\aap},
  title		= {UltraCompCAT: A comprehensive catalogue of ultra-compact
		  and short orbital period X-ray binaries},
  year		= {2023},
  month		= sep,
  pages		= {A186},
  volume	= {677},
  archiveprefix	= {arXiv},
  doi		= {10.1051/0004-6361/202346797},
  eid		= {A186},
  eprint	= {2305.07691},
  primaryclass	= {astro-ph.HE},
  url		= {https://ui.adsabs.harvard.edu/abs/2023A&A...677A.186A}
}

@InCollection{Bahramian2023,
  author    = {Bahramian, Arash and Degenaar, Nathalie},
  booktitle = {Handbook of {X}-ray and {Gamma}-ray {Astrophysics}},
  title     = {Low-{Mass} {X}-ray {Binaries}},
  year      = {2023},
  month     = feb,
  note      = {ADS Bibcode: 2023hxga.book..120B},
  pages     = {120},
  doi       = {10.1007/978-981-16-4544-0_94-1},
  url       = {https://ui.adsabs.harvard.edu/abs/2023hxga.book..120B},
  urldate   = {2026-04-29},
}

@Article{Balbus1991,
  author	= {Balbus, Steven A. and Hawley, John F.},
  journal	= {\apj},
  title		= {A Powerful Local Shear Instability in Weakly Magnetized
		  Disks. I. Linear Analysis},
  year		= {1991},
  month		= {Jul},
  pages		= {214},
  volume	= {376},
  doi		= {10.1086/170270},
  url		= {https://ui.adsabs.harvard.edu/abs/1991ApJ...376..214B}
}

@Article{Bandyopadhyay1997,
  author	= {Bandyopadhyay, R. and Shahbaz, T. and Charles, P. A. and
		  van Kerkwijk, M. H. and Naylor, T.},
  journal	= {\mnras},
  title		= {Infrared spectroscopy of low-mass X-ray binaries},
  year		= {1997},
  month		= {Mar},
  number	= {4},
  pages		= {718-724},
  volume	= {285},
  doi		= {10.1093/mnras/285.4.718},
  url		= {https://ui.adsabs.harvard.edu/abs/1997MNRAS.285..718B}
}

@Article{Bandyopadhyay1999,
  author	= {{Bandyopadhyay}, R.~M. and {Shahbaz}, T. and {Charles},
		  P.~A. and {Naylor}, T.},
  journal	= {\mnras},
  month		= jun,
  pages		= {417-426},
  title		= {{Infrared spectroscopy of low-mass X-ray binaries - II}},
  volume	= {306},
  year		= {1999},
  doi		= {10.1046/j.1365-8711.1999.02547.x}
}

@Article{Beals1929,
  author	= {Beals, C. S.},
  journal	= {\mnras},
  month		= dec,
  pages		= {202-212},
  title		= {On the nature of Wolf-Rayet emission},
  volume	= {90},
  year		= {1929},
  doi		= {10.1093/mnras/90.2.202},
  url		= {http://adsabs.harvard.edu/abs/1929MNRAS..90..202B}
}

@Article{Beals1931,
  author	= {{Beals}, C.~S.},
  title		= {{The contours of emission bands in nov{\ae} and Wolf-Rayet
		  stars}},
  journal	= {\mnras},
  year		= {1931},
  volume	= {91},
  pages		= {966-977},
  month		= {Jun},
  adsurl	= {https://ui.adsabs.harvard.edu/abs/1931MNRAS..91..966B},
  doi		= {10.1093/mnras/91.9.966}
}

@Article{Beals1953,
  author	= {Beals, C. S.},
  journal	= {Publications of the Dominion Astrophysical Observatory
		  Victoria},
  title		= {The {Spectra} of the {P} {Cygni} {Stars}},
  year		= {1953},
  issn		= {0078-6950},
  month		= jan,
  note		= {ADS Bibcode: 1953PDAO....9....1B},
  pages		= {1},
  volume	= {9},
  url		= {https://ui.adsabs.harvard.edu/abs/1953PDAO....9....1B},
  urldate	= {2025-02-18}
}

@Article{Begelman1983,
  author	= {{Begelman}, M.~C. and {McKee}, C.~F. and {Shields},
		  G.~A.},
  title		= "{Compton heated winds and coronae above accretion disks.
		  I. Dynamics.}",
  journal	= {\apj},
  year		= 1983,
  month		= aug,
  volume	= {271},
  pages		= {70-88},
  doi		= {10.1086/161178},
  adsurl	= {https://ui.adsabs.harvard.edu/abs/1983ApJ...271...70B}
}

@Article{Begelman1983b,
  author	= {{Begelman}, M.~C. and {McKee}, C.~F.},
  title		= "{Compton heated winds and coronae above accretion disks.
		  II. Radiativetransfer and observable consequences.}",
  journal	= {\apj},
  year		= 1983,
  month		= aug,
  volume	= {271},
  pages		= {89-112},
  doi		= {10.1086/161179},
  adsurl	= {https://ui.adsabs.harvard.edu/abs/1983ApJ...271...89B}
}

@Article{Belloni2005,
  author	= {{Belloni}, T. and {Homan}, J. and {Casella}, P. and {van
		  der Klis}, M. and {Nespoli}, E. and {Lewin}, W.~H.~G. and
		  {Miller}, J.~M. and {M{\'e}ndez}, M.},
  title		= {{The evolution of the timing properties of the black-hole
		  transient GX 339-4 during its 2002/2003 outburst}},
  year		= {2005},
  volume	= {440},
  month		= sep,
  pages		= {207-222},
  doi		= {10.1051/0004-6361:20042457},
  eprint	= {arXiv:astro-ph/0504577},
  adsurl	= {http://ads.nao.ac.jp/abs/2005A%26A...440..207B},
  journal	= {\aap}
}

@Article{Belloni2010,
  author	= {{Belloni}, T.~M.},
  title		= {{States and Transitions in Black Hole Binaries}},
  year		= {2010},
  editor	= {{T.~Belloni}},
  series	= {Lecture Notes in Physics, Berlin Springer Verlag},
  volume	= {794},
  pages		= {53-+},
  doi		= {10.1007/978-3-540-76937-8_3},
  adsurl	= {http://adsabs.harvard.edu/abs/2010LNP...794...53B},
  booktitle	= {Lecture Notes in Physics, Berlin Springer Verlag},
  journal	= {Lecture Notes in Physics, Berlin Springer Verlag}
}

@Article{Bianchi2017,
  author	= {Bianchi, S. and Ponti, G. and Mu{\~n}oz-Darias, T. and
		  Petrucci, P.-O.},
  title		= {Photoionization instability of the Fe K absorbing plasma
		  in the neutron star transient AX J1745.6-2901},
  journal	= {\mnras},
  year		= {2017},
  volume	= {472},
  pages		= {2454-2461},
  month		= dec,
  archiveprefix	= {arXiv},
  doi		= {10.1093/mnras/stx2173},
  eprint	= {1709.00860},
  primaryclass	= {astro-ph.HE},
  url		= {http://adsabs.harvard.edu/abs/2017MNRAS.472.2454B}
}

@Article{Bozzo2016,
  author	= {Bozzo, E. and Pjanka, P. and Romano, P. and Papitto, A.
		  and Ferrigno, C. and Motta, S. and Zdziarski, A. A. and
		  Pintore, F. and Di Salvo, T. and Burderi, L. and Lazzati,
		  D. and Ponti, G. and Pavan, L.},
  journal	= {\aap},
  title		= {IGR J17451-3022: A dipping and eclipsing low mass X-ray
		  binary},
  year		= {2016},
  month		= may,
  pages		= {A42},
  volume	= {589},
  archiveprefix	= {arXiv},
  doi		= {10.1051/0004-6361/201527501},
  eid		= {A42},
  eprint	= {1603.03353},
  primaryclass	= {astro-ph.HE},
  url		= {https://ui.adsabs.harvard.edu/abs/2016A&A...589A..42B}
}

@Article{Buisson2021,
  author    = {Buisson, D J K and Altamirano, D and Armas Padilla, M and Arzoumanian, Z and Bult, P and Castro Segura, N and Charles, P A and Degenaar, N and Díaz Trigo, M and van den Eijnden, J and Fogantini, F and Gandhi, P and Gendreau, K and Hare, J and Homan, J and Knigge, C and Malacaria, C and Mendez, M and Muñoz Darias, T and Ng, M and Özbey Arabacı, M and Remillard, R and Strohmayer, T E and Tombesi, F and Tomsick, J A and Vincentelli, F and Walton, D J},
  journal   = {Monthly Notices of the Royal Astronomical Society},
  title     = {Dips and eclipses in the X-ray binary Swift J1858.6–0814 observed with NICER},
  year      = {2021},
  issn      = {1365-2966},
  month     = mar,
  number    = {4},
  pages     = {5600--5610},
  volume    = {503},
  doi       = {10.1093/mnras/stab863},
  publisher = {Oxford University Press (OUP)},
}

@Article{Casares1991,
  author	= {{Casares}, J. and {Charles}, P.~A. and {Jones}, D.~H.~P.
		  and {Rutten}, R.~G.~M. and {Callanan}, P.~J.},
  journal	= {\mnras},
  month		= jun,
  pages		= {712-725},
  title		= {{Optical studies of V404 Cyg, the X-ray transient GS2023 +
		  338. I - The 1989 outburst and decline}},
  volume	= {250},
  year		= {1991}
}

@Article{Casares2015,
  author	= {Casares, J.},
  journal	= {The Astrophysical Journal},
  title		= {A {FWHM}-{K2} {Correlation} in {Black} {Hole}
		  {Transients}},
  year		= {2015},
  issn		= {0004-637X},
  month		= jul,
  note		= {ADS Bibcode: 2015ApJ...808...80C},
  pages		= {80},
  volume	= {808},
  doi		= {10.1088/0004-637X/808/1/80},
  publisher	= {IOP},
  url		= {https://ui.adsabs.harvard.edu/abs/2015ApJ...808...80C},
  urldate	= {2025-02-18}
}

@Article{Casares2019,
  author	= {Casares, J. and Mu{\~n}oz-Darias, T. and Mata S{\'a}nchez,
		  D. and Charles, P. A. and Torres, M. A. P. and Armas
		  Padilla, M. and Fender, R. P. and Garc{\'\i}a-Rojas, J.},
  journal	= {\mnras},
  title		= {Accretion and outflow in V404 Cyg},
  year		= {2019},
  month		= sep,
  number	= {1},
  pages		= {1356-1365},
  volume	= {488},
  archiveprefix	= {arXiv},
  doi		= {10.1093/mnras/stz1793},
  eprint	= {1907.00005},
  primaryclass	= {astro-ph.HE},
  url		= {https://ui.adsabs.harvard.edu/abs/2019MNRAS.488.1356C}
}

@Article{Castor1970,
  author	= {Castor, J. I.},
  journal	= {Monthly Notices of the Royal Astronomical Society},
  title		= {Spectral line formation in {Wolf}-{Rayet} envelopes.},
  year		= {1970},
  issn		= {0035-8711},
  month		= jan,
  note		= {ADS Bibcode: 1970MNRAS.149..111C},
  pages		= {111},
  volume	= {149},
  doi		= {10.1093/mnras/149.2.111},
  publisher	= {OUP},
  url		= {https://ui.adsabs.harvard.edu/abs/1970MNRAS.149..111C},
  urldate	= {2025-02-17}
}

@Article{Castor1975,
  author	= {Castor, J. I. and Abbott, D. C. and Klein, R. I.},
  journal	= {The Astrophysical Journal},
  title		= {Radiation-driven winds in {Of} stars.},
  year		= {1975},
  issn		= {0004-637X},
  month		= jan,
  note		= {ADS Bibcode: 1975ApJ...195..157C},
  pages		= {157--174},
  volume	= {195},
  doi		= {10.1086/153315},
  publisher	= {IOP},
  url		= {https://ui.adsabs.harvard.edu/abs/1975ApJ...195..157C},
  urldate	= {2025-02-18}
}

@Article{Castor1979,
  author	= {Castor, J. I. and Lamers, H. J. G. L. M.},
  journal	= {\apjs},
  month		= apr,
  pages		= {481-511},
  title		= {An atlas of theoretical P Cygni profiles},
  volume	= {39},
  year		= {1979},
  doi		= {10.1086/190583},
  url		= {http://adsabs.harvard.edu/abs/1979ApJS...39..481C}
}

@Article{Castrosegura2022,
  author	= {Castro Segura, N. and Knigge, C. and Long, K. S. and
		  Altamirano, D. and Armas Padilla, M. and Bailyn, C. and
		  Buckley, D. A. H. and Buisson, D. J. K. and Casares, J. and
		  Charles, P. and Combi, J. A. and C{\'u}neo, V. A. and
		  Degenaar, N. D. and del Palacio, S. and D{\'\i}az Trigo, M.
		  and Fender, R. and Gandhi, P. and Georganti, M. and
		  Guti{\'e}rrez, C. and Hernandez Santisteban, J. V. and
		  Jim{\'e}nez-Ibarra, F. and Matthews, J. and M{\'e}ndez, M.
		  and Middleton, M. and Mu{\~n}oz-Darias, T. and {\"O}zbey
		  Arabac{\i}, M. and Pahari, M. and Rhodes, L. and Russell,
		  T. D. and Scaringi, S. and van den Eijnden, J. and
		  Vasilopoulos, G. and Vincentelli, F. M. and Wiseman, P.},
  journal	= {\nat},
  month		= mar,
  number	= {7899},
  pages		= {52-57},
  title		= {A persistent ultraviolet outflow from an accreting neutron
		  star binary transient},
  volume	= {603},
  year		= {2022},
  doi		= {10.1038/s41586-021-04324-2},
  url		= {https://ui.adsabs.harvard.edu/abs/2022Natur.603...52C}
}

@Article{Chakravorty2013,
  author	= {Chakravorty, S. and Lee, J. C. and Neilsen, J.},
  journal	= {\mnras},
  title		= {The effects of thermodynamic stability on wind properties
		  in different low-mass black hole binary states},
  year		= {2013},
  month		= nov,
  pages		= {560-569},
  volume	= {436},
  archiveprefix	= {arXiv},
  doi		= {10.1093/mnras/stt1593},
  eprint	= {1308.4574},
  primaryclass	= {astro-ph.HE},
  url		= {http://adsabs.harvard.edu/abs/2013MNRAS.436..560C}
}

@Article{Charles2019,
  author	= {{Charles}, Phil and {Matthews}, James H. and {Buckley},
		  David A.~H. and {Gandhi}, Poshak and {Kotze}, Enrico and
		  {Paice}, John},
  journal	= {\mnras},
  month		= {Oct},
  number	= {1},
  pages		= {L47-L52},
  title		= {{Hot, dense He II outflows during the 2017 outburst of the
		  X-ray transient Swift J1357.2-0933}},
  volume	= {489},
  year		= {2019},
  doi		= {10.1093/mnrasl/slz120}
}

@Article{Chaty2003,
  author	= {Chaty, S. and Charles, P. A. and Mart{\'{\i}}, J. and
		  Mirabel, I. F. and Rodr{\'{\i}}guez, L. F. and Shahbaz, T.},
  journal	= {\mnras},
  month		= jul,
  pages		= {169-174},
  title		= {Optical and near-infrared observations of the microquasar
		  V4641 Sgr during the 1999 September outburst},
  volume	= {343},
  year		= {2003},
  doi		= {10.1046/j.1365-8711.2003.06651.x},
  url		= {http://adsabs.harvard.edu/abs/2003MNRAS.343..169C}
}

@Article{Corral-santana2013,
  author	= {{Corral-Santana}, J.~M. and {Casares}, J. and
		  {Munoz-Darias}, T. and {Rodriguez-Gil}, P. and {Shahbaz},
		  T. and {Torres}, M.~A. and {Zurita}, C. and {Tyndall},
		  A.~A.},
  title		= {{A black hole nova obscured by an inner disk torus.}},
  year		= {2013},
  volume	= {339},
  pages		= {1048-1051},
  eprint	= {1303.0034},
  adsurl	= {http://adsabs.harvard.edu/abs/2013Sci...339.1048C},
  archiveprefix	= {arXiv},
  journal	= {Science},
  primaryclass	= {astro-ph.GA}
}

@Article{Cuneo2020,
  author	= {C{\'u}neo, V. A. and Mu{\~n}oz-Darias, T. and
		  S{\'a}nchez-Sierras, J. and Jim{\'e}nez-Ibarra, F. and
		  Armas Padilla, M. and Buckley, D. A. H. and Casares, J. and
		  Charles, P. and Corral-Santana, J. M. and Fender, R. and
		  Fern{\'a}ndez-Ontiveros, J. A. and Mata S{\'a}nchez, D. and
		  Panizo-Espinar, G. and Ponti, G. and Torres, M. A. P.},
  journal	= {\mnras},
  month		= aug,
  number	= {1},
  pages		= {25-32},
  title		= {Discovery of optical outflows and inflows in the black
		  hole candidate GRS 1716-249},
  volume	= {498},
  year		= {2020},
  doi		= {10.1093/mnras/staa2241},
  url		= {https://ui.adsabs.harvard.edu/abs/2020MNRAS.498...25C}
}

@Article{Degenaar2014,
  author	= {Degenaar, N. and Maitra, D. and Cackett, E. M. and
		  Reynolds, M. T. and Miller, J. M. and Reis, R. C. and King,
		  A. L. and G{\"u}ltekin, K. and Bailyn, C. D. and Buxton, M.
		  M. and MacDonald, R. K. D. and Fabian, A. C. and Fox, D. B.
		  and Rykoff, E. S.},
  journal	= {\apj},
  title		= {Multi-wavelength Coverage of State Transitions in the New
		  Black Hole X-Ray Binary Swift J1910.2-0546},
  year		= {2014},
  month		= apr,
  number	= {2},
  pages		= {122},
  volume	= {784},
  archiveprefix	= {arXiv},
  doi		= {10.1088/0004-637X/784/2/122},
  eid		= {122},
  eprint	= {1403.0939},
  primaryclass	= {astro-ph.HE},
  url		= {https://ui.adsabs.harvard.edu/abs/2014ApJ...784..122D}
}

@Article{DiazTrigo2006,
  author	= {{D{\'{\i}}az Trigo}, M. and {Parmar}, A.~N. and {Boirin},
		  L. and {M{\'e}ndez}, M. and {Kaastra}, J.~S.},
  journal	= {\aap},
  month		= jan,
  pages		= {179-195},
  title		= {{Spectral changes during dipping in low-mass X-ray
		  binaries due to highly-ionized absorbers}},
  volume	= {445},
  year		= {2006},
  doi		= {10.1051/0004-6361:20053586}
}

@Article{DiazTrigo2012,
  author	= {{D{\'\i}az Trigo}, M. and {Sidoli}, L. and {Boirin}, L.
		  and {Parmar}, A.~N.},
  journal	= {\aap},
  month		= jul,
  pages		= {A50},
  title		= {{XMM-Newton observations of GX 13 + 1: correlation between
		  photoionised absorption and broad line emission}},
  volume	= {543},
  year		= {2012},
  doi		= {10.1051/0004-6361/201219049},
  eid		= {A50}
}

@ARTICLE{DiazTrigo2014,
       author = {{D{\'\i}az Trigo}, M. and {Migliari}, S. and {Miller-Jones}, J.~C.~A. and {Guainazzi}, M.},
        title = "{XMM-Newton observations reveal the disappearance of the wind in 4U 1630-47}",
      journal = {\aap},
         year = 2014,
        month = nov,
       volume = {571},
          eid = {A76},
        pages = {A76},
          doi = {10.1051/0004-6361/201424554},
archivePrefix = {arXiv},
       eprint = {1409.3406},
 primaryClass = {astro-ph.HE},
       adsurl = {https://ui.adsabs.harvard.edu/abs/2014A&A...571A..76D}
}

@Article{DiazTrigo2016,
  author	= {D{\'{\i}}az Trigo, M. and Boirin, L.},
  title		= {Accretion disc atmospheres and winds in low-mass X-ray
		  binaries},
  journal	= {Astronomische Nachrichten},
  year		= {2016},
  volume	= {337},
  pages		= {368},
  month		= may,
  archiveprefix	= {arXiv},
  doi		= {10.1002/asna.201612315},
  eprint	= {1510.03576},
  primaryclass	= {astro-ph.HE},
  url		= {http://adsabs.harvard.edu/abs/2016AN....337..368D}
}

@Article{Done2007,
  author	= {Done, Chris and Gierlinski, Marek and Kubota, Aya},
  journal	= {\aapr},
  title		= {Modelling the behaviour of accretion flows in X-ray
		  binaries. Everything you always wanted to know about
		  accretion but were afraid to ask},
  year		= {2007},
  month		= {Dec},
  number	= {1},
  pages		= {1-66},
  volume	= {15},
  archiveprefix	= {arXiv},
  doi		= {10.1007/s00159-007-0006-1},
  eprint	= {0708.0148},
  primaryclass	= {astro-ph},
  url		= {https://ui.adsabs.harvard.edu/abs/2007A\&ARv..15....1D}
}

@Article{Dubus2001,
  author	= {Dubus, G. and Kim, R. S. J. and Menou, K. and Szkody, P.
		  and Bowen, D. V.},
  journal	= {\apj},
  title		= {Optical Spectroscopy of the X-Ray Transient XTE J1118+480
		  in Outburst},
  year		= {2001},
  month		= may,
  pages		= {307-320},
  volume	= {553},
  doi		= {10.1086/320648},
  eprint	= {astro-ph/0009148},
  url		= {http://adsabs.harvard.edu/abs/2001ApJ...553..307D}
}

@Article{Dubus2019,
  author	= {Guillaume Dubus and Chris Done and Bailey E. Tetarenko and
		  Jean-Marie Hameury},
  title		= {The impact of thermal winds on the outburst lightcurves of
		  black hole X-ray binaries},
  journal	= {Astronomy {\&} Astrophysics},
  year		= {2019},
  volume	= {632},
  pages		= {A40},
  month		= {nov},
  doi		= {10.1051/0004-6361/201936333},
  publisher	= {{EDP} Sciences}
}

@Article{Dunn2010,
  author	= {{Dunn}, R.~J.~H. and {Fender}, R.~P. and {K{\"o}rding},
		  E.~G. and {Belloni}, T. and {Cabanac}, C.},
  title		= {{A global spectral study of black hole X-ray binaries}},
  year		= {2010},
  volume	= {403},
  month		= mar,
  pages		= {61-82},
  doi		= {10.1111/j.1365-2966.2010.16114.x},
  eprint	= {0912.0142},
  adsurl	= {http://ads.nao.ac.jp/abs/2010MNRAS.403...61D},
  archiveprefix	= {arXiv},
  journal	= {\mnras},
  primaryclass	= {astro-ph.HE}
}

@Article{Fender2004,
  author	= {{Fender}, R.~P. and {Belloni}, T.~M. and {Gallo}, E.},
  title		= {{Towards a unified model for black hole X-ray binary
		  jets}},
  year		= {2004},
  volume	= {355},
  month		= dec,
  pages		= {1105-1118},
  doi		= {10.1111/j.1365-2966.2004.08384.x},
  eprint	= {arXiv:astro-ph/0409360},
  adsurl	= {http://ads.nao.ac.jp/abs/2004MNRAS.355.1105F},
  journal	= {\mnras}
}

@Article{Fender2012,
  author	= {{Fender}, R. and {Belloni}, T.},
  title		= {{Stellar-Mass Black Holes and Ultraluminous X-ray
		  Sources}},
  year		= {2012},
  volume	= {337},
  month		= aug,
  pages		= {540-},
  doi		= {10.1126/science.1221790},
  eprint	= {1208.1138},
  adsurl	= {http://ads.nao.ac.jp/abs/2012Sci...337..540F},
  archiveprefix	= {arXiv},
  journal	= {Science},
  primaryclass	= {astro-ph.HE}
}

@InProceedings{Fender2016,
  author	= {Fender, R. and Mu{\~n}oz-Darias, T.},
  title		= {The Balance of Power: Accretion and Feedback in Stellar
		  Mass Black Holes},
  booktitle	= {Lecture Notes in Physics, Berlin Springer Verlag},
  year		= {2016},
  editor	= {{Haardt}, F. and {Gorini}, V. and {Moschella}, U. and
		  {Treves}, A. and {Colpi}, M.},
  volume	= {905},
  series	= {Lecture Notes in Physics, Berlin Springer Verlag},
  pages		= {65},
  archiveprefix	= {arXiv},
  doi		= {10.1007/978-3-319-19416-5_3},
  eprint	= {1505.03526},
  primaryclass	= {astro-ph.HE},
  url		= {http://adsabs.harvard.edu/abs/2016LNP...905...65F}
}

@Article{Fijma2023,
  author	= {Fijma, S. and Castro Segura, N. and Degenaar, N. and
		  Knigge, C. and Higginbottom, N. and Hern{\'a}ndez
		  Santisteban, J.~V. and Maccarone, T.~J.},
  journal	= {\mnras},
  title		= {A transient ultraviolet outflow in the short-period X-ray
		  binary UW CrB},
  year		= {2023},
  month		= nov,
  number	= {1},
  pages		= {L149-L154},
  volume	= {526},
  archiveprefix	= {arXiv},
  doi		= {10.1093/mnrasl/slad125},
  eprint	= {2305.10793},
  primaryclass	= {astro-ph.HE},
  url		= {https://ui.adsabs.harvard.edu/abs/2023MNRAS.526L.149F}
}

@Book{Frank1992,
  author	= {{Frank}, J. and {King}, A. and {Raine}, D.},
  title		= {{Accretion Power in Astrophysics}},
  year		= {1992},
  publisher	= {Accretion Power in Astrophysics, ISBN 0521408636,
		  Cambridge University Press, 1992.},
  adsurl	= {http://adsabs.harvard.edu/abs/1992apa..book.....F}
}

@Article{Gatuzz2019,
  author	= {Gatuzz, E. and D{\'{\i}}az Trigo, M. and Miller-Jones, J.
		  C. A. and Migliari, S.},
  title		= {Chandra high-resolution spectra of 4U 1630-47: the
		  disappearance of the wind},
  journal	= {\mnras},
  year		= {2019},
  volume	= {482},
  pages		= {2597-2611},
  month		= jan,
  archiveprefix	= {arXiv},
  doi		= {10.1093/mnras/sty2850},
  eprint	= {1810.09464},
  primaryclass	= {astro-ph.HE},
  url		= {http://adsabs.harvard.edu/abs/2019MNRAS.482.2597G}
}

@InProceedings{Gilfanov2010,
  author	= {{Gilfanov}, M.},
  editor	= {{T.~Belloni}},
  title		= {{X-Ray Emission from Black-Hole Binaries}},
  booktitle	= {Lecture Notes in Physics, Berlin Springer Verlag},
  year		= {2010},
  volume	= {794},
  series	= {Lecture Notes in Physics, Berlin Springer Verlag},
  month		= mar,
  pages		= {17-+},
  doi		= {10.1007/978-3-540-76937-8_2},
  eprint	= {0909.2567},
  adsurl	= {http://ads.nao.ac.jp/abs/2010LNP...794...17G},
  archiveprefix	= {arXiv},
  primaryclass	= {astro-ph.HE}
}

@Article{Hasinger1989,
  author	= {{Hasinger}, G. and {van der Klis}, M.},
  title		= {{Two patterns of correlated X-ray timing and spectral
		  behaviour in low-mass X-ray binaries}},
  year		= {1989},
  volume	= {225},
  month		= nov,
  pages		= {79-96},
  adsurl	= {http://adsabs.harvard.edu/abs/1989A%26A...225...79H},
  journal	= {\aap}
}

@Article{Higginbottom2017,
  author	= {{Higginbottom}, N. and {Proga}, D. and {Knigge}, C. and
		  {Long}, K.~S.},
  title		= "{Thermal Disk Winds in X-Ray Binaries: Realistic Heating
		  and Cooling Rates Give Rise to Slow, but Massive,
		  Outflows}",
  journal	= {\apj},
  year		= 2017,
  month		= feb,
  volume	= {836},
  number	= {1},
  eid		= {42},
  pages		= {42},
  doi		= {10.3847/1538-4357/836/1/42},
  archiveprefix	= {arXiv},
  eprint	= {1612.08996},
  primaryclass	= {astro-ph.HE},
  adsurl	= {https://ui.adsabs.harvard.edu/abs/2017ApJ...836...42H}
}

@Article{Higginbottom2018,
  author	= {{Higginbottom}, Nick and {Knigge}, Christian and {Long},
		  Knox S. and {Matthews}, James H. and {Sim}, Stuart A. and
		  {Hewitt}, Henrietta A.},
  title		= "{Radiation-hydrodynamic simulations of thermally driven
		  disc winds in X-ray binaries: a direct comparison to GRO
		  J1655-40}",
  journal	= {\mnras},
  year		= 2018,
  month		= sep,
  volume	= {479},
  number	= {3},
  pages		= {3651-3662},
  doi		= {10.1093/mnras/sty1599},
  archiveprefix	= {arXiv},
  eprint	= {1806.04887},
  primaryclass	= {astro-ph.HE},
  adsurl	= {https://ui.adsabs.harvard.edu/abs/2018MNRAS.479.3651H}
}

@Article{Higginbottom2019,
  author	= {Higginbottom, Nick and Knigge, Christian and Long, Knox S.
		  and Matthews, James H. and Parkinson, Edward J.},
  journal	= {\mnras},
  title		= {The luminosity dependence of thermally driven disc winds
		  in low-mass X-ray binaries},
  year		= {2019},
  month		= {Apr},
  number	= {4},
  pages		= {4635-4644},
  volume	= {484},
  archiveprefix	= {arXiv},
  doi		= {10.1093/mnras/stz310},
  eprint	= {1901.09684},
  primaryclass	= {astro-ph.HE},
  url		= {https://ui.adsabs.harvard.edu/abs/2019MNRAS.484.4635H}
}

@Article{Higginbottom2020,
  author	= {Higginbottom, Nick and Knigge, Christian and Sim, Stuart
		  A. and Long, Knox S. and Matthews, James H. and Hewitt,
		  Henrietta A. and Parkinson, Edward J. and Mangham, Sam W.},
  journal	= {\mnras},
  title		= {Thermal and radiation driving can produce observable disc
		  winds in hard-state X-ray binaries},
  year		= {2020},
  month		= mar,
  number	= {4},
  pages		= {5271-5279},
  volume	= {492},
  archiveprefix	= {arXiv},
  doi		= {10.1093/mnras/staa209},
  eprint	= {2001.08547},
  primaryclass	= {astro-ph.HE},
  url		= {https://ui.adsabs.harvard.edu/abs/2020MNRAS.492.5271H}
}

@Article{Higginbottom2024,
  author	= {{Higginbottom}, Nick and {Scepi}, Nicolas and {Knigge},
		  Christian and {Long}, Knox S. and {Matthews}, James H. and
		  {Sim}, Stuart A.},
  title		= "{State-of-the-art simulations of line-driven accretion
		  disc winds: realistic radiation hydrodynamics leads to
		  weaker outflows}",
  journal	= {\mnras},
  year		= 2024,
  month		= jan,
  volume	= {527},
  number	= {3},
  pages		= {9236-9249},
  doi		= {10.1093/mnras/stad3830},
  archiveprefix	= {arXiv},
  eprint	= {2312.06042},
  primaryclass	= {astro-ph.HE},
  adsurl	= {https://ui.adsabs.harvard.edu/abs/2024MNRAS.527.9236H}
}

@Article{Higgionbottom2015,
  author	= {{Higginbottom}, Nick and {Proga}, Daniel},
  title		= "{Coronae and Winds from Irradiated Disks in X-Ray
		  Binaries}",
  journal	= {\apj},
  year		= 2015,
  month		= jul,
  volume	= {807},
  number	= {1},
  eid		= {107},
  pages		= {107},
  doi		= {10.1088/0004-637X/807/1/107},
  archiveprefix	= {arXiv},
  eprint	= {1504.03328},
  primaryclass	= {astro-ph.HE},
  adsurl	= {https://ui.adsabs.harvard.edu/abs/2015ApJ...807..107H}
}

@Article{Homan2001,
  author	= {{Homan}, J. and {Wijnands}, R. and {van der Klis}, M. and
		  {Belloni}, T. and {van Paradijs}, J. and {Klein-Wolt}, M.
		  and {Fender}, R. and {M\'endez}, M.},
  title		= {{Correlated X-Ray Spectral and Timing Behavior of the
		  Black Hole Candidate XTE J1550-564: A New Interpretation of
		  Black Hole States}},
  year		= {2001},
  volume	= {132},
  month		= feb,
  pages		= {377-402},
  doi		= {10.1086/318954},
  eprint	= {arXiv:astro-ph/0001163},
  adsurl	= {http://adsabs.harvard.edu/abs/2001ApJS..132..377H},
  journal	= {\apjs}
}

@Article{Homan2010,
  author	= {{Homan}, J. and {van der Klis}, M. and {Fridriksson},
		  J.~K. and {Remillard}, R.~A. and {Wijnands}, R. and
		  {M{\'e}ndez}, M. and {Lin}, D. and {Altamirano}, D. and
		  {Casella}, P. and {Belloni}, T.~M. and {Lewin}, W.~H.~G.},
  title		= {{XTE J1701-462 and Its Implications for the Nature of
		  Subclasses in Low-magnetic-field Neutron Star Low-mass
		  X-ray Binaries}},
  year		= {2010},
  volume	= {719},
  month		= aug,
  pages		= {201-212},
  doi		= {10.1088/0004-637X/719/1/201},
  eprint	= {1005.3210},
  adsurl	= {http://ads.nao.ac.jp/abs/2010ApJ...719..201H},
  archiveprefix	= {arXiv},
  journal	= {\apj},
  primaryclass	= {astro-ph.HE}
}

@Article{Homan2016,
  author	= {Homan, J. and Neilsen, J. and Allen, J. L. and
		  Chakrabarty, D. and Fender, R. and Fridriksson, J. K. and
		  Remillard, R. A. and Schulz, N.},
  journal	= {\apjl},
  month		= oct,
  pages		= {L5},
  title		= {Evidence for Simultaneous Jets and Disk Winds in Luminous
		  Low-mass X-Ray Binaries},
  volume	= {830},
  year		= {2016},
  doi		= {10.3847/2041-8205/830/1/L5},
  eid		= {L5},
  url		= {http://adsabs.harvard.edu/abs/2016ApJ...830L...5H}
}

@Article{Igo2020,
  author	= {Igo, Z. and Parker, M.~L. and Matzeu, G.~A. and Alston, W.
		  and Alvarez Crespo, N. and F{\"u}rst, F. and Buisson,
		  D.~J.~K. and Lobban, A. and Joyce, A.~M. and Mallick, L.
		  and Schartel, N. and Santos-Lle{\'o}, M.},
  journal	= {\mnras},
  title		= {Searching for ultra-fast outflows in AGN using variability
		  spectra},
  year		= {2020},
  month		= mar,
  number	= {1},
  pages		= {1088-1108},
  volume	= {493},
  archiveprefix	= {arXiv},
  doi		= {10.1093/mnras/staa265},
  eprint	= {2001.08208},
  primaryclass	= {astro-ph.HE},
  url		= {https://ui.adsabs.harvard.edu/abs/2020MNRAS.493.1088I}
}

@Article{Iijima2003,
  author	= {Iijima, T. and Esenoglu, H. H.},
  title		= {Spectral evolution of Nova (V1494) Aql and its high
		  velocity jets},
  journal	= {\aap},
  year		= {2003},
  volume	= {404},
  pages		= {997-1009},
  month		= jun,
  doi		= {10.1051/0004-6361:20030528},
  url		= {http://adsabs.harvard.edu/abs/2003A%26A...404..997I}
}

@Article{Jimenez-ibarra2019b,
  author	= {Jim{\'e}nez-Ibarra, F. and Mu{\~n}oz-Darias, T. and
		  Casares, J. and Armas Padilla, M. and Corral-Santana, J. M.},
  journal	= {\mnras},
  title		= {An equatorial outflow in the black hole optical dipper
		  Swift J1357.2-0933},
  year		= {2019},
  month		= nov,
  pages		= {3420-3426},
  volume	= {489},
  archiveprefix	= {arXiv},
  doi		= {10.1093/mnras/stz2393},
  eprint	= {1908.00356},
  primaryclass	= {astro-ph.HE},
  url		= {https://ui.adsabs.harvard.edu/abs/2019MNRAS.489.3420J}
}

@Article{Kafka2004,
  author	= {Kafka, S. and Honeycutt, R. K.},
  title		= {Detecting Outflows from Cataclysmic Variables in the
		  Optical},
  journal	= {\aj},
  year		= {2004},
  volume	= {128},
  pages		= {2420-2429},
  month		= nov,
  doi		= {10.1086/424618},
  url		= {http://adsabs.harvard.edu/abs/2004AJ....128.2420K}
}

@Article{Kallman1982,
  author	= {Kallman, T. R. and McCray, R.},
  journal	= {\apjs},
  title		= {X-ray nebular models},
  year		= {1982},
  month		= dec,
  pages		= {263-317},
  volume	= {50},
  doi		= {10.1086/190828},
  url		= {https://ui.adsabs.harvard.edu/abs/1982ApJS...50..263K}
}

@Article{King2015,
  author	= {{King}, A.~L. and {Miller}, J.~M. and {Raymond}, J. and
		  {Reynolds}, M.~T. and {Morningstar}, W.},
  journal	= {\apjl},
  title		= {{High-resolution Chandra HETG Spectroscopy of V404 Cygni
		  in Outburst}},
  year		= {2015},
  month		= nov,
  pages		= {L37},
  volume	= {813},
  adsurl	= {http://ads.nao.ac.jp/abs/2015ApJ...813L..37K},
  archiveprefix	= {arXiv},
  doi		= {10.1088/2041-8205/813/2/L37},
  eid		= {L37},
  eprint	= {1508.01181},
  primaryclass	= {astro-ph.HE}
}

@Article{King2015a,
  author	= {King, Andrew and Pounds, Ken},
  journal	= {\araa},
  title		= {Powerful Outflows and Feedback from Active Galactic
		  Nuclei},
  year		= {2015},
  month		= {Aug},
  pages		= {115-154},
  volume	= {53},
  archiveprefix	= {arXiv},
  doi		= {10.1146/annurev-astro-082214-122316},
  eprint	= {1503.05206},
  primaryclass	= {astro-ph.GA},
  url		= {https://ui.adsabs.harvard.edu/abs/2015ARA\&A..53..115K}
}

@Article{Koljonen2023,
  author        = {{Koljonen}, K.~I.~I. and {Long}, K.~S. and {Matthews}, J.~H. and {Knigge}, C.},
  journal       = {\mnras},
  title         = {{The origin of optical emission lines in the soft state of X-ray binary outbursts: the case of MAXI J1820+070}},
  year          = {2023},
  month         = may,
  number        = {3},
  pages         = {4190-4206},
  volume        = {521},
  adsurl        = {https://ui.adsabs.harvard.edu/abs/2023MNRAS.521.4190K},
  archiveprefix = {arXiv},
  doi           = {10.1093/mnras/stad809},
  eprint        = {2303.09242},
  primaryclass  = {astro-ph.HE},
}

@Article{Krolik1981,
  author	= {Krolik, J. H. and McKee, C. F. and Tarter, C. B.},
  journal	= {\apj},
  title		= {Two-phase models of quasar emission line regions.},
  year		= {1981},
  month		= oct,
  pages		= {422-442},
  volume	= {249},
  doi		= {10.1086/159303},
  url		= {https://ui.adsabs.harvard.edu/abs/1981ApJ...249..422K}
}

@Article{Kudritzki2000,
  author	= {Kudritzki, R.-P. and Puls, J.},
  title		= {Winds from Hot Stars},
  journal	= {\araa},
  year		= {2000},
  volume	= {38},
  pages		= {613-666},
  doi		= {10.1146/annurev.astro.38.1.613},
  url		= {http://adsabs.harvard.edu/abs/2000ARA%26A..38..613K}
}

@Article{Kuulkers1994,
  author	= {{Kuulkers}, E. and {van der Klis}, M. and {Oosterbroek},
		  T. and {Asai}, K. and {Dotani}, T. and {van Paradijs}, J.
		  and {Lewin}, W.~H.~G.},
  journal	= {\aap},
  title		= {{Spectral and correlated timing behaviour of GX 5-1}},
  year		= {1994},
  month		= sep,
  pages		= {795-821},
  volume	= {289},
  adsurl	= {http://adsabs.harvard.edu/abs/1994A%26A...289..795K}
}

@Article{Lasota2001,
  author	= {Lasota, J.-P.},
  title		= {The disc instability model of dwarf novae and low-mass
		  X-ray binary transients},
  journal	= {\nar},
  year		= {2001},
  volume	= {45},
  pages		= {449-508},
  month		= jun,
  doi		= {10.1016/S1387-6473(01)00112-9},
  eprint	= {astro-ph/0102072}
}

@Article{Lee2002,
  author	= {{Lee}, J.~C. and {Reynolds}, C.~S. and {Remillard}, R. and
		  {Schulz}, N.~S. and {Blackman}, E.~G. and {Fabian}, A.~C.},
  title		= {{High-Resolution Chandra HETGS and Rossi X-Ray Timing
		  Explorer Observations of GRS 1915+105: A Hot Disk
		  Atmosphere and Cold Gas Enriched in Iron and Silicon}},
  year		= {2002},
  volume	= {567},
  month		= mar,
  pages		= {1102-1111},
  doi		= {10.1086/338588},
  eprint	= {astro-ph/0208187},
  adsurl	= {http://ads.nao.ac.jp/abs/2002ApJ...567.1102L},
  journal	= {\apj}
}

@Article{Lin2007,
  author	= {{Lin}, D. and {Remillard}, R.~A. and {Homan}, J.},
  title		= {{Evaluating Spectral Models and the X-Ray States of
		  Neutron Star X-Ray Transients}},
  year		= {2007},
  volume	= {667},
  month		= oct,
  pages		= {1073-1086},
  doi		= {10.1086/521181},
  eprint	= {astro-ph/0702089},
  adsurl	= {http://ads.nao.ac.jp/abs/2007ApJ...667.1073L},
  journal	= {\apj}
}

@Article{Lindstrom2005,
  author	= {{Lindstr{\o}m}, C. and {Griffin}, J. and {Kiss}, L.~L. and
		  {Uemura}, M. and {Derekas}, A. and {M{\'e}sz{\'a}ros}, S.
		  and {Sz{\'e}kely}, P.},
  journal	= {\mnras},
  month		= nov,
  pages		= {882-890},
  title		= {{New clues on outburst mechanisms and improved
		  spectroscopic elements of the black hole binary V4641
		  Sagittarii$^{*}$}},
  volume	= {363},
  year		= {2005},
  doi		= {10.1111/j.1365-2966.2005.09483.x}
}

@Article{Marsh1988,
  author	= {{Marsh}, T.~R. and {Horne}, K.},
  journal	= {\mnras},
  month		= nov,
  pages		= {269-286},
  title		= {{Images of accretion discs. II - Doppler tomography}},
  volume	= {235},
  year		= {1988}
}

@Article{Matasanchez2015,
  author	= {Mata S{\'a}nchez, D. and Mu{\~n}oz-Darias, T. and Casares,
		  J. and Steeghs, D. and Ramos Almeida, C. and Acosta Pulido,
		  J. A.},
  title		= {Mass constraints to Sco X-1 from Bowen fluorescence and
		  deep near-infrared spectroscopy},
  journal	= {\mnras},
  year		= {2015},
  volume	= {449},
  pages		= {L1-L5},
  month		= apr,
  archiveprefix	= {arXiv},
  doi		= {10.1093/mnrasl/slv002},
  eprint	= {1501.02269},
  primaryclass	= {astro-ph.HE}
}

@Article{Matasanchez2018,
  author	= {Mata S{\'a}nchez, D. and Mu{\~n}oz-Darias, T. and Casares,
		  J. and Charles, P. A. and Armas Padilla, M. and
		  Fern{\'a}ndez-Ontiveros, J. A. and Jim{\'e}nez-Ibarra, F.
		  and Jonker, P. G. and Linares, M. and Torres, M. A. P. and
		  Shaw, A. W. and Rodr{\'{\i}}guez-Gil, P. and van Grunsven,
		  T. and Blay, P. and Caballero-Garc{\'{\i}}a, M. D. and
		  Castro-Tirado, A. and Chinchilla, P. and Farina, C. and
		  Ferragamo, A. and Lopez-Martinez, F. and Rubi{\~n}o-Martin,
		  J. A. and Su{\'a}rez-Andr{\'e}s, L.},
  journal	= {\mnras},
  month		= dec,
  pages		= {2646-2665},
  title		= {The 1989 and 2015 outbursts of V404 Cygni: a global study
		  of wind-related optical features},
  volume	= {481},
  year		= {2018},
  doi		= {10.1093/mnras/sty2402},
  url		= {http://adsabs.harvard.edu/abs/2018MNRAS.481.2646M}
}

@Article{Matasanchez2022,
  author	= {Mata S{\'a}nchez, D. and Mu{\~n}oz-Darias, T. and
		  C{\'u}neo, V. A. and Armas Padilla, M. and
		  S{\'a}nchez-Sierras, J. and Panizo-Espinar, G. and Casares,
		  J. and Corral-Santana, J. M. and Torres, M. A. P.},
  journal	= {\apjl},
  month		= feb,
  number	= {2},
  pages		= {L10},
  title		= {Hard-state Optical Wind during the Discovery Outburst of
		  the Black Hole X-Ray Dipper MAXI J1803-298},
  volume	= {926},
  year		= {2022},
  doi		= {10.3847/2041-8213/ac502f},
  eid		= {L10},
  url		= {https://ui.adsabs.harvard.edu/abs/2022ApJ...926L..10M}
}

@Article{Matasanchez2023,
  author	= {Mata S{\'a}nchez, D. and Mu{\~n}oz-Darias, T. and Casares,
		  J. and Huertas-Company, M. and Panizo-Espinar, G.},
  journal	= {\mnras},
  title		= {Ask the machine: systematic detection of wind-type
		  outflows in low-mass X-ray binaries},
  year		= {2023},
  month		= sep,
  number	= {1},
  pages		= {338-350},
  volume	= {524},
  archiveprefix	= {arXiv},
  doi		= {10.1093/mnras/stad1895},
  eprint	= {2306.12475},
  primaryclass	= {astro-ph.HE},
  url		= {https://ui.adsabs.harvard.edu/abs/2023MNRAS.524..338M}
}

@Article{Matasanchez2024a,
  author	= {Mata S{\'a}nchez, D. and Mu{\~n}oz-Darias, T. and Armas
		  Padilla, M. and Casares, J. and Torres, M.~A.~P.},
  journal	= {\aap},
  month		= feb,
  pages		= {L1},
  title		= {Evidence for inflows and outflows in the nearby black hole
		  transient Swift J1727.8{\ensuremath{-}}162},
  volume	= {682},
  year		= {2024},
  doi		= {10.1051/0004-6361/202348754},
  eid		= {L1},
  url		= {https://ui.adsabs.harvard.edu/abs/2024A&A...682L...1M}
}

@Article{Matthews2015,
  author	= {Matthews, J. H. and Knigge, C. and Long, K. S. and Sim, S.
		  A. and Higginbottom, N.},
  journal	= {\mnras},
  title		= {The impact of accretion disc winds on the optical spectra
		  of cataclysmic variables},
  year		= {2015},
  month		= jul,
  pages		= {3331-3344},
  volume	= {450},
  archiveprefix	= {arXiv},
  doi		= {10.1093/mnras/stv867},
  eprint	= {1504.05590},
  primaryclass	= {astro-ph.SR},
  url		= {http://adsabs.harvard.edu/abs/2015MNRAS.450.3331M}
}

@Article{Maury1897,
  author	= {{Maury}, Antonia C. and {Pickering}, Edward C.},
  journal	= {Annals of Harvard College Observatory},
  month		= jan,
  pages		= {1-128},
  title		= {{Spectra of bright stars photographed with the 11-inch
		  Draper Telescope as part of the Henry Draper Memorial.}},
  volume	= {28},
  year		= {1897}
}

@Article{Migliari2006,
  author	= {{Migliari}, S. and {Fender}, R.~P.},
  title		= {{Jets in neutron star X-ray binaries: a comparison with
		  black holes}},
  year		= {2006},
  volume	= {366},
  month		= feb,
  pages		= {79-91},
  doi		= {10.1111/j.1365-2966.2005.09777.x},
  eprint	= {astro-ph/0510698},
  adsurl	= {http://ads.nao.ac.jp/abs/2006MNRAS.366...79M},
  journal	= {\mnras}
}

@Article{Miller2006,
  author	= {Miller, J. M. and Raymond, J. and Fabian, A. and Steeghs,
		  D. and Homan, J. and Reynolds, C. and van der Klis, M. and
		  Wijnands, R.},
  title		= {The magnetic nature of disk accretion onto black holes},
  journal	= {\nat},
  year		= {2006},
  volume	= {441},
  pages		= {953-955},
  month		= jun,
  doi		= {10.1038/nature04912},
  eprint	= {astro-ph/0605390}
}

@Article{Miller2011,
  author	= {Miller, Jon M. and Maitra, Dipankar and Cackett, Edward M.
		  and Bhattacharyya, Sudip and Strohmayer, Tod E.},
  journal	= {\apjl},
  title		= {A Fast X-ray Disk Wind in the Transient Pulsar IGR
		  J17480-2446 in Terzan 5},
  year		= {2011},
  month		= apr,
  number	= {1},
  pages		= {L7},
  volume	= {731},
  archiveprefix	= {arXiv},
  doi		= {10.1088/2041-8205/731/1/L7},
  eid		= {L7},
  eprint	= {1101.2377},
  primaryclass	= {astro-ph.HE},
  url		= {https://ui.adsabs.harvard.edu/abs/2011ApJ...731L...7M}
}

@Article{Miller2014,
  author	= {Miller, J. M. and Raymond, J. and Kallman, T. R. and
		  Maitra, D. and Fabian, A. C. and Proga, D. and Reynolds, C.
		  S. and Reynolds, M. T. and Degenaar, N. and King, A. L. and
		  Cackett, E. M. and Kennea, J. A. and Beardmore, A.},
  journal	= {\apj},
  title		= {Chandra Spectroscopy of MAXI J1305-704: Detection of an
		  Infalling Black Hole Disk Wind?},
  year		= {2014},
  month		= jun,
  pages		= {53},
  volume	= {788},
  archiveprefix	= {arXiv},
  doi		= {10.1088/0004-637X/788/1/53},
  eid		= {53},
  eprint	= {1306.2915},
  primaryclass	= {astro-ph.HE},
  url		= {https://ui.adsabs.harvard.edu/abs/2014ApJ...788...53M}
}

@Article{Miller2020,
  author	= {Miller, J.~M. and Zoghbi, A. and Raymond, J. and
		  Balakrishnan, M. and Brenneman, L. and Cackett, E. and
		  Draghis, P. and Fabian, A.~C. and Gallo, E. and Kaastra, J.
		  and Kallman, T. and Kammoun, E. and Motta, S.~E. and Proga,
		  D. and Reynolds, M.~T. and Trueba, N.},
  journal	= {\apj},
  title		= {An Obscured, Seyfert 2-like State of the Stellar-mass
		  Black Hole GRS 1915+105 Caused by Failed Disk Winds},
  year		= {2020},
  month		= nov,
  number	= {1},
  pages		= {30},
  volume	= {904},
  archiveprefix	= {arXiv},
  doi		= {10.3847/1538-4357/abbb31},
  eid		= {30},
  eprint	= {2007.07005},
  primaryclass	= {astro-ph.HE},
  url		= {https://ui.adsabs.harvard.edu/abs/2020ApJ...904...30M}
}

@Article{Mirabel1999,
  author	= {Mirabel, I. F. and Rodriguez, L. F.},
  journal	= {Ann. Rev. Astron. Astrophys.},
  title		= {{Sources of relativistic jets in the galaxy}},
  year		= {1999},
  pages		= {409-443},
  volume	= {37},
  archiveprefix	= {arXiv},
  doi		= {10.1146/annurev.astro.37.1.409},
  eprint	= {astro-ph/9902062},
  primaryclass	= {astro-ph},
  slaccitation	= {%%CITATION = ASTRO-PH/9902062;%%}
}

@Article{Miyamoto1992,
  author	= {{Miyamoto}, S. and {Kitamoto}, S. and {Iga}, S. and
		  {Negoro}, H. and {Terada}, K.},
  title		= {{Canonical time variations of X-rays from black hole
		  candidates in the low-intensity state}},
  year		= {1992},
  volume	= {391},
  month		= may,
  pages		= {L21-L24},
  doi		= {10.1086/186389},
  adsurl	= {http://ads.nao.ac.jp/abs/1992ApJ...391L..21M},
  journal	= {\apjl}
}

@Article{Miyamoto1993,
  author	= {{Miyamoto}, S. and {Iga}, S. and {Kitamoto}, S. and
		  {Kamado}, Y.},
  journal	= {\apjl},
  title		= {{Another canonical time variation of X-rays from black
		  hole candidates in the very high flare state?}},
  year		= {1993},
  month		= jan,
  pages		= {L39-L42},
  volume	= {403},
  adsurl	= {http://ads.nao.ac.jp/abs/1993ApJ...403L..39M},
  doi		= {10.1086/186716}
}

@Article{Motta2017b,
  author	= {Motta, S. E. and Kajava, J. J. E. and
		  S{\'a}nchez-Fern{\'a}ndez, C. and Giustini, M. and
		  Kuulkers, E.},
  journal	= {\mnras},
  title		= {The black hole binary V404 Cygni: a highly accreting
		  obscured AGN analogue},
  year		= {2017},
  month		= jun,
  pages		= {981-993},
  volume	= {468},
  archiveprefix	= {arXiv},
  doi		= {10.1093/mnras/stx466},
  eprint	= {1607.02255},
  primaryclass	= {astro-ph.HE},
  url		= {http://adsabs.harvard.edu/abs/2017MNRAS.468..981M}
}

@Article{Motta2021,
  author	= {Motta, S. E. and {Kajava}, J.~J.~E. and {Giustini}, M.
		  and {Williams}, D.~R.~A. and {Del Santo}, M. and {Fender},
		  R. and {Green}, D.~A. and {Heywood}, I. and {Rhodes}, L.
		  and {Segreto}, A. and {Sivakoff}, G. and {Woudt}, P.~A.},
  journal	= {\mnras},
  title		= {{Observations of a radio-bright, X-ray obscured GRS
		  1915+105}},
  year		= {2021},
  month		= may,
  number	= {1},
  pages		= {152-161},
  volume	= {503},
  adsurl	= {https://ui.adsabs.harvard.edu/abs/2021MNRAS.503..152M},
  archiveprefix	= {arXiv},
  doi		= {10.1093/mnras/stab511},
  eprint	= {2101.01187},
  primaryclass	= {astro-ph.HE}
}

@Article{Munoz-darias2011,
  author	= {{Mu{\~n}oz-Darias}, T. and {Motta}, S. and {Belloni},
		  T.~M.},
  title		= {{Fast variability as a tracer of accretion regimes in
		  black hole transients}},
  journal	= {\mnras},
  year		= {2011},
  volume	= {410},
  pages		= {679-684},
  month		= jan,
  adsurl	= {http://ads.nao.ac.jp/abs/2011MNRAS.410..679M},
  archiveprefix	= {arXiv},
  doi		= {10.1111/j.1365-2966.2010.17476.x},
  eprint	= {1008.0558},
  primaryclass	= {astro-ph.HE}
}

@Article{Munoz-darias2013b,
  author	= {{Mu{\~n}oz-Darias}, T. and {Coriat}, M. and {Plant}, D.~S.
		  and {Ponti}, G. and {Fender}, R.~P. and {Dunn}, R.~J.~H.},
  title		= {{Inclination and relativistic effects in the outburst
		  evolution of black hole transients}},
  journal	= {\mnras},
  year		= {2013},
  volume	= {432},
  pages		= {1330-1337},
  month		= jun,
  adsurl	= {http://adsabs.harvard.edu/abs/2013MNRAS.432.1330M},
  archiveprefix	= {arXiv},
  doi		= {10.1093/mnras/stt546},
  eprint	= {1304.2072},
  primaryclass	= {astro-ph.HE}
}

@Article{Munoz-darias2014,
  author	= {{Mu{\~n}oz-Darias}, T. and {Fender}, R.~P. and {Motta},
		  S.~E. and {Belloni}, T.~M.},
  title		= {{Black hole-like hysteresis and accretion states in
		  neutron star low-mass X-ray binaries}},
  journal	= {\mnras},
  year		= {2014},
  volume	= {443},
  pages		= {3270-3283},
  month		= oct,
  adsurl	= {http://adsabs.harvard.edu/abs/2014MNRAS.443.3270M},
  archiveprefix	= {arXiv},
  doi		= {10.1093/mnras/stu1334},
  eprint	= {1407.1318},
  primaryclass	= {astro-ph.HE}
}

@Article{Munoz-darias2016,
  author	= {Mu{\~n}oz-Darias, T. and {Casares}, J. and {Mata
		  S{\'a}nchez}, D. and {Fender}, R.~P. and {Armas Padilla},
		  M. and {Linares}, M. and {Ponti}, G. and {Charles}, P.~A.
		  and {Mooley}, K.~P. and {Rodriguez}, J.},
  journal	= {\nat},
  month		= jun,
  pages		= {75-78},
  title		= {{Regulation of black-hole accretion by a disk wind during
		  a violent outburst of V404 Cygni}},
  volume	= {534},
  year		= {2016},
  doi		= {10.1038/nature17446}
}

@Article{Munoz-darias2017,
  author	= {Mu{\~n}oz-Darias, T. and Casares, J. and Mata S{\'a}nchez,
		  D. and Fender, R. P. and Armas Padilla, M. and Mooley, K.
		  and Hardy, L. and Charles, P. A. and Ponti, G. and Motta,
		  S. E. and Dhillon, V. S. and Gandhi, P. and
		  Jim{\'e}nez-Ibarra, F. and Butterley, T. and Carey, S. and
		  Grainge, K. J. B. and Hickish, J. and Littlefair, S. P. and
		  Perrott, Y. C. and Razavi-Ghods, N. and Rumsey, C. and
		  Scaife, A. M. M. and Scott, P. F. and Titterington, D. J.
		  and Wilson, R. W.},
  journal	= {\mnras},
  month		= feb,
  pages		= {L124-L128},
  title		= {Flares, wind and nebulae: the 2015 December mini-outburst
		  of V404 Cygni},
  volume	= {465},
  year		= {2017},
  doi		= {10.1093/mnrasl/slw222}
}

@Article{Munoz-darias2018,
  author	= {Mu{\~n}oz-Darias, T. and Torres, M. A. P. and Garcia, M.
		  R.},
  journal	= {\mnras},
  month		= sep,
  pages		= {3987-3995},
  title		= {The low-luminosity accretion disc wind of the black hole
		  transient V4641 Sagittarii},
  volume	= {479},
  year		= {2018},
  doi		= {10.1093/mnras/sty1711},
  url		= {http://adsabs.harvard.edu/abs/2018MNRAS.479.3987M}
}

@Article{Munoz-darias2019,
  author	= {Mu{\~n}oz-Darias, T. and Jim{\'e}nez-Ibarra, F. and
		  Panizo-Espinar, G. and Casares, J. and Mata S{\'a}nchez, D.
		  and Ponti, G. and Fender, R. P. and Buckley, D. A. H. and
		  Garnavich, P. and Torres, M. A. P. and Armas Padilla, M.
		  and Charles, P. A. and Corral-Santana, J. M. and Kajava, J.
		  J. E. and Kotze, E. J. and Littlefield, C. and
		  S{\'a}nchez-Sierras, J. and Steeghs, D. and Thomas, J.},
  journal	= {\apjl},
  month		= jul,
  pages		= {L4},
  title		= {Hard-state Accretion Disk Winds from Black Holes: The
		  Revealing Case of MAXI J1820+070},
  volume	= {879},
  year		= {2019},
  doi		= {10.3847/2041-8213/ab2768},
  eid		= {L4},
  url		= {https://ui.adsabs.harvard.edu/abs/2019ApJ...879L...4M}
}

@Article{Munoz-darias2020,
  author	= {Mu{\~n}oz-Darias, T. and Armas Padilla, M. and
		  Jim{\'e}nez-Ibarra, F. and Panizo-Espinar, G. and Casares,
		  J. and Altamirano, D. and Buisson, D. J. K. and Castro
		  Segura, N. and C{\'u}neo, V. A. and Degenaar, N. and
		  Fogantini, F. A. and Knigge, C. and Mata S{\'a}nchez, D.
		  and {\"O}zbey Arabaci, M. and S{\'a}nchez-Sierras, J. and
		  Torres, M. A. P. and van den Eijnden, J. and Vincentelli, F. M.},
  journal	= {\apjl},
  month		= apr,
  number	= {1},
  pages		= {L19},
  title		= {The Changing-look Optical Wind of the Flaring X-Ray
		  Transient Swift J1858.6-0814},
  volume	= {893},
  year		= {2020},
  doi		= {10.3847/2041-8213/ab8381},
  eid		= {L19},
  url		= {https://ui.adsabs.harvard.edu/abs/2020ApJ...893L..19M}
}

@Article{Munoz-darias2022,
  author	= {Mu{\~n}oz-Darias, Teo and Ponti, Gabriele},
  journal	= {\aap},
  title		= {Simultaneous X-ray and optical spectroscopy of V404 Cygni
		  supports the multi-phase nature of X-ray binary accretion
		  disc winds},
  year		= {2022},
  month		= aug,
  pages		= {A104},
  volume	= {664},
  archiveprefix	= {arXiv},
  doi		= {10.1051/0004-6361/202243769},
  eid		= {A104},
  eprint	= {2205.14162},
  primaryclass	= {astro-ph.HE},
  url		= {https://ui.adsabs.harvard.edu/abs/2022A&A...664A.104M}
}

@Article{Murray1996,
  author	= {Murray, N. and Chiang, J.},
  journal	= {\nat},
  title		= {Wind-dominated optical line emission from accretion disks
		  around luminous cataclysmic variable stars},
  year		= {1996},
  month		= aug,
  number	= {6594},
  pages		= {789-791},
  volume	= {382},
  doi		= {10.1038/382789a0},
  url		= {https://ui.adsabs.harvard.edu/abs/1996Natur.382..789M}
}

@Article{Neilsen2009,
  author	= {{Neilsen}, J. and {Lee}, J.~C.},
  title		= {{Accretion disk winds as the jet suppression mechanism in
		  the microquasar GRS 1915+105}},
  journal	= {\nat},
  year		= {2009},
  volume	= {458},
  pages		= {481-484},
  month		= mar,
  adsurl	= {http://ads.nao.ac.jp/abs/2009Natur.458..481N},
  doi		= {10.1038/nature07680}
}

@Article{Neilsen2011,
  author	= {{Neilsen}, J. and {Remillard}, R.~A. and {Lee}, J.~C.},
  title		= {{The Physics of the ''Heartbeat'' State of GRS 1915+105}},
  year		= {2011},
  volume	= {737},
  eid		= {69},
  month		= aug,
  pages		= {69},
  doi		= {10.1088/0004-637X/737/2/69},
  eprint	= {1106.0298},
  adsurl	= {http://ads.nao.ac.jp/abs/2011ApJ...737...69N},
  archiveprefix	= {arXiv},
  journal	= {\apj},
  primaryclass	= {astro-ph.HE}
}

@Article{Paice2018,
  author	= {{Paice}, J.~A. and {Gandhi}, P. and {Dhillon}, V.~S. and
		  {Marsh}, T.~R. and {Green}, M. and {Breedt}, E.},
  title		= {{Blue Oscillations and Rapid Red Flares in Swift
		  J1858.6-0814 Observed with ULTRACAM/NTT}},
  journal	= {The Astronomer's Telegram},
  year		= {2018},
  volume	= {12197},
  pages		= {1},
  month		= {Nov},
  adsurl	= {https://ui.adsabs.harvard.edu/abs/2018ATel12197....1P}
}

@Article{Panizoespinar2021,
  author	= {Panizo-Espinar, G. and Mu{\~n}oz-Darias, T. and Armas
		  Padilla, M. and Jim{\'e}nez-Ibarra, F. and Casares, J. and
		  Mata S{\'a}nchez, D.},
  journal	= {\aap},
  month		= jun,
  pages		= {A135},
  title		= {Optical nebular emission following the most luminous
		  outburst of Aquila X-1},
  volume	= {650},
  year		= {2021},
  doi		= {10.1051/0004-6361/202140323},
  eid		= {A135},
  url		= {https://ui.adsabs.harvard.edu/abs/2021A&A...650A.135P}
}

@Article{Panizoespinar2022,
  author	= {Panizo-Espinar, G. and Armas Padilla, M. and
		  Mu{\~n}oz-Darias, T. and Koljonen, K. I. I. and C{\'u}neo,
		  V. A. and S{\'a}nchez-Sierras, J. and Mata S{\'a}nchez, D.
		  and Casares, J. and Corral-Santana, J. and Fender, R. P.
		  and Jim{\'e}nez-Ibarra, F. and Ponti, G. and Steeghs, D.
		  and Torres, M. A. P.},
  journal	= {\aap},
  month		= aug,
  pages		= {A100},
  title		= {Discovery of optical and infrared accretion disc wind
		  signatures in the black hole candidate MAXI J1348-630},
  volume	= {664},
  year		= {2022},
  doi		= {10.1051/0004-6361/202243426},
  eid		= {A100},
  url		= {https://ui.adsabs.harvard.edu/abs/2022A&A...664A.100P}
}

@Article{Panizoespinar2024,
  author	= {Panizo-Espinar, G. and Mu{\~n}oz-Darias, T. and Armas
		  Padilla, M. and Jim{\'e}nez-Ibarra, F. and Mata
		  S{\'a}nchez, D. and Yanes-Rizo, I.~V. and Alabarta, K. and
		  Baglio, M.~C. and Caruso, E. and Casares, J. and
		  Corral-Santana, J.~M. and Lewis, F. and Russell, D.~M. and
		  Saikia, P. and S{\'a}nchez-Sierras, J. and Shahbaz, T. and
		  Torres, M.~A.~P. and Vincentelli, F.},
  journal	= {\aap},
  title		= {The omnipresent flux-dependent optical dips of the black
		  hole transient Swift J1357.2{\ensuremath{-}}0933},
  year		= {2024},
  month		= feb,
  pages		= {A19},
  volume	= {682},
  archiveprefix	= {arXiv},
  doi		= {10.1051/0004-6361/202347955},
  eid		= {A19},
  eprint	= {2311.03460},
  primaryclass	= {astro-ph.HE},
  url		= {https://ui.adsabs.harvard.edu/abs/2024A&A...682A..19P}
}

@Article{Parra2024,
  author	= {Parra, M. and Petrucci, P. O. and Bianchi, S. and
		  Gianolli, V. E. and Ursini, F. and Ponti, G.},
  journal	= {\aap},
  title		= {The current state of disk wind observations in BHLMXBs
		  through X-ray absorption lines in the iron band},
  year		= {2024},
  month		= jan,
  pages		= {A49},
  volume	= {681},
  archiveprefix	= {arXiv},
  doi		= {10.1051/0004-6361/202346920},
  eid		= {A49},
  eprint	= {2308.00691},
  primaryclass	= {astro-ph.HE},
  url		= {https://ui.adsabs.harvard.edu/abs/2024A&A...681A..49P}
}

@Article{Payne1930,
  author	= {Payne, Cecilia H.},
  journal	= {Harvard College Observatory Bulletin},
  title		= {On the {Spectra} of the {Wolf}-{Rayet} {Stars}},
  year		= {1930},
  issn		= {0891-3943},
  month		= mar,
  note		= {ADS Bibcode: 1930BHarO.874...23P},
  pages		= {23--30},
  volume	= {874},
  url		= {https://ui.adsabs.harvard.edu/abs/1930BHarO.874...23P},
  urldate	= {2025-02-21}
}

@Article{Ponti2012,
  author	= {{Ponti}, G. and {Fender}, R.~P. and {Begelman}, M.~C. and
		  {Dunn}, R.~J.~H. and {Neilsen}, J. and {Coriat}, M.},
  journal	= {\mnras},
  month		= may,
  pages		= {L11},
  title		= {{Ubiquitous equatorial accretion disc winds in black hole
		  soft states}},
  volume	= {422},
  year		= {2012},
  doi		= {10.1111/j.1745-3933.2012.01224.x}
}

@Article{Ponti2014,
  author	= {{Ponti}, G. and {Mu{\~n}oz-Darias}, T. and {Fender},
		  R.~P.},
  title		= {{A connection between accretion state and Fe K absorption
		  in an accreting neutron star: black hole-like soft-state
		  winds?}},
  year		= {2014},
  volume	= {444},
  month		= oct,
  pages		= {1829-1834},
  doi		= {10.1093/mnras/stu1742},
  eprint	= {1407.4468},
  adsurl	= {http://ads.nao.ac.jp/abs/2014MNRAS.444.1829P},
  archiveprefix	= {arXiv},
  journal	= {\mnras},
  primaryclass	= {astro-ph.HE}
}

@Article{Ponti2015,
  author	= {{Ponti}, G. and {Bianchi}, S. and {Mu{\~n}oz-Darias}, T.
		  and {De Marco}, B. and {Dwelly}, T. and {Fender}, R.~P. and
		  {Nandra}, K. and {Rea}, N. and {Mori}, K. and {Haggard}, D.
		  and {Heinke}, C.~O. and {Degenaar}, N. and {Aramaki}, T.
		  and {Clavel}, M. and {Goldwurm}, A. and {Hailey}, C.~J. and
		  {Israel}, G.~L. and {Morris}, M.~R. and {Rushton}, A. and
		  {Terrier}, R.},
  title		= {{On the Fe K absorption - accretion state connection in
		  the Galactic Centre neutron star X-ray binary AX
		  J1745.6-2901}},
  journal	= {\mnras},
  year		= {2015},
  volume	= {446},
  pages		= {1536-1550},
  month		= jan,
  adsurl	= {http://adsabs.harvard.edu/abs/2015MNRAS.446.1536P},
  archiveprefix	= {arXiv},
  doi		= {10.1093/mnras/stu1853},
  eprint	= {1409.3224},
  primaryclass	= {astro-ph.HE}
}

@Article{Ponti2016,
  author	= {Ponti, G. and Bianchi, S. and Mu{\~n}oz-Darias, T. and De,
		  K. and Fender, R. and Merloni, A.},
  journal	= {Astronomische Nachrichten},
  title		= {High ionisation absorption in low mass X-ray binaries},
  year		= {2016},
  month		= may,
  pages		= {512-517},
  volume	= {337},
  archiveprefix	= {arXiv},
  doi		= {10.1002/asna.201612339},
  eprint	= {1510.08902},
  primaryclass	= {astro-ph.HE},
  url		= {http://adsabs.harvard.edu/abs/2016AN....337..512P}
}

@Article{Ponti2017,
  author	= {Ponti, G. and De, K. and Mu{\~n}oz-Darias, T. and Stella,
		  L. and Nandra, K.},
  journal	= {\mnras},
  title		= {The puzzling orbital period evolution of the LMXB AX
		  J1745.6-2901},
  year		= {2017},
  month		= jan,
  number	= {1},
  pages		= {840-849},
  volume	= {464},
  archiveprefix	= {arXiv},
  doi		= {10.1093/mnras/stw2317},
  eprint	= {1511.02855},
  primaryclass	= {astro-ph.HE},
  url		= {https://ui.adsabs.harvard.edu/abs/2017MNRAS.464..840P}
}

@Article{Ponti2018,
  author	= {Ponti, Gabriele and Bianchi, Stefano and Mu{\~n}oz-Darias,
		  Teo and Nandra, Kirpal},
  journal	= {\mnras},
  title		= {Measuring masses in low mass X-ray binaries via X-ray
		  spectroscopy: the case of MXB 1659-298},
  year		= {2018},
  month		= nov,
  number	= {1},
  pages		= {L94-L99},
  volume	= {481},
  archiveprefix	= {arXiv},
  doi		= {10.1093/mnrasl/sly120},
  eprint	= {1807.04757},
  primaryclass	= {astro-ph.HE},
  url		= {https://ui.adsabs.harvard.edu/abs/2018MNRAS.481L..94P}
}

@ARTICLE{Poutanen2007,
       author = {{Poutanen}, Juri and {Lipunova}, Galina and {Fabrika}, Sergei and {Butkevich}, Alexey G. and {Abolmasov}, Pavel},
        title = "{Supercritically accreting stellar mass black holes as ultraluminous X-ray sources}",
      journal = {\mnras},
         year = 2007,
        month = may,
       volume = {377},
       number = {3},
        pages = {1187-1194},
          doi = {10.1111/j.1365-2966.2007.11668.x},
archivePrefix = {arXiv},
       eprint = {astro-ph/0609274},
 primaryClass = {astro-ph},
       adsurl = {https://ui.adsabs.harvard.edu/abs/2007MNRAS.377.1187P}
}

@Article{Prinja1994,
  author	= {Prinja, R. K. and Fullerton, A. W.},
  journal	= {\apj},
  month		= may,
  pages		= {345-356},
  title		= {Low-velocity variability in the stellar wind of HD 152408
		  (O8: Iafpe)},
  volume	= {426},
  year		= {1994},
  doi		= {10.1086/174070},
  url		= {http://adsabs.harvard.edu/abs/1994ApJ...426..345P}
}

@Article{Proga2000,
  author	= {{Proga}, Daniel and {Stone}, James M. and {Kallman},
		  Timothy R.},
  title		= "{Dynamics of Line-driven Disk Winds in Active Galactic
		  Nuclei}",
  journal	= {\apj},
  year		= 2000,
  month		= nov,
  volume	= {543},
  number	= {2},
  pages		= {686-696},
  doi		= {10.1086/317154},
  archiveprefix	= {arXiv},
  eprint	= {astro-ph/0005315},
  primaryclass	= {astro-ph},
  adsurl	= {https://ui.adsabs.harvard.edu/abs/2000ApJ...543..686P}
}

@Article{Proga2002,
  author	= {{Proga}, Daniel and {Kallman}, Timothy R.},
  title		= "{On the Role of the Ultraviolet and X-Ray Radiation in
		  Driving a Disk Wind in X-Ray Binaries}",
  journal	= {\apj},
  year		= 2002,
  month		= jan,
  volume	= {565},
  number	= {1},
  pages		= {455-470},
  doi		= {10.1086/324534},
  archiveprefix	= {arXiv},
  eprint	= {astro-ph/0109064},
  primaryclass	= {astro-ph},
  adsurl	= {https://ui.adsabs.harvard.edu/abs/2002ApJ...565..455P}
}

@Article{Rahoui2014,
  author	= {Rahoui, F. and Coriat, M. and Lee, J. C.},
  journal	= {\mnras},
  month		= aug,
  pages		= {1610-1618},
  title		= {Optical and near-infrared spectroscopy of the black hole
		  GX 339-4 - II. The spectroscopic content in the low/hard
		  and high/soft states},
  volume	= {442},
  year		= {2014},
  doi		= {10.1093/mnras/stu977},
  url		= {http://adsabs.harvard.edu/abs/2014MNRAS.442.1610R}
}

@Article{Rahoui2017,
  author	= {Rahoui, F. and Tomsick, J. A. and Gandhi, P. and Casella,
		  P. and F{\"u}rst, F. and Natalucci, L. and Rossi, A. and
		  Shaw, A. W. and Testa, V. and Walton, D. J.},
  journal	= {\mnras},
  title		= {The nova-like nebular optical spectrum of V404 Cygni at
		  the beginning of the 2015 outburst decay},
  year		= {2017},
  month		= mar,
  pages		= {4468-4481},
  volume	= {465},
  archiveprefix	= {arXiv},
  doi		= {10.1093/mnras/stw2890},
  eprint	= {1611.02278},
  primaryclass	= {astro-ph.HE},
  url		= {http://adsabs.harvard.edu/abs/2017MNRAS.465.4468R}
}

@Article{Sala2007,
  author	= {{Sala}, G. and {Greiner}, J. and {Ajello}, M. and
		  {Bottacini}, E. and {Haberl}, F.},
  journal	= {\aap},
  title		= {{XMM-Newton and INTEGRAL observations of the black hole
		  candidate <ASTROBJ>XTE J1817-330</ASTROBJ>}},
  year		= {2007},
  month		= oct,
  pages		= {561-568},
  volume	= {473},
  adsurl	= {http://ads.nao.ac.jp/abs/2007A%26A...473..561S},
  archiveprefix	= {arXiv},
  doi		= {10.1051/0004-6361:20077360},
  eprint	= {0707.4155}
}

@Article{Sanchezsierras2020,
  author	= {S{\'a}nchez-Sierras, J. and Mu{\~n}oz-Darias, T.},
  journal	= {\aap},
  month		= aug,
  pages		= {L3},
  title		= {Near-infrared emission lines trace the state-independent
		  accretion disc wind of the black hole transient MAXI
		  J1820+070},
  volume	= {640},
  year		= {2020},
  doi		= {10.1051/0004-6361/202038406},
  eid		= {L3},
  url		= {https://ui.adsabs.harvard.edu/abs/2020A&A...640L...3S}
}

@Article{Shakura1973,
  author	= {{Shakura}, N.~I. and {Sunyaev}, R.~A.},
  title		= "{Black holes in binary systems. Observational
		  appearance.}",
  journal	= {\aap},
  year		= 1973,
  month		= jan,
  volume	= {24},
  pages		= {337-355},
  adsurl	= {https://ui.adsabs.harvard.edu/abs/1973A&A....24..337S}
}

@Article{Shidatsu2013,
  author	= {Shidatsu, M. and Ueda, Y. and Nakahira, S. and Done, C.
		  and Morihana, K. and Sugizaki, M. and Mihara, T. and Hori,
		  T. and Negoro, H. and Kawai, N. and Yamaoka, K. and
		  Ebisawa, K. and Matsuoka, M. and Serino, M. and Yoshikawa,
		  T. and Nagayama, T. and Matsunaga, N.},
  journal	= {\apj},
  title		= {The Accretion Disk and Ionized Absorber of the 9.7 hr
		  Dipping Black Hole Binary MAXI J1305-704},
  year		= {2013},
  month		= dec,
  number	= {1},
  pages		= {26},
  volume	= {779},
  archiveprefix	= {arXiv},
  doi		= {10.1088/0004-637X/779/1/26},
  eid		= {26},
  eprint	= {1310.0019},
  primaryclass	= {astro-ph.HE},
  url		= {https://ui.adsabs.harvard.edu/abs/2013ApJ...779...26S}
}

@Article{Shidatsu2016,
  author	= {Shidatsu, M. and Done, C. and Ueda, Y.},
  title		= {An Optically Thick Disk Wind in GRO J1655-40?},
  journal	= {\apj},
  year		= {2016},
  volume	= {823},
  pages		= {159},
  month		= jun,
  archiveprefix	= {arXiv},
  doi		= {10.3847/0004-637X/823/2/159},
  eid		= {159},
  eprint	= {1604.04346},
  primaryclass	= {astro-ph.HE},
  url		= {http://adsabs.harvard.edu/abs/2016ApJ...823..159S}
}

@Article{Shidatsu2019,
  author	= {{Shidatsu}, Megumi and {Done}, Chris},
  title		= "{Application of the Thermal Wind Model to Absorption
		  Features in the Black Hole X-Ray Binary H1743-322}",
  journal	= {\apj},
  year		= 2019,
  month		= nov,
  volume	= {885},
  number	= {2},
  eid		= {112},
  pages		= {112},
  doi		= {10.3847/1538-4357/ab46b3},
  archiveprefix	= {arXiv},
  eprint	= {1906.02469},
  primaryclass	= {astro-ph.HE},
  adsurl	= {https://ui.adsabs.harvard.edu/abs/2019ApJ...885..112S}
}

@Article{Soria1999,
  author	= {Soria, Roberto and Wu, Kinwah and Johnston, Helen M.},
  journal	= {\mnras},
  month		= nov,
  number	= {1},
  pages		= {71-77},
  title		= {Optical spectroscopy of GX 339-4 during the high-soft and
		  low-hard states - I},
  volume	= {310},
  year		= {1999},
  doi		= {10.1046/j.1365-8711.1999.02933.x},
  url		= {https://ui.adsabs.harvard.edu/abs/1999MNRAS.310...71S}
}

@Article{Soria2000,
  author	= {Soria, R. and Wu, K. and Hunstead, R. W.},
  journal	= {\apj},
  month		= aug,
  pages		= {445-462},
  title		= {Optical Spectroscopy of GRO J1655-40},
  volume	= {539},
  year		= {2000},
  doi		= {10.1086/309194},
  url		= {http://adsabs.harvard.edu/abs/2000ApJ...539..445S}
}

@Article{Tetarenko2016,
  author	= {Tetarenko, B. E. and Sivakoff, G. R. and Heinke, C. O. and
		  Gladstone, J. C.},
  journal	= {\apjs},
  title		= {WATCHDOG: A Comprehensive All-sky Database of Galactic
		  Black Hole X-ray Binaries},
  year		= {2016},
  month		= feb,
  pages		= {15},
  volume	= {222},
  archiveprefix	= {arXiv},
  doi		= {10.3847/0067-0049/222/2/15},
  eid		= {15},
  eprint	= {1512.00778},
  primaryclass	= {astro-ph.HE},
  url		= {http://adsabs.harvard.edu/abs/2016ApJS..222...15T}
}

@Article{Tetarenko2018,
  author	= {Tetarenko, B. E. and Lasota, J.-P. and Heinke, C. O. and
		  Dubus, G. and Sivakoff, G. R.},
  journal	= {\nat},
  title		= {Strong disk winds traced throughout outbursts in
		  black-hole X-ray binaries},
  year		= {2018},
  month		= feb,
  pages		= {69-72},
  volume	= {554},
  archiveprefix	= {arXiv},
  doi		= {10.1038/nature25159},
  eprint	= {1801.07203},
  primaryclass	= {astro-ph.HE},
  url		= {http://adsabs.harvard.edu/abs/2018Natur.554...69T}
}

@Article{Tomaru2019,
  author	= {{Tomaru}, Ryota and {Done}, Chris and {Ohsuga}, Ken and
		  {Nomura}, Mariko and {Takahashi}, Tadayuki},
  title		= "{The thermal-radiative wind in low-mass X-ray binary
		  H1743-322: radiation hydrodynamic simulations}",
  journal	= {\mnras},
  year		= 2019,
  month		= dec,
  volume	= {490},
  number	= {3},
  pages		= {3098-3111},
  doi		= {10.1093/mnras/stz2738},
  archiveprefix	= {arXiv},
  eprint	= {1905.11763},
  primaryclass	= {astro-ph.HE},
  adsurl	= {https://ui.adsabs.harvard.edu/abs/2019MNRAS.490.3098T}
}

@Article{Tomaru2020,
  author	= {{Tomaru}, Ryota and {Done}, Chris and {Ohsuga}, Ken and
		  {Odaka}, Hirokazu and {Takahashi}, Tadayuki},
  title		= "{The thermal-radiative wind in low-mass X-ray binary
		  H1743-322 - II. Iron line predictions from Monte Carlo
		  radiation transfer}",
  journal	= {\mnras},
  year		= 2020,
  month		= may,
  volume	= {494},
  number	= {3},
  pages		= {3413-3421},
  doi		= {10.1093/mnras/staa961},
  archiveprefix	= {arXiv},
  eprint	= {1911.01660},
  primaryclass	= {astro-ph.HE},
  adsurl	= {https://ui.adsabs.harvard.edu/abs/2020MNRAS.494.3413T}
}

@Article{Tomaru2020b,
  author	= {{Tomaru}, Ryota and {Done}, Chris and {Ohsuga}, Ken and
		  {Odaka}, Hirokazu and {Takahashi}, Tadayuki},
  title		= "{The thermal-radiative wind in the neutron star low-mass
		  X-ray binary GX 13 + 1}",
  journal	= {\mnras},
  year		= 2020,
  month		= oct,
  volume	= {497},
  number	= {4},
  pages		= {4970-4980},
  doi		= {10.1093/mnras/staa2254},
  archiveprefix	= {arXiv},
  eprint	= {2007.14607},
  primaryclass	= {astro-ph.HE},
  adsurl	= {https://ui.adsabs.harvard.edu/abs/2020MNRAS.497.4970T}
}

@Article{Tomaru2023,
 author = {{Tomaru}, Ryota and {Done}, Chris and {Mao}, Junjie},
        title = "{What powers the wind from the black hole accretion disc in GRO J1655-40?}",
      journal = {\mnras},
         year = 2023,
        month = jan,
       volume = {518},
       number = {2},
        pages = {1789-1801},
          doi = {10.1093/mnras/stac3210},
archivePrefix = {arXiv},
       eprint = {2204.08802},
 primaryClass = {astro-ph.HE},
       adsurl = {https://ui.adsabs.harvard.edu/abs/2023MNRAS.518.1789T}
}

@Article{Tomaru2023b,
  author	= {{Tomaru}, Ryota and {Done}, Chris and {Odaka}, Hirokazu
		  and {Tanimoto}, Atsushi},
  title		= "{A different view of wind in X-ray binaries: the accretion
		  disc corona source 2S 0921-630}",
  journal	= {\mnras},
  year		= 2023,
  month		= aug,
  volume	= {523},
  number	= {3},
  pages		= {3441-3449},
  doi		= {10.1093/mnras/stad1637},
  archiveprefix	= {arXiv},
  eprint	= {2302.12638},
  primaryclass	= {astro-ph.HE},
  adsurl	= {https://ui.adsabs.harvard.edu/abs/2023MNRAS.523.3441T}
}

@Article{Tomaru2024,
  author	= {{Tomaru}, Ryota and {Done}, Chris and {Odaka}, Hirokazu},
  title		= "{X-ray polarization properties of thermal-radiative disc
		  winds in binary systems}",
  journal	= {\mnras},
  year		= 2024,
  month		= jan,
  volume	= {527},
  number	= {3},
  pages		= {7047-7054},
  doi		= {10.1093/mnras/stad3649},
  archiveprefix	= {arXiv},
  eprint	= {2308.07237},
  primaryclass	= {astro-ph.HE},
  adsurl	= {https://ui.adsabs.harvard.edu/abs/2024MNRAS.527.7047T}
}

@Article{Ueda1998,
  author	= {Ueda, Y. and Inoue, H. and Tanaka, Y. and Ebisawa, K. and
		  Nagase, F. and Kotani, T. and Gehrels, N.},
  title		= {Detection of Absorption-Line Features in the X-Ray Spectra
		  of the Galactic Superluminal Source GRO J1655-40},
  journal	= {\apj},
  year		= {1998},
  volume	= {492},
  pages		= {782-787},
  month		= jan,
  doi		= {10.1086/305063},
  url		= {http://adsabs.harvard.edu/abs/1998ApJ...492..782U}
}

@Article{Vanderklis2006,
  author	= {{van der Klis}, M.},
  title		= {{Rapid X-ray Variability}},
  year		= {2006},
  editor	= {{Lewin, W.~H.~G.~\& van der Klis, M.}},
  month		= apr,
  pages		= {39-112},
  adsurl	= {http://adsabs.harvard.edu/abs/2006csxs.book...39V},
  journal = {Cambridge Astrophysics Series},
  booktitle	= {Compact stellar X-ray sources}
}

@Article{Vincentelli2023,
  author	= {Vincentelli, F. M. and Neilsen, J. and Tetarenko, A. J.
		  and Cavecchi, Y. and Castro Segura, N. and del Palacio, S.
		  and van den Eijnden, J. and Vasilopoulos, G. and
		  Altamirano, D. and Armas Padilla, M. and Bailyn, C. D. and
		  Belloni, T. and Buisson, D. J. K. and C{\'u}neo, V. A. and
		  Degenaar, N. and Knigge, C. and Long, K. S. and
		  Jim{\'e}nez-Ibarra, F. and Milburn, J. and Mu{\~n}oz
		  Darias, T. and {\"O}zbey Arabac{\i}, M. and Remillard, R.
		  and Russell, T.},
  journal	= {\nat},
  title		= {A shared accretion instability for black holes and neutron
		  stars},
  year		= {2023},
  month		= mar,
  number	= {7950},
  pages		= {45-49},
  volume	= {615},
  archiveprefix	= {arXiv},
  doi		= {10.1038/s41586-022-05648-3},
  eprint	= {2303.00020},
  primaryclass	= {astro-ph.HE},
  url		= {https://ui.adsabs.harvard.edu/abs/2023Natur.615...45V}
}

@Article{Wang2024,
  author	= {Wang, Jingyi and Kara, Erin and García, Javier A. and
		  Altamirano, Diego and Belloni, Tomaso and Steiner, James F.
		  and van der Klis, Michiel and Ingram, Adam and Mastroserio,
		  Guglielmo and Connors, Riley and Lucchini, Matteo and
		  Dauser, Thomas and Neilsen, Joseph and Lewin, Collin and
		  Remillard, Ron A. and Homan, Jeroen},
  journal	= {The Astrophysical Journal},
  title		= {The 2022 {Outburst} of {IGR} {J17091}?3624: {Connecting}
		  the {Exotic} {GRS} 1915+105 to {Standard} {Black} {Hole}
		  {X}-{Ray} {Binaries}},
  year		= {2024},
  issn		= {0004-637X},
  month		= mar,
  note		= {ADS Bibcode: 2024ApJ...963...14W},
  pages		= {14},
  volume	= {963},
  doi		= {10.3847/1538-4357/ad1595},
  publisher	= {IOP},
  shorttitle	= {The 2022 {Outburst} of {IGR} {J17091}?3624},
  url		= {https://ui.adsabs.harvard.edu/abs/2024ApJ...963...14W},
  urldate	= {2025-02-19}
}

@Article{Waters2021,
  author	= {{Waters}, Tim and {Proga}, Daniel and {Dannen}, Randall},
  title		= "{Multiphase AGN Winds from X-Ray-irradiated Disk
		  Atmospheres}",
  journal	= {\apj},
  year		= 2021,
  month		= jun,
  volume	= {914},
  number	= {1},
  eid		= {62},
  pages		= {62},
  doi		= {10.3847/1538-4357/abfbe6},
  archiveprefix	= {arXiv},
  eprint	= {2101.09273},
  primaryclass	= {astro-ph.GA},
  adsurl	= {https://ui.adsabs.harvard.edu/abs/2021ApJ...914...62W}
}

@Article{Woods1996,
  author	= {{Woods}, D. Tod and {Klein}, Richard I. and {Castor}, John
		  I. and {McKee}, Christopher F. and {Bell}, John B.},
  title		= "{X-Ray--heated Coronae and Winds from Accretion Disks:
		  Time-dependent Two-dimensional Hydrodynamics with Adaptive
		  Mesh Refinement}",
  journal	= {\apj},
  year		= 1996,
  month		= apr,
  volume	= {461},
  pages		= {767},
  doi		= {10.1086/177101},
  adsurl	= {https://ui.adsabs.harvard.edu/abs/1996ApJ...461..767W}
}

@ARTICLE{xrism_pds456,
              author = {{Xrism Collaboration} and {Audard}, Marc and {Awaki}, Hisamitsu and {Ballhausen}, Ralf and {Bamba}, Aya and {Behar}, Ehud and {Boissay-Malaquin}, Rozenn and {Brenneman}, Laura and {Brown}, Gregory V. and {Corrales}, Lia and {Costantini}, Elisa and {Cumbee}, Renata and {Trigo}, Mar{\'\i}a D{\'\i}az and {Done}, Chris and {Dotani}, Tadayasu and {Ebisawa}, Ken and {Eckart}, Megan and {Eckert}, Dominique and {Enoto}, Teruaki and {Eguchi}, Satoshi and {Ezoe}, Yuichiro and {Foster}, Adam and {Fujimoto}, Ryuichi and {Fujita}, Yutaka and {Fukazawa}, Yasushi and {Fukushima}, Kotaro and {Furuzawa}, Akihiro and {Gallo}, Luigi and {Garc{\'\i}a}, Javier A. and {Gu}, Liyi and {Guainazzi}, Matteo and {Hagino}, Kouichi and {Hamaguchi}, Kenji and {Hatsukade}, Isamu and {Hayashi}, Katsuhiro and {Hayashi}, Takayuki and {Hell}, Natalie and {Hodges-Kluck}, Edmund and {Hornschemeier}, Ann and {Ichinohe}, Yuto and {Ishida}, Manabu and {Ishikawa}, Kumi and {Ishisaki}, Yoshitaka and {Kaastra}, Jelle and {Kallman}, Timothy and {Kara}, Erin and {Katsuda}, Satoru and {Kanemaru}, Yoshiaki and {Kelley}, Richard and {Kilbourne}, Caroline and {Kitamoto}, Shunji and {Kobayashi}, Shogo and {Kohmura}, Takayoshi and {Kubota}, Aya and {Leutenegger}, Maurice and {Loewenstein}, Michael and {Maeda}, Yoshitomo and {Markevitch}, Maxim and {Matsumoto}, Hironori and {Matsushita}, Kyoko and {McCammon}, Dan and {McNamara}, Brian and {Mernier}, Fran{\c{c}}ois and {Miller}, Eric D. and {Miller}, Jon M. and {Mitsuishi}, Ikuyuki and {Mizumoto}, Misaki and {Mizuno}, Tsunefumi and {Mori}, Koji and {Mukai}, Koji and {Murakami}, Hiroshi and {Mushotzky}, Richard and {Nakajima}, Hiroshi and {Nakazawa}, Kazuhiro and {Ness}, Jan-Uwe and {Nobukawa}, Kumiko and {Nobukawa}, Masayoshi and {Noda}, Hirofumi and {Odaka}, Hirokazu and {Ogawa}, Shoji and {Ogorzalek}, Anna and {Okajima}, Takashi and {Ota}, Naomi and {Paltani}, Stephane and {Petre}, Robert and {Plucinsky}, Paul and {Porter}, Frederick Scott and {Pottschmidt}, Katja and {Sato}, Kosuke and {Sato}, Toshiki and {Sawada}, Makoto and {Seta}, Hiromi and {Shidatsu}, Megumi and {Simionescu}, Aurora and {Smith}, Randall and {Suzuki}, Hiromasa and {Szymkowiak}, Andrew and {Takahashi}, Hiromitsu and {Takeo}, Mai and {Tamagawa}, Toru and {Tamura}, Keisuke and {Tanaka}, Takaaki and {Tanimoto}, Atsushi and {Tashiro}, Makoto and {Terada}, Yukikatsu and {Terashima}, Yuichi and {Tsuboi}, Yohko and {Tsujimoto}, Masahiro and {Tsunemi}, Hiroshi and {Tsuru}, Takeshi G. and {Uchida}, Hiroyuki and {Uchida}, Nagomi and {Uchida}, Yuusuke and {Uchiyama}, Hideki and {Ueda}, Yoshihiro and {Uno}, Shinichiro and {Vink}, Jacco and {Watanabe}, Shin and {Williams}, Brian J. and {Yamada}, Satoshi and {Yamada}, Shinya and {Yamaguchi}, Hiroya and {Yamaoka}, Kazutaka and {Yamasaki}, Noriko and {Yamauchi}, Makoto and {Yamauchi}, Shigeo and {Yaqoob}, Tahir and {Yoneyama}, Tomokage and {Yoshida}, Tessei and {Yukita}, Mihoko and {Zhuravleva}, Irina and {Braito}, Valentina and {Cond{\`o}}, Pierpaolo and {Fukumura}, Keigo and {Gonzalez}, Adam and {Luminari}, Alfredo and {Miyamoto}, Aiko and {Mizukawa}, Ryuki and {Reeves}, James and {Sato}, Riki and {Tombesi}, Francesco and {Xu}, Yerong},
        title = "{Structured ionized winds shooting out from a quasar at relativistic speeds}",
      journal = {\nat},
         year = 2025,
        month = may,
       volume = {641},
       number = {8065},
        pages = {1132-1136},
          doi = {10.1038/s41586-025-08968-2},
archivePrefix = {arXiv},
       eprint = {2505.09171},
 primaryClass = {astro-ph.HE},
       adsurl = {https://ui.adsabs.harvard.edu/abs/2025Natur.641.1132X}
}

@Article{Menzel1929,
  author    = {Menzel, Donald H.},
  journal   = {Publications of the Astronomical Society of the Pacific},
  title     = {The {Wolf}-{Rayet} {Stars}},
  year      = {1929},
  issn      = {0004-6280},
  month     = dec,
  note      = {ADS Bibcode: 1929PASP...41..344M},
  pages     = {344},
  volume    = {41},
  doi       = {10.1086/123970},
  publisher = {IOP},
  url       = {https://ui.adsabs.harvard.edu/abs/1929PASP...41..344M},
  urldate   = {2025-02-24},
}

@Article{Matthews2023,
  author    = {Matthews, James H. and Strong-Wright, Jago and Knigge, Christian and Hewett, Paul and Temple, Matthew J. and Long, Knox S. and Rankine, Amy L. and Stepney, Matthew and Banerji, Manda and Richards, Gordon T.},
  journal   = {Monthly Notices of the Royal Astronomical Society},
  title     = {A disc wind model for blueshifts in quasar broad emission lines},
  year      = {2023},
  issn      = {0035-8711},
  month     = dec,
  note      = {ADS Bibcode: 2023MNRAS.526.3967M},
  pages     = {3967--3986},
  volume    = {526},
  doi       = {10.1093/mnras/stad2895},
  publisher = {OUP},
  url       = {https://ui.adsabs.harvard.edu/abs/2023MNRAS.526.3967M},
  urldate   = {2025-02-25},
}

@Article{Richards2011,
  author    = {Richards, Gordon T. and Kruczek, Nicholas E. and Gallagher, S. C. and Hall, Patrick B. and Hewett, Paul C. and Leighly, Karen M. and Deo, Rajesh P. and Kratzer, Rachael M. and Shen, Yue},
  journal   = {The Astronomical Journal},
  title     = {Unification of {Luminous} {Type} 1 {Quasars} through {C} {IV} {Emission}},
  year      = {2011},
  issn      = {0004-6256},
  month     = may,
  note      = {ADS Bibcode: 2011AJ....141..167R},
  pages     = {167},
  volume    = {141},
  doi       = {10.1088/0004-6256/141/5/167},
  publisher = {IOP},
  url       = {https://ui.adsabs.harvard.edu/abs/2011AJ....141..167R},
  urldate   = {2025-02-25},
}

@Article{Richards2002,
  author     = {Richards, Gordon T. and Vanden Berk, Daniel E. and Reichard, Timothy A. and Hall, Patrick B. and Schneider, Donald P. and SubbaRao, Mark and Thakar, Anirudda R. and York, Donald G.},
  journal    = {The Astronomical Journal},
  title      = {Broad {Emission}-{Line} {Shifts} in {Quasars}: {An} {Orientation} {Measure} for {Radio}-{Quiet} {Quasars}?},
  year       = {2002},
  issn       = {0004-6256},
  month      = jul,
  note       = {ADS Bibcode: 2002AJ....124....1R},
  pages      = {1--17},
  volume     = {124},
  doi        = {10.1086/341167},
  publisher  = {IOP},
  shorttitle = {Broad {Emission}-{Line} {Shifts} in {Quasars}},
  url        = {https://ui.adsabs.harvard.edu/abs/2002AJ....124....1R},
  urldate    = {2025-02-25},
}

@ARTICLE{Blandford1982,
       author = {{Blandford}, R.~D. and {Payne}, D.~G.},
        title = "{Hydromagnetic flows from accretion disks and the production of radio jets.}",
      journal = {\mnras},
         year = 1982,
        month = jun,
       volume = {199},
        pages = {883-903},
          doi = {10.1093/mnras/199.4.883},
       adsurl = {https://ui.adsabs.harvard.edu/abs/1982MNRAS.199..883B}
}

@Article{Carotenuto2021,
  author     = {Carotenuto, F. and Corbel, S. and Tremou, E. and Russell, T. D. and Tzioumis, A. and Fender, R. P. and Woudt, P. A. and Motta, S. E. and Miller-Jones, J. C. A. and Chauhan, J. and Tetarenko, A. J. and Sivakoff, G. R. and Heywood, I. and Horesh, A. and van der Horst, A. J. and Koerding, E. and Mooley, K. P.},
  journal    = {Monthly Notices of the Royal Astronomical Society},
  title      = {The black hole transient {MAXI} {J1348}-630: evolution of the compact and transient jets during its 2019/2020 outburst},
  year       = {2021},
  issn       = {0035-8711},
  month      = jun,
  note       = {ADS Bibcode: 2021MNRAS.504..444C},
  pages      = {444--468},
  volume     = {504},
  doi        = {10.1093/mnras/stab864},
  publisher  = {OUP},
  shorttitle = {The black hole transient {MAXI} {J1348}-630},
  url        = {https://ui.adsabs.harvard.edu/abs/2021MNRAS.504..444C},
  urldate    = {2025-03-18},
}

@Article{Morris1990,
  author    = {Morris, Simon L. and Liebert, James and Stocke, John T. and Gioia, Isabella M. and Schild, Rudy E. and Wolter, Anna},
  journal   = {The Astrophysical Journal},
  title     = {{MS} 1603.6+2600, an {Unusual} {X}-{Ray} {Selected} {Binary} {System} at {High} {Galactic} {Latitude}},
  year      = {1990},
  issn      = {0004-637X},
  month     = dec,
  note      = {ADS Bibcode: 1990ApJ...365..686M},
  pages     = {686},
  volume    = {365},
  doi       = {10.1086/169523},
  publisher = {IOP},
  url       = {https://ui.adsabs.harvard.edu/abs/1990ApJ...365..686M},
  urldate   = {2025-03-18},
}

@Article{Vincentelli2025,
  author        = {{Vincentelli}, Federico M. and {Mu{\~n}oz-Darias}, Teo},
  journal       = {\aap},
  title         = {{Accretion disc winds imprint distinct signatures in the optical variability spectrum of black hole transients}},
  year          = {2025},
  month         = mar,
  pages         = {A181},
  volume        = {695},
  adsurl        = {https://ui.adsabs.harvard.edu/abs/2025A&A...695A.181V},
  archiveprefix = {arXiv},
  doi           = {10.1051/0004-6361/202452634},
  eid           = {A181},
  eprint        = {2501.04087},
  primaryclass  = {astro-ph.HE},
}

@ARTICLE{Datta2024,
       author = {{Datta}, Sudeb Ranjan and {Chakravorty}, Susmita and {Ferreira}, Jonathan and {Petrucci}, Pierre-Olivier and {Kallman}, Timothy R. and {Jacquemin-Ide}, Jonatan and {Zimniak}, Nathan and {Wilms}, Joern and {Bianchi}, Stefano and {Parra}, Maxime and {Clavel}, Ma{\"\i}ca},
        title = "{Impact of disc magnetisation on MHD disc wind signature}",
      journal = {\aap},
         year = 2024,
        month = jul,
       volume = {687},
          eid = {A2},
        pages = {A2},
          doi = {10.1051/0004-6361/202349129},
archivePrefix = {arXiv},
       eprint = {2403.13077},
 primaryClass = {astro-ph.HE},
       adsurl = {https://ui.adsabs.harvard.edu/abs/2024A&A...687A...2D}
}

@ARTICLE{Chakravorty2023,
       author = {{Chakravorty}, Susmita and {Petrucci}, Pierre-Olivier and {Datta}, Sudeb Ranjan and {Ferreira}, Jonathan and {Wilms}, Joern and {Jacquemin-Ide}, Jonatan and {Clavel}, Maica and {Marcel}, Gregoire and {Rodriguez}, Jerome and {Malzac}, Julien and {Belmont}, Renaud and {Corbel}, Stephane and {Coriat}, Mickael and {Henri}, Gilles and {Parra}, Maxime},
        title = "{Absorption lines from magnetically driven winds in X-ray binaries - II. High resolution observational signatures expected from future X-ray observatories}",
      journal = {\mnras},
         year = 2023,
        month = jan,
       volume = {518},
       number = {1},
        pages = {1335-1351},
          doi = {10.1093/mnras/stac2835},
archivePrefix = {arXiv},
       eprint = {2209.15127},
 primaryClass = {astro-ph.HE},
       adsurl = {https://ui.adsabs.harvard.edu/abs/2023MNRAS.518.1335C}
}

@ARTICLE{Fukumura2017,
       author = {{Fukumura}, Keigo and {Kazanas}, Demosthenes and {Shrader}, Chris and {Behar}, Ehud and {Tombesi}, Francesco and {Contopoulos}, Ioannis},
        title = "{Magnetic origin of black hole winds across the mass scale}",
      journal = {Nature Astronomy},
         year = 2017,
        month = mar,
       volume = {1},
          eid = {0062},
        pages = {0062},
          doi = {10.1038/s41550-017-0062},
archivePrefix = {arXiv},
       eprint = {1702.02197},
 primaryClass = {astro-ph.HE},
       adsurl = {https://ui.adsabs.harvard.edu/abs/2017NatAs...1E..62F}
}

@ARTICLE{Fukumura2010,
       author = {{Fukumura}, Keigo and {Kazanas}, Demosthenes and {Contopoulos}, Ioannis and {Behar}, Ehud},
        title = "{Magnetohydrodynamic Accretion Disk Winds as X-ray Absorbers in Active Galactic Nuclei}",
      journal = {\apj},
         year = 2010,
        month = may,
       volume = {715},
       number = {1},
        pages = {636-650},
          doi = {10.1088/0004-637X/715/1/636},
archivePrefix = {arXiv},
       eprint = {0910.3001},
 primaryClass = {astro-ph.HE},
       adsurl = {https://ui.adsabs.harvard.edu/abs/2010ApJ...715..636F}
}

@ARTICLE{Contopoulos1994,
       author = {{Contopoulos}, J. and {Lovelace}, R.~V.~E.},
        title = "{Magnetically Driven Jets and Winds: Exact Solutions}",
      journal = {\apj},
         year = 1994,
        month = jul,
       volume = {429},
        pages = {139},
          doi = {10.1086/174307},
       adsurl = {https://ui.adsabs.harvard.edu/abs/1994ApJ...429..139C}
}

@ARTICLE{Ferreira1997,
       author = {{Ferreira}, J.},
        title = "{Magnetically-driven jets from Keplerian accretion discs.}",
      journal = {\aap},
         year = 1997,
        month = mar,
       volume = {319},
        pages = {340-359},
          doi = {10.48550/arXiv.astro-ph/9607057},
archivePrefix = {arXiv},
       eprint = {astro-ph/9607057},
 primaryClass = {astro-ph},
       adsurl = {https://ui.adsabs.harvard.edu/abs/1997A&A...319..340F}
}

@ARTICLE{Ferreira1995,
       author = {{Ferreira}, J. and {Pelletier}, G.},
        title = "{Magnetized accretion-ejection structures. III. Stellar and extragalactic jets as weakly dissipative disk outflows.}",
      journal = {\aap},
         year = 1995,
        month = mar,
       volume = {295},
        pages = {807},
       adsurl = {https://ui.adsabs.harvard.edu/abs/1995A&A...295..807F}
}

@ARTICLE{Chakravorty2016,
       author = {{Chakravorty}, S. and {Petrucci}, P. -O. and {Ferreira}, J. and {Henri}, G. and {Belmont}, R. and {Clavel}, M. and {Corbel}, S. and {Rodriguez}, J. and {Coriat}, M. and {Drappeau}, S. and {Malzac}, J.},
        title = "{Absorption lines from magnetically driven winds in X-ray binaries}",
      journal = {\aap},
         year = 2016,
        month = may,
       volume = {589},
          eid = {A119},
        pages = {A119},
          doi = {10.1051/0004-6361/201527163},
archivePrefix = {arXiv},
       eprint = {1512.09149},
 primaryClass = {astro-ph.HE},
       adsurl = {https://ui.adsabs.harvard.edu/abs/2016A&A...589A.119C}
}

@ARTICLE{Fukumura2015,
       author = {{Fukumura}, Keigo and {Tombesi}, Francesco and {Kazanas}, Demosthenes and {Shrader}, Chris and {Behar}, Ehud and {Contopoulos}, Ioannis},
        title = "{Magnetically Driven Accretion Disk Winds and Ultra-fast Outflows in PG 1211+143}",
      journal = {\apj},
         year = 2015,
        month = may,
       volume = {805},
       number = {1},
          eid = {17},
        pages = {17},
          doi = {10.1088/0004-637X/805/1/17},
archivePrefix = {arXiv},
       eprint = {1503.04074},
 primaryClass = {astro-ph.HE},
       adsurl = {https://ui.adsabs.harvard.edu/abs/2015ApJ...805...17F}
}

@Article{Erkal2022,
  author   = {Erkal, J. and Manara, C. F. and Schneider, P. C. and Vincenzi, M. and Nisini, B. and Coffey, D. and Alcalá, J. M. and Fedele, D. and Antoniucci, S.},
  journal  = {Astronomy and Astrophysics},
  title = {The He I $\lambda$10830\,{\AA} line as a probe of winds and accretion in young stars in Lupus and Upper Scorpius},
  year     = {2022},
  issn     = {0004-6361},
  month    = oct,
  note     = {ADS Bibcode: 2022A\&A...666A.188E},
  pages    = {A188},
  volume   = {666},
  doi      = {10.1051/0004-6361/202244254},
  url      = {https://ui.adsabs.harvard.edu/abs/2022A&A...666A.188E},
  urldate  = {2025-04-10},
}

@PhdThesis{Matthews2016,
  author     = {Matthews, James H.},
  title      = {Disc {Winds} {Matter}: {Modelling} {Accretion} {And} {Outflow} {On} {All} {Scales}},
  year       = {2016},
  month      = jun,
  note       = {ADS Bibcode: 2016PhDT.......348M},
  doi        = {10.5281/zenodo.1256805},
  shorttitle = {Disc {Winds} {Matter}},
  url        = {https://ui.adsabs.harvard.edu/abs/2016PhDT.......348M},
  urldate    = {2025-04-10},
}

@Article{Rahoui2012,
  author    = {Rahoui, F. and Coriat, M. and Corbel, S. and Cadolle Bel, M. and Tomsick, J. A. and Lee, J. C. and Rodriguez, J. and Russell, D. M. and Migliari, S.},
  journal   = {Monthly Notices of the Royal Astronomical Society},
  title     = {Optical and near-infrared spectroscopy of the black hole {GX} 339-4 - {I}. {A} focus on the continuum in the low/hard and high/soft states},
  year      = {2012},
  issn      = {0035-8711},
  month     = may,
  note      = {ADS Bibcode: 2012MNRAS.422.2202R},
  pages     = {2202--2212},
  volume    = {422},
  doi       = {10.1111/j.1365-2966.2012.20763.x},
  publisher = {OUP},
  url       = {https://ui.adsabs.harvard.edu/abs/2012MNRAS.422.2202R},
  urldate   = {2025-04-15},
}

@InProceedings{Puls1998,
  author  = {Puls, J. and Kudritzki, R. -P. and Santolaya-Rey, A. E. and Herrero, A. and Owocki, S. P. and McCarthy, J. K.},
  title   = {Spectral {Diagnostics} of {Blue} {Stars} with {Winds}},
  year    = {1998},
  month   = jan,
  note    = {ADS Bibcode: 1998ASPC..131..245P},
  pages   = {245},
  volume  = {131},
  url     = {https://ui.adsabs.harvard.edu/abs/1998ASPC..131..245P},
  urldate = {2025-04-15},
}

@Article{Petrov2014,
  author     = {Petrov, Blagovest and Vink, Jorick S. and Gräfener, Götz},
  journal    = {Astronomy and Astrophysics},
  title      = {On the H$\alpha$ behaviour of blue supergiants: rise and fall over the bi-stability jump},
  year       = {2014},
  issn       = {0004-6361},
  month      = may,
  note       = {ADS Bibcode: 2014A\&A...565A..62P},
  pages      = {A62},
  volume     = {565},
  doi        = {10.1051/0004-6361/201322754},
  shorttitle = {On the {Hα} behaviour of blue supergiants},
  url        = {https://ui.adsabs.harvard.edu/abs/2014A&A...565A..62P},
  urldate    = {2025-04-15},
}

@ARTICLE{Kallman2001,
       author = {{Kallman}, T. and {Bautista}, M.},
        title = "{Photoionization and High-Density Gas}",
      journal = {\apjs},
         year = 2001,
        month = mar,
       volume = {133},
       number = {1},
        pages = {221-253},
          doi = {10.1086/319184},
       adsurl = {https://ui.adsabs.harvard.edu/abs/2001ApJS..133..221K}
}

@ARTICLE{Ferland2017,
       author = {{Ferland}, G.~J. and {Chatzikos}, M. and {Guzm{\'a}n}, F. and {Lykins}, M.~L. and {van Hoof}, P.~A.~M. and {Williams}, R.~J.~R. and {Abel}, N.~P. and {Badnell}, N.~R. and {Keenan}, F.~P. and {Porter}, R.~L. and {Stancil}, P.~C.},
        title = "{The 2017 Release Cloudy}",
      journal = {\rmxaa},
         year = 2017,
        month = oct,
       volume = {53},
        pages = {385-438},
          doi = {10.48550/arXiv.1705.10877},
archivePrefix = {arXiv},
       eprint = {1705.10877},
 primaryClass = {astro-ph.GA},
       adsurl = {https://ui.adsabs.harvard.edu/abs/2017RMxAA..53..385F}
}

@ARTICLE{Gunasekera2023,
       author = {{Gunasekera}, Chamani M. and {van Hoof}, Peter A.~M. and {Chatzikos}, Marios and {Ferland}, Gary J.},
        title = "{The 23.01 Release of Cloudy}",
      journal = {Research Notes of the American Astronomical Society},
         year = 2023,
        month = nov,
       volume = {7},
       number = {11},
          eid = {246},
        pages = {246},
          doi = {10.3847/2515-5172/ad0e75},
archivePrefix = {arXiv},
       eprint = {2311.10163},
 primaryClass = {astro-ph.GA},
       adsurl = {https://ui.adsabs.harvard.edu/abs/2023RNAAS...7..246G}
}

@ARTICLE{Pudritz1986,
       author = {{Pudritz}, R.~E. and {Norman}, C.~A.},
        title = "{Bipolar Hydromagnetic Winds from Disks around Protostellar Objects}",
      journal = {\apj},
         year = 1986,
        month = feb,
       volume = {301},
        pages = {571},
          doi = {10.1086/163924},
       adsurl = {https://ui.adsabs.harvard.edu/abs/1986ApJ...301..571P}
}

@ARTICLE{Eggum1988,
       author = {{Eggum}, G.~E. and {Coroniti}, F.~V. and {Katz}, J.~I.},
        title = "{Radiation Hydrodynamic Calculation of Super-Eddington Accretion Disks}",
      journal = {\apj},
         year = 1988,
        month = jul,
       volume = {330},
        pages = {142},
          doi = {10.1086/166462},
       adsurl = {https://ui.adsabs.harvard.edu/abs/1988ApJ...330..142E}
}

@ARTICLE{Ohsuga2005,
       author = {{Ohsuga}, Ken and {Mori}, Masao and {Nakamoto}, Taishi and {Mineshige}, Shin},
        title = "{Supercritical Accretion Flows around Black Holes: Two-dimensional, Radiation Pressure-dominated Disks with Photon Trapping}",
      journal = {\apj},
         year = 2005,
        month = jul,
       volume = {628},
       number = {1},
        pages = {368-381},
          doi = {10.1086/430728},
archivePrefix = {arXiv},
       eprint = {astro-ph/0504168},
 primaryClass = {astro-ph},
       adsurl = {https://ui.adsabs.harvard.edu/abs/2005ApJ...628..368O}
}

@ARTICLE{Sadowski2014,
       author = {{Sadowski}, Aleksander and {Narayan}, Ramesh and {McKinney}, Jonathan C. and {Tchekhovskoy}, Alexander},
        title = "{Numerical simulations of super-critical black hole accretion flows in general relativity}",
      journal = {\mnras},
         year = 2014,
        month = mar,
       volume = {439},
       number = {1},
        pages = {503-520},
          doi = {10.1093/mnras/stt2479},
archivePrefix = {arXiv},
       eprint = {1311.5900},
 primaryClass = {astro-ph.HE},
       adsurl = {https://ui.adsabs.harvard.edu/abs/2014MNRAS.439..503S}
}

@ARTICLE{Jiang2014,
       author = {{Jiang}, Yan-Fei and {Stone}, James M. and {Davis}, Shane W.},
        title = "{A Global Three-dimensional Radiation Magneto-hydrodynamic Simulation of Super-Eddington Accretion Disks}",
      journal = {\apj},
         year = 2014,
        month = dec,
       volume = {796},
       number = {2},
          eid = {106},
        pages = {106},
          doi = {10.1088/0004-637X/796/2/106},
archivePrefix = {arXiv},
       eprint = {1410.0678},
 primaryClass = {astro-ph.HE},
       adsurl = {https://ui.adsabs.harvard.edu/abs/2014ApJ...796..106J}
}

@ARTICLE{Proga1998,
       author = {{Proga}, Daniel and {Stone}, James M. and {Drew}, Janet E.},
        title = "{Radiation-driven winds from luminous accretion discs}",
      journal = {\mnras},
         year = 1998,
        month = apr,
       volume = {295},
       number = {3},
        pages = {595-617},
          doi = {10.1046/j.1365-8711.1998.01337.x},
archivePrefix = {arXiv},
       eprint = {astro-ph/9710305},
 primaryClass = {astro-ph},
       adsurl = {https://ui.adsabs.harvard.edu/abs/1998MNRAS.295..595P}
}

@ARTICLE{Nomura2016,
       author = {{Nomura}, Mariko and {Ohsuga}, Ken and {Takahashi}, Hiroyuki R. and {Wada}, Keiichi and {Yoshida}, Tessei},
        title = "{Radiation hydrodynamic simulations of line-driven disk winds for ultra-fast outflows}",
      journal = {\pasj},
         year = 2016,
        month = feb,
       volume = {68},
       number = {1},
          eid = {16},
        pages = {16},
          doi = {10.1093/pasj/psv124},
archivePrefix = {arXiv},
       eprint = {1511.08815},
 primaryClass = {astro-ph.HE},
       adsurl = {https://ui.adsabs.harvard.edu/abs/2016PASJ...68...16N}
}

@ARTICLE{Boirin2004,
   author = {{Boirin}, L. and {Parmar}, A.~N. and {Barret}, D. and {Paltani}, S. and
    {Grindlay}, J.~E.},
    title = "{Discovery of X-ray absorption features from the dipping low-mass X-ray binary XB 1916-053 with XMM-Newton}",
  journal = {\aap},
     year = 2004,
    month = may,
   volume = 418,
    pages = {1061-1072},
   adsurl = {http://cdsads.u-strasbg.fr/cgi-bin/nph-bib_query?bibcode=2004A%26A...418.1061B&amp;db_key=AST},
}

@ARTICLE{Juett2006,
   author = {{Juett}, A. and {Chakrabarty}, D.},
    title = "{Detection of Highly Ionized Metal Absorption Lines in the Ultracompact X-Ray Dipper 4U 1916-05}",
  journal = {\apj},
     year = 2006,
    month = jul,
   volume = 646,
    pages = {493-498},
   adsurl = {http://adsabs.harvard.edu/abs/2006ApJ...646..493J},
}

@ARTICLE{Iaria2006,
       author = {{Iaria}, R. and {Di Salvo}, T. and {Lavagetto}, G. and {Robba}, N.~R. and {Burderi}, L.},
        title = "{Chandra Observation of the Persistent Emission from the Dipping Source XB 1916-053}",
      journal = {\apj},
         year = 2006,
        month = aug,
       volume = {647},
       number = {2},
        pages = {1341-1348},
          doi = {10.1086/505616},
archivePrefix = {arXiv},
       eprint = {astro-ph/0605055},
 primaryClass = {astro-ph},
       adsurl = {https://ui.adsabs.harvard.edu/abs/2006ApJ...647.1341I}
}

@ARTICLE{Zhang2014,
       author = {{Zhang}, Zhongli and {Makishima}, Kazuo and {Sakurai}, Soki and {Sasano}, Makoto and {Ono}, Ko},
        title = "{Probing the accretion scheme of the dipping X-ray binary 4U 1915-05 with Suzaku}",
      journal = {\pasj},
         year = 2014,
        month = dec,
       volume = {66},
       number = {6},
          eid = {120},
        pages = {120},
          doi = {10.1093/pasj/psu117},
archivePrefix = {arXiv},
       eprint = {1409.2091},
 primaryClass = {astro-ph.HE},
       adsurl = {https://ui.adsabs.harvard.edu/abs/2014PASJ...66..120Z}
}

@ARTICLE{Gambino2019,
       author = {{Gambino}, A.~F. and {Iaria}, R. and {Di Salvo}, T. and {Mazzola}, S.~M. and {Marino}, A. and {Burderi}, L. and {Riggio}, A. and {Sanna}, A. and {D'Amico}, N.},
        title = "{Spectral analysis of the dipping LMXB system XB 1916-053}",
      journal = {\aap},
         year = 2019,
        month = may,
       volume = {625},
          eid = {A92},
        pages = {A92},
          doi = {10.1051/0004-6361/201832984},
archivePrefix = {arXiv},
       eprint = {1904.05770},
 primaryClass = {astro-ph.HE},
       adsurl = {https://ui.adsabs.harvard.edu/abs/2019A&A...625A..92G}
}

@ARTICLE{Trueba2020,
       author = {{Trueba}, Nicolas and {Miller}, J.~M. and {Fabian}, A.~C. and {Kaastra}, J. and {Kallman}, T. and {Lohfink}, A. and {Proga}, D. and {Raymond}, J. and {Reynolds}, C. and {Reynolds}, M. and {Zoghbi}, A.},
        title = "{A Redshifted Inner Disk Atmosphere and Transient Absorbers in the Ultracompact Neutron Star X-Ray Binary 4U 1916-053}",
      journal = {\apjl},
         year = 2020,
        month = aug,
       volume = {899},
       number = {1},
          eid = {L16},
        pages = {L16},
          doi = {10.3847/2041-8213/aba9de},
archivePrefix = {arXiv},
       eprint = {2008.01083},
 primaryClass = {astro-ph.HE},
       adsurl = {https://ui.adsabs.harvard.edu/abs/2020ApJ...899L..16T}
}

@ARTICLE{Gavriil2012,
       author = {{Gavriil}, Fotis P. and {Strohmayer}, Tod E. and {Bhattacharyya}, Sudip},
        title = "{An Fe XXIV Absorption Line in the Persistent Spectrum of the Dipping Low-mass X-Ray Binary 1A 1744-361}",
      journal = {\apj},
         year = 2012,
        month = jul,
       volume = {753},
       number = {1},
          eid = {2},
        pages = {2},
          doi = {10.1088/0004-637X/753/1/2},
archivePrefix = {arXiv},
       eprint = {0909.1607},
 primaryClass = {astro-ph.HE},
       adsurl = {https://ui.adsabs.harvard.edu/abs/2012ApJ...753....2G}
}

@ARTICLE{Mondal2024,
       author = {{Mondal}, Aditya S. and {Raychaudhuri}, B. and {Dewangan}, G.~C.},
        title = "{Relativistic X-ray reflection and highly ionized absorption in the spectrum of NS LMXB 1A 1744-361}",
      journal = {\mnras},
         year = 2024,
        month = jan,
       volume = {527},
       number = {2},
        pages = {2362-2370},
          doi = {10.1093/mnras/stad3326},
archivePrefix = {arXiv},
       eprint = {2309.12637},
 primaryClass = {astro-ph.HE},
       adsurl = {https://ui.adsabs.harvard.edu/abs/2024MNRAS.527.2362M}
}

@ARTICLE{Ng2024,
       author = {{Ng}, Mason and {Hughes}, Andrew K. and {Homan}, Jeroen and {Miller}, Jon M. and {Pike}, Sean N. and {Altamirano}, Diego and {Bult}, Peter and {Chakrabarty}, Deepto and {Buisson}, D.~J.~K. and {Coughenour}, Benjamin M. and {Fender}, Rob and {Guillot}, Sebastien and {G{\"u}ver}, Tolga and {Jaisawal}, Gaurava K. and {Jaodand}, Amruta D. and {Malacaria}, Christian and {Miller-Jones}, James C.~A. and {Sanna}, Andrea and {Sivakoff}, Gregory R. and {Strohmayer}, Tod E. and {Tomsick}, John A. and {van den Eijnden}, Jakob},
        title = "{X-Ray and Radio Monitoring of the Neutron Star Low-mass X-Ray Binary 1A 1744-361: Quasiperiodic Oscillations, Transient Ejections, and a Disk Atmosphere}",
      journal = {\apj},
         year = 2024,
        month = may,
       volume = {966},
       number = {2},
          eid = {232},
        pages = {232},
          doi = {10.3847/1538-4357/ad35bd},
archivePrefix = {arXiv},
       eprint = {2310.01511},
 primaryClass = {astro-ph.HE},
       adsurl = {https://ui.adsabs.harvard.edu/abs/2024ApJ...966..232N}
}

@ARTICLE{Boirin2005,
    author = {{Boirin}, L. and {M\'endez}, M. {D{\'i}az Trigo}, M. and {Parmar}, A. N. and {Kaastra}, J.},
    title = "{A highly-ionised absorber in the X-ray binary 4U\,1323$-$619: a
    new explanation of the dipping phenomena}",
    journal = {\aap},
    year = 2005,
    month = feb,
    volume = {436},
    pages={195},
    adsurl = {unknown}
}

@ARTICLE{Church2005,
   author = {{Church}, M.~J. and {Reed}, D. and {Dotani}, T. and {Ba{\l}uci{\'n}ska-Church}, M. and
    {Smale}, A.~P.},
    title = "{Discovery of absorption features of the accretion disc corona and systematic acceleration of the X-ray burst rate in XB1323-619}",
  journal = {\mnras},
     year = 2005,
    month = jun,
   volume = 359,
    pages = {1336-1344},
      doi = {10.1111/j.1365-2966.2005.08728.x},
   adsurl = {http://adsabs.harvard.edu/cgi-bin/nph-bib_query?bibcode=2005MNRAS.359.1336C&db_key=AST}
}

@ARTICLE{Balucinska-Church2009,
       author = {{Ba{\l}uci{\'n}ska-Church}, M. and {Dotani}, T. and {Hirotsu}, T. and {Church}, M.~J.},
        title = "{Neutral absorber dips in the periodic burster LMXB XB 1323-619 from Suzaku}",
      journal = {\aap},
         year = 2009,
        month = jun,
       volume = {500},
       number = {2},
        pages = {873-882},
          doi = {10.1051/0004-6361/200811215},
archivePrefix = {arXiv},
       eprint = {0905.0618},
 primaryClass = {astro-ph.HE},
       adsurl = {https://ui.adsabs.harvard.edu/abs/2009A&A...500..873B}
}

@ARTICLE{Raman2018,
       author = {{Raman}, Gayathri and {Maitra}, Chandreyee and {Paul}, Biswajit},
        title = "{Observation of variable pre-eclipse dips and disc windsin the eclipsing LMXB XTE J1710-281}",
      journal = {\mnras},
         year = 2018,
        month = jul,
       volume = {477},
       number = {4},
        pages = {5358-5366},
          doi = {10.1093/mnras/sty918},
archivePrefix = {arXiv},
       eprint = {1804.06073},
 primaryClass = {astro-ph.HE},
       adsurl = {https://ui.adsabs.harvard.edu/abs/2018MNRAS.477.5358R}
}

@ARTICLE{Trueba2022,
       author = {{Trueba}, Nicolas and {Miller}, J.~M. and {Fabian}, A.~C. and {Kaastra}, J. and {Kallman}, T. and {Lohfink}, A. and {Ludlam}, R.~M. and {Proga}, D. and {Raymond}, J. and {Reynolds}, C. and {Reynolds}, M. and {Zoghbi}, A.},
        title = "{A Spectroscopic Angle on Central Engine Size Scales in Accreting Neutron Stars}",
      journal = {\apj},
         year = 2022,
        month = feb,
       volume = {925},
       number = {2},
          eid = {113},
        pages = {113},
          doi = {10.3847/1538-4357/ac3766},
archivePrefix = {arXiv},
       eprint = {2111.04764},
 primaryClass = {astro-ph.HE},
       adsurl = {https://ui.adsabs.harvard.edu/abs/2022ApJ...925..113T}
}

@ARTICLE{Cottam2001,
    author = {{Cottam}, J. and {Kahn}, S.~M. and {Brinkman}, A.~C. and {den Herder}, J.~W. and
    {Erd}, C.},
    title = "{High-resolution spectroscopy of the low-mass X-ray binary <ASTROBJ>EXO 0748-67</ASTROBJ>}",
    journal = {\aap},
    year = 2001,
    month = jan,
    volume = 365,
    pages = {L277-L281},
    adsurl = {http://adsabs.harvard.edu/cgi-bin/nph-bib_query?bibcode=2001A%26A...365L.277C&amp;db_key=AST},
}

@ARTICLE{Jimenez2003,
    author = {{Jimenez-Garate}, M.~A. and {Schulz}, N.~S. and {Marshall}, H.~L.
    },
    title = "{Discrete X-Ray Signatures of a Photoionized Plasma above the Accretion Disk of the Neutron Star EXO 0748-676}",
    journal = {\apj},
    year = 2003,
    month = jun,
    volume = 590,
    pages = {432-444},
    adsurl = {http://adsabs.harvard.edu/cgi-bin/nph-bib_query?bibcode=2003ApJ...590..432J&amp;db_key=AST},
}

@ARTICLE{vanPeet2009,
       author = {{van Peet}, J.~C.~A. and {Costantini}, E. and {M{\'e}ndez}, M. and {Paerels}, F.~B.~S. and {Cottam}, J.},
        title = "{Properties of the ionised plasma in the vicinity of the neutron-star X-ray binary EXO 0748-676}",
      journal = {\aap},
         year = 2009,
        month = apr,
       volume = {497},
       number = {3},
        pages = {805-813},
          doi = {10.1051/0004-6361/200811181},
archivePrefix = {arXiv},
       eprint = {0902.4470},
 primaryClass = {astro-ph.GA},
       adsurl = {https://ui.adsabs.harvard.edu/abs/2009A&A...497..805V}
}

@ARTICLE{Psaradaki2018,
       author = {{Psaradaki}, I. and {Costantini}, E. and {Mehdipour}, M. and {D{\'\i}az Trigo}, M.},
        title = "{Modelling the disc atmosphere of the low mass X-ray binary EXO 0748-676}",
      journal = {\aap},
         year = 2018,
        month = dec,
       volume = {620},
          eid = {A129},
        pages = {A129},
          doi = {10.1051/0004-6361/201834000},
archivePrefix = {arXiv},
       eprint = {1809.06864},
 primaryClass = {astro-ph.HE},
       adsurl = {https://ui.adsabs.harvard.edu/abs/2018A&A...620A.129P}
}

@ARTICLE{Bhattacharya2024,
       author = {{Bhattacharya}, Sayantan and {Bhattacharyya}, Sudip and {Shaw}, Gargi},
        title = "{XMM-Newton High-resolution Spectroscopy of EXO 0748{\textendash}676 after Its Reemergence from a Long Quiescence}",
      journal = {\apjl},
         year = 2024,
        month = dec,
       volume = {977},
       number = {1},
          eid = {L17},
        pages = {L17},
          doi = {10.3847/2041-8213/ad9337},
archivePrefix = {arXiv},
       eprint = {2408.02715},
 primaryClass = {astro-ph.HE},
       adsurl = {https://ui.adsabs.harvard.edu/abs/2024ApJ...977L..17B}
}

@ARTICLE{Boirin2003,
    author = {{Boirin}, L. and {Parmar}, A.~N.},
    title = "{Discovery of narrow X-ray absorption features from the low-mass X-ray binary X 1254-690 with XMM-Newton}",
    journal = {\aap},
    year = 2003,
    month = sep,
    volume = 407,
    pages = {1079-1084},
    adsurl = {http://cdsads.u-strasbg.fr/cgi-bin/nph-bib_query?bibcode=2003A%26A...407.1079B&db_key=AST},
}

@ARTICLE{Iaria2007,
   author = {{Iaria}, R. and {di Salvo}, T. and {Lavagetto}, G. and {D'A{\'{\i}}}, A. and 
	{Robba}, N.~R.},
    title = "{Chandra observation of the dipping source XB 1254-690}",
  journal = {\aap},
   eprint = {astro-ph/0612592},
     year = 2007,
    month = mar,
   volume = 464,
    pages = {291-297},
      doi = {10.1051/0004-6361:20065644},
   adsurl = {http://cdsads.u-strasbg.fr/cgi-bin/nph-bib_query?bibcode=2007A%26A...464..291I&db_key=AST}
}

@ARTICLE{DiazTrigo2009,
   author = {{D{\'{\i}}az Trigo}, M. and {Parmar}, A.~N. and {Boirin}, L. and 
	{Motch}, C. and {Talavera}, A. and {Balman}, S.},
    title = "{Variations in the dip properties of the low-mass X-ray binary XB 1254-690 observed with XMM-Newton and INTEGRAL}",
  journal = {\aap},
archivePrefix = "arXiv",
   eprint = {0810.0432},
     year = 2009,
    month = jan,
   volume = 493,
    pages = {145-157},
      doi = {10.1051/0004-6361:200810154},
   adsurl = {http://adsabs.harvard.edu/abs/2009A%26A...493..145D}
}

@ARTICLE{Sidoli2001,
     author = {{Sidoli}, L. and {Oosterbroek}, T. and {Parmar}, A.~N. and {Lumb}, D. and
     {Erd}, C.},
     title = "{An XMM-Newton study of the X-ray binary MXB 1659-298 and the discovery of narrow X-ray absorption lines}",
     journal = {\aap},
     year = 2001,
     month = nov,
     volume = 379,
     pages = {540-550},
     url = {http://cdsads.u-strasbg.fr/cgi-bin/nph-bib_query?bibcode=2001A%26A...379..540S&db_key=AST},
 }

@ARTICLE{Ponti2019,
       author = {{Ponti}, G. and {Bianchi}, S. and {De Marco}, B. and {Bahramian}, A. and {Degenaar}, N. and {Heinke}, C.~O.},
        title = "{Evolution of the disc atmosphere in the X-ray binary MXB 1659-298, during its 2015-17 outburst}",
      journal = {\mnras},
         year = 2019,
        month = jul,
       volume = {487},
       number = {1},
        pages = {858-870},
          doi = {10.1093/mnras/stz1245},
archivePrefix = {arXiv},
       eprint = {1905.01308},
 primaryClass = {astro-ph.HE},
       adsurl = {https://ui.adsabs.harvard.edu/abs/2019MNRAS.487..858P}
}

@ARTICLE{Iaria2019,
       author = {{Iaria}, R. and {Mazzola}, S.~M. and {Bassi}, T. and {Gambino}, A.~F. and {Marino}, A. and {Di Salvo}, T. and {Sanna}, A. and {Riggio}, A. and {Burderi}, L. and {D'Amico}, N.},
        title = "{Broadband spectral analysis of MXB 1659-298 in its soft and hard state}",
      journal = {\aap},
         year = 2019,
        month = oct,
       volume = {630},
          eid = {A138},
        pages = {A138},
          doi = {10.1051/0004-6361/201833982},
archivePrefix = {arXiv},
       eprint = {1807.11431},
 primaryClass = {astro-ph.HE},
       adsurl = {https://ui.adsabs.harvard.edu/abs/2019A&A...630A.138I}
}

@ARTICLE{Hyodo2009,
       author = {{Hyodo}, Yoshiaki and {Ueda}, Yoshihiro and {Yuasa}, Takayuki and {Maeda}, Yoshitomo and {Makishima}, Kazuo and {Koyama}, Katsuji},
        title = "{Timing and Spectral Study of AXJ1745.6-2901 with Suzaku}",
      journal = {\pasj},
         year = 2009,
        month = jan,
       volume = {61},
        pages = {S99},
          doi = {10.1093/pasj/61.sp1.S99},
       adsurl = {https://ui.adsabs.harvard.edu/abs/2009PASJ...61S..99H}
}

@ARTICLE{Parmar2002,
     author = {{Parmar}, A.~N. and {Oosterbroek}, T. and {Boirin}, L. and {Lumb}, D.
     },
     title = "{Discovery of narrow X-ray absorption features from the dipping low-mass X-ray binary X 1624-490 with XMM-Newton}",
     journal = {\aap},
     year = 2002,
     month = may,
     volume = 386,
     pages = {910-915},
     url = {http://cdsads.u-strasbg.fr/cgi-bin/nph-bib_query?bibcode=2002A%26A...386..910P&db_key=AST},
 }

@ARTICLE{Iaria2007b,
       author = {{Iaria}, R. and {Lavagetto}, G. and {D'A{\'\i}}, A. and {di Salvo}, T. and {Robba}, N.~R.},
        title = "{Chandra observation of the Big Dipper X 1624-490}",
      journal = {\aap},
         year = 2007,
        month = feb,
       volume = {463},
       number = {1},
        pages = {289-295},
          doi = {10.1051/0004-6361:20065862},
archivePrefix = {arXiv},
       eprint = {astro-ph/0612269},
 primaryClass = {astro-ph},
       adsurl = {https://ui.adsabs.harvard.edu/abs/2007A&A...463..289I}
}

@ARTICLE{Xiang2009,
   author = {{Xiang}, J. and {Lee}, J.~C. and {Nowak}, M.~A. and {Wilms}, J. and 
	{Schulz}, N.~S.},
    title = "{The Accretion Disk Corona and Disk Atmosphere of 4U 1624-490  as Viewed by the Chandra-High Energy Transmission Grating Spectrometer}",
  journal = {\apj},
archivePrefix = "arXiv",
   eprint = {0905.3925},
 primaryClass = "astro-ph.HE",
     year = 2009,
    month = aug,
   volume = 701,
    pages = {984-993},
      doi = {10.1088/0004-637X/701/2/984},
   adsurl = {http://adsabs.harvard.edu/abs/2009ApJ...701..984X}
}

@ARTICLE{Schulz2002,
    author = {{Schulz}, N.~S. and {Brandt}, W.~N.},
    title = "{Variability of the X-Ray P Cygni Line Profiles from Circinus X-1 near Zero Phase}",
    journal = {\apj},
    year = 2002,
    month = jun,
    volume = 572,
    pages = {971-983},
    adsurl = {http://cdsads.u-strasbg.fr/cgi-bin/nph-bib_query?bibcode=2002ApJ...572..971S&db_key=AST},
}

@ARTICLE{Schulz2008,
   author = {{Schulz}, N.~S. and {Kallman}, T.~E. and {Galloway}, D.~K. and 
	{Brandt}, W.~N.},
    title = "{The Variable Warm Absorber in Circinus X-1}",
  journal = {\apj},
archivePrefix = "arXiv",
   eprint = {0709.3336},
     year = 2008,
    month = jan,
   volume = 672,
    pages = {1091-1102},
      doi = {10.1086/523809},
   adsurl = {http://adsabs.harvard.edu/abs/2008ApJ...672.1091S}
}

@ARTICLE{Iaria2008,
       author = {{Iaria}, R. and {D'A{\'\i}}, A. and {Lavagetto}, G. and {Di Salvo}, T. and {Robba}, N.~R. and {Burderi}, L.},
        title = "{Chandra Observation of Cir X-1 near the Periastron Passage: Evidence for an X-Ray Jet?}",
      journal = {\apj},
         year = 2008,
        month = feb,
       volume = {673},
       number = {2},
        pages = {1033-1043},
          doi = {10.1086/524311},
       adsurl = {https://ui.adsabs.harvard.edu/abs/2008ApJ...673.1033I}
}

@ARTICLE{Schulz2020,
       author = {{Schulz}, N.~S. and {Kallman}, T.~E. and {Heinz}, S. and {Sell}, P. and {Jonker}, P. and {Brandt}, W.~N.},
        title = "{Origins of X-Ray Line Emissions in Circinus X-1 at Very Low X-Ray Flux}",
      journal = {\apj},
         year = 2020,
        month = mar,
       volume = {891},
       number = {2},
          eid = {150},
        pages = {150},
          doi = {10.3847/1538-4357/ab6dc8},
archivePrefix = {arXiv},
       eprint = {2001.05638},
 primaryClass = {astro-ph.HE},
       adsurl = {https://ui.adsabs.harvard.edu/abs/2020ApJ...891..150S}
}

@ARTICLE{Tominaga2023,
       author = {{Tominaga}, Mayu and {Tsujimoto}, Masahiro and {Ebisawa}, Ken and {Enoto}, Teruaki and {Hayasaki}, Kimitake},
        title = "{X-Ray Spectral Variations of Circinus X-1 Observed with NICER throughout an Entire Orbital Cycle}",
      journal = {\apj},
         year = 2023,
        month = nov,
       volume = {958},
       number = {1},
          eid = {52},
        pages = {52},
          doi = {10.3847/1538-4357/ad0034},
archivePrefix = {arXiv},
       eprint = {2310.07158},
 primaryClass = {astro-ph.HE},
       adsurl = {https://ui.adsabs.harvard.edu/abs/2023ApJ...958...52T}
}

@ARTICLE{Tsujimoto2025,
              author = {{Tsujimoto}, Masahiro and {Enoto}, Teruaki and {D{\'\i}az Trigo}, Mar{\'\i}a and {Hell}, Natalie and {Chakraborty}, Priyanka and {Leutenegger}, Maurice A. and {Loewenstein}, Michael and {Pradhan}, Pragati and {Shidatsu}, Megumi and {Takahashi}, Hiromitsu and {Yaqoob}, Tahir},
        title = "{Outflowing photoionized plasma in Circinus X-1 using the high-resolution X-ray spectrometer Resolve onboard XRISM and the radiative transfer code cloudy}",
      journal = {\pasj},
         year = 2025,
        month = sep,
       volume = {77},
        pages = {S72-S85},
          doi = {10.1093/pasj/psaf022},
archivePrefix = {arXiv},
       eprint = {2503.08254},
 primaryClass = {astro-ph.HE},
       adsurl = {https://ui.adsabs.harvard.edu/abs/2025PASJ...77S..72T}
}

@ARTICLE{Kotani2000,
       author = {{Kotani}, Taro and {Ebisawa}, Ken and {Dotani}, Tadayasu and {Inoue}, Hajime and {Nagase}, Fumiaki and {Tanaka}, Yasuo and {Ueda}, Yoshihiro},
        title = "{ASCA Observations of the Absorption Line Features from the Superluminal Jet Source GRS 1915+105}",
      journal = {\apj},
         year = 2000,
        month = aug,
       volume = {539},
       number = {1},
        pages = {413-423},
          doi = {10.1086/309200},
archivePrefix = {arXiv},
       eprint = {astro-ph/0003237},
 primaryClass = {astro-ph},
       adsurl = {https://ui.adsabs.harvard.edu/abs/2000ApJ...539..413K}
}

@ARTICLE{Ueda2001,
    author = {{Ueda}, Y. and {Asai}, K. and {Yamaoka}, K. and {Dotani}, T. and
    {Inoue}, H.},
    title = "{Discovery of an Iron K Absorption Line in the Low-Mass X-Ray Binary GX 13+1}",
    journal = {\apjl},
    year = 2001,
    month = aug,
    volume = 556,
    pages = {L87-L90},
    adsurl = {http://cdsads.u-strasbg.fr/cgi-bin/nph-bib_query?bibcode=2001ApJ...556L..87U&db_key=AST},
}

@ARTICLE{Sidoli2002,
    author = {{Sidoli}, L. and {Parmar}, A.~N. and {Oosterbroek}, T. and {Lumb}, D.
    },
    title = "{Discovery of complex narrow X-ray absorption features from the low-mass X-ray binary GX 13+1 with XMM-Newton}",
    journal = {\aap},
    year = 2002,
    month = apr,
    volume = 385,
    pages = {940-946},
    adsurl = {http://cdsads.u-strasbg.fr/cgi-bin/nph-bib_query?bibcode=2002A%26A...385..940S&db_key=AST},
}

@ARTICLE{Ueda2004,
   author = {{Ueda}, Y. and {Murakami}, H. and {Yamaoka}, K. and {Dotani}, T. and
    {Ebisawa}, K.},
    title = "{Chandra High-Resolution Spectroscopy of the Absorption-Line Features in the Low-Mass X-Ray Binary GX 13+1}",
  journal = {\apj},
     year = 2004,
    month = jul,
   volume = 609,
    pages = {325-334},
   adsurl = {http://adsabs.harvard.edu/cgi-bin/nph-bib_query?bibcode=2004ApJ...609..325U&db_key=AST}
}

@ARTICLE{Madej2014,
   author = {{Madej}, O.~K. and {Jonker}, P.~G. and {D{\'{\i}}az Trigo}, M. and 
	{Mi{\v s}kovi{\v c}ov{\'a}}, I.},
    title = "{Variable Doppler shifts of the thermal wind absorption lines in low-mass X-ray binaries}",
  journal = {\mnras},
archivePrefix = "arXiv",
   eprint = {1311.0874},
 primaryClass = "astro-ph.HE",
     year = 2014,
    month = feb,
   volume = 438,
    pages = {145-155},
      doi = {10.1093/mnras/stt2119},
   adsurl = {http://adsabs.harvard.edu/abs/2014MNRAS.438..145M}
}

@ARTICLE{Allen2018,
   author = {{Allen}, J.~L. and {Schulz}, N.~S. and {Homan}, J. and {Neilsen}, J. and 
	{Nowak}, M.~A. and {Chakrabarty}, D.},
    title = "{The Disk Wind in the Neutron Star Low-mass X-Ray Binary GX 13+1}",
  journal = {\apj},
archivePrefix = "arXiv",
   eprint = {1806.08800},
 primaryClass = "astro-ph.HE",
     year = 2018,
    month = jul,
   volume = 861,
      eid = {26},
    pages = {26},
      doi = {10.3847/1538-4357/aac2d1},
   adsurl = {http://adsabs.harvard.edu/abs/2018ApJ...861...26A}
}

@ARTICLE{Tomaru2018,
       author = {{Tomaru}, Ryota and {Done}, Chris and {Odaka}, Hirokazu and {Watanabe}, Shin and {Takahashi}, Tadayuki},
        title = "{Monte Carlo simulations of the detailed iron absorption line profiles from thermal winds in X-ray binaries}",
      journal = {\mnras},
         year = 2018,
        month = may,
       volume = {476},
       number = {2},
        pages = {1776-1784},
          doi = {10.1093/mnras/sty336},
archivePrefix = {arXiv},
       eprint = {1802.07019},
 primaryClass = {astro-ph.HE},
       adsurl = {https://ui.adsabs.harvard.edu/abs/2018MNRAS.476.1776T}
}

@ARTICLE{Rogantini2025,
       author = {{Rogantini}, Daniele and {Homan}, Jeroen and {Plotkin}, Richard M. and {van den Berg}, Maureen and {Miller-Jones}, James and {Neilsen}, Joey and {Chakrabarty}, Deepto and {Fender}, Rob P. and {Schulz}, Norbert},
        title = "{A persistent disk wind and variable jet outflow in the neutron-star low-mass X-ray binary GX 13+1}",
      journal = {arXiv e-prints},
         year = 2025,
        month = apr,
          eid = {arXiv:2504.05452},
        pages = {arXiv:2504.05452},
          doi = {10.48550/arXiv.2504.05452},
archivePrefix = {arXiv},
       eprint = {2504.05452},
 primaryClass = {astro-ph.HE},
       adsurl = {https://ui.adsabs.harvard.edu/abs/2025arXiv250405452R}
}

@ARTICLE{Younes2015,
       author = {{Younes}, G. and {Kouveliotou}, C. and {Grefenstette}, B.~W. and {Tomsick}, J.~A. and {Tennant}, A. and {Finger}, M.~H. and {F{\"u}rst}, F. and {Pottschmidt}, K. and {Bhalerao}, V. and {Boggs}, S.~E. and {Boirin}, L. and {Chakrabarty}, D. and {Christensen}, F.~E. and {Craig}, W.~W. and {Degenaar}, N. and {Fabian}, A.~C. and {Gandhi}, P. and {G{\"o}{\u{g}}{\"u}{\c{s}}}, E. and {Hailey}, C.~J. and {Harrison}, F.~A. and {Kennea}, J.~A. and {Miller}, J.~M. and {Stern}, D. and {Zhang}, W.~W.},
        title = "{Simultaneous NuSTAR/Chandra Observations of the Bursting Pulsar GRO J1744-28 during Its Third Reactivation}",
      journal = {\apj},
         year = 2015,
        month = may,
       volume = {804},
       number = {1},
          eid = {43},
        pages = {43},
          doi = {10.1088/0004-637X/804/1/43},
archivePrefix = {arXiv},
       eprint = {1502.05982},
 primaryClass = {astro-ph.HE},
       adsurl = {https://ui.adsabs.harvard.edu/abs/2015ApJ...804...43Y}
}

@ARTICLE{Miller2016,
       author = {{Miller}, J.~M. and {Raymond}, J. and {Cackett}, E. and {Grinberg}, V. and {Nowak}, M.},
        title = "{An Ultra-fast X-Ray Disk Wind in the Neutron Star Binary GX 340+0}",
      journal = {\apjl},
         year = 2016,
        month = may,
       volume = {822},
       number = {1},
          eid = {L18},
        pages = {L18},
          doi = {10.3847/2041-8205/822/1/L18},
archivePrefix = {arXiv},
       eprint = {1604.03329},
 primaryClass = {astro-ph.HE},
       adsurl = {https://ui.adsabs.harvard.edu/abs/2016ApJ...822L..18M}
}

@ARTICLE{Nowak2019,
       author = {{Nowak}, Michael A. and {Paizis}, Adamantia and {Jaisawal}, Gaurava Kumar and {Chenevez}, J{\'e}r{\^o}me and {Chaty}, Sylvain and {Fortin}, Francis and {Rodriguez}, J{\'e}r{\^o}me and {Wilms}, J{\"o}rn},
        title = "{Chandra-HETGS Characterization of an Outflowing Wind in the Accreting Millisecond Pulsar IGR J17591-2342}",
      journal = {\apj},
         year = 2019,
        month = mar,
       volume = {874},
       number = {1},
          eid = {69},
        pages = {69},
          doi = {10.3847/1538-4357/ab0a71},
archivePrefix = {arXiv},
       eprint = {1902.09577},
 primaryClass = {astro-ph.HE},
       adsurl = {https://ui.adsabs.harvard.edu/abs/2019ApJ...874...69N}
}

@ARTICLE{Manca2023,
       author = {{Manca}, A. and {Gambino}, A.~F. and {Sanna}, A. and {Jaisawal}, G.~K. and {Di Salvo}, T. and {Iaria}, R. and {Mazzola}, S.~M. and {Marino}, A. and {Anitra}, A. and {Bozzo}, E. and {Riggio}, A. and {Burderi}, L.},
        title = "{Spectral analysis of the AMXP IGR J17591-2342 during its 2018 outburst}",
      journal = {\mnras},
         year = 2023,
        month = feb,
       volume = {519},
       number = {2},
        pages = {2309-2320},
          doi = {10.1093/mnras/stac3707},
archivePrefix = {arXiv},
       eprint = {2212.07157},
 primaryClass = {astro-ph.HE},
       adsurl = {https://ui.adsabs.harvard.edu/abs/2023MNRAS.519.2309M}
}

@ARTICLE{Kallman2003,
       author = {{Kallman}, T.~R. and {Angelini}, L. and {Boroson}, B. and {Cottam}, J.},
        title = "{Chandra and XMM Observations of the Accretion Disk Corona Source 2S 0921-63}",
      journal = {\apj},
         year = 2003,
        month = feb,
       volume = {583},
       number = {2},
        pages = {861-877},
          doi = {10.1086/345475},
archivePrefix = {arXiv},
       eprint = {astro-ph/0209010},
 primaryClass = {astro-ph},
       adsurl = {https://ui.adsabs.harvard.edu/abs/2003ApJ...583..861K}
}

@ARTICLE{Yoneyama2023,
       author = {{Yoneyama}, Tomokage and {Dotani}, Tadayasu},
        title = "{X-ray spectroscopy of the accretion disk corona source 2S 0921-630 with Suzaku archival data}",
      journal = {\pasj},
         year = 2023,
        month = feb,
       volume = {75},
       number = {1},
        pages = {30-36},
          doi = {10.1093/pasj/psac086},
archivePrefix = {arXiv},
       eprint = {2210.10792},
 primaryClass = {astro-ph.HE},
       adsurl = {https://ui.adsabs.harvard.edu/abs/2023PASJ...75...30Y}
}

@ARTICLE{Ji2011,
       author = {{Ji}, L. and {Schulz}, N.~S. and {Nowak}, M.~A. and {Canizares}, C.~R.},
        title = "{Implications of X-ray Line Variations for 4U1822-371}",
      journal = {\apj},
         year = 2011,
        month = mar,
       volume = {729},
       number = {2},
          eid = {102},
        pages = {102},
          doi = {10.1088/0004-637X/729/2/102},
archivePrefix = {arXiv},
       eprint = {1007.3839},
 primaryClass = {astro-ph.HE},
       adsurl = {https://ui.adsabs.harvard.edu/abs/2011ApJ...729..102J}
}

@ARTICLE{Iaria2013,
       author = {{Iaria}, R. and {Di Salvo}, T. and {D'A{\`\i}}, A. and {Burderi}, L. and {Mineo}, T. and {Riggio}, A. and {Papitto}, A. and {Robba}, N.~R.},
        title = "{X-ray spectroscopy of the ADC source X1822-371 with Chandra and XMM-Newton}",
      journal = {\aap},
         year = 2013,
        month = jan,
       volume = {549},
          eid = {A33},
        pages = {A33},
          doi = {10.1051/0004-6361/201015305},
archivePrefix = {arXiv},
       eprint = {1210.0874},
 primaryClass = {astro-ph.HE},
       adsurl = {https://ui.adsabs.harvard.edu/abs/2013A&A...549A..33I}
}

@ARTICLE{Sasano2014,
       author = {{Sasano}, Makoto and {Makishima}, Kazuo and {Sakurai}, Soki and {Zhang}, Zhongli and {Enoto}, Teruaki},
        title = "{Suzaku view of the neutron star in the dipping source 4U 1822-37}",
      journal = {\pasj},
         year = 2014,
        month = apr,
       volume = {66},
       number = {2},
          eid = {35},
        pages = {35},
          doi = {10.1093/pasj/psu002},
archivePrefix = {arXiv},
       eprint = {1311.4618},
 primaryClass = {astro-ph.HE},
       adsurl = {https://ui.adsabs.harvard.edu/abs/2014PASJ...66...35S}
}

@ARTICLE{Jimenez-Garate2002,
       author = {{Jimenez-Garate}, M.~A. and {Hailey}, C.~J. and {den Herder}, J.~W. and {Zane}, S. and {Ramsay}, G.},
        title = "{High-Resolution X-Ray Spectroscopy of Hercules X-1 with the XMM-Newton Reflection Grating Spectrometer: CNO Element Abundance Measurements and Density Diagnostics of a Photoionized Plasma}",
      journal = {\apj},
         year = 2002,
        month = oct,
       volume = {578},
       number = {1},
        pages = {391-404},
          doi = {10.1086/342348},
archivePrefix = {arXiv},
       eprint = {astro-ph/0206181},
 primaryClass = {astro-ph},
       adsurl = {https://ui.adsabs.harvard.edu/abs/2002ApJ...578..391J}
}

@ARTICLE{Jimenez-Garate2005,
       author = {{Jimenez-Garate}, M.~A. and {Raymond}, J.~C. and {Liedahl}, D.~A. and {Hailey}, C.~J.},
        title = "{Identification of an Extended Accretion Disk Corona in the Hercules X-1 Low State: Moderate Optical Depth, Precise Density Determination, and Verification of CNO Abundances}",
      journal = {\apj},
         year = 2005,
        month = jun,
       volume = {625},
       number = {2},
        pages = {931-950},
          doi = {10.1086/426702},
archivePrefix = {arXiv},
       eprint = {astro-ph/0411780},
 primaryClass = {astro-ph},
       adsurl = {https://ui.adsabs.harvard.edu/abs/2005ApJ...625..931J}
}

@ARTICLE{Ji2009,
       author = {{Ji}, L. and {Schulz}, N. and {Nowak}, M. and {Marshall}, H.~L. and {Kallman}, T.},
        title = "{The Photoionized Accretion Disk in Her X-1}",
      journal = {\apj},
         year = 2009,
        month = aug,
       volume = {700},
       number = {2},
        pages = {977-988},
          doi = {10.1088/0004-637X/700/2/977},
archivePrefix = {arXiv},
       eprint = {0905.3773},
 primaryClass = {astro-ph.HE},
       adsurl = {https://ui.adsabs.harvard.edu/abs/2009ApJ...700..977J}
}

@ARTICLE{Kosec2020,
       author = {{Kosec}, P. and {Fabian}, A.~C. and {Pinto}, C. and {Walton}, D.~J. and {Dyda}, S. and {Reynolds}, C.~S.},
        title = "{An ionized accretion disc wind in Hercules X-1}",
      journal = {\mnras},
         year = 2020,
        month = jan,
       volume = {491},
       number = {3},
        pages = {3730-3750},
          doi = {10.1093/mnras/stz3200},
archivePrefix = {arXiv},
       eprint = {1910.08337},
 primaryClass = {astro-ph.HE},
       adsurl = {https://ui.adsabs.harvard.edu/abs/2020MNRAS.491.3730K}
}

@ARTICLE{Kosec2023,
       author = {{Kosec}, P. and {Kara}, E. and {Fabian}, A.~C. and {F{\"u}rst}, F. and {Pinto}, C. and {Psaradaki}, I. and {Reynolds}, C.~S. and {Rogantini}, D. and {Walton}, D.~J. and {Ballhausen}, R. and {Canizares}, C. and {Dyda}, S. and {Staubert}, R. and {Wilms}, J.},
        title = "{Vertical wind structure in an X-ray binary revealed by a precessing accretion disk}",
      journal = {Nature Astronomy},
         year = 2023,
        month = jun,
       volume = {7},
        pages = {715-723},
          doi = {10.1038/s41550-023-01929-7},
archivePrefix = {arXiv},
       eprint = {2304.05490},
 primaryClass = {astro-ph.HE},
       adsurl = {https://ui.adsabs.harvard.edu/abs/2023NatAs...7..715K}
}

@ARTICLE{Degenaar2016,
       author = {{Degenaar}, N. and {Altamirano}, D. and {Parker}, M. and {Miller-Jones}, J.~C.~A. and {Miller}, J.~M. and {Heinke}, C.~O. and {Wijnands}, R. and {Ludlam}, R. and {Parikh}, A. and {Hessels}, J.~W.~T. and {Gusinskaia}, N. and {Deller}, A.~T. and {Fabian}, A.~C.},
        title = "{Disc reflection and a possible disc wind during a soft X-ray state in the neutron star low-mass X-ray binary 1RXS J180408.9-342058}",
      journal = {\mnras},
         year = 2016,
        month = oct,
       volume = {461},
       number = {4},
        pages = {4049-4058},
          doi = {10.1093/mnras/stw1593},
archivePrefix = {arXiv},
       eprint = {1607.01780},
 primaryClass = {astro-ph.HE},
       adsurl = {https://ui.adsabs.harvard.edu/abs/2016MNRAS.461.4049D}
}

@ARTICLE{Narayan2003,
       author = {{Narayan}, Ramesh and {Igumenshchev}, Igor V. and {Abramowicz}, Marek A.},
        title = "{Magnetically Arrested Disk: an Energetically Efficient Accretion Flow}",
      journal = {\pasj},
         year = 2003,
        month = dec,
       volume = {55},
        pages = {L69-L72},
          doi = {10.1093/pasj/55.6.L69},
archivePrefix = {arXiv},
       eprint = {astro-ph/0305029},
 primaryClass = {astro-ph},
       adsurl = {https://ui.adsabs.harvard.edu/abs/2003PASJ...55L..69N}
}

@BOOK{Chandrasekhar1961, 
       author = {{Chandrasekhar}, Subrahmanyan},
        title = "{Hydrodynamic and hydromagnetic stability}",
         year = 1961,
       adsurl = {https://ui.adsabs.harvard.edu/abs/1961hhs..book.....C}
}

@ARTICLE{Done2018,
       author = {{Done}, Chris and {Tomaru}, Ryota and {Takahashi}, Tadayuki},
        title = "{Thermal winds in stellar mass black hole and neutron star binary systems}",
      journal = {\mnras},
         year = 2018,
        month = jan,
       volume = {473},
       number = {1},
        pages = {838-848},
          doi = {10.1093/mnras/stx2400},
archivePrefix = {arXiv},
       eprint = {1612.09377},
 primaryClass = {astro-ph.HE},
       adsurl = {https://ui.adsabs.harvard.edu/abs/2018MNRAS.473..838D}
}

@ARTICLE{elmellah2018,
       author = {{El Mellah}, I. and {Sundqvist}, J.~O. and {Keppens}, R.},
        title = "{Accretion from a clumpy massive-star wind in supergiant X-ray binaries}",
      journal = {\mnras},
         year = 2018,
        month = apr,
       volume = {475},
       number = {3},
        pages = {3240-3252},
          doi = {10.1093/mnras/stx3211},
archivePrefix = {arXiv},
       eprint = {1711.08709},
 primaryClass = {astro-ph.HE},
       adsurl = {https://ui.adsabs.harvard.edu/abs/2018MNRAS.475.3240E}
}

@ARTICLE{Kubota2019,
       author = {{Kubota}, Aya and {Done}, Chris},
        title = "{Modelling the spectral energy distribution of super-Eddington quasars}",
      journal = {\mnras},
         year = 2019,
        month = oct,
       volume = {489},
       number = {1},
        pages = {524-533},
          doi = {10.1093/mnras/stz2140},
archivePrefix = {arXiv},
       eprint = {1905.02920},
 primaryClass = {astro-ph.GA},
       adsurl = {https://ui.adsabs.harvard.edu/abs/2019MNRAS.489..524K}
}

@ARTICLE{Higginbottom2014,
       author = {{Higginbottom}, Nick and {Proga}, Daniel and {Knigge}, Christian and {Long}, Knox S. and {Matthews}, James H. and {Sim}, Stuart A.},
        title = "{Line-driven Disk Winds in Active Galactic Nuclei: The Critical Importance of Ionization and Radiative Transfer}",
      journal = {\apj},
         year = 2014,
        month = jul,
       volume = {789},
       number = {1},
          eid = {19},
        pages = {19},
          doi = {10.1088/0004-637X/789/1/19},
archivePrefix = {arXiv},
       eprint = {1402.1849},
 primaryClass = {astro-ph.GA},
       adsurl = {https://ui.adsabs.harvard.edu/abs/2014ApJ...789...19H}
}

@ARTICLE{Hori2018,
       author = {{Hori}, Takafumi and {Ueda}, Yoshihiro and {Done}, Chris and {Shidatsu}, Megumi and {Kubota}, Aya},
        title = "{Evolution of Thermally Driven Disk Wind in the Black Hole Binary 4U 1630-47 Observed with Suzaku and NuSTAR}",
      journal = {\apj},
         year = 2018,
        month = dec,
       volume = {869},
       number = {2},
          eid = {183},
        pages = {183},
          doi = {10.3847/1538-4357/aaea5e},
       adsurl = {https://ui.adsabs.harvard.edu/abs/2018ApJ...869..183H}
}

@ARTICLE{Lovelace1991,
       author = {{Lovelace}, R.~V.~E. and {Berk}, H.~L. and {Contopoulos}, J.},
        title = "{Magnetically Driven Jets and Winds}",
      journal = {\apj},
         year = 1991,
        month = oct,
       volume = {379},
        pages = {696},
          doi = {10.1086/170544},
       adsurl = {https://ui.adsabs.harvard.edu/abs/1991ApJ...379..696L}
}

@ARTICLE{Mizumoto2021,
       author = {{Mizumoto}, Misaki and {Nomura}, Mariko and {Done}, Chris and {Ohsuga}, Ken and {Odaka}, Hirokazu},
        title = "{UV line-driven disc wind as the origin of UltraFast Outflows in AGN}",
      journal = {\mnras},
         year = 2021,
        month = may,
       volume = {503},
       number = {1},
        pages = {1442-1458},
          doi = {10.1093/mnras/staa3282},
archivePrefix = {arXiv},
       eprint = {2003.01137},
 primaryClass = {astro-ph.HE},
       adsurl = {https://ui.adsabs.harvard.edu/abs/2021MNRAS.503.1442M}
}

@ARTICLE{Miller2006b,
       author = {{Miller}, J.~M. and {Raymond}, J. and {Homan}, J. and {Fabian}, A.~C. and {Steeghs}, D. and {Wijnands}, R. and {Rupen}, M. and {Charles}, P. and {van der Klis}, M. and {Lewin}, W.~H.~G.},
        title = "{Simultaneous Chandra and RXTE Spectroscopy of the Microquasar H1743-322: Clues to Disk Wind and Jet Formation from a Variable Ionized Outflow}",
      journal = {\apj},
         year = 2006,
        month = jul,
       volume = {646},
       number = {1},
        pages = {394-406},
          doi = {10.1086/504673},
archivePrefix = {arXiv},
       eprint = {astro-ph/0406272},
 primaryClass = {astro-ph},
       adsurl = {https://ui.adsabs.harvard.edu/abs/2006ApJ...646..394M}
}

@ARTICLE{Kallman2004,
       author = {{Kallman}, T.~R. and {Palmeri}, P. and {Bautista}, M.~A. and {Mendoza}, C. and {Krolik}, J.~H.},
        title = "{Photoionization Modeling and the K Lines of Iron}",
      journal = {\apjs},
         year = 2004,
        month = dec,
       volume = {155},
       number = {2},
        pages = {675-701},
          doi = {10.1086/424039},
archivePrefix = {arXiv},
       eprint = {astro-ph/0405210},
 primaryClass = {astro-ph},
       adsurl = {https://ui.adsabs.harvard.edu/abs/2004ApJS..155..675K}
}

@ARTICLE{Kubota2007,
       author = {{Kubota}, Aya and {Dotani}, Tadayasu and {Cottam}, Jean and {Kotani}, Taro and {Done}, Chris and {Ueda}, Yoshihiro and {Fabian}, Andrew C. and {Yasuda}, Tomonori and {Takahashi}, Hiromitsu and {Fukazawa}, Yasushi and {Yamaoka}, Kazutaka and {Makishima}, Kazuo and {Yamada}, Shinya and {Kohmura}, Takayoshi and {Angelini}, Lorella},
        title = "{Suzaku Discovery of Iron Absorption Lines in Outburst Spectra of the X-Ray Transient 4U 1630-472}",
      journal = {\pasj},
         year = 2007,
        month = jan,
       volume = {59},
        pages = {185-198},
          doi = {10.1093/pasj/59.sp1.S185},
archivePrefix = {arXiv},
       eprint = {astro-ph/0610496},
 primaryClass = {astro-ph},
       adsurl = {https://ui.adsabs.harvard.edu/abs/2007PASJ...59S.185K}
}

@ARTICLE{Kallman2009,
       author = {{Kallman}, T.~R. and {Bautista}, M.~A. and {Goriely}, Stephane and {Mendoza}, Claudio and {Miller}, Jon M. and {Palmeri}, Patrick and {Quinet}, Pascal and {Raymond}, John},
        title = "{Spectrum Synthesis Modeling of the X-Ray Spectrum of GRO J1655-40 Taken During the 2005 Outburst}",
      journal = {\apj},
         year = 2009,
        month = aug,
       volume = {701},
       number = {2},
        pages = {865-884},
          doi = {10.1088/0004-637X/701/2/865},
archivePrefix = {arXiv},
       eprint = {0905.4206},
 primaryClass = {astro-ph.HE},
       adsurl = {https://ui.adsabs.harvard.edu/abs/2009ApJ...701..865K}
}

@Article{Neilsen2023,
  author        = {{Neilsen}, Joey and {Degenaar}, Nathalie},
  journal       = {arXiv e-prints},
  title         = {{High-Resolution Spectroscopy of X-ray Binaries}},
  year          = {2023},
  month         = apr,
  pages         = {arXiv:2304.05412},
  adsurl        = {https://ui.adsabs.harvard.edu/abs/2023arXiv230405412N},
  archiveprefix = {arXiv},
  doi           = {10.48550/arXiv.2304.05412},
  eid           = {arXiv:2304.05412},
  eprint        = {2304.05412},
  primaryclass  = {astro-ph.HE},
}

@Article{Spencer1979,
  author   = {Spencer, R. E.},
  journal  = {Nature},
  title    = {A radio jet in {SS433}},
  year     = {1979},
  issn     = {0028-0836},
  month    = nov,
  note     = {ADS Bibcode: 1979Natur.282..483S},
  pages    = {483--484},
  volume   = {282},
  doi      = {10.1038/282483a0},
  url      = {https://ui.adsabs.harvard.edu/abs/1979Natur.282..483S},
  urldate  = {2025-06-06},
}

@Article{Proga2002a,
  author    = {Proga, Daniel and Kallman, Timothy R. and Drew, Janet E. and Hartley, Louise E.},
  journal   = {The Astrophysical Journal},
  title     = {Resonance {Line} {Profile} {Calculations} {Based} on {Hydrodynamical} {Models} of {Cataclysmic} {Variable} {Winds}},
  year      = {2002},
  issn      = {0004-637X},
  month     = jun,
  note      = {ADS Bibcode: 2002ApJ...572..382P},
  pages     = {382--391},
  volume    = {572},
  doi       = {10.1086/340339},
  publisher = {IOP},
  url       = {https://ui.adsabs.harvard.edu/abs/2002ApJ...572..382P},
  urldate   = {2025-06-09},
}

@ARTICLE{Behar2001,
       author = {{Behar}, Ehud and {Sako}, Masao and {Kahn}, Steven M.},
        title = "{Soft X-Ray Absorption by Fe$^{0+}$ to Fe$^{15+}$ in Active Galactic Nuclei}",
      journal = {\apj},
         year = 2001,
        month = dec,
       volume = {563},
       number = {2},
        pages = {497-504},
          doi = {10.1086/323966},
archivePrefix = {arXiv},
       eprint = {astro-ph/0109314},
 primaryClass = {astro-ph},
       adsurl = {https://ui.adsabs.harvard.edu/abs/2001ApJ...563..497B}
}

@ARTICLE{Degenaar2017,
       author = {{Degenaar}, N. and {Pinto}, C. and {Miller}, J.~M. and {Wijnands}, R. and {Altamirano}, D. and {Paerels}, F. and {Fabian}, A.~C. and {Chakrabarty}, D.},
        title = "{An in-depth study of a neutron star accreting at low Eddington rate: on the possibility of a truncated disc and an outflow}",
      journal = {\mnras},
         year = 2017,
        month = jan,
       volume = {464},
       number = {1},
        pages = {398-409},
          doi = {10.1093/mnras/stw2355},
archivePrefix = {arXiv},
       eprint = {1609.04816},
 primaryClass = {astro-ph.HE},
       adsurl = {https://ui.adsabs.harvard.edu/abs/2017MNRAS.464..398D}
}

@ARTICLE{vandenEijnden2018,
       author = {{van den Eijnden}, J. and {Degenaar}, N. and {Pinto}, C. and {Patruno}, A. and {Wette}, K. and {Messenger}, C. and {Hern{\'a}ndez Santisteban}, J.~V. and {Wijnands}, R. and {Miller}, J.~M. and {Altamirano}, D. and {Paerels}, F. and {Chakrabarty}, D. and {Fabian}, A.~C.},
        title = "{The very faint X-ray binary IGR J17062-6143: a truncated disc, no pulsations, and a possible outflow}",
      journal = {\mnras},
         year = 2018,
        month = apr,
       volume = {475},
       number = {2},
        pages = {2027-2044},
          doi = {10.1093/mnras/stx3224},
       adsurl = {https://ui.adsabs.harvard.edu/abs/2018MNRAS.475.2027V}
}

@ARTICLE{Miller2025,
       author = {{Miller}, Jon M. and {Mizumoto}, Misaki and {Shidatsu}, Megumi and {Ballhausen}, Ralf and {Behar}, Ehud and {Diaz Trigo}, Maria and {Done}, Chris and {Dotani}, Tadayasu and {Garcia}, Javier and {Kallman}, Timothy and {Kobayashi}, Shogo B. and {Kubota}, Aya and {Smith}, Randall and {Takahashi}, Hiromitsu and {Tashiro}, Makoto and {Ueda}, Yoshihiro and {Vink}, Jacco and {Yamada}, Shinya and {Watanabe}, Shin and {Iizuka}, Ryo and {Terada}, Yukikatsu and {Baluta}, Chris and {Kanemaru}, Yoshiaki and {Ogawa}, Shoji and {Yoshida}, Tessei and {Hayashi}, Katsuhiro},
        title = "{XRISM Spectroscopy of the Stellar-Mass Black Hole 4U 1630-472 in Outburst}",
      journal = {arXiv e-prints},
         year = 2025,
        month = jun,
          eid = {arXiv:2506.07319},
        pages = {arXiv:2506.07319},
          doi = {10.48550/arXiv.2506.07319},
archivePrefix = {arXiv},
       eprint = {2506.07319},
 primaryClass = {astro-ph.HE},
       adsurl = {https://ui.adsabs.harvard.edu/abs/2025arXiv250607319M}
}

@ARTICLE{Gatuzz2020,
       author = {{Gatuzz}, E. and {D{\'\i}az Trigo}, M. and {Miller-Jones}, J.~C.~A. and {Migliari}, S.},
        title = "{Simultaneous detection of an intrinsic absorber and a compact jet emission in the X-ray binary IGR J17091-3624 during a hard accretion state}",
      journal = {\mnras},
         year = 2020,
        month = feb,
       volume = {491},
       number = {4},
        pages = {4857-4868},
          doi = {10.1093/mnras/stz3385},
archivePrefix = {arXiv},
       eprint = {1912.02180},
 primaryClass = {astro-ph.HE},
       adsurl = {https://ui.adsabs.harvard.edu/abs/2020MNRAS.491.4857G}
}

@Article{Cuneo2023,
  author   = {C\'uneo, V. A. and Mu\~noz-Darias, T. and Jiménez-Ibarra, F. and Panizo-Espinar, G. and Sánchez-Sierras, J. and Armas Padilla, M. and Casares, J. and Mata Sánchez, D. and Torres, M. A. P. and Vincentelli, F. and Ambrifi, A.},
  journal  = {Astronomy and Astrophysics},
  title    = {Unveiling optical signatures of outflows in accreting white dwarfs},
  year     = {2023},
  issn     = {0004-6361},
  month    = nov,
  note     = {ADS Bibcode: 2023A\&A...679A..85C},
  pages    = {A85},
  volume   = {679},
  doi      = {10.1051/0004-6361/202347265},
  url      = {https://ui.adsabs.harvard.edu/abs/2023A&A...679A..85C},
  urldate  = {2025-06-17},
}

@Article{Blundell2011,
  author     = {Blundell, Katherine M. and Schmidtobreick, Linda and Trushkin, Sergei},
  journal    = {Monthly Notices of the Royal Astronomical Society},
  title      = {{SS433}'s accretion disc, wind and jets: before, during and after a major flare},
  year       = {2011},
  issn       = {0035-8711},
  month      = nov,
  note       = {ADS Bibcode: 2011MNRAS.417.2401B},
  pages      = {2401--2410},
  volume     = {417},
  doi        = {10.1111/j.1365-2966.2011.18785.x},
  publisher  = {OUP},
  shorttitle = {{SS433}'s accretion disc, wind and jets},
  url        = {https://ui.adsabs.harvard.edu/abs/2011MNRAS.417.2401B},
  urldate    = {2025-06-17},
}

@Article{Fabrika2004,
  author        = {{Fabrika}, S.},
  journal       = {\apspr},
  title         = {{The jets and supercritical accretion disk in SS433}},
  year          = {2004},
  month         = jan,
  pages         = {1-152},
  volume        = {12},
  adsurl        = {https://ui.adsabs.harvard.edu/abs/2004ASPRv..12....1F},
  archiveprefix = {arXiv},
  doi           = {10.48550/arXiv.astro-ph/0603390},
  eprint        = {astro-ph/0603390},
  primaryclass  = {astro-ph},
}

@Article{Folha2001,
  author    = {Folha, D. F. M. and Emerson, J. P.},
  journal   = {Astronomy and Astrophysics},
  title     = {Near infrared hydrogen lines as diagnostic of accretion and winds in {T} {Tauri} stars},
  year      = {2001},
  issn      = {0004-6361},
  month     = jan,
  note      = {ADS Bibcode: 2001A\&A...365...90F},
  pages     = {90--109},
  volume    = {365},

  doi       = {10.1051/0004-6361:20000018},
  publisher = {EDP},
  url       = {https://ui.adsabs.harvard.edu/abs/2001A&A...365...90F},
  urldate   = {2025-06-23},
}

@Article{Miller2007,
  author   = {Miller, J. M.},
  journal  = {Annual Review of Astronomy and Astrophysics},
  title    = {Relativistic {X}-{Ray} {Lines} from the {Inner} {Accretion} {Disks} {Around} {Black} {Holes}},
  year     = {2007},
  issn     = {0066-4146},
  month    = sep,
  note     = {ADS Bibcode: 2007ARA\&A..45..441M},
  pages    = {441--479},
  volume   = {45},
  doi      = {10.1146/annurev.astro.45.051806.110555},
  url      = {https://ui.adsabs.harvard.edu/abs/2007ARA&A..45..441M},
  urldate  = {2025-06-24},
}

@Article{Rozanska2014,
  author   = {Różańska, A. and Madej, J. and Bagińska, P. and Hryniewicz, K. and Handzlik, B.},
  journal  = {Astronomy and Astrophysics},
  title    = {Disk emission and atmospheric absorption lines in black hole candidate {4U} 1630-472},
  year     = {2014},
  issn     = {0004-6361},
  month    = feb,
  note     = {ADS Bibcode: 2014A\&A...562A..81R},
  pages    = {A81},
  volume   = {562},
  doi      = {10.1051/0004-6361/201321567},
  url      = {https://ui.adsabs.harvard.edu/abs/2014A&A...562A..81R},
  urldate  = {2025-06-24},
}

@Article{Miceli2024,
  author   = {Miceli, C. and Mata Sánchez, D. and Anitra, A. and Muñoz-Darias, T. and Di Salvo, T. and Iaria, R. and Marino, A. and Leone, W. and Del Santo, M. and Armas-Padilla, M. and Degenaar, N. and Miller, J. M. and Reynolds, M.},
  journal  = {Astronomy and Astrophysics},
  title    = {Soft-state optical spectroscopy of the black hole {MAXI} {J1305}-704},
  year     = {2024},
  issn     = {0004-6361},
  month    = apr,
  note     = {ADS Bibcode: 2024A\&A...684A..67M},
  pages    = {A67},
  volume   = {684},
  doi      = {10.1051/0004-6361/202348482},
  url      = {https://ui.adsabs.harvard.edu/abs/2024A&A...684A..67M},
  urldate  = {2025-06-24},
}

@Article{Avakyan2023,
  author     = {Avakyan, A. and Neumann, M. and Zainab, A. and Doroshenko, V. and Wilms, J. and Santangelo, A.},
  journal    = {Astronomy and Astrophysics},
  title      = {{XRBcats}: {Galactic} low-mass {X}-ray binary catalogue},
  year       = {2023},
  issn       = {0004-6361},
  month      = jul,
  note       = {ADS Bibcode: 2023A\&A...675A.199A},
  pages      = {A199},
  volume     = {675},
  doi        = {10.1051/0004-6361/202346522},
  publisher  = {EDP},
  shorttitle = {{XRBcats}},
  url        = {https://ui.adsabs.harvard.edu/abs/2023A&A...675A.199A},
  urldate    = {2025-06-26},
}

@Article{Corral-Santana2016,
  author        = {{Corral-Santana}, J.~M. and {Casares}, J. and {Mu{\~n}oz-Darias}, T. and {Bauer}, F.~E. and {Mart{\'{\i}}nez-Pais}, I.~G. and {Russell}, D.~M.},
  journal       = {\aap},
  title         = {{BlackCAT: A catalogue of stellar-mass black holes in X-ray transients}},
  year          = {2016},
  month         = mar,
  pages         = {A61},
  volume        = {587},
  adsurl        = {http://ads.nao.ac.jp/abs/2016A%26A...587A..61C},
  archiveprefix = {arXiv},
  doi           = {10.1051/0004-6361/201527130},
  eid           = {A61},
  eprint        = {1510.08869},
  primaryclass  = {astro-ph.HE},
}

@Article{Matthews2025,
  author     = {Matthews, James H. and Long, Knox S. and Knigge, Christian and Sim, Stuart A. and Parkinson, Edward J. and Higginbottom, Nick and Mangham, Samuel W. and Scepi, Nicolas and Wallis, Austen and Hewitt, Henrietta A. and Mosallanezhad, Amin},
  journal    = {Monthly Notices of the Royal Astronomical Society},
  title      = {{SIROCCO}: a publicly available {Monte} {Carlo} ionization and radiative transfer code for astrophysical outflows},
  year       = {2025},
  issn       = {0035-8711},
  month      = jan,
  note       = {ADS Bibcode: 2025MNRAS.536..879M},
  pages      = {879--904},
  volume     = {536},
  doi        = {10.1093/mnras/stae2677},
  publisher  = {OUP},
  shorttitle = {{SIROCCO}},
  url        = {https://ui.adsabs.harvard.edu/abs/2025MNRAS.536..879M},
  urldate    = {2025-06-27},
}

@ARTICLE{Zhang2025,
       author = {{Zhang}, Lizhong and {Stone}, James M. and {Mullen}, Patrick D. and {Davis}, Shane W. and {Jiang}, Yan-Fei and {White}, Christopher J.},
        title = "{Radiation GRMHD Models of Accretion onto Stellar-Mass Black Holes: I. Survey of Eddington Ratios}",
      journal = {arXiv e-prints},
         year = 2025,
        month = jun,
          eid = {arXiv:2506.02289},
        pages = {arXiv:2506.02289},
          doi = {10.48550/arXiv.2506.02289},
archivePrefix = {arXiv},
       eprint = {2506.02289},
 primaryClass = {astro-ph.HE},
       adsurl = {https://ui.adsabs.harvard.edu/abs/2025arXiv250602289Z}
}

@Article{Buisson2020,
  author     = {Buisson, D. J. K. and Altamirano, D. and Díaz Trigo, M. and Mendez, M. and Armas Padilla, M. and Castro Segura, N. and Degenaar, N. D. and van den Eijnden, J. and Fogantini, F. A. and Gandhi, P. and Knigge, C. and Muñoz-Darias, T. and Özbey Arabacı, M. and Vincentelli, F. M.},
  journal    = {Monthly Notices of the Royal Astronomical Society},
  title      = {Soft {X}-ray emission lines in the {X}-ray binary {Swift} {J1858}.6-0814 observed with {XMM}-{Newton} {Reflection} {Grating} {Spectrometer}: disc atmosphere or wind?},
  year       = {2020},
  issn       = {0035-8711},
  month      = oct,
  note       = {ADS Bibcode: 2020MNRAS.498...68B},
  pages      = {68--76},
  volume     = {498},
  doi        = {10.1093/mnras/staa2258},
  publisher  = {OUP},
  shorttitle = {Soft {X}-ray emission lines in the {X}-ray binary {Swift} {J1858}.6-0814 observed with {XMM}-{Newton} {Reflection} {Grating} {Spectrometer}},
  url        = {https://ui.adsabs.harvard.edu/abs/2020MNRAS.498...68B},
  urldate    = {2025-06-30},
}

@Article{CastroSegura2024,
  author    = {Castro Segura, N. and Knigge, C. and Matthews, J. H. and Vincentelli, F. M. and Charles, P. and Long, K. S. and Altamirano, D. and Buckley, D. A. H. and Modiano, D. and Torres, M. A. P. and Buisson, D. J. K. and Fijma, S. and Alabarta, K. and Degenaar, N. and Georganti, M. and Baglio, M. C.},
  journal   = {Monthly Notices of the Royal Astronomical Society},
  title     = {Shedding far-ultraviolet light on the donor star and evolutionary state of the neutron-star {LMXB} {Swift} {J1858}.6-0814},
  year      = {2024},
  issn      = {0035-8711},
  month     = jan,
  note      = {ADS Bibcode: 2024MNRAS.527.2508C},
  pages     = {2508--2522},
  volume    = {527},
  doi       = {10.1093/mnras/stad3109},
  publisher = {OUP},
  url       = {https://ui.adsabs.harvard.edu/abs/2024MNRAS.527.2508C},
  urldate   = {2025-06-30},
}

@Article{King2012,
  author    = {King, A. L. and Miller, J. M. and Raymond, J. and Fabian, A. C. and Reynolds, C. S. and Kallman, T. R. and Maitra, D. and Cackett, E. M. and Rupen, M. P.},
  journal   = {The Astrophysical Journal},
  title     = {An {Extreme} {X}-{Ray} {Disk} {Wind} in the {Black} {Hole} {Candidate} {IGR} {J17091}-3624},
  year      = {2012},
  issn      = {0004-637X},
  month     = feb,
  note      = {ADS Bibcode: 2012ApJ...746L..20K},
  pages     = {L20},
  volume    = {746},
  doi       = {10.1088/2041-8205/746/2/L20},
  publisher = {IOP},
  url       = {https://ui.adsabs.harvard.edu/abs/2012ApJ...746L..20K},
  urldate   = {2025-07-15},
}

@Article{Nitindala2025,
  author    = {Nitindala, Anagha P. and Veledina, Alexandra and Poutanen, Juri},
  journal   = {Astronomy and Astrophysics},
  title     = {X-ray polarization from accretion disk winds},
  year      = {2025},
  issn      = {0004-6361},
  month     = feb,
  note      = {ADS Bibcode: 2025A\&A...694A.230N},
  pages     = {A230},
  volume    = {694},
  doi       = {10.1051/0004-6361/202453188},
  publisher = {EDP},
  url       = {https://ui.adsabs.harvard.edu/abs/2025A&A...694A.230N},
  urldate   = {2025-07-17},
}

@Article{Weisskopf2022,
  author     = {Weisskopf, Martin C. and Soffitta, Paolo and Baldini, Luca and Ramsey, Brian D. and O'Dell, Stephen L. and Romani, Roger W. and Matt, Giorgio and Deininger, William D. and Baumgartner, Wayne H. and Bellazzini, Ronaldo and Costa, Enrico and Kolodziejczak, Jeffery J. and Latronico, Luca and Marshall, Herman L. and Muleri, Fabio and Bongiorno, Stephen D. and Tennant, Allyn and Bucciantini, Niccolo and Dovciak, Michal and Marin, Frederic and Marscher, Alan and Poutanen, Juri and Slane, Pat and Turolla, Roberto and Kalinowski, William and Di Marco, Alessandro and Fabiani, Sergio and Minuti, Massimo and La Monaca, Fabio and Pinchera, Michele and Rankin, John and Sgro', Carmelo and Trois, Alessio and Xie, Fei and Alexander, Cheryl and Allen, D. Zachery and Amici, Fabrizio and Andersen, Jason and Antonelli, Angelo and Antoniak, Spencer and Attinà, Primo and Barbanera, Mattia and Bachetti, Matteo and Baggett, Randy M. and Bladt, Jeff and Brez, Alessandro and Bonino, Raffaella and Boree, Christopher and Borotto, Fabio and Breeding, Shawn and Brienza, Daniele and Bygott, H. Kyle and Caporale, Ciro and Cardelli, Claudia and Carpentiero, Rita and Castellano, Simone and Castronuovo, Marco and Cavalli, Luca and Cavazzuti, Elisabetta and Ceccanti, Marco and Centrone, Mauro and Citraro, Saverio and D'Amico, Fabio and D'Alba, Elisa and Di Gesu, Laura and Del Monte, Ettore and Dietz, Kurtis L. and Di Lalla, Niccolo' and Persio, Giuseppe Di and Dolan, David and Donnarumma, Immacolata and Evangelista, Yuri and Ferrant, Kevin and Ferrazzoli, Riccardo and Ferrie, MacKenzie and Footdale, Joseph and Forsyth, Brent and Foster, Michelle and Garelick, Benjamin and Gunji, Shuichi and Gurnee, Eli and Head, Michael and Hibbard, Grant and Johnson, Samantha and Kelly, Erik and Kilaru, Kiranmayee and Lefevre, Carlo and Roy, Shelley Le and Loffredo, Pasqualino and Lorenzi, Paolo and Lucchesi, Leonardo and Maddox, Tyler and Magazzu, Guido and Maldera, Simone and Manfreda, Alberto and Mangraviti, Elio and Marengo, Marco and Marrocchesi, Alessandra and Massaro, Francesco and Mauger, David and McCracken, Jeffrey and McEachen, Michael and Mize, Rondal and Mereu, Paolo and Mitchell, Scott and Mitsuishi, Ikuyuki and Morbidini, Alfredo and Mosti, Federico and Nasimi, Hikmat and Negri, Barbara and Negro, Michela and Nguyen, Toan and Nitschke, Isaac and Nuti, Alessio and Onizuka, Mitch and Oppedisano, Chiara and Orsini, Leonardo and Osborne, Darren and Pacheco, Richard and Paggi, Alessandro and Painter, Will and Pavelitz, Steven D. and Pentz, Christina and Piazzolla, Raffaele and Perri, Matteo and Pesce-Rollins, Melissa and Peterson, Colin and Pilia, Maura and Profeti, Alessandro and Puccetti, Simonetta and Ranganathan, Jaganathan and Ratheesh, Ajay and Reedy, Lee and Root, Noah and Rubini, Alda and Ruswick, Stephanie and Sanchez, Javier and Sarra, Paolo and Santoli, Francesco and Scalise, Emanuele and Sciortino, Andrea and Schroeder, Christopher and Seek, Tim and Sosdian, Kalie and Spandre, Gloria and Speegle, Chet O. and Tamagawa, Toru and Tardiola, Marcello and Tobia, Antonino and Thomas, Nicholas E. and Valerie, Robert and Vimercati, Marco and Walden, Amy L. and Weddendorf, Bruce and Wedmore, Jeffrey and Welch, David and Zanetti, Davide and Zanetti, Francesco},
  journal    = {Journal of Astronomical Telescopes, Instruments, and Systems},
  title      = {The {Imaging} {X}-{Ray} {Polarimetry} {Explorer} ({IXPE}): {Pre}-{Launch}},
  year       = {2022},
  month      = apr,
  note       = {ADS Bibcode: 2022JATIS...8b6002W},
  pages      = {026002},
  volume     = {8},
  doi        = {10.1117/1.JATIS.8.2.026002},
  shorttitle = {The {Imaging} {X}-{Ray} {Polarimetry} {Explorer} ({IXPE})},
  url        = {https://ui.adsabs.harvard.edu/abs/2022JATIS...8b6002W},
  urldate    = {2025-07-17},
}

@InProceedings{Schipani2018,
  author     = {Schipani, P. and Campana, S. and Claudi, R. and Käufl, H. U. and Accardo, M. and Aliverti, M. and Baruffolo, A. and Ben Ami, S. and Biondi, F. and Brucalassi, A. and Capasso, G. and Cosentino, R. and D'Alessio, F. and D'Avanzo, P. and Hershko, O. and Gardiol, D. and Kuncarayacti, H. and Munari, M. and Rubin, A. and Scuderi, S. and Vitali, F. and Achrén, J. and Araiza-Duran, J. Antonio and Arcavi, I. and Bianco, A. and Cappellaro, E. and Colapietro, M. and Della Valle, M. and Diner, O. and D'Orsi, S. and Fantinel, D. and Fynbo, J. and Gal-Yam, A. and Genoni, M. and Hirvonen, M. and Kotilainen, J. and Kumar, T. and Landoni, M. and Lehti, J. and Li Causi, G. and Loreggia, D. and Marafatto, L. and Mattila, S. and Pariani, G. and Pignata, G. and Rappaport, M. and Ricci, D. and Riva, M. and Salasnich, B. and Zanmar Sanchez, R. and Smartt, S. and Turatto, M.},
  title      = {{SOXS}: a wide band spectrograph to follow up transients},
  year       = {2018},
  address    = {eprint: arXiv:1807.08828},
  month      = jul,
  note       = {ADS Bibcode: 2018SPIE10702E..0FS},
  pages      = {107020F},
  volume     = {10702},
  doi        = {10.1117/12.2307349},
  shorttitle = {{SOXS}},
  url        = {https://ui.adsabs.harvard.edu/abs/2018SPIE10702E..0FS},
  urldate    = {2025-07-17},
}

@Article{Kennedy2025,
  author   = {Kennedy, Mark R. and Callanan, P. and Garnavich, P. M. and Breton, R. P. and Brown, A. J. and Segura, N. Castro and Dhillon, V. S. and Dyer, M. J. and Fijma, S. and Garbutt, J. and Green, M. J. and Hakala, P. and Jiminez-Ibarra, F. and Kerry, P. and Littlefair, S. and Munday, J. and Mason, P. A. and Mata-Sanchez, D. and Munoz-Darias, T. and Parsons, S. and Pelisoli, I. and Sahman, D.},
  journal  = {The Open Journal of Astrophysics},
  title    = {Analysis of optical spectroscopy and photometry of the type {I} {X}-ray bursting system {UW} {CrB}},
  year     = {2025},
  issn     = {2565-6120},
  month    = jun,
  note     = {ADS Bibcode: 2025OJAp....8E..71K},
  pages    = {71},
  volume   = {8},
  doi      = {10.33232/001c.140729},
  url      = {https://ui.adsabs.harvard.edu/abs/2025OJAp....8E..71K},
  urldate  = {2025-07-21},
}

@ARTICLE{Costantini12,
       author = {{Costantini}, E. and {Pinto}, C. and {Kaastra}, J.~S. and {in't Zand}, J.~J.~M. and {Freyberg}, M.~J. and {Kuiper}, L. and {M{\'e}ndez}, M. and {de Vries}, C.~P. and {Waters}, L.~B.~F.~M.},
        title = "{XMM-Newton observation of 4U 1820-30. Broad band spectrum and the contribution of the cold interstellar medium}",
      journal = {\aap},
         year = 2012,
        month = mar,
       volume = {539},
          eid = {A32},
        pages = {A32},
          doi = {10.1051/0004-6361/201117818},
archivePrefix = {arXiv},
       eprint = {1112.4349},
 primaryClass = {astro-ph.GA},
       adsurl = {https://ui.adsabs.harvard.edu/abs/2012A&A...539A..32C}
}

@Article{Prabhakar2023,
  author     = {Prabhakar, Geethu and Mandal, Samir and Bhuvana, G. R. and Nandi, Anuj},
  journal    = {Monthly Notices of the Royal Astronomical Society},
  title      = {Wideband study of the brightest black hole {X}-ray binary {4U} 1543-47 in the 2021 outburst: signature of disc-wind regulated accretion},
  year       = {2023},
  issn       = {0035-8711},
  month      = apr,
  note       = {ADS Bibcode: 2023MNRAS.520.4889P},
  pages      = {4889--4901},
  volume     = {520},
  doi        = {10.1093/mnras/stad080},
  publisher  = {OUP},
  shorttitle = {Wideband study of the brightest black hole {X}-ray binary {4U} 1543-47 in the 2021 outburst},
  url        = {https://ui.adsabs.harvard.edu/abs/2023MNRAS.520.4889P},
  urldate    = {2025-07-23},
}

@Article{Xu2020,
  author     = {Xu, Yanjun and Harrison, Fiona A. and Tomsick, John A. and Walton, Dominic J. and Barret, Didier and García, Javier A. and Hare, Jeremy and Parker, Michael L.},
  journal    = {The Astrophysical Journal},
  title      = {Studying the {Reflection} {Spectra} of the {New} {Black} {Hole} {X}-{Ray} {Binary} {Candidate} {MAXI} {J1631}-479 {Observed} by {NuSTAR}: {A} {Variable} {Broad} {Iron} {Line} {Profile}},
  year       = {2020},
  issn       = {0004-637X},
  month      = apr,
  note       = {ADS Bibcode: 2020ApJ...893...30X},
  pages      = {30},
  volume     = {893},
  doi        = {10.3847/1538-4357/ab7dc0},
  publisher  = {IOP},
  shorttitle = {Studying the {Reflection} {Spectra} of the {New} {Black} {Hole} {X}-{Ray} {Binary} {Candidate} {MAXI} {J1631}-479 {Observed} by {NuSTAR}},
  url        = {https://ui.adsabs.harvard.edu/abs/2020ApJ...893...30X},
  urldate    = {2025-07-23},
}

@Article{Zhang2024,
  author    = {Zhang, Zuobin and Bambi, Cosimo and Liu, Honghui and Jiang, Jiachen and Shi, Fangzheng and Zhang, Yuexin and Young, Andrew J. and Tomsick, John A. and Coughenour, Benjamin M. and Zhou, Menglei},
  journal   = {The Astrophysical Journal},
  title     = {Variable {Ionized} {Disk} {Winds} in {MAXI} {J1803}‑298 {Revealed} by {NICER}},
  year      = {2024},
  issn      = {0004-637X},
  month     = nov,
  note      = {ADS Bibcode: 2024ApJ...975...22Z},
  pages     = {22},
  volume    = {975},
  doi       = {10.3847/1538-4357/ad7b29},
  publisher = {IOP},
  url       = {https://ui.adsabs.harvard.edu/abs/2024ApJ...975...22Z},
  urldate   = {2025-07-23},
}

@Article{DelSanto2023,
  author    = {Del Santo, M. and Pinto, C. and Marino, A. and D'Aì, A. and Petrucci, P. -O. and Malzac, J. and Ferreira, J. and Pintore, F. and Motta, S. E. and Russell, T. D. and Segreto, A. and Sanna, A.},
  journal   = {Monthly Notices of the Royal Astronomical Society},
  title     = {An ultrafast outflow in the black hole candidate {MAXI} {J1810}-222?},
  year      = {2023},
  issn      = {0035-8711},
  month     = jul,
  note      = {ADS Bibcode: 2023MNRAS.523L..15D},
  pages     = {L15--L20},
  volume    = {523},
  doi       = {10.1093/mnrasl/slad048},
  publisher = {OUP},
  url       = {https://ui.adsabs.harvard.edu/abs/2023MNRAS.523L..15D},
  urldate   = {2025-07-23},
}

@Article{Fabian2020,
  author     = {Fabian, A. C. and Buisson, D. J. and Kosec, P. and Reynolds, C. S. and Wilkins, D. R. and Tomsick, J. A. and Walton, D. J. and Gandhi, P. and Altamirano, D. and Arzoumanian, Z. and Cackett, E. M. and Dyda, S. and Garcia, J. A. and Gendreau, K. C. and Grefenstette, B. W. and Homan, J. and Kara, E. and Ludlam, R. M. and Miller, J. M. and Steiner, J. F.},
  journal    = {Monthly Notices of the Royal Astronomical Society},
  title      = {The soft state of the black hole transient source {MAXI} {J1820}+070: emission from the edge of the plunge region?},
  year       = {2020},
  issn       = {0035-8711},
  month      = apr,
  note       = {ADS Bibcode: 2020MNRAS.493.5389F},
  pages      = {5389--5396},
  volume     = {493},
  doi        = {10.1093/mnras/staa564},
  publisher  = {OUP},
  shorttitle = {The soft state of the black hole transient source {MAXI} {J1820}+070},
  url        = {https://ui.adsabs.harvard.edu/abs/2020MNRAS.493.5389F},
  urldate    = {2025-07-24},
}

@Article{Miller2015,
  author    = {Miller, J. M. and Fabian, A. C. and Kaastra, J. and Kallman, T. and King, A. L. and Proga, D. and Raymond, J. and Reynolds, C. S.},
  journal   = {The Astrophysical Journal},
  title     = {Powerful, {Rotating} {Disk} {Winds} from {Stellar}-mass {Black} {Holes}},
  year      = {2015},
  issn      = {0004-637X},
  month     = dec,
  note      = {ADS Bibcode: 2015ApJ...814...87M},
  pages     = {87},
  volume    = {814},
  doi       = {10.1088/0004-637X/814/2/87},
  publisher = {IOP},
  url       = {https://ui.adsabs.harvard.edu/abs/2015ApJ...814...87M},
  urldate   = {2025-07-24},
}

@Article{Chakraborty2021,
  author     = {Chakraborty, Sudip and Ratheesh, Ajay and Bhattacharyya, Sudip and Tomsick, John A. and Tombesi, Francesco and Fukumura, Keigo and Jaisawal, Gaurava K.},
  journal    = {Monthly Notices of the Royal Astronomical Society},
  title      = {{NuSTAR} monitoring of {MAXI} {J1348}-630: evidence of high density disc reflection},
  year       = {2021},
  issn       = {0035-8711},
  month      = nov,
  note       = {ADS Bibcode: 2021MNRAS.508..475C},
  pages      = {475--488},
  volume     = {508},
  doi        = {10.1093/mnras/stab2530},
  publisher  = {OUP},
  shorttitle = {{NuSTAR} monitoring of {MAXI} {J1348}-630},
  url        = {https://ui.adsabs.harvard.edu/abs/2021MNRAS.508..475C},
  urldate    = {2025-07-24},
}

@Article{Wu2023,
  author    = {Wu, Hanji and Wang, Wei and Sai, Na},
  journal   = {Journal of High Energy Astrophysics},
  title     = {Accretion disk wind during the outburst of the stellar-mass black hole {MAXI} {J1348}-630},
  year      = {2023},
  issn      = {2214-4048},
  month     = mar,
  note      = {ADS Bibcode: 2023JHEAp..37...25W},
  pages     = {25--33},
  volume    = {37},
  doi       = {10.1016/j.jheap.2022.12.001},
  publisher = {Elsevier},
  url       = {https://ui.adsabs.harvard.edu/abs/2023JHEAp..37...25W},
  urldate   = {2025-07-24},
}

@Article{Neilsen2014,
  author    = {Neilsen, Joseph and Coriat, Mickaël and Fender, Rob and Lee, Julia C. and Ponti, Gabriele and Tzioumis, Anastasios K. and Edwards, Philip G. and Broderick, Jess W.},
  journal   = {The Astrophysical Journal},
  title     = {A {Link} between {X}-{Ray} {Emission} {Lines} and {Radio} {Jets} in {4U} 1630-47?},
  year      = {2014},
  issn      = {0004-637X},
  month     = mar,
  note      = {ADS Bibcode: 2014ApJ...784L...5N},
  pages     = {L5},
  volume    = {784},
  doi       = {10.1088/2041-8205/784/1/L5},
  publisher = {IOP},
  url       = {https://ui.adsabs.harvard.edu/abs/2014ApJ...784L...5N},
  urldate   = {2025-07-24},
}

@Article{Reeves2009,
  author    = {Reeves, J. N. and O'Brien, P. T. and Braito, V. and Behar, E. and Miller, L. and Turner, T. J. and Fabian, A. C. and Kaspi, S. and Mushotzky, R. and Ward, M.},
  journal   = {The Astrophysical Journal},
  title     = {A {Compton}-thick {Wind} in the {High}-luminosity {Quasar}, {PDS} 456},
  year      = {2009},
  issn      = {0004-637X},
  month     = aug,
  note      = {ADS Bibcode: 2009ApJ...701..493R},
  pages     = {493--507},
  volume    = {701},
  doi       = {10.1088/0004-637X/701/1/493},
  publisher = {IOP},
  url       = {https://ui.adsabs.harvard.edu/abs/2009ApJ...701..493R},
}

@ARTICLE{Wang2021,
       author = {{Wang}, Yanan and {Ji}, Long and {Garc{\'\i}a}, Javier A. and {Dauser}, Thomas and {M{\'e}ndez}, Mariano and {Mao}, Junjie and {Tao}, L. and {Altamirano}, Diego and {Maggi}, Pierre and {Zhang}, S.~N. and {Ge}, M.~Y. and {Zhang}, L. and {Qu}, J.~L. and {Zhang}, S. and {Ma}, X. and {Lu}, F.~J. and {Li}, T.~P. and {Huang}, Y. and {Zheng}, S.~J. and {Chang}, Z. and {Tuo}, Y.~L. and {Song}, L.~M. and {Xu}, Y.~P. and {Chen}, Y. and {Liu}, C.~Z. and {Bu}, Q.~C. and {Cai}, C. and {Cao}, X.~L. and {Chen}, L. and {Chen}, T.~X. and {Chen}, Y.~P. and {Cui}, W.~W. and {Du}, Y.~Y. and {Gao}, G.~H. and {Gu}, Y.~D. and {Guan}, J. and {Guo}, C.~C. and {Han}, D.~W. and {Huo}, J. and {Jia}, S.~M. and {Jiang}, W.~C. and {Jin}, J. and {Kong}, L.~D. and {Li}, B. and {Li}, C.~K. and {Li}, G. and {Li}, W. and {Li}, X. and {Li}, X.~B. and {Li}, X.~F. and {Li}, Z.~W. and {Liang}, X.~H. and {Liao}, J.~Y. and {Liu}, H.~W. and {Liu}, X.~J. and {Lu}, X.~F. and {Luo}, Q. and {Luo}, T. and {Meng}, B. and {Nang}, Y. and {Nie}, J.~Y. and {Ou}, G. and {Sai}, N. and {Shang}, R.~C. and {Song}, X.~Y. and {Sun}, L. and {Tan}, Y. and {Wang}, W.~S. and {Wang}, Y.~D. and {Wang}, Y.~S. and {Wen}, X.~Y. and {Wu}, B.~B. and {Wu}, B.~Y. and {Wu}, M. and {Xiao}, G.~C. and {Xiao}, S. and {Xiong}, S.~L. and {Yang}, S. and {Yang}, Y.~J. and {Yi}, Q.~B. and {Yin}, Q.~Q. and {You}, Y. and {Zhang}, F. and {Zhang}, H.~M. and {Zhang}, J. and {Zhang}, W.~C. and {Zhang}, W. and {Zhang}, Y.~F. and {Zhao}, H.~S. and {Zhao}, X.~F. and {Zhou}, D.~K.},
        title = "{A Variable Ionized Disk Wind in the Black Hole Candidate EXO 1846-031}",
      journal = {\apj},
         year = 2021,
        month = jan,
       volume = {906},
       number = {1},
          eid = {11},
        pages = {11},
          doi = {10.3847/1538-4357/abc55e},
archivePrefix = {arXiv},
       eprint = {2010.14662},
 primaryClass = {astro-ph.HE},
       adsurl = {https://ui.adsabs.harvard.edu/abs/2021ApJ...906...11W}
}

@ARTICLE{King2013,
       author = {{King}, A.~L. and {Miller}, J.~M. and {Raymond}, J. and {Fabian}, A.~C. and {Reynolds}, C.~S. and {G{\"u}ltekin}, K. and {Cackett}, E.~M. and {Allen}, S.~W. and {Proga}, D. and {Kallman}, T.~R.},
        title = "{Regulation of Black Hole Winds and Jets across the Mass Scale}",
      journal = {\apj},
         year = 2013,
        month = jan,
       volume = {762},
       number = {2},
          eid = {103},
        pages = {103},
          doi = {10.1088/0004-637X/762/2/103},
archivePrefix = {arXiv},
       eprint = {1205.4222},
 primaryClass = {astro-ph.HE},
       adsurl = {https://ui.adsabs.harvard.edu/abs/2013ApJ...762..103K}
}

@ARTICLE{Miller2012,
       author = {{Miller}, J.~M. and {Raymond}, J. and {Fabian}, A.~C. and {Reynolds}, C.~S. and {King}, A.~L. and {Kallman}, T.~R. and {Cackett}, E.~M. and {van der Klis}, M. and {Steeghs}, D.~T.~H.},
        title = "{The Disk-wind-Jet Connection in the Black Hole H 1743-322}",
      journal = {\apjl},
         year = 2012,
        month = nov,
       volume = {759},
       number = {1},
          eid = {L6},
        pages = {L6},
          doi = {10.1088/2041-8205/759/1/L6},
archivePrefix = {arXiv},
       eprint = {1208.4514},
 primaryClass = {astro-ph.HE},
       adsurl = {https://ui.adsabs.harvard.edu/abs/2012ApJ...759L...6M}
}

@ARTICLE{Blum2010,
       author = {{Blum}, J.~L. and {Miller}, J.~M. and {Cackett}, E. and {Yamaoka}, K. and {Takahashi}, H. and {Raymond}, J. and {Reynolds}, C.~S. and {Fabian}, A.~C.},
        title = "{Suzaku Observations of the Black Hole H1743-322 in Outburst}",
      journal = {\apj},
         year = 2010,
        month = apr,
       volume = {713},
       number = {2},
        pages = {1244-1248},
          doi = {10.1088/0004-637X/713/2/1244},
archivePrefix = {arXiv},
       eprint = {1003.3417},
 primaryClass = {astro-ph.HE},
       adsurl = {https://ui.adsabs.harvard.edu/abs/2010ApJ...713.1244B}
}

@ARTICLE{King2014,
       author = {{King}, Ashley L. and {Walton}, Dominic J. and {Miller}, Jon M. and {Barret}, Didier and {Boggs}, Steven E. and {Christensen}, Finn E. and {Craig}, William W. and {Fabian}, Andy C. and {F{\"u}rst}, Felix and {Hailey}, Charles J. and {Harrison}, Fiona A. and {Krivonos}, Roman and {Mori}, Kaya and {Natalucci}, Lorenzo and {Stern}, Daniel and {Tomsick}, John A. and {Zhang}, William W.},
        title = "{The Disk Wind in the Rapidly Spinning Stellar-mass Black Hole 4U 1630-472 Observed with NuSTAR}",
      journal = {\apjl},
         year = 2014,
        month = mar,
       volume = {784},
       number = {1},
          eid = {L2},
        pages = {L2},
          doi = {10.1088/2041-8205/784/1/L2},
archivePrefix = {arXiv},
       eprint = {1401.3646},
 primaryClass = {astro-ph.HE},
       adsurl = {https://ui.adsabs.harvard.edu/abs/2014ApJ...784L...2K}
}

@ARTICLE{Wang2016,
       author = {{Wang}, Yanan and {M{\'e}ndez}, Mariano},
        title = "{The XMM-Newton spectra of the 2012 outburst of the black hole candidate 4U 1630-47 revisited}",
      journal = {\mnras},
         year = 2016,
        month = feb,
       volume = {456},
       number = {2},
        pages = {1579-1586},
          doi = {10.1093/mnras/stv2814},
archivePrefix = {arXiv},
       eprint = {1512.00512},
 primaryClass = {astro-ph.HE},
       adsurl = {https://ui.adsabs.harvard.edu/abs/2016MNRAS.456.1579W}
}

@ARTICLE{Pahari2018,
       author = {{Pahari}, Mayukh and {Bhattacharyya}, Sudip and {Rao}, A.~R. and {Bhattacharya}, Dipankar and {Vadawale}, Santosh V. and {Dewangan}, Gulab C. and {McHardy}, I.~M. and {Gandhi}, Poshak and {Corbel}, St{\'e}phane and {Schulz}, Norbert S. and {Altamirano}, Diego},
        title = "{AstroSat and Chandra View of the High Soft State of 4U 1630-47 (4U 1630-472): Evidence of the Disk Wind and a Rapidly Spinning Black Hole}",
      journal = {\apj},
         year = 2018,
        month = nov,
       volume = {867},
       number = {2},
          eid = {86},
        pages = {86},
          doi = {10.3847/1538-4357/aae53b},
archivePrefix = {arXiv},
       eprint = {1810.01275},
 primaryClass = {astro-ph.HE},
       adsurl = {https://ui.adsabs.harvard.edu/abs/2018ApJ...867...86P}
}

@ARTICLE{Zeegers2019,
       author = {{Zeegers}, S.~T. and {Costantini}, E. and {Rogantini}, D. and {de Vries}, C.~P. and {Mutschke}, H. and {Mohr}, P. and {de Groot}, F. and {Tielens}, A.~G.~G.~M.},
        title = "{Dust absorption and scattering in the silicon K-edge}",
      journal = {\aap},
         year = 2019,
        month = jul,
       volume = {627},
          eid = {A16},
        pages = {A16},
          doi = {10.1051/0004-6361/201935050},
archivePrefix = {arXiv},
       eprint = {1905.06560},
 primaryClass = {astro-ph.GA},
       adsurl = {https://ui.adsabs.harvard.edu/abs/2019A&A...627A..16Z}
}

@ARTICLE{Trueba2019,
       author = {{Trueba}, N. and {Miller}, J.~M. and {Kaastra}, J. and {Zoghbi}, A. and {Fabian}, A.~C. and {Kallman}, T. and {Proga}, D. and {Raymond}, J.},
        title = "{A Comprehensive Chandra Study of the Disk Wind in the Black Hole Candidate 4U 1630-472}",
      journal = {\apj},
         year = 2019,
        month = dec,
       volume = {886},
       number = {2},
          eid = {104},
        pages = {104},
          doi = {10.3847/1538-4357/ab4f70},
archivePrefix = {arXiv},
       eprint = {1910.11243},
 primaryClass = {astro-ph.HE},
       adsurl = {https://ui.adsabs.harvard.edu/abs/2019ApJ...886..104T}
}

@ARTICLE{DiazTrigo2007,
       author = {{D{\'\i}az Trigo}, M. and {Parmar}, A.~N. and {Miller}, J. and {Kuulkers}, E. and {Caballero-Garc{\'\i}a}, M.~D.},
        title = "{XMM-Newton and INTEGRAL spectroscopy of the microquasar GRO J1655-40 during its 2005 outburst}",
      journal = {\aap},
         year = 2007,
        month = feb,
       volume = {462},
       number = {2},
        pages = {657-666},
          doi = {10.1051/0004-6361:20065406},
archivePrefix = {arXiv},
       eprint = {astro-ph/0610873},
 primaryClass = {astro-ph},
       adsurl = {https://ui.adsabs.harvard.edu/abs/2007A&A...462..657D}
}

@ARTICLE{Miller2008,
       author = {{Miller}, J.~M. and {Raymond}, J. and {Reynolds}, C.~S. and {Fabian}, A.~C. and {Kallman}, T.~R. and {Homan}, J.},
        title = "{The Accretion Disk Wind in the Black Hole GRO J1655-40}",
      journal = {\apj},
         year = 2008,
        month = jun,
       volume = {680},
       number = {2},
        pages = {1359-1377},
          doi = {10.1086/588521},
archivePrefix = {arXiv},
       eprint = {0802.2026},
 primaryClass = {astro-ph},
       adsurl = {https://ui.adsabs.harvard.edu/abs/2008ApJ...680.1359M}
}

@ARTICLE{Reis2009,
       author = {{Reis}, R.~C. and {Fabian}, A.~C. and {Ross}, R.~R. and {Miller}, J.~M.},
        title = "{Determining the spin of two stellar-mass black holes from disc reflection signatures}",
      journal = {\mnras},
         year = 2009,
        month = may,
       volume = {395},
       number = {3},
        pages = {1257-1264},
          doi = {10.1111/j.1365-2966.2009.14622.x},
       adsurl = {https://ui.adsabs.harvard.edu/abs/2009MNRAS.395.1257R}
}

@ARTICLE{Zhang2012,
       author = {{Zhang}, Shuang-Nan and {Liao}, Jinyuan and {Yao}, Yangsen},
        title = "{Measuring the black hole masses in accreting X-ray binaries by detecting the Doppler orbital motion of their accretion disc wind absorption lines}",
      journal = {\mnras},
         year = 2012,
        month = apr,
       volume = {421},
       number = {4},
        pages = {3550-3556},
          doi = {10.1111/j.1365-2966.2012.20579.x},
archivePrefix = {arXiv},
       eprint = {1201.3451},
 primaryClass = {astro-ph.HE},
       adsurl = {https://ui.adsabs.harvard.edu/abs/2012MNRAS.421.3550Z}
}

@ARTICLE{Neilsen2012,
       author = {{Neilsen}, Joseph and {Homan}, Jeroen},
        title = "{A Hybrid Magnetically/Thermally Driven Wind in the Black Hole GRO J1655-40?}",
      journal = {\apj},
         year = 2012,
        month = may,
       volume = {750},
       number = {1},
          eid = {27},
        pages = {27},
          doi = {10.1088/0004-637X/750/1/27},
archivePrefix = {arXiv},
       eprint = {1202.6053},
 primaryClass = {astro-ph.HE},
       adsurl = {https://ui.adsabs.harvard.edu/abs/2012ApJ...750...27N}
}

\end{document}